\documentclass[prd,preprint,tightenlines,preprintnumbers,aps,eqsecnum,nofootinbib,
superscriptaddress,showpacs,floatfix]{revtex4-1}

\usepackage{amsmath}
\usepackage{amssymb}
\usepackage{bm}
\usepackage{graphicx}
\usepackage{hyperref}

\usepackage{breakcites}
\newcommand{\ZVbb}{\ensuremath{Z_{V^4_{bb}}}}
\newcommand{\ZVll}{\ensuremath{Z_{V^4_{ss}}}}
\newcommand{\mph}{\ensuremath{\phantom{-}}}
\newcommand{\coarse}{$a\approx 0.12$~fm\ }
\newcommand{\fine}{$a\approx 0.09$~fm\ }
\newcommand{\superfine}{$a\approx 0.06$~fm\ }
\newcommand{\ultrafine}{$a\approx 0.045$~fm\ }

\newcommand{\order}{\mathcal{O}} 
\newcommand{\vp}{\ensuremath{E}}

\newcommand{\bra}[1]{\langle #1|}
\newcommand{\ket}[1]{|#1\rangle}

\begin{document}
\title{\boldmath\texorpdfstring{$B\to Kl^+l^-$}{B to Kl+l-} decay form factors from three-flavor lattice QCD}
\author{Jon A.~Bailey}
\affiliation{Department of Physics and Astronomy, Seoul National University, Seoul, South Korea}

\author{A.~Bazavov}
\affiliation{Physics Department, Brookhaven National Laboratory, Upton, NY, USA}

\author{C.~Bernard}
\affiliation{Department of Physics, Washington University, St.~Louis, Missouri, USA}

\author{C.~M.~Bouchard}
\altaffiliation[Present address:~]%
{Physics Department, College of William and Mary, Williamsburg, VA, USA}
\affiliation{Department of Physics, The Ohio State University, Columbus, OH, USA}

\author{C.~DeTar}
\affiliation{Department of Physics and Astronomy, University of Utah, Salt Lake City, Utah, USA}

\author{Daping~Du}
\affiliation{Department of Physics, Syracuse University, Syracuse, New York, USA}

\author{A.~X.~El-Khadra}
\affiliation{Department of Physics, University of Illinois, Urbana, Illinois, USA}

\author{J.~Foley}
\affiliation{Department of Physics and Astronomy, University of Utah, Salt Lake City, Utah, USA}

\author{E.~D.~Freeland}
\affiliation{School of the Art Institute of Chicago, Chicago, Illinois, USA}

\author{E.~G\'amiz}
\affiliation{CAFPE and Departamento de F\'{\i}sica Te\'orica y del Cosmos, Universidad de Granada,
Granada, Spain}

\author{Steven~Gottlieb}
\affiliation{Department of Physics, Indiana University, Bloomington, Indiana, USA}

\author{U.~M.~Heller}
\affiliation{American Physical Society, Ridge, New York, USA}

\author{R.~D.~Jain}
\affiliation{Department of Physics, University of Illinois, Urbana, Illinois, USA}

\author{J.~Komijani}
\affiliation{Department of Physics, Washington University, St.~Louis, Missouri, USA}

\author{A.~S.~Kronfeld}
\affiliation{Fermi National Accelerator Laboratory, Batavia, Illinois, USA}
\affiliation{Institute for Advanced Study, Technische Universit\"at M\"unchen, Garching, Germany}

\author{J.~Laiho}
\affiliation{Department of Physics, Syracuse University, Syracuse, New York, USA}

\author{L.~Levkova}
\affiliation{Department of Physics and Astronomy, University of Utah, Salt Lake City, Utah, USA}

\author{Yuzhi~Liu}
\affiliation{Department of Physics, University of Colorado, Boulder, CO, USA}

\author{P.~B.~Mackenzie}
\affiliation{Fermi National Accelerator Laboratory, Batavia, Illinois, USA}

\author{Y.~Meurice}
\affiliation{Department of Physics and Astronomy, University of Iowa, Iowa City, IA, USA}

\author{E.~T.~Neil}
\affiliation{Department of Physics, University of Colorado, Boulder, CO 80309, USA}
\affiliation{RIKEN-BNL Research Center, Brookhaven National Laboratory, Upton, NY 11973, USA}

\author{Si-Wei Qiu}
\altaffiliation[Present address:~]%
{Laboratory of Biological Modeling, NIDDK, NIH, Bethesda, Maryland, USA}
\affiliation{Department of Physics and Astronomy, University of Utah, \\ Salt Lake City, Utah, USA}

\author{J.~N.~Simone}
\affiliation{Fermi National Accelerator Laboratory, Batavia, Illinois, USA}

\author{R.~Sugar}
\affiliation{Department of Physics, University of California, Santa Barbara, California, USA}

\author{D.~Toussaint}
\affiliation{Department of Physics, University of Arizona, Tucson, Arizona, USA}

\author{R.S.~Van~de~Water}
\affiliation{Fermi National Accelerator Laboratory, Batavia, Illinois, USA}

\author{Ran Zhou}
\email{zhouran@fnal.gov}
\affiliation{Fermi National Accelerator Laboratory, Batavia, Illinois, USA}

\collaboration{Fermilab Lattice and MILC Collaborations}
\noaffiliation

\begin{abstract}
We compute the form factors for the $B \to Kl^+l^-$ semileptonic decay process in 
lattice QCD using gauge-field ensembles with 2+1 flavors of sea quark, 
generated by the MILC Collaboration. The ensembles span lattice spacings from 0.12 
to 0.045 fm and have multiple sea-quark masses to help control the chiral extrapolation.
The asqtad improved staggered action is used for the light valence and sea quarks, and 
the clover action with the Fermilab interpretation is used for the heavy $b$ quark.
We present results for the form factors $f_+(q^2)$, $f_0(q^2)$, and $f_T(q^2)$, 
where $q^2$ is the momentum transfer, together with a comprehensive 
examination of systematic errors. Lattice QCD determines the form factors for 
a limited range of~$q^2$, and we use the model-independent $z$~expansion to cover 
the whole kinematically allowed range. We present our final form-factor 
results as coefficients of the $z$~expansion and the correlations between
them, where the errors on the coefficients include statistical and all systematic 
uncertainties.
We use this complete description of the form factors to test QCD predictions of 
the form factors at high and low~$q^2$. 
\end{abstract}

\date{\today} % replace with posting date when done

\maketitle

\section{Introduction}

Flavor-changing neutral-current interactions (FCNC) place important constraints on physics beyond the
Standard Model.
In the Standard Model, tree-level FCNC contributions vanish by the Glashow-Iliopolous-Maiani (GIM) mechanism.
Even at the one-loop level, the GIM mechanism suppresses these amplitudes, as do factors of the
Cabibbo-Kobayashi-Maskawa (CKM) mixing matrix.
Thus, new-physics effects may be substantially larger than the small Standard-Model contribution and, hence,
observable.
In this paper, we present an unquenched lattice-QCD calculation of the amplitudes for the FCNC process
$B\to{}Kl^+l^-$.
Within the Standard Model and beyond, three form factors can arise, and we present results for all three.
This work is part of a larger program by the Fermilab Lattice and MILC collaborations  to
calculate form factors for exclusive semileptonic $B$ decays needed to test the Standard Model and search
for new physics, all of which use the same lattice actions and parameters.
It builds upon our previous work on charged-current semileptonic $B$ decays,
$B\to\pi{}l\nu$~\cite{Bailey:2008wp,Lattice:2015tia} and $B\to
D^{(*)}l\nu$~\cite{Bernard:2008dn,Bailey:2014tva,Lattice:2015rga}, which are used to determine the CKM
matrix elements $|V_{ub}|$ and $|V_{cb}|$~\cite{Agashe:2014kda}.
It is also part of a suite of form factors needed for searching for new physics in rare semileptonic
$B$-decay processes such as $B\to\pi l^+ l^-$~\cite{Bailey:2015nbd}, 
$B\to D\tau\nu$~\cite{Bailey:2012rr} and $B_s\to\mu^+\mu^-$~\cite{Bailey:2012jg}.

Experimental research on rare $B$-meson decays is active~\cite{Hurth:2010tk,Antonelli:2009ws}.
The BaBar, Belle, and CDF collaborations have measured the differential branching ratio, the
forward-backward asymmetry and other observables for both $B \to Kl^+l^-$ and $B \to K^*l^+l^-$
decays~\cite{Aubert:2008ps,Wei:2009zv,Aaltonen:2011cn,Aaltonen:2011qs,Lees:2012tva}.
The LHCb Collaboration has reported more precise results for the $B^0 \to K^0 l^+l^-$ and 
$B^{\pm}\to K^\pm l^+l^-$ decays~\cite{Aaij:2012cq,Aaij:2012vr,Aaij:2014pli}.
The high-intensity $B$ factories will also have results in the future~\cite{Hewett:2012ns}.
Thus, it is timely to improve the precision of the theoretical calculation of these processes.
Recently, the HPQCD Collaboration published the first three-flavor lattice calculation of $B \to K l^+
l^-$~\cite{Bouchard:2013eph}, also analyzing the phenomenological implications~\cite{Bouchard:2013mia}.
Three-flavor results for the modes with vector mesons in the final state, 
$B\to K^*l^+l^-$ and $B_s\to\phi l^+l^-$, have also been presented~\cite{Horgan:2013hoa}.

The theoretical description of the $B\to K^{(*)}l^+l^-$ process is based on the 
operator-product expansion, which leads to a low-energy effective Hamiltonian~\cite{Grinstein:1988me,Buras:1993xp,Bobeth:1999mk,%
Altmannshofer:2008dz}. Amplitudes are expressed in terms of Wilson coefficients that encode the 
high-scale physics and hadronic matrix elements that capture the low-energy nonperturbative QCD contributions. 
Hadronic matrix elements of local operators can be parameterized in terms of form factors. 
The uncertainties in the form factors are an important source of error in the theoretical predictions of the observables mentioned above.
In order to calculate the form factors, one requires knowledge of nonperturbative QCD dynamics, 
and lattice QCD is the tool of choice. We focus on $B\to Kl^+l^-$, rather than $B\to K^*l^+l^-$, because the former 
is a ``gold-plated'' decay with a stable hadron (under strong interactions) in the final state.  In the vector-meson 
case, the $K^*$ is unstable, and the resonance would have to be distinguished from $K\pi$ states.

The goal of this work is to improve our knowledge of the $B\to Kl^+l^-$ form factors.
We use the three-flavor gauge-field ensembles generated by the MILC Collaboration with dynamical up, down,
and strange quarks.
We extrapolate our lattice simulation data to the physical light-quark masses and continuum using SU(2)
chiral perturbation theory formulated for the process $B \to K l^+ l^-$.
Because the strange-quark mass is integrated out of the SU(2) theory, the use of SU(2) $\chi$PT, rather than
SU(3), improves the convergence of the chiral expansion, thereby reducing the systematic uncertainty due to
the chiral-continuum extrapolation.
On currently available lattices, we directly obtain the form factors at large momentum transfer (low
recoil), $q^2\gtrsim17~\textrm{GeV}^2$.
Following Ref.~\cite{Bailey:2008wp}, we use the $z$~expansion to extend the lattice-QCD calculation to the
full range of $q^2$.
Compared with the work of the HPQCD Collaboration~\cite{Bouchard:2013mia,Bouchard:2013eph}, we use twice 
as many ensembles, covering a larger range of lattice spacings and using lighter sea-quark masses. 
In particular, the smallest lattice spacing and the smallest light-quark mass in our ensemble set are 
smaller by a factor of two compared to the set used by HPQCD.  In addition, we use the 
Sheikholeslami-Wohlert (SW) action~\cite{Sheikholeslami:1985ij} with the Fermilab
interpretation~\cite{ElKhadra:1996mp} for the $b$-quark, while the HPQCD Collaboration uses a nonrelativistic
QCD $b$ quark~\cite{Lepage:1992tx}.
As discussed below, details of the chiral-continuum extrapolation and the $z$~expansion also differ.

This paper is organized as follows.
In Sec.~\ref{sec:lattices}, we define the form factors for the $B\to Kl^+l^-$ decay.
We then describe the lattice ensembles used in our simulations.
We also discuss the formalism for the light and heavy quarks.
In Sec~\ref{sec:results}, we present the numerical analysis.
This section has four parts.
We first present results for the simulated $K$ and $B$ meson masses.
Next, we determine the lattice form factors from two-point and three-point correlation functions.
We then correct our form-factor data to account for the slight difference between the simulated $b$-quark
mass and the physical $b$-quark mass.
Last, we extrapolate the lattice simulation results to the chiral and continuum limits with SU(2)
heavy meson rooted staggered chiral perturbation theory (HMrS$\chi$PT).
In Sec.~\ref{sec:sys_errs}, we analyze the systematic errors in our calculation and give a complete error
budget for the range of momenta $q^2 \gtrsim 17~{\rm GeV}^2$ accessible in our numerical simulations.
In Sec.~\ref{sec:zexp}, we extrapolate our form factors from low to high recoil using the
$z$~expansion~\cite{Boyd:1994tt,Arnesen:2005ez,Bourrely:2008za,Bharucha:2010im}.
We present our final results for $f_+(q^2)$, $f_0(q^2)$, and $f_T(q^2)$, including statistical and all
systematic errors, as coefficients of the $z$~expansion and the correlations between them; this provides a
complete description of our form factors valid over the entire kinematic range.
In Sec.~\ref{sec:QCDTests}, we use these results to test predictions for the form 
factor from heavy-quark symmetry at high~$q^2$ and from QCD factorization
at low~$q^2$. Finally, 
we compare our form factors with other lattice-QCD and 
light-cone-sum-rule results, and present an outlook for future work, in Sec.~\ref{sec:conclusion}.

Preliminary results have been reported in Refs.~\cite{Zhou:2011be,Zhou:2012sn,Zhou:2013uu}.
Here we present a full analysis that includes the tensor-current form factor
and complete systematic error budgets.

\section{Lattice Calculation}
\label{sec:lattices}

In this section, we present the methods and ingredients used in this work.
We give the definitions of the form factors for the $B\to Kl^+l^-$ process and their relation to physical
observables in Sec.~\ref{subsec:methodology}.
We then describe the lattice actions and parameters used for gluon and fermion fields in our simulation in
Sec.~\ref{subsec:actions-and-parameters}.
Finally, we define the lattice currents in Sec.~\ref{subsec:op_renorm}.

\subsection{Matrix elements and form factors}
\label{subsec:methodology}

An operator-product expansion analysis of the $B\to Kl^+l^-$ decay in the Standard Model shows that two
currents, a vector current $\mathcal{V}^\mu=\bar{s}\gamma^\mu b$ and a tensor current
$\mathcal{T}^{\mu\nu}=i\bar{s}\sigma^{\mu\nu}b$, contribute to the $b\to s$ process at lowest
order~\cite{Hurth:2010tk}.
In general Standard Model extensions, a scalar current $\mathcal{S}=\bar{s}b$ can also arise.
The matrix elements of the vector, tensor, and scalar current are characterized by three form factors
$f_+(q^2)$, $f_0(q^2)$, and $f_T(q^2)$, which are defined~via
\begin{align}
    \bra{K} \bar{s}\gamma^\mu b \ket{B} &= f_+(q^2)\left(p^\mu+k^\mu-\frac{M_B^2-M_K^2}{q^2}q^\mu \right)
        +f_0(q^2)\frac{M_B^2-M_K^2}{q^2}q^\mu,
    \label{eq:def.f+f0}\\
    \bra{K} i\bar{s}\sigma^{\mu\nu} b\ket{B} &= \frac{2f_T(q^2)}{M_B+M_K} (p^\mu k^\nu-p^\nu k^\mu), 
    \label{eq:def.fT} \\
    \bra{K}\bar{s} b\ket{B} &= \frac{M_B^2-M_K^2}{m_b-m_s}f_0(q^2),
    \label{eq:def.f0} 
\end{align}
where $p$ and $k$ are the $B$-meson and kaon momenta, respectively, and $q=p-k$ is the 
momentum carried off by the leptons. 
The Ward identity relating the matrix element of a vector current to that of the corresponding 
scalar current ensures that $f_0$ is the same in Eqs.~(\ref{eq:def.f+f0}) and~(\ref{eq:def.f0}).

For the analysis that follows, it is convenient to write the vector-current matrix element~as
\begin{equation}
    \bra{K} \bar{s}\gamma^\mu b \ket{B} = \sqrt{2M_B}
        \left[v^\mu f_\parallel(E_K) + k_\perp^\mu f_\perp(E_K)\right],
\end{equation}
where $v^\mu=p^\mu/M_B$ is the four-velocity of the $B$ meson, $k_\perp^\mu=k^\mu-(k\cdot v)v^\mu$, and
$E_K=v\cdot k$ is the kaon energy in the $B$-meson rest frame.
From energy-momentum conservation, $q^2=M_B^2+M_K^2-2M_BE_K$.
We obtain $f_\parallel(E_K)$ and $f_\perp(E_K)$ from the temporal and spatial components 
of the matrix element of the vector current:
\begin{align}
    f_\parallel(E_K) &= \frac{\bra{K} \bar{s} \gamma^0 b \ket{B}}{\sqrt{2M_B}}, \\
    f_\perp(E_K)     &= \frac{\bra{K} \bar{s}\gamma^i b \ket{B}}{\sqrt{2M_B}k^i}.
\end{align}
Similarly, we obtain the tensor form factor $f_T$ from
\begin{align}
        f_T(q^2) &= \frac{M_B+M_K}{\sqrt{2M_B}}\frac{\bra{K} \bar{s}\sigma^{0i}b\ket{B}}{\sqrt{2M_B}k^i}. 
    \label{eq:contift}
\end{align}
Finally, the vector and scalar form factors $f_+$ and $f_0$ can be obtained from
\begin{align}
    f_+(q^2) &= \frac{1}{\sqrt{2M_B}}\left [ f_\parallel(E_K)+(M_B-E_K)f_\perp(E_K) \right],
    \label{eq:contif+} \\
    f_0(q^2) &= \frac{\sqrt{2M_B}}{M_B^2-M_K^2}
        \left[ (M_B-E_K)f_\parallel(E_K)+(E_K^2-M_K^2)f_\perp(E_K) \right]. %,
    \label{eq:contif0} % \\
\end{align}
Equations~(\ref{eq:contif+})--(\ref{eq:contif0}) satisfy the kinematic 
constraint, $f_+(0)=f_0(0)$, automatically.
At low recoil, the form factor $f_\perp$ gives the dominant contribution to $f_+$.

Physical observables can be described in terms of the form factors, if we neglect non-factorizable 
contributions. For example, the Standard-Model differential decay rate for $B\to Kl^+l^-$
is~\cite{Grinstein:1988me,Becirevic:2012fy,Ali:2013zfa}
\begin{align}
    \frac{d\Gamma}{dq^2} &= \frac{G_F^2\alpha^2|V_{tb}V_{ts}^*|^2}{2^7\pi^5} |\bm{k}|\beta_+
        \left\{ \vphantom{\left|2C_7^\text{eff}\frac{m_b+m_s}{M_B+M_K} f_T(q^2)\right|^2}
        \frac{2}{3}|\bm{k}|^2\beta_+^2 \left|C_{10}^\text{eff} f_+(q^2)\right|^2 \right.
    + 
        \frac{m_l^2(M_B^2-M_K^2)^2}{q^2M_B^2}
        \left|C_{10}^\text{eff} f_0(q^2)\right|^2
    \nonumber \\ & { } + \left.
        |\bm{k}|^2 \left[1 - \frac{1}{3}\beta_+^2 \right]
        \left| C_9^\text{eff} f_+(q^2) + 2C_7^\text{eff}\frac{m_b+m_s}{M_B+M_K} f_T(q^2) \right|^2
        \right\},
    \label{eq:dGamma}
\end{align}
where $G_F$, $\alpha$, and $V_{tq}$ are the Fermi constant, the (QED) fine structure constant, and CKM
matrix elements, respectively, $|\bm{k}|=\sqrt{E_K^2-M_K^2}$ is the kaon momentum in the $B$-meson rest
frame, and $\beta_+^2=1-4m_l^2/q^2$, with $m_l$ being the lepton mass.
The $C_i^\text{eff}$ are effective Wilson coefficients~\cite{Buras:1993xp}; we follow the notation of
Ref.~\cite{Altmannshofer:2008dz} in Eq.~(\ref{eq:dGamma}).
When $q^2$ corresponds to a charmonium resonance, further contributions must be added to
Eq.~(\ref{eq:dGamma}).
Beyond the Standard Model, the expression can become more complicated, but $f_+(q^2)$, $f_T(q^2)$,
and~$f_0(q^2)$ still suffice.

\subsection{Actions and parameters}
\label{subsec:actions-and-parameters}

Our calculations employ the $N_f=2+1$ flavor gauge configurations generated 
by the MILC Collaboration~\cite{Bernard:2001av,Aubin:2004wf}, which include the effects of 
dynamical $u$, $d$, and $s$ quarks. The one-loop improved 
L\"{u}scher-Weisz action is used for the gluon fields, which leads to lattice
artifacts of $\order(\alpha_s a^2)$~\cite{Luscher:1984xn}.
(The gluon-loop correction is included~\cite{Luscher:1985zq}, but not that of the quark 
loop~\cite{Hao:2007iz}.) 

For light quarks ($u$, $d$ and $s$), these configurations employ the $a^2$ tadpole-improved staggered action
(asqtad)~\cite{Blum:1996uf,Lepage:1997id,Lagae:1998pe,Lepage:1998vj,Orginos:1998ue,Orginos:1999cr,%
Bernard:1999xx}, leading to discretization errors of $\order(\alpha_s a^2)$ and
$\order(a^4)$~\cite{Bazavov:2009bb}.
The sea quarks are simulated with the fourth root of the staggered fermion determinant.
Several theoretical and numerical analyses support the idea that this procedure yields
continuum QCD as the lattice spacing $a\to0$~\cite{Bazavov:2009bb,Shamir:2004zc,Shamir:2006nj,Lee:1999zxa,Bernard:2006zw,%
Bernard:2007ma,Prelovsek:2005rf,Bernard:2007qf,Aubin:2008wk,Sharpe:2004is,Durr:2005ax,Sharpe:2006re,%
Kronfeld:2007ek,Donald:2011if}.

\begin{table}[tp]
    \centering
    \caption{Parameters of the QCD gauge-field ensembles and light valence-quark masses 
used in this work, lattice spacing $a$, lattice size $N_s^3\times N_t$, 
sea-quark masses $am_l^\prime$ and $am_h^\prime$,  light-valence mass $am_l$, daughter mass $am_h$, 
the number of configurations and sources denoted as  
$N_{\rm conf}\times N_{\rm src}$, and the box size times the pion mass. 
On all ensembles but one, we use the same light valence- and sea-quark mass.
(The only exception is on the \fine ensemble with $m_l^\prime=0.0465$, where
the light valence-quark mass is 0.0047 instead of 0.00465.) On the \coarse and \fine ensembles we 
also use the same valence and sea strange-quark mass. On the \superfine and \ultrafine ensembles,  
we use slightly different valence strange-quark masses than in the sea; the valence masses are 
tuned to be closer to the physical value. The values of $M_\pi L$ are taken from 
Refs.~\cite{Bazavov:2009bb,Bailey:2014tva}. 
The gauge-field configurations can be downloaded using the digital object 
identifier (DOI) links provided in Refs.~\cite{asqtad_C0.2ms,asqtad_C0.2ms_b,asqtad_C0.15ms,asqtad_C0.1ms,asqtad_F0.2ms,asqtad_F0.2ms_b,asqtad_F0.2ms_c,asqtad_F0.15ms,asqtad_F0.1ms_a,asqtad_F0.1ms_b,asqtad_F0.05ms,asqtad_SF0.2ms_a,asqtad_SF0.2ms_b,asqtad_SF0.1ms_a,asqtad_SF0.1ms_b,asqtad_UF0.2ms}.}
    \label{tab:ensembles}
    \begin{tabular}{lr@{$\times$}lllllr@{$\times$}lc}
    \hline\hline  
    ~$\approx a$~(fm)~~~~~~~ & $N_s^3$ & $N_t$~~~ & $am_l'\qquad~$ & $am_h'\qquad$ & $am_l\qquad$ & $am_h\quad$
        & $N_{\rm conf}$ & $N_{\rm src}$ & ~~~$M_\pi L$~~~\\
    \hline
    0.12~\cite{asqtad_C0.2ms,asqtad_C0.2ms_b}
           & $20^3$ & $64$ & 0.01    & 0.05   & 0.01  & 0.05   & $2259$ & $4$ & 4.5 \\
    0.12~\cite{asqtad_C0.15ms}
           & $20^3$ & $64$ & 0.007   & 0.05   & 0.007 & 0.05   & $2110$ & $4$ & 3.8 \\
    0.12~\cite{asqtad_C0.1ms}
           & $24^3$ & $64$ & 0.005   & 0.05   & 0.005 & 0.05   & $2099$ & $4$ & 3.8 \\
    \hline
    0.09~\cite{asqtad_F0.2ms,asqtad_F0.2ms_b,asqtad_F0.2ms_c}  
          & $28^3$ & $96$ & 0.0062  & 0.031  & 0.0062  & 0.031  & $1931$ & $4$ & 4.1 \\
    0.09~\cite{asqtad_F0.15ms}  
          & $32^3$ & $96$ & 0.00465  & 0.031  & 0.0047 & 0.031  &  $984$ & $4$ & 4.1 \\
    0.09~\cite{asqtad_F0.1ms_a,asqtad_F0.1ms_b}  
          & $40^3$ & $96$ & 0.0031  & 0.031  & 0.0031  & 0.031  & $1015$ & $4$ & 4.2 \\
    0.09~\cite{asqtad_F0.05ms}    
          & $64^3$ & $94$ & 0.00155 & 0.031  & 0.00155 & 0.031  &  $791$ & $4$ & 4.8 \\
    \hline
    0.06~\cite{asqtad_SF0.2ms_a,asqtad_SF0.2ms_b}    
          & $48^3$ &$144$ & 0.0036  & 0.0180 & 0.0036  & 0.0188 &  $673$ & $4$ & 4.5 \\
    0.06~\cite{asqtad_SF0.1ms_a,asqtad_SF0.1ms_b}
          & $64^3$ &$144$ & 0.0018  & 0.0180 & 0.0018  & 0.0188 &  $827$ & $4$ & 4.3 \\
    \hline
    0.045~\cite{asqtad_UF0.2ms} 
          & $64^3$ &$192$ & 0.0028  & 0.0140 & 0.0028  & 0.0130 &  $801$ & $4$ & 4.6 \\
    \hline  
    \hline
    \end{tabular}
\end{table}

Table~\ref{tab:ensembles} summarizes the properties of the ensembles used in this work.
We use the asqtad ensembles at four lattice spacings: $a\approx 0.12$~fm, $a\approx0.09$~fm,
$a\approx 0.06$~fm, and $a\approx 0.045$~fm.
The volumes of the lattices are large enough ($M_\pi L \gtrsim 4$) to suppress finite-volume effects.
The strange sea-quark mass is tuned to be close to its physical value.
The light-to-strange sea-quark mass ratios range from $am_l^\prime/am_h^\prime=0.2$ down to~0.05, to
facilitate reliable chiral extrapolations.
On the \coarse and \fine ensembles, we use unitary data, with the light and strange valence-quark masses set
equal to the corresponding sea-quark masses, with one exception.
On the \superfine and \ultrafine ensembles, however, we use valence strange-quark masses that are closer to
the physical value and, thus, differ slightly from the strange-quark mass in the sea.

On each configuration, we compute the correlation functions starting at four different source locations, to
increase the available statistics.
We first translate the gauge field by a different random four-vector on each configuration and then fix the
spatial source locations at $\bm{x}=\bm{0}$ and the temporal source locations at $t=0$, $N_t/4$, $N_t/2$,
and $3N_t/4$.
The correlation between the results from different source locations is weak.
The random translation of the gauge field reduces autocorrelations between successive configurations.

For the heavy $b$ quark, we use the Sheikholeslami-Wohlert (SW) action~\cite{Sheikholeslami:1985ij} with the
Fermilab interpretation~\cite{ElKhadra:1996mp}.
The lattice action and currents are matched to the continuum QCD action via HQET~\cite{Kronfeld:2000ck}.
The heavy-quark action can be systematically improved to arbitrarily high orders in $1/m_b$, or,
equivalently, $a$, by including higher-dimensional operators in the lattice
action~\cite{ElKhadra:1996mp,Kronfeld:2000ck,Oktay:2008ex} and
currents~\cite{Harada:2001fi,Harada:2001fj,Bailey:2014jga}.
In this work, we remove the leading discretization errors in the action by tuning the
hopping parameter $\kappa$ and clover coefficient $c_\text{SW}$.
We fix the bare $b$-quark mass by tuning the value of $\kappa_b$ to reproduce the spin-averaged $B_s$ meson
kinetic mass as in Ref.~\cite{Bailey:2014tva}.
We use the tadpole-improved tree-level value for $c_\text{SW}=u_0^{-3}$, where $u_0$ is obtained from the
fourth root of the plaquette.
We also remove the leading discretization error in the vector and tensor currents; see
Sec.~\ref{subsec:op_renorm}.
The values of the parameters for $b$ quarks used in our simulations are listed in
Table~\ref{tab:heavy_quark_para}.

\begin{table}[tp]
 \centering
 \caption{
Parameters used in the simulation of the heavy $b$ quark~\cite{Bailey:2014tva}.  
  We list the clover coefficient $c_\text{SW}$, input $b$-quark hopping parameter 
  $\kappa_b'$, and rotation coefficient $d_1$.}
 \label{tab:heavy_quark_para}
 \begin{tabular}{ccccc}
  \hline  
  \hline  
  $\approx a$~(fm) & $am_l^\prime$  & ~~~$c_\text{SW}$~~~ & ~~~~$\kappa_b'$~~~~ & ~~~~~~$d_1$~~~~~~\\
  \hline
  0.12 & 0.01   & 1.531 & 0.0901 & 0.093340 \\
  0.12 & 0.007  & 1.530 & 0.0901 & 0.093320 \\
  0.12 & 0.005  & 1.530 & 0.0901 &  0.093320\\
  \hline
  0.09 & 0.0062  & 1.476 & 0.0979 & 0.096765 \\
  0.09 & 0.00465 & 1.477 & 0.0977 &  0.096708 \\
  0.09 & 0.0031  & 1.478 & 0.0976 & 0.096690 \\
  0.09 & 0.00155 & 1.4784 & 0.0976 & 0.096700\\
  \hline
  0.06 & 0.0036  & 1.4287 & 0.1052 & 0.096300 \\
  0.06 & 0.0018  & 1.4298 & 0.1052 & 0.096300 \\
  \hline  
  0.045 & 0.0028 & 1.3943 & 0.1143 & 0.08864 \\
  \hline
  \hline
 \end{tabular}
\end{table}

To extrapolate the form factors calculated on the lattice to the continuum limit, we need a unified scale to
compare the results from different spacings and convert to physical units.
We do so with the scale $r_1$ which is defined such that
$r_1^2F(r_1)=1.0$~\cite{Sommer:1993ce,Bernard:2000gd}.
Here $F(r)$ is the force between static quarks at distance $r$.
We first determine the relative scale $r_1/a$ on each ensemble, and then interpolate $r_1/a$ with a smooth
function of the gauge coupling $\beta$; the smoothed $r_1$ values are independent of the light sea-quark
mass.
(The explicit form of the smoothing function is given in Ref.~\cite{Bailey:2014tva}.) %
In this paper, we choose a mass-independent scheme for $r_1/a$, so that it is  the same for all sea masses with the
same approximate lattice spacing.
We use the values of $r_1/a$ to convert all lattice quantities to $r_1$ units.
We can then combine results from different ensembles and perform a chiral-continuum
extrapolation.
The physical value $r_1=0.3117(22)$~fm~\cite{Bazavov:2009bb,Bazavov:2011aa}
is determined by requiring that
the continuum limit of the pion decay constant at the physical quark masses takes the PDG
value~\cite{Agashe:2014kda}. 
The RBC-UKQCD Collaboration also reported the physical value $r_1=0.323(8)(4)$~fm in
Ref.~\cite{Arthur:2012opa}. This result is consistent with the one we use, but less
precise.
The values of $r_1/a$ used in this work are provided in Table~\ref{tab:r1}.

\begin{table}[tp]
  \centering
  \caption{Relative scales $r_1/a$ used in this work, for corresponding values of 
    $\beta$~\cite{Bazavov:2009bb,Bailey:2014tva}.
    The statistical and systematic errors on $r_1/a$ are both 0.1--0.3\%~\cite{Bailey:2014tva}.
We also list  the Goldstone pion mass ($M_\pi$) and root-mean-square (RMS) pion 
mass ($M_\pi^{\rm RMS}$) here.}
  \label{tab:r1}
  \begin{tabular}{cccccc}
    \hline\hline  
  $\approx a$~(fm) & ~$am_l'/am_h'$~ & ~~~~$\beta$~~~~ & ~~$r_1/a$~~ & ~$M_\pi$(MeV)~ & ~$M_\pi^{\rm RMS}$(MeV)~ \\
  \hline
  0.12 & 0.01/0.05  & 6.760 & 2.739 & 389 & 532 \\
  0.12 & 0.007/0.05 & 6.760 & 2.739 & 327 & 488 \\
  0.12 & 0.005/0.05 & 6.760 & 2.739 & 277 & 456 \\
  \hline
  0.09 & 0.0062/0.031  & 7.090 & 3.789 & 354 & 413 \\
  0.09 & 0.00465/0.031 & 7.085 & 3.772 & 307 & 374 \\
  0.09 & 0.0031/0.031  & 7.080 & 3.755 & 249 & 329 \\
  0.09 & 0.00155/0.031 & 7.075 & 3.738 & 177 & 277 \\
  \hline
  0.06 & 0.0036/0.018  & 7.470 & 5.353 & 316 & 340 \\
  0.06 & 0.0018/0.018  & 7.460 & 5.307 & 224 & 255 \\
  \hline
  0.045 & 0.0028/0.014 & 7.810 & 7.208 & 324 & 331  \\
  \hline  
  \hline
 \end{tabular}
\end{table}

\subsection{Definition of currents}
\label{subsec:op_renorm}

We define the current operators on the lattice as in Refs.~\cite{Wingate:2002fh,Bailey:2008wp}:
\begin{align}
    V_\xi^\mu(x)&=\bar{\Psi}_\alpha(x)\gamma^\mu_{\alpha\beta}\Omega_{\beta \xi}(x)\chi(x), 
    \label{eq:Vlat} \\
    T_\xi^{\mu\nu}(x)&=\bar{\Psi}_\alpha(x)\sigma^{\mu\nu}_{\alpha\beta}\Omega_{\beta \xi}(x)\chi(x),
    \label{eq:Tlat}
\end{align}
where the matrix $\Omega=\gamma_4^{x_4/a}\gamma_1^{x_1/a}\gamma_2^{x_2/a}\gamma_3^{x_3/a}$
and $\chi(x)$ is the one-component staggered fermion field.
The clover $b$-quark field is rotated to remove discretization 
errors of order~$a$ from the lattice current~\cite{ElKhadra:1996mp}:
\begin{equation}
    \Psi = \left(1+ad_1\bm{\gamma} \cdot \bm{D}_\text{lat}\right)\psi ,
    \label{eq:rotated_psi}
\end{equation}
where $\psi$ is the field in the Fermilab action (for the $b$ quark), 
$\bm{D}_\text{lat}$ is the symmetric, nearest-neighbor, covariant
difference operator, and $d_1$ is adjusted to remove discretization errors.
In practice, we set the rotation coefficient $d_1$ to its tadpole-improved tree-level value: 
\begin{equation}
    d_1 = \frac{1}{u_0} \left( \frac{1}{2+m_0a}-\frac{1}{2(1+m_0a)} \right) \ ,
\end{equation}
where $m_0a$ is the bare lattice $b$-quark mass. 
The index $\xi$ in Eqs.~(\ref{eq:Vlat}) and~(\ref{eq:Tlat}) corresponds to taste, 
and it is contracted with another taste index in the heavy-light operators coupling 
the $B$ meson to the vacuum~\cite{Wingate:2002fh}.

To calculate the form factors on the lattice, we have to define currents with the correct 
continuum limit. As in earlier work~\cite{Bailey:2008wp, Kronfeld:2000ck}, we define
\begin{align}
    \mathcal{V}^\mu     &  \doteq 
        Z_{V^\mu}      V^\mu , \\
    \mathcal{T}^{\mu\nu} &  \doteq 
        Z_{T^{\mu\nu}} T^{\mu\nu},
\end{align}
where \{$\mathcal{V}$, $\mathcal{T}$\} and \{$V$, $T$\} are the continuum and 
lattice current operators, respectively.
We use a mostly nonperturbative renormalization procedure to obtain the $Z$ factors~\cite{ElKhadra:2001rv},
\begin{equation}
    Z_J = \rho_J \sqrt{\ZVbb \ZVll},
    \label{eq:rhodef}
\end{equation}
where $\ZVbb$ and $\ZVll$ are computed nonperturbatively, and the remaining factor $\rho_J$ is 
calculated at one-loop order in mean-field improved lattice perturbation theory~\cite{Harada:2001fi}. 

The light-light renormalization factor \ZVll\ is calculated nonperturbatively
from the charge normalization condition of a $\bar{c}s$ meson:
\begin{equation}
    \ZVll^{-1} = \int d^3x \langle D_s | V^4_{ss}(x) |D_s \rangle
    \label{eq:charge-normalization-condition}
\end{equation}
as in Ref.~\cite{Bazavov:2011aa}, but with random color wall sources and higher statistics,
leading to the values listed in Table~\ref{tab:Z}.
The result for \ZVll\ is insensitive to the mass of the spectator quark 
in the correlation function, so we use a heavy charm quark to improve the 
statistical errors. 
The heavy-heavy renormalization factor \ZVbb\ is computed analogously 
from the charge normalization condition of the $B$ meson  using 
data generated for our $B\to D l \nu$ analysis~\cite{Qiu:2011ur}.
We compute \ZVbb\ on the same jackknife samples as the form factors 
and propagate the statistical error directly throughout the remainder of the analysis.
The values of \ZVbb\ are shown in Table~\ref{tab:Z}.

The remaining factor $\rho_J$ (here, $J=V^\mu$, $T^{\mu\nu}$) is close to
unity~\cite{Harada:2001fi,ElKhadra:2007qe}, because most of the radiative corrections, particularly those
from tadpole diagrams, cancel among the $Z$ factors in Eq.~(\ref{eq:rhodef}).
We expand the factor $\rho_J$ perturbatively as
\begin{equation}
    \rho_J = 1+\alpha_s(q^*) \rho^{[1]}_{J}+\order(\alpha_s^2) ,
    \label{eq:def_of_rho}
\end{equation}
where $\alpha_s$ is the QCD coupling~\cite{Lepage:1992xa,Mason:2005zx}.
Details of the one-loop perturbative calculation will be given in a separate
publication~\cite{rho_in_prep:2012}; the values used here are listed in 
Table~\ref{tab:rho}.
In practice, we evaluate the coupling in the $V$-scheme~\cite{Brodsky:1982gc,Lepage:1992xa} 
at the scale $q^* = 2/a$ in mean-field improved lattice perturbation theory.  
For $\rho_{V^4}$ we find that the one-loop corrections are less than 1\%, while 
for $\rho_{V^i}$ they range from $1.5\%$ to $2.6\%$. The tensor current is scale dependent, 
and we renormalize it at the scale $\mu=m_b$ (where according to the Fermilab prescription 
$m_2=m_b$). We find that for $\rho_{T}$ the corrections range from $3\%$ to $6\%$.

Because $\rho_J$ is computed separately from the correlation functions, we used it to 
introduce a blinding procedure (as in many $B$ physics experiments) to reduce subjective bias.
Those of us carrying out the perturbative calculation~\cite{rho_in_prep:2012}  multiplied $\rho_J$ by a 
constant prefactor. Only after we finalized the choices made in our analysis, including 
tests and estimates of systematic uncertainties, was the prefactor revealed to the rest 
of the collaboration and removed from the results reported here.

\begin{table}[tp]
 \centering
 \caption{The flavor-conserving renormalization factors \ZVll\ and \ZVbb\ used in this work. 
     Errors shown are statistical. }
 \label{tab:Z}
 \begin{tabular}{lllllllll}
  \hline  
  \hline  
  $\approx a$~(fm) & $am_l'\qquad$ & $am_h'\quad$~ & ~~~$\kappa_b^\prime$~~~~ &  ~~$\ZVll$~~~~~ & ~~~~$\ZVbb$~~~~  \\
  \hline
  0.12 & 0.01 & 0.05   & 0.0901 & 1.741(3) &  0.5065(57) \\
  0.12 & 0.007 & 0.05  & 0.0901 & 1.741(3) & 0.5119(75) \\
  0.12 & 0.005 & 0.05  & 0.0901 & 1.741(3) & 0.5026(71) \\
  \hline
  0.09 & 0.0062 & 0.031  & 0.0979 & 1.777(5)  & 0.4482(57)\\
  0.09 & 0.00465 & 0.031 & 0.0977 & 1.776(5)  & 0.4694(100)\\
  0.09 & 0.0031 & 0.031  & 0.0976 & 1.776(5)  & 0.4608(94)\\
  0.09 & 0.00155 & 0.031 & 0.0976 & 1.776(5)  & 0.4491(116)\\
  \hline
  0.06 & 0.0036 & 0.018  & 0.1052 & 1.808(6)  & 0.4196(101)\\
  0.06 & 0.0018 & 0.018  & 0.1052 & 1.807(7)  & 0.4100(103)\\
  \hline
  0.045 & 0.0028 & 0.014 & 0.1143 & 1.841(6)  & 0.3564(65)\\
  \hline  
  \hline
 \end{tabular}
\end{table}

\begin{table}[tp]
    \centering
    \caption{Matching factors $\rho_{V^4}$, $\rho_{V^1}$, and $\rho_{T}$ calculated 
      at one loop in tadpole-improved lattice perturbation theory.
      Here, $\rho_T$ brings $f_T$ to the $\overline{\textrm{MS}}$ scheme 
      at $\mu=m_2$, and $m_2$ should be interpreted as the pole mass. }
    \label{tab:rho}
    \begin{tabular}{lllllllc}
    \hline\hline  
    $\approx a$~(fm) & ~$am_l'$~~~~ & ~$am_h'$~~ & ~$am_h$~~ & ~~~$\kappa_b^\prime$~~~~~ & 
        ~~$\rho_{V^4}$~~~~~ & ~~$\rho_{V^1}$~~~~ & $\rho_T(\mu=m_2)$\\
    \hline
    0.12  & 0.010   & 0.050 & 0.050  & 0.0901 & 1.0071 & 0.9737 & 1.0334 \\
    0.12  & 0.007   & 0.050 & 0.050  & 0.0901 & 1.0071 & 0.9737 & 1.0333 \\
    0.12  & 0.005   & 0.050 & 0.050  & 0.0901 & 1.0072 & 0.9738 & 1.0333 \\
    \hline
    0.09  & 0.0062  & 0.031 & 0.031  & 0.0979 & 0.9997 & 0.9759 & 1.0366 \\
    0.09  & 0.00465 & 0.031 & 0.031  & 0.0977 & 0.9998 & 0.9759 & 1.0364 \\
    0.09  & 0.0031  & 0.031 & 0.031  & 0.0976 & 0.9999 & 0.9758 & 1.0364 \\
    0.09  & 0.00155 & 0.031 & 0.031  & 0.0976 & 0.9999 & 0.9757 & 1.0364 \\
    \hline
    0.06  & 0.0036  & 0.018 & 0.0188 & 0.1052 & 0.9956 & 0.9792 & 1.0432 \\
    0.06  & 0.0018  & 0.018 & 0.0188 & 0.1052 & 0.9956 & 0.9792 & 1.0433 \\
    \hline
    0.045 & 0.0028  & 0.014 & 0.013  & 0.1143 & 0.9943 & 0.9843 & 1.0588 \\
    \hline \hline
    \end{tabular}
\end{table}

\section{Analysis}
\label{sec:results}

In this section, we present our form-factor analysis and results.
In Sec.~\ref{subsec:mass_fit}, we obtain the $B$-meson and kaon masses and 
energies by fitting two-point correlation functions. In 
Sec.~\ref{subsec:form_factors}, we extract the lattice form factors from 
ratios of three-point over two-point correlation functions.
In Sec.~\ref{subsec:kappa_b_shift}, we slightly shift  the full set of lattice form-factor data 
from the simulated $\kappa_b^\prime$ to the physical value.
In Sec.~\ref{subsec:chiral_fit}, we carry out the chiral-continuum 
extrapolation by fitting the form factors to the expression derived in
heavy meson rooted staggered chiral perturbation theory (HMrS$\chi$PT).

\subsection{\texorpdfstring{\boldmath $B$ and $K$}{B and K} meson masses}
\label{subsec:mass_fit}

We extract meson masses and energies from two-point correlation functions defined at Euclidean time $t$:
\begin{equation}
    C_2(t;\bm{k}) = \sum_x 
        \langle \mathcal{O}_P(\bm{x},t) \mathcal{O}_P^\dagger(\bm{0},0) \rangle 
                {e^{-i\bm{k}\cdot \bm{x}}},
\end{equation}
where the subscript $P$ denotes the $K$ or $B$ pseudoscalar meson in the interpolating 
operator. For the kaon we use a local interpolating operator.
For the $B$ meson we use the wave function for bottomonium given by the Richardson potential
model~\cite{Richardson:1978bt} as explained in Refs.~\cite{Menscher:2005,Bernard:2010fr,Neil:2011ku}.
We generate correlators with kaon three-momenta $\bm{k}=2\pi(0,0,0)/L$, $2\pi(1,0,0)/L$,
$2\pi(1,1,0)/L$, and $2\pi(1,1,1)/L$.

\begin{figure}
    \centering
    \includegraphics[scale=0.8]{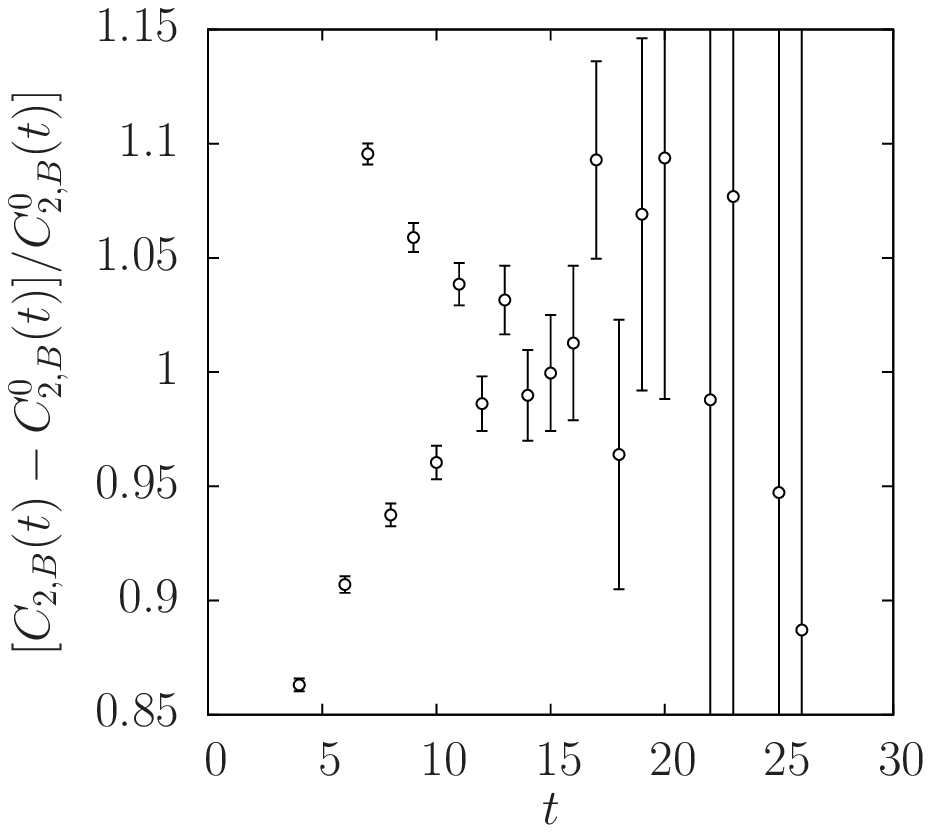}
    \hfill
    \includegraphics[scale=0.8]{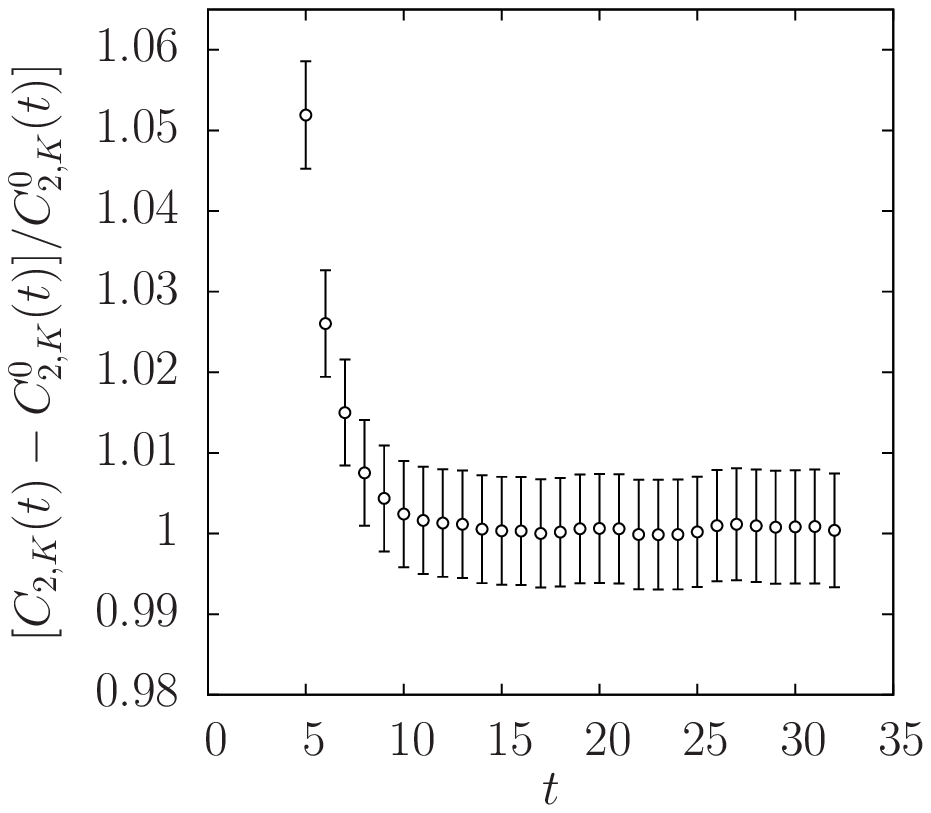}
    \caption{Scaled correlator $[C_2(t) - C_2^{(0)}(t)]/C_2^{(0)}(t)$ as a function 
      of time $t$ on the $am_l^\prime/am_h^\prime=0.005/0.05$ \coarse ensemble at the unitary point. 
      The oscillating opposite-parity-state contribution is clearly visible in the $B$-meson 
      correlator (left), but it is small in the zero-momentum kaon correlator 
      (right). }
    \label{fig:2pt_state_choice}
\end{figure}

The meson masses and energies are extracted from the large-$t$ behavior of the two-point correlation
functions.
By inserting a complete set of states, two-point correlation functions can be decomposed into a sum of
energy levels as 
\begin{equation}
    C_2(t;\bm{k}) = \sum_m (-1)^{m(t+1)}\frac{|\bra{0}\mathcal{O}_P\ket{P(m)}|^2}{2E_P^{(m)}} e^{-E_P^{(m)}t}.
\label{eq:2pt_fit_function}
\end{equation}
The amplitudes of terms with odd $m$ oscillate in time as $(-1)^{m(t+1)}$ and are due to
opposite-parity-state contributions to staggered correlators. 
Figure~\ref{fig:2pt_state_choice} shows sample kaon and $B$-meson scaled
correlators $[C_2(t)-C_2^{(0)}(t)]/C_2^{(0)}(t)$ on the \coarse ensemble with 
$m_l^\prime = 0.1m_h^\prime$ and  momentum $\bm{k}=\bm{0}$, where
\begin{equation}
    C_2^{(0)}(t)  = \frac{|\bra{0}\mathcal{O}_P\ket{P(0)}|^2}{2E_P^{(0)}} e^{-E_P^{(0)} t} 
\end{equation}
is the ground-state contribution determined by our fit. The opposite-parity-state 
contribution is insignificant for the zero-momentum kaon but is visible 
for the $B$ meson. We employ a simple strategy to fit the two-point 
correlators because the statistical errors in the kaon and $B$-meson energies 
contribute little to the errors in form factors, which stem primarily from the 
three-point correlators. For the kaon correlators, we perform two-state 
fits that include the ground state and a same-parity excited state. For the 
$B$-meson correlators, we perform three-state fits including the ground state, 
its excited state, and the lowest-lying opposite-parity state.

We use a  single-elimination jackknife analysis to estimate the statistical errors in this
work. We first average the correlation functions generated from the four sources 
at 0, $N_t/4$, $N_t/2$, and $3N_t/4$. 
We fit $C_2(t)$ in an interval $t\in[t_\text{min}, t_\text{max}]$, taking correlation
from time slice to time slice into account. In general, we 
choose $t_\text{max}$ so that the fractional error in the correlator remains below 4\%.
We choose $t_\text{min}$ such that we obtain a good correlated $p$~value.
We use the same interval $[t_\text{min}, t_\text{max}]$ for all kaon or $B$-meson fits 
at a given lattice spacing, and use similar physical distances 
for $[t_\text{min}, t_\text{max}]$ on the four lattice spacings.
These fit ranges are given in Table~\ref{tab:fit_range}. We use
a 2+1-state fit for the $B$ meson in this paper and find consistent
results with the 1+1-state, larger $t_\text{min}$ fit of
Ref.~\cite{Lattice:2015tia}.

\begin{table}
 \centering
 \caption{Fit ranges $[t_\text{min},t_\text{max}]$ used in the kaon and $B$-meson mass and energy fits.}
 \label{tab:fit_range}
 \begin{tabular}{ccc}
  \hline\hline
  $\approx a$~(fm) & ~~kaon~~  & ~~$B$ meson ~~\\
  \hline
  0.12  & [7,30]  & [3,15] \\
  0.09  & [10,35] & [5,20] \\
  0.06  & [17,60] & [7,30] \\
  0.045 & [20,90] & [8,40] \\
  \hline\hline
 \end{tabular}
\end{table}

\begin{figure}
    \centering
    \includegraphics[scale=0.8]{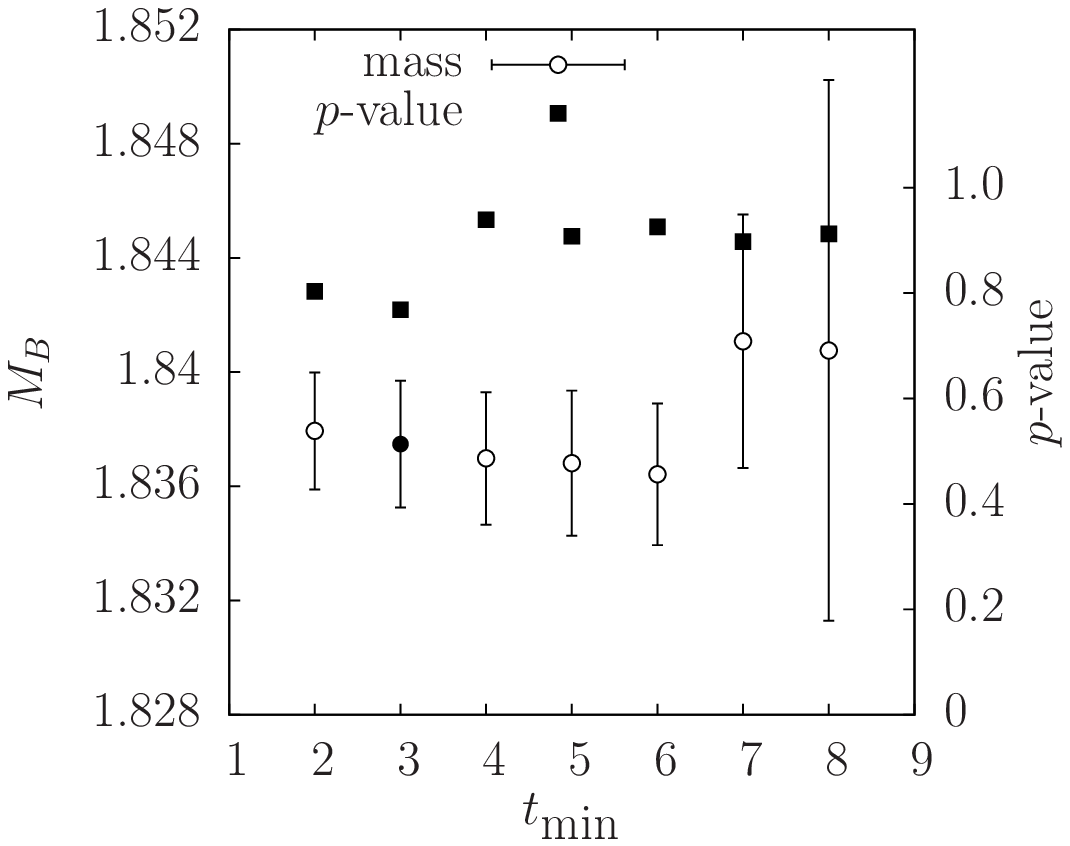} \\[2em]
    \includegraphics[scale=0.8]{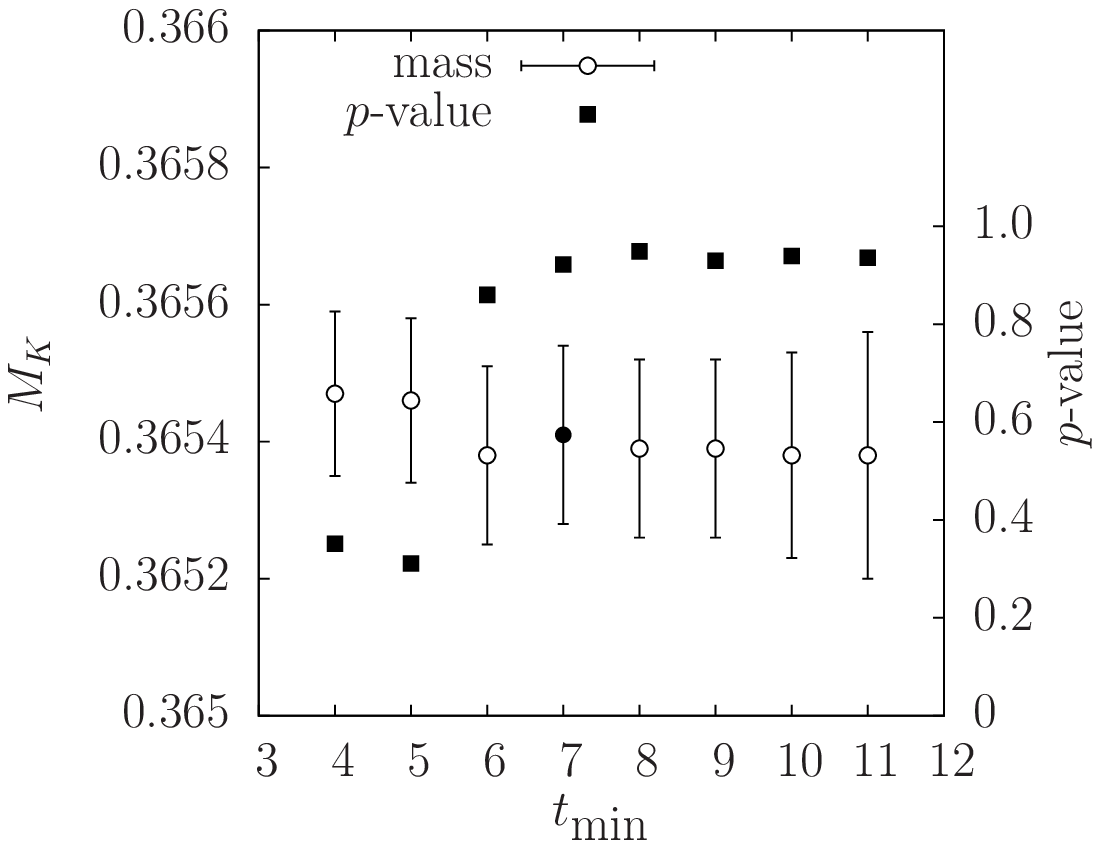}
    \caption{$B$-meson (upper) and kaon (lower) mass vs.~$t_\text{min}$ on the \coarse, 
      $m_l^\prime = 0.1m_h^\prime$ ensemble for fixed $t_{\rm max} = 15$ and 30, respectively.
      The left and right vertical axes show the fitted mass and the $p$~value (confidence level) of the fit,
      respectively. The filled circles show the values of $t_{\min}$ selected for the analysis.}
    \label{fig:mass-fit}
\end{figure}

Figure~\ref{fig:mass-fit} shows sample $B$-meson and kaon correlator fits 
versus $t_\text{min}$ for fixed $t_\text{max}$ on the same \coarse ensemble
as in Fig.~\ref{fig:2pt_state_choice}.  The fit 
results and errors are stable versus $t_{\min}$, and show no evidence of 
residual excited-state contamination.

\begin{figure}
    \centering
    \includegraphics[scale=0.8]{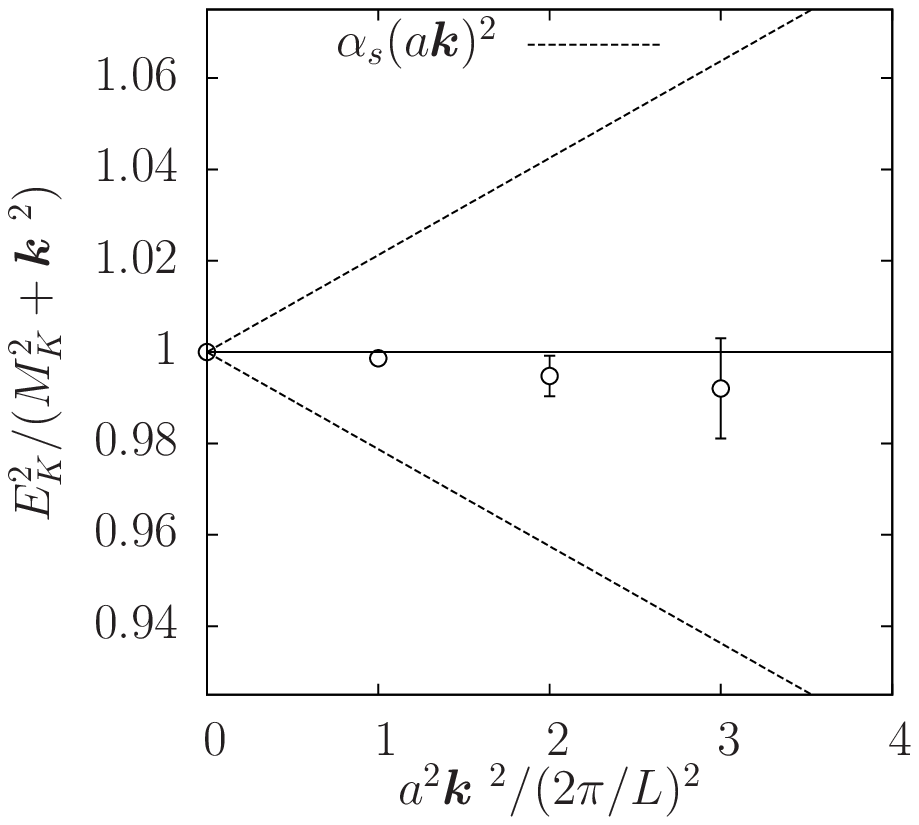}
    \caption{$E_K^2/(M_K^2+\bm{k}^2)$ vs.~kaon momentum in units of $2\pi/L$ on the \coarse, 
      $m_l^\prime = 0.1m_h^\prime$ ensemble.
      The continuum dispersion relation is well respected through momentum $2\pi(1,1,1)/L$.
      The dashed lines show a power-counting estimate for the size of the momentum-dependent discretization 
      error for comparison.}
    \label{fig:EP}
\end{figure}

For kaons with nonzero momentum, we can either extract the energy from two-point correlation functions with
nonzero momentum, or we can use the kaon mass from the zero-momentum correlator and the 
continuum dispersion relation, $E^2=M^2+\bm{k}^2$.
Figure~\ref{fig:EP} shows a comparison of the kaon energy calculated from the continuum dispersion relation
and from directly fitting the nonzero momentum two-point correlation functions on the ensemble discussed
above.
We do not observe any statistically significant deviations from the continuum dispersion relation.  
Further, while the statistical errors grow with increasing momentum, the 
kaon energies are consistent with a continuum dispersion relation within a 2\% statistical accuracy even 
at our largest simulated lattice kaon momentum.
Therefore, we use the continuum dispersion relation to obtain the 
kaon energies at nonzero lattice momenta because this yields smaller statistical errors than the 
direct fit. 

\begin{figure}
    \centering
    \includegraphics[scale=0.8]{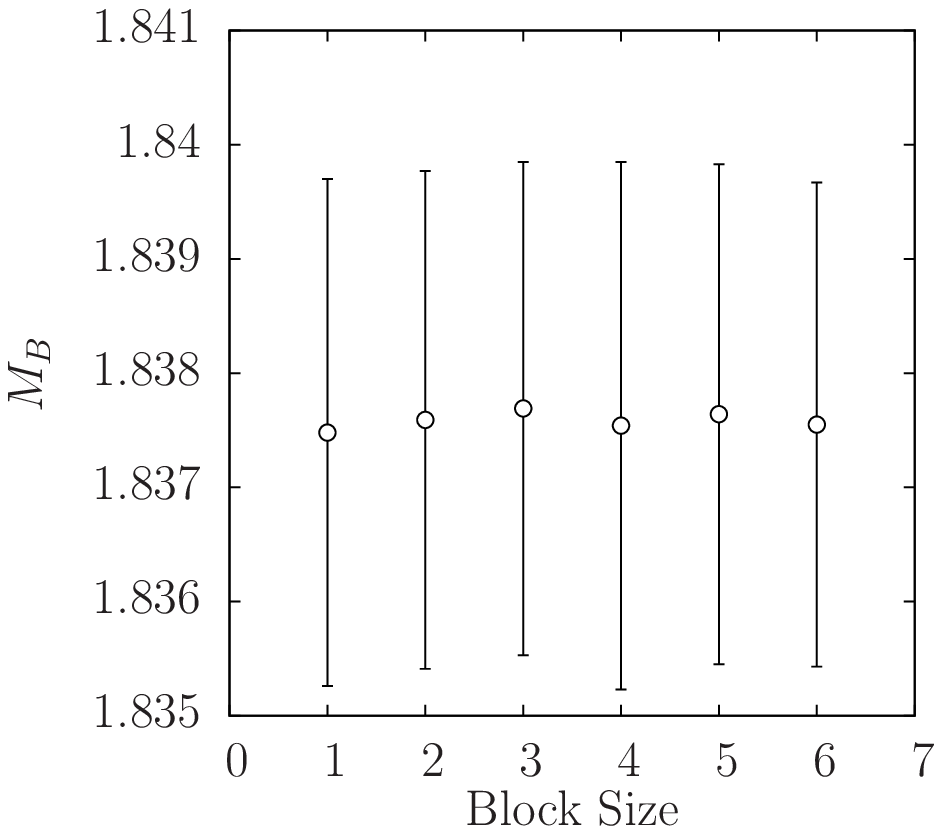}\hfill
    \includegraphics[scale=0.8]{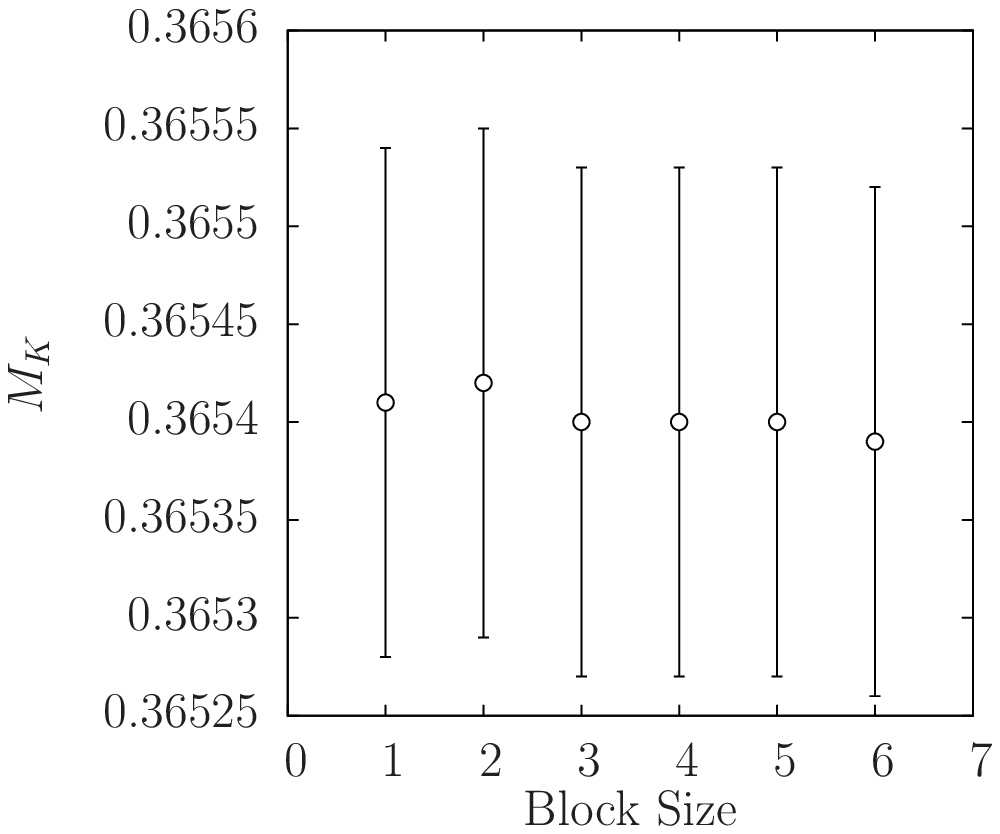} \\ \quad \\
    \includegraphics[scale=0.8]{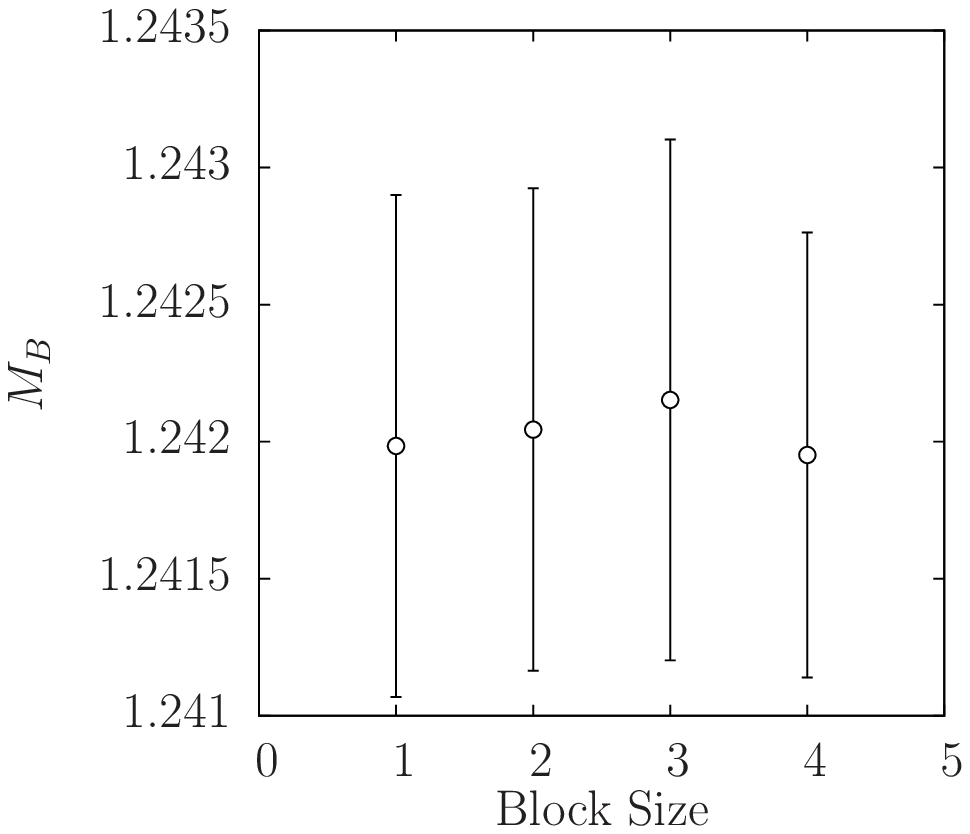}\hfill
    \includegraphics[scale=0.8]{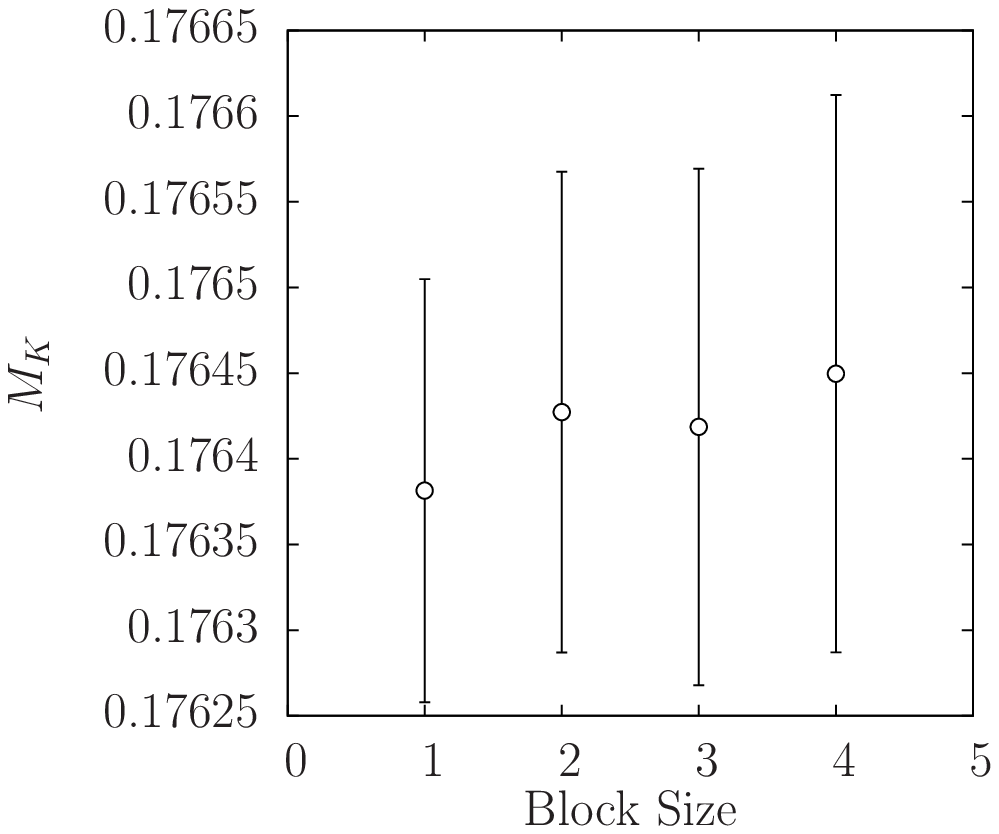}
    \caption{$M_B$ (left column)  and $M_K$ (right column) vs. block
      size on the \coarse~(top row) and \superfine~(bottom row),
      $m_l^\prime=0.1m_h^\prime$ ensemble. The fit results are
      stable as the block 
      size increases.}
    \label{fig:mass.autocorr}
\end{figure}

The meson propagators from consecutive gauge-field configurations are, in
principle, correlated, so we look for possible autocorrelations by studying 
the effect of the block size on our fit results.
We perform this test on every ensemble. As illustrated in Figure~\ref{fig:mass.autocorr} 
for two of the ensembles, the central values  and errors are stable with increasing 
the block size, so we do not block the data or inflate the statistical errors in 
our analysis.

\subsection{Extracting form factors}
\label{subsec:form_factors}

We extract the lattice form factors $f_\parallel$, $f_\perp$, and $f_T$
from the ratio of three-point to two-point correlation functions.
The three-point functions are defined as
\begin{align}
    C^\mu_3(t,T; \bm{k}) & = \sum_{\bm{x},\bm{y}} e^{i \bm{k} \cdot \bm{y}} 
        \langle \mathcal{O}_K (0,\bm{0})\, 
        V^\mu (t,\bm{y})\, \mathcal{O}^\dagger_B (T,\bm{x}) \rangle,
    \label{eq:C3mu} \\
    C^{\mu\nu}_3(t,T; \bm{k}) & = \sum_{\bm{x},\bm{y}} e^{i \bm{k} \cdot \bm{y}} 
        \langle \mathcal{O}_K (0,\bm{0})\, 
        T^{\mu\nu} (t,\bm{y})\, \mathcal{O}^\dagger_B (T,\bm{x}) \rangle ,
    \label{eq:C3munu}
\end{align}
where the kaon source is at time slice 0 and the $B$-meson sink is at time slice $T$.
The source-sink separations $T$ are given in Table~\ref{tab:3pt_fit_range}.
Because we calculate the form factors in the $B$-meson rest frame, only the kaon has nonzero
momentum~$\bm{k}$.

By inserting two complete sets of states, the three-point correlation function $C^\mu_3$ can 
be decomposed into sums over energy levels as 
\begin{align}
    C^\mu_3(t,T;\bm{k}) &= \sum_{m,n} (-1)^{m(t+1)} (-1)^{n(T-t-1)}
        A_{mn}^{\mu}e^{-E_K^{(m)}t}e^{-M_B^{(n)}(T-t)},
\end{align}
where 
\begin{align}
    A_{mn}^{\mu} &= \frac{\bra{0}\mathcal{O}_K\ket{K^{(m)}}}{2E_K^{(m)}} 
    \bra{K^{(m)}}V^\mu\ket{B^{(n)}}\frac{\bra{B^{(n)}}\mathcal{O}_B\ket{0}}{2M_B^{(n)}}.
    \label{eq:amplitude-not-amputated}
\end{align}
The contributions from the first few terms dominate $C^\mu_3$ at 
times sufficiently far from both the source and sink.
A similar decomposition applies to $C^{\mu\nu}_3$. 

We use the averages introduced in Ref.~\cite{Bailey:2008wp} to suppress the 
contribution from oscillating states in correlation functions. 
We average the value of the two-point correlator on successive time slices:
\begin{align}
        \bar{C}_{2} (t) &\equiv \frac{e^{-M_P^{(0)}t}}{4} \left[ \frac{C_{2} (t)}{e^{-M_P^{(0)}t}} + 
            \frac{2 C_{2} (t+1)}{e^{-M_P^{(0)}(t+1)}} + \frac{C_{2} (t+2)}{e^{-M_P^{(0)}(t+2)}}  \right] 
            \nonumber\\
            &= \frac{Z_P^2}{2 M_P^{(0)}} e^{-M_P^{(0)}t} + \order(\Delta M_P^2),
\label{eq:2pt_ave}
\end{align}
where $Z_P = \langle0|\mathcal{O}_P|P\rangle$ is the ground-state amplitude of the kaon or $B$~meson, and
$\Delta M_P$ is the energy difference between the ground and first oscillating state.
For three-point functions, we also average the value of the correlator for two neighboring sink locations
$T$ and $T+1$:
\begin{align}
    \bar{C}_{3}^{\mu(\nu)} (t,T; \bm{k}) \equiv &\,  \frac{1}{8} \bigg[  
        e^{-E_K^{(0)}t} \, e^{-M_B^{(0)}(T-t)} \bigg] \times
        \left[ \frac{C_3^{\mu(\nu)} (t,T; \bm{k})}{e^{-E_K^{(0)}t} e^{-M_B^{(0)}(T-t)}} + 
           \frac{C_3^{\mu(\nu)}(t,T+1; \bm{k})}{e^{-E_K^{(0)}(t)} e^{-M_B^{(0)}(T+1-t)}} \right.
   \nonumber \\
        & + \frac{2 \, C_3^{\mu(\nu)} (t+1,T; \bm{k})}{e^{-E_K^{(0)}(t+1)} e^{-M_B^{(0)}(T-t-1)}} +  
           \frac{2 \, C_3^{\mu(\nu)} (t+1,T+1; \bm{k})}{e^{-E_K^{(0)}(t+1)} e^{-M_B^{(0)}(T-t)}}
   \nonumber \\
        & + \left. \frac{C_3^{\mu(\nu)} (t+2,T; \bm{k})}{e^{-E_K^{(0)}(t+2)} e^{-M_B^{(0)}(T-t-2)}} +  
           \frac{C_3^{\mu(\nu)} (t+2,T+1; \bm{k})}{e^{-E_K^{(0)}(t+2)} e^{-M_B^{(0)}(T-t-1)}} \right] \\
        = &\, A_{00}e^{-E_K^{(0)}t} \, e^{-M_B^{(0)}(T-t)} + 
            (-1)^{T+1} A_{11} e^{- E_K^{(1)}t} e^{-M_B^{(1)}(T-t)} \left( \frac{\Delta M_B}{2} \right) 
    \nonumber \\
        & + \order(\Delta E_K^2,\, \Delta E_K \Delta M_B,\, \Delta M_B^2) .
\label{eq:3pt_ave}
\end{align}
We then form the ratios
\begin{equation}
    \bar{R}^{\mu(\nu)} (t,T; \bm{k}) \equiv 
        \frac{\bar{C}_{3}^{\mu(\nu)} (t,T; \bm{k})}{\sqrt{\bar{C}_{2}^{K} (t; \bm{k})\bar{C}_{2}^{B} (T-t)}}
        \sqrt{\frac{2 E_K^{(0)}}{e^{-E_K^{(0)}t} \, e^{-M_B^{(0)}(T-t)}}} ,
\label{eq:2pt_3pt_ratio}
\end{equation}
where $E^{(0)}_K$ and $M_B^{(0)}$ are obtained from fits to Eq.~(\ref{eq:2pt_fit_function}) with 
$E^{(0)}_K = \sqrt{M_K^{(0)} + \bm{k}^2}$.
From Eqs.~(\ref{eq:2pt_ave}) and~(\ref{eq:3pt_ave}), the ratio $\bar{R}^{\mu(\nu)}$ contains 
a $t$-independent term proportional to the desired matrix element,
and other higher-order terms from the excited states.

\begin{figure}
    \includegraphics[scale=0.75]{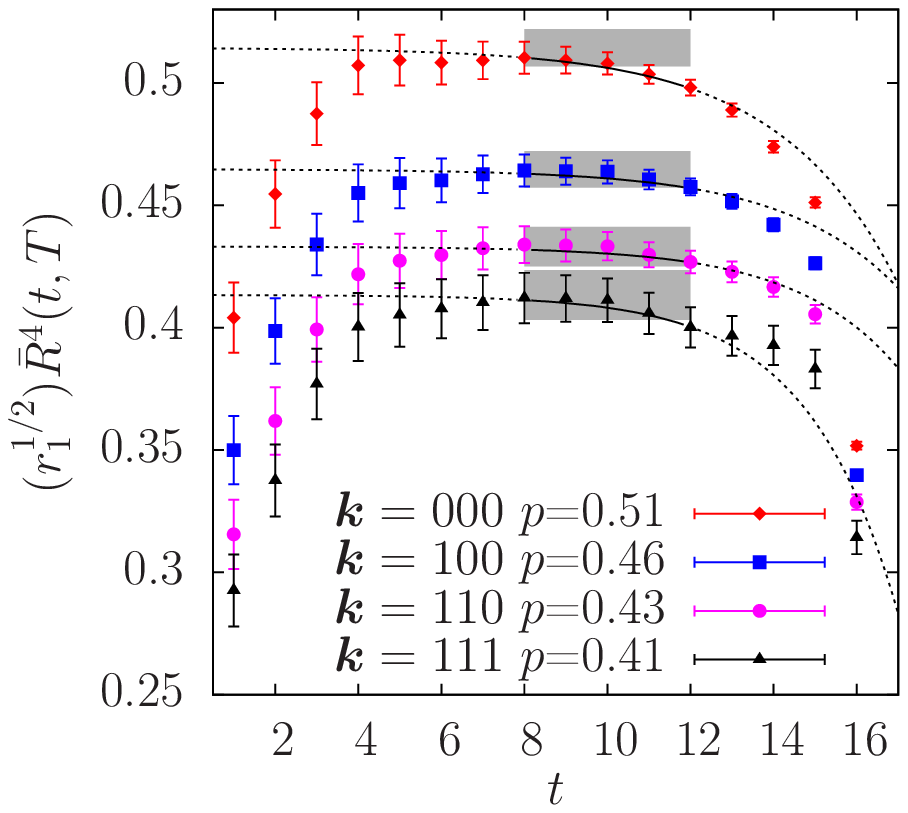} \\[1em]
    \includegraphics[scale=0.75]{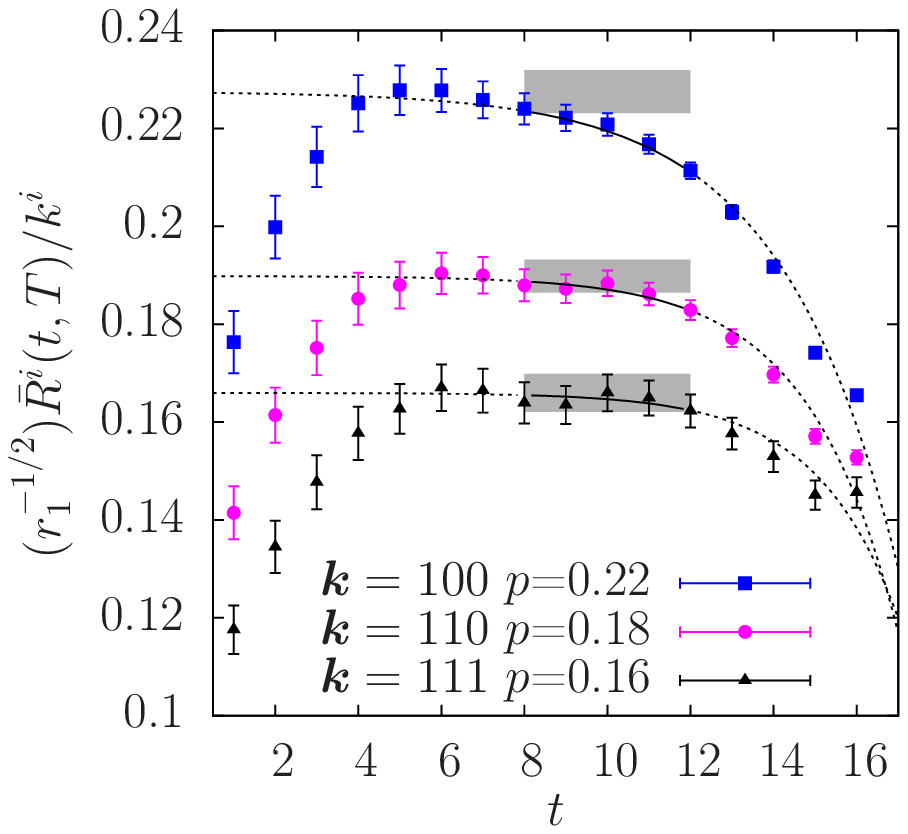} \\[1em]
    \includegraphics[scale=0.75]{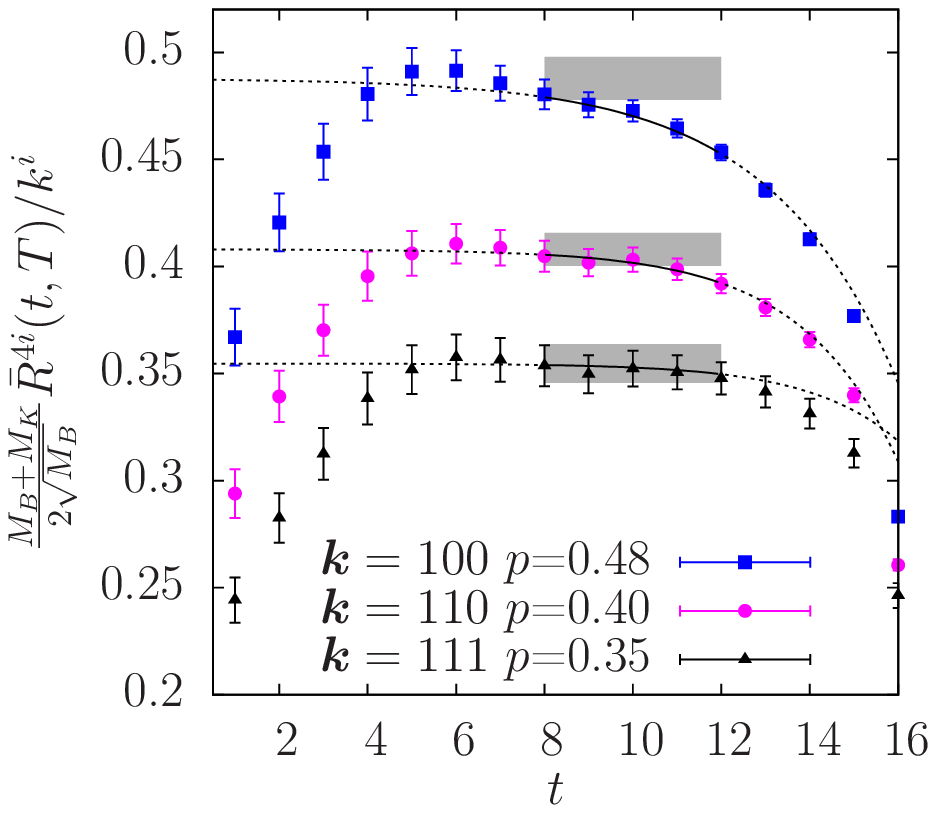}
    \caption{Form-factor ratio $\bar{R}^{\mu(\nu)}$ fits on the \coarse, 
$m_l^\prime = 0.1m_h^\prime$ ensemble. From top to bottom, the three plots show the ratios for the 
temporal vector, spatial vector, and tensor currents. In the top plot, the data sets correspond to 
lattice kaon momenta $\bm{k}=2\pi(0,0,0)/L$, $2\pi(1,0,0)/L$, $2\pi(1,1,0)/L$ and $2\pi(1,1,1)/L$;
nonzero momentum is required to extract the form factors in  the bottom two plots, so there are 
only three sets of data in each of them. The gray horizontal bands 
show the fit results with statistical errors for $C_0^{\mu(\nu)}$ in Eq.~(\ref{eq:form_fit}).
The black solid and dashed curves show the fit result within and extended beyond 
the fit range, respectively.}
    \label{fig:form_fit}
\end{figure}

We show an example of the ratio $\bar{R}^{\mu(\nu)}$ on the \coarse, 
$m_l' = 0.1m_s'$ ensemble in Fig.~\ref{fig:form_fit}. There is a short plateau region 
in the middle between $0$ and $T$, with kaon excited-state contributions visible
on the left and $B$-meson excited-state contributions visible on the right. The $B$-meson 
excited-state contributions, however, are smaller as indicated by the less dramatic falloff 
of the correlator on the right-hand side. We therefore choose to fit the correlator 
closer to the $B$-meson side including the contribution from a single 
$B$-meson excited state, but sufficiently far from the kaon 
that we can neglect kaon excited states. The fit function is given by: 
\begin{equation}
\bar{R}^{\mu(\nu)}(t,T; \bm{k})=D_0^{\mu(\nu)}\left [1-D_1 e^{-\Delta M_B(T-t)}\right ] \,
\label{eq:form_fit} 
\end{equation}
where $D_0^{\mu(\nu)}$, $D_1$, and $\Delta M_B$ are fit parameters.  
Although the second term in Eq.~(\ref{eq:form_fit}) models all excited
states, we expect $\Delta M_B$ to be close to the mass difference
of the first excited state.

\begin{figure}
    \includegraphics[scale=0.75]{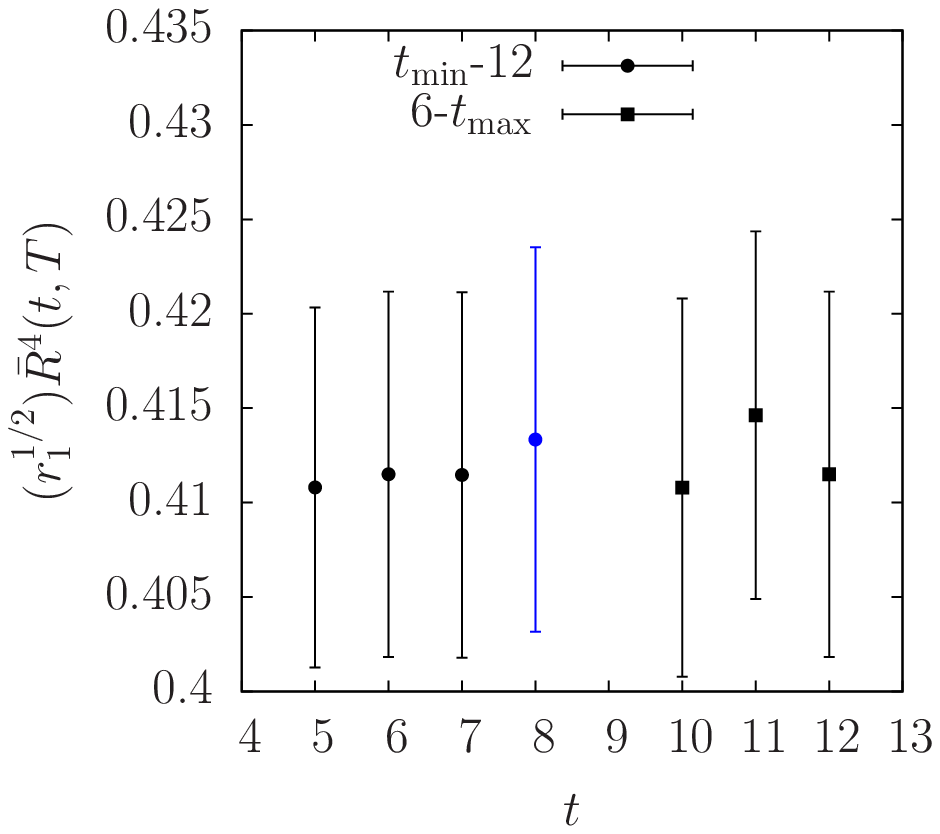} \\[1em]
    \includegraphics[scale=0.75]{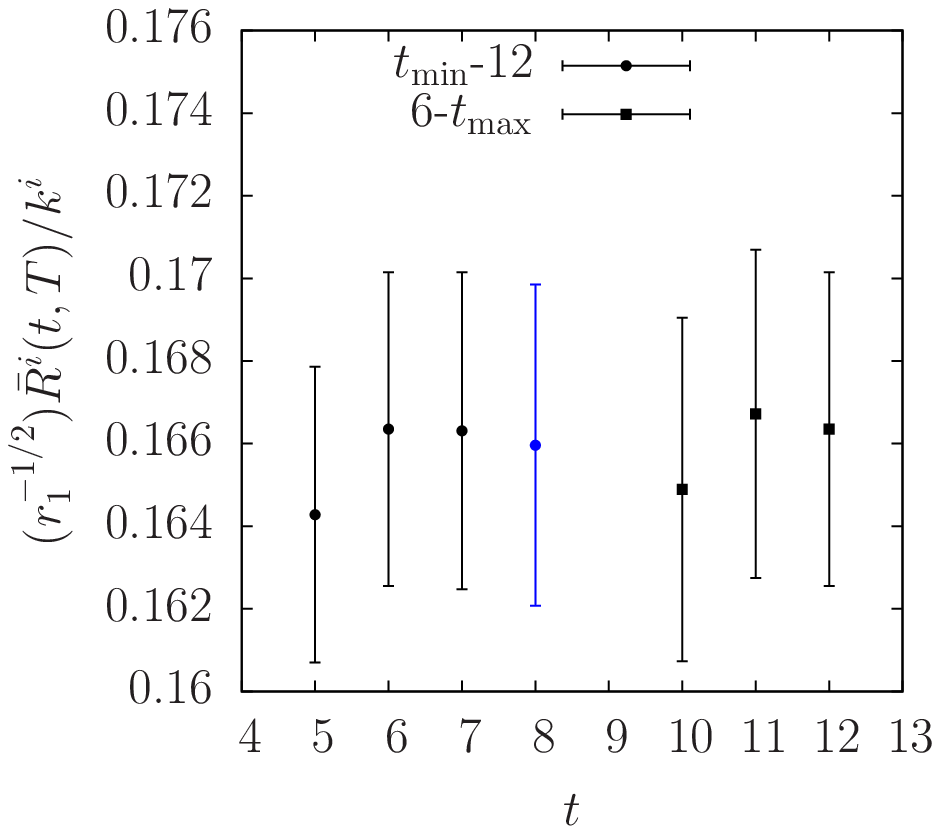} \\[1em]
    \includegraphics[scale=0.75]{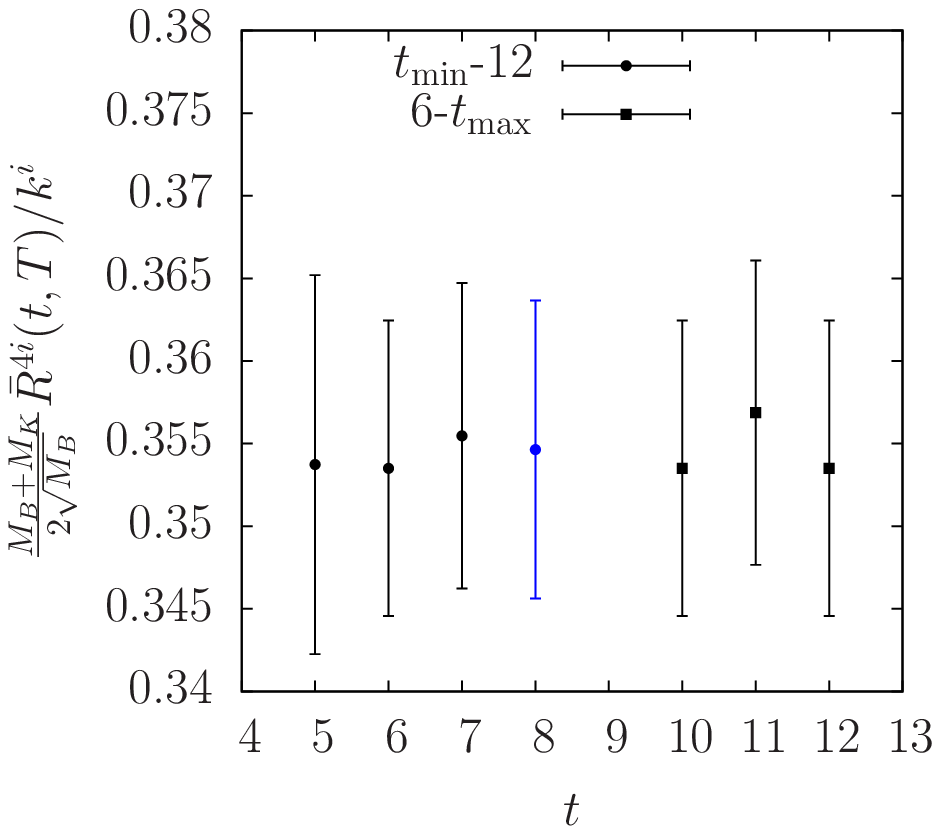}
    \caption{Fit results of $\bar{R}^{\mu(\nu)}$ from different fit ranges on the \coarse, 
$m_l^\prime = 0.1m_h^\prime$ ensemble with lattice kaon momentum $p=\frac{2\pi}{L}(1, 1, 1)$. 
From top to bottom, the three plots show the ratios for the temporal vector, spatial vector, and tensor currents. 
We vary the fit range by changing $t_{\rm {min}}$ and $t_{\rm {max}}$. The blue data point denotes
the result from the fit range used in this paper.}
    \label{fig:form_fit_stability}
\end{figure}

We employ a correlated, constrained fit~\cite{Lepage:2001ym,Morningstar:2001je} to Eq.~(\ref{eq:form_fit}), 
with priors determined as follows. For the prior 
on $D_0^{\mu(\nu)}$, we select a point from the middle of the plateau region
and use its central value with the error inflated by a factor of two.
For $D_1$, we use a prior of central value zero and width one. 
For $\Delta M_B$,  we use the central value and width of $M_B^{(1)}-M_B^{(0)}$ 
obtained from the corresponding two-point correlator fit. 
We minimize the augmented $\chi_{\rm aug}^2$~\cite{Morningstar:2001je}
\begin{equation}
    \chi^2_\text{aug} = \chi^2+\sum_i \frac{(P^{(i)}-\tilde{P}^{(i)})^2}{\sigma_i^2} ,
\label{eq:aug_chi2}
\end{equation}
where $P^{(i)}$ is the $i$th fit parameter, and $\tilde{P}^{(i)}$ and $\sigma_i$ are the prior central value
and width.
We measure the goodness of fit using the $\chi^2_\text{aug}/$dof or $p$~value, obtaining $p$ from
$\chi^2_\text{aug}$ and the number of degrees of freedom equal to the sum of the number of data points and
prior constraints minus the number of fit parameters.
We choose the fit interval $[t_{\rm min}, t_{\rm max}]$ such that we obtain a good $p$ value, using the same
fit range for all momenta on the same ensemble.
We select approximately the same physical fit ranges on the ensembles with different lattice spacings.
Figure~\ref{fig:form_fit} shows sample fits of the three form-factor ratios on the \coarse, $m_l' = 0.1m_s'$
ensemble.
Figure~\ref{fig:form_fit_stability} shows an example of the stability of the fit result against the variations 
of the fit range. 
We choose the preferred fit range to be $[t_\text{min},t_\text{max}]=[8,12]$, where we find a good $p$~value.
The fit ranges and source-sink separations used on other ensembles are given in 
Table~\ref{tab:3pt_fit_range}. 

\begin{figure}
    \centering
    \includegraphics[height=0.275\textheight]{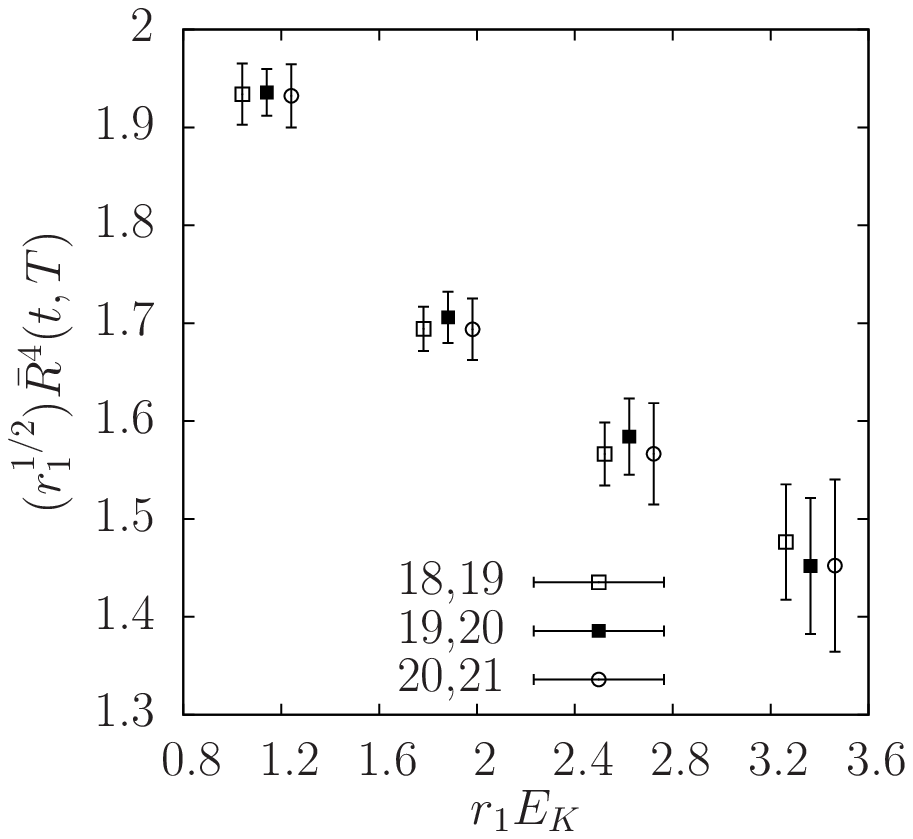} \\[1em]
    \includegraphics[height=0.275\textheight]{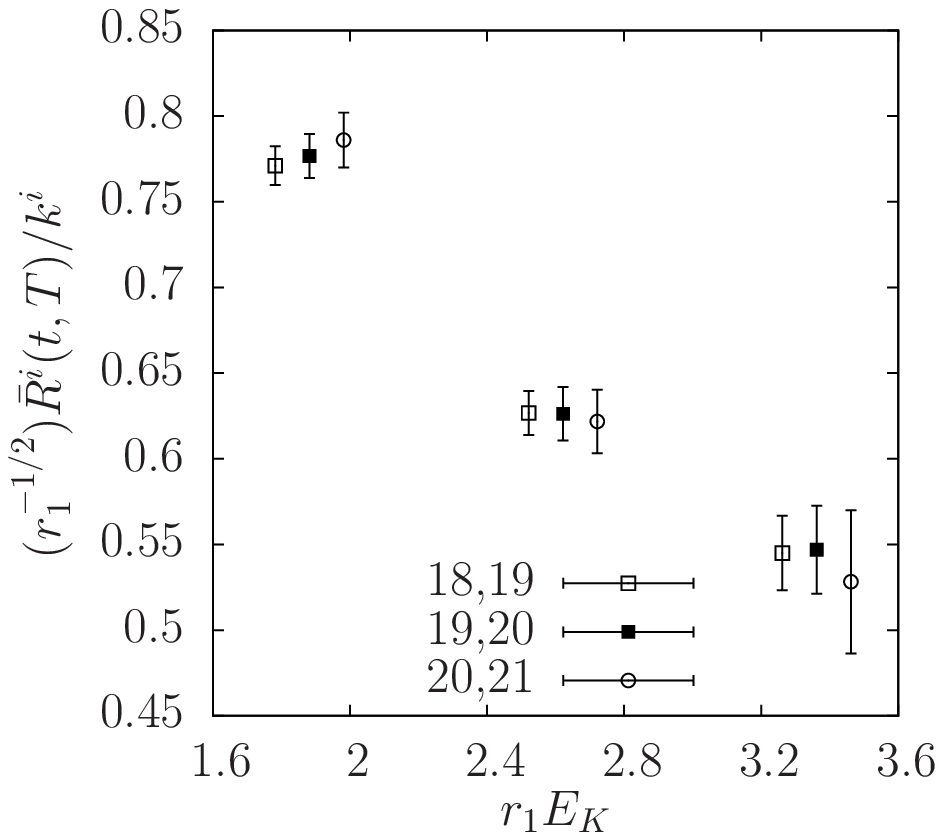} \\[1em]
    \includegraphics[height=0.275\textheight]{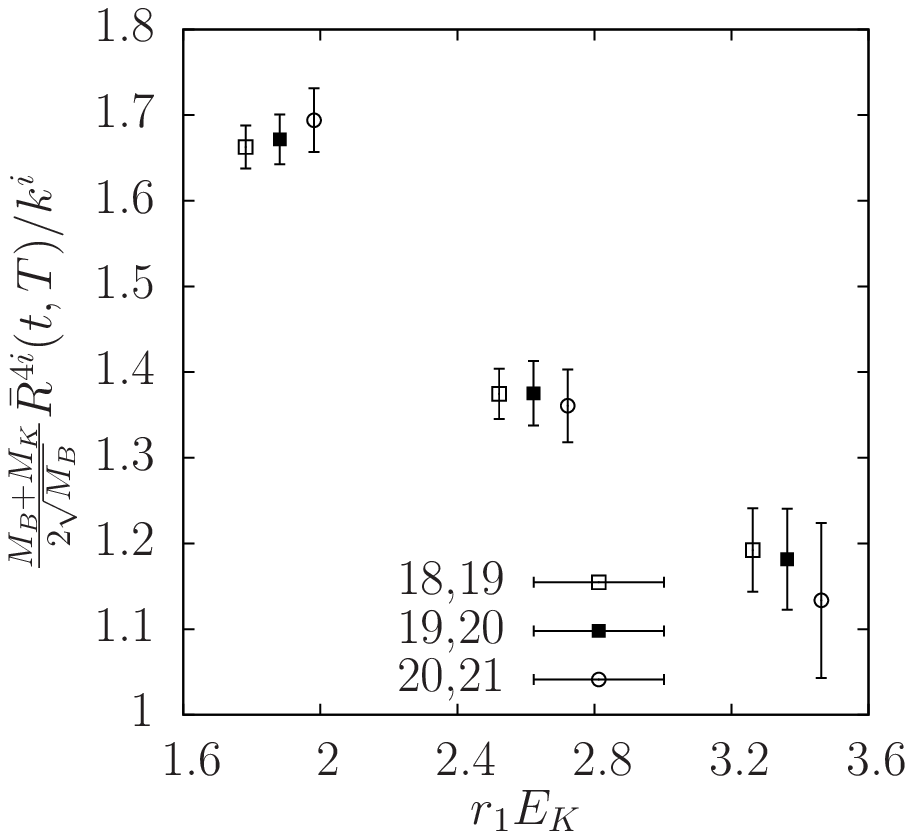}
    \caption{Form-factor ratio $\bar{R}^{\mu(\nu)}$ fits on the $a\approx 0.12$~fm, 
$m_l^\prime/m_h^\prime=0.007/0.05$ ensemble. From top to bottom, the three plots 
show the ratios for the temporal vector $f_\parallel$, spatial vector 
$f_\perp$, and tensor $f_T$ currents. The fit results 
for different pairs of source-sink separations $T,T+1$ are shown as 
a function of $E_K$. The results from larger sink combinations are 
slightly displaced to the right for clarity.}
    \label{fig:source-sink}
\end{figure}

To study the effects of residual excited-state contamination, we generated three-point
correlators on the  $a \approx 0.12$~fm, $m_l^\prime = 0.14m_s^\prime$ 
ensemble with several source-sink 
separations $T = 18, 19, 20, 21$. We repeat the correlator fits with three sink-location 
combinations $(T,T+1)$=$(18,19)$, $(19,20)$, and~$(20,21)$, and the results are 
shown in Fig.~\ref{fig:source-sink} for four different momenta. 
We find no statistically significant differences for all operators and momenta except
for $f_\perp$ and $f_T$ at $\bm{p}=2\pi(1,0,0)/L$. 
These differences, however, are still sufficiently small that 
increasing the error on all $\bm{p}=2\pi(1,0,0)/L$ points in the chiral-continuum fit
does not change the physical form-factor results.

\begin{table}[tp]
\centering
    \caption{Pairs of source-sink separations $T, T+1$ and fit ranges used in the $\bar{R}^{\mu(\nu)}$ fits.}
    \label{tab:3pt_fit_range}
    \begin{tabular}{ccc}
        \hline\hline  
        $\approx a$~(fm) & ~~$T, T+1$~~ & $[t_\text{min}, t_\text{max}]$ \\
        \hline
        0.12  & 18, 19 &  [8, 12] \\
        0.09  & 25, 26 &  [10, 16] \\
        0.06  & 36, 37 &  [16, 24] \\
        0.045 & 48, 49 &  [20, 32] \\
        \hline\hline
    \end{tabular}
\end{table}

The fit parameters $C_0^{\mu(\nu)}$ are proportional to the matrix elements 
$\bra{K^0}J\ket{B^0}$. The lattice form factors are obtained as
\begin{align}
    f_\parallel^{\rm lat}(E_K) & = {D_0^4}(\bm{k}), \\
    f_\perp^{\rm lat}(E_K)     & = \frac{{D_0^i}(\bm{k})}{k^i}, \\
    f_T^{\rm lat}(E_K)         & = \frac{M_B+M_K}{\sqrt{2M_B}} \label{eq:latft}
        \frac{{D_0^{4i}}(\bm{k})}{k^i}.
\end{align}
The factor $(M_B+M_K)/\sqrt{2M_B}$ in $f_T$ in Eq.~(\ref{eq:latft}), which stems from 
Eq.~(\ref{eq:contift}), is evaluated with the physical meson masses to 
avoid introducing $m_q$ dependence not captured in the $\chi$PT formula.

\subsection{\texorpdfstring{\boldmath $b$-quark}{b-quark} mass correction}
\label{subsec:kappa_b_shift}

The $b$-quark hopping parameter used in our simulations $\kappa_b^\prime$  
differs slightly from the physical value $\kappa_b$
because our production runs started before a more precise 
tuning of the $b$-quark hopping parameter $\kappa_b$ was completed. 
For our desired accuracy, we need to apply a correction. To this end, we have carried out 
runs with multiple values of $\kappa_b^\prime$ on the \coarse ensemble with 
$m_l^\prime/m_h^\prime=0.2$. In addition to the production value of $\kappa_b^\prime=0.0901$, 
we repeated the run with $\kappa_b^\prime = 0.0820$ and 0.0860, allowing us to 
bracket the physical value $\kappa_b=0.0868$. The form factors depend on 
the $b$-quark kinetic mass $m_2^\prime$. At the tree level
\begin{equation}
        \frac{1}{m_2^\prime a} = \frac{2}{m_0^\prime a(2+m_0^\prime a)} + \frac{1}{1+m_0^\prime a},
\end{equation}
where
\begin{equation}
m_0^\prime a = \frac{1}{2u_0}\left (\frac{1}{\kappa^\prime}-\frac{1}{\kappa_\text{crit}}\right ).
\end{equation}
The values of $u_0$ and $\kappa_\text{crit}$ are given in Table~\ref{tab:phy_kappa}.
Following Ref.~\cite{Lattice:2015tia}, we expand the form factor in $m_2^{-1}$ about $m_2'$: 
\begin{equation}
  f(m_2^\prime  , E_K)  = f(m_2, E_K)\left [1- \frac{\partial{\ln{f}}}{\partial {\ln{m_2}}} \left ( \frac{m_2}{m_2^\prime}-1\right )\right ] \ ,
\label{eq:kappa-shift}
\end{equation}
where $m_2$ denotes the physical $b$-quark kinetic mass. We determine the slope, 
$\frac{\partial{\ln{f}}}{\partial{\ln{m_2}}}$, 
in our companion work on the semileptonic decay $B\to\pi l\nu$~\cite{Lattice:2015tia}. Because
the slope depends mildly on the daughter-quark mass, and the 
daughter-quark mass is tuned close to its physical value in our calculation, we neglect 
the daughter-quark dependence of the slope in this work. Finally, we quote 
$\frac{\partial{\ln{f}}}{\partial{\ln{m_2}}}$ of $f_\parallel$, $f_\perp$, and $f_T$ 
at the simulated daughter-quark mass as 0.115(9), 0.139(13), and 0.126(13)~\cite{Lattice:2015tia}. 
We find relative shifts due to b-quark mass tuning of about 0.5\%-1.5\% on the 
different ensembles.

\begin{table}[tp]
 \centering
 \caption{
The simulation $\kappa_b'$ and physical $\kappa_b$~\cite{Bailey:2014tva}.  We also include 
$\kappa_\text{crit}$ and $u_0$ from the plaquette in this table for 
convenience, because they are used in the calculation of the $b$-quark kinetic mass.}
 \label{tab:phy_kappa}
 \begin{tabular}{cccccc}
  \hline  
  \hline  
  $\approx a$~(fm) & $am_l'/am_h'$ & ~~~~~~$\kappa_b'$~~~~~~ &  
~~~~~~$\kappa_b$~~~~~~ 
& ~~~~~~$\kappa_\text{crit}$~~~~~~ & ~~~~~~$u_0$~~~~~~ \\
  \hline
  0.12  & 0.01/0.05     & 0.0901 & 0.0868(9)(3) & 0.14091  & 0.8677  \\
  0.12  & 0.007/0.05    & 0.0901 & 0.0868(9)(3) & 0.14095  & 0.8678  \\
  0.12  & 0.005/0.05    & 0.0901 & 0.0868(9)(3) & 0.14096  & 0.8678  \\
  \hline
  0.09  & 0.0062/0.031  & 0.0979 & 0.0967(7)(3) & 0.139119 & 0.8782  \\
  0.09  & 0.00465/0.031 & 0.0977 & 0.0966(7)(3) & 0.139134 & 0.8781  \\
  0.09  & 0.0031/0.031  & 0.0976 & 0.0965(7)(3) & 0.139173 & 0.8779  \\
  0.09  & 0.00155/0.031 & 0.0976 & 0.0964(7)(3) & 0.139190 & 0.877805  \\
  \hline 
  0.06  & 0.0036/0.018  & 0.1052 & 0.1052(5)(2) & 0.137632 & 0.88788  \\
  0.06  & 0.0018/0.018  & 0.1052 & 0.1050(5)(2) & 0.137678 & 0.88764  \\
  \hline
  0.045 & 0.0028/0.014  & 0.1143 & 0.1116(3)(2) & 0.136640 & 0.89511  \\
  \hline  
  \hline
 \end{tabular}
\end{table}

\subsection{Chiral-continuum extrapolations}
\label{subsec:chiral_fit}

The lattice form factors are computed numerically on ensembles with  
degenerate up-/down-quark masses that are heavier than 
the value in nature, as well as at nonzero 
lattice spacing. To obtain physical results, we first compute the form factors on 
several lattice spacings with varying up/down-quark masses and 
close-to-physical strange-quark masses, and
then extrapolate to the physical light-quark mass and 
continuum (and interpolate to the physical strange-quark mass)
using heavy meson rooted staggered chiral perturbation theory 
(HMrS$\chi$PT)~\cite{Aubin:2005aq,Aubin:2007mc}.

For the chiral-continuum extrapolation we use an HMrS$\chi$PT formula valid to 
leading order in $1/m_b$ and next-to-leading order (NLO) in the light-quark masses, kaon 
energy, and lattice spacing, supplemented by next-to-next-to leading order (NNLO) analytical 
terms.  We have tested both SU(3) HMrS$\chi$PT~\cite{Aubin:2007mc}, which includes the 
effects of dynamical pions, kaons, and $\eta$ mesons, and SU(2) HMrS$\chi$PT, in which 
the mesons with strange quarks are integrated out. In addition, we also consider 
hard-kaon HMrS$\chi$PT, which applies to semileptonic decays with energetic kaons. 
We find that NLO SU(3) HMrS$\chi$PT even supplemented with NNLO analytical terms, does not provide a good description of 
the data for $f_\parallel$~\cite{Zhou:2011be,Zhou:2012sn,Zhou:2013uu},
and the $p$~value of the fit is $10^{-9}$.
On the other hand, SU(2) HMrS$\chi$PT describes the data well even at NLO.
We therefore choose SU(2) HMrS$\chi$PT to perform the chiral-continuum extrapolations.

The kaon energies in our numerical simulations are much larger than the rest mass of the 
physical kaon. Therefore standard HMrS$\chi$PT, which is derived for the situation in which 
the kaon momenta are soft, may not provide a good description of our data throughout the 
available kinematic range.  We therefore also consider hard-kaon HMrS$\chi$PT, which applies 
for semileptonic decays with energetic kaons. Recently, Bijnens and Jemos derived the 
continuum NLO hard-kaon (pion) HM$\chi$PT formulae for both $B \to K$ and $B\to \pi$ 
processes~\cite{Bijnens:2010ws,Bijnens:2010jg}. We derive the corresponding NLO staggered SU(2) 
and SU(3) hard kaon (pion) HMrS$\chi$PT formulae in Appendix~\ref{app.su2}. It turns 
out that the chiral logarithms in NLO hard-kaon SU(2) HMrS$\chi$PT are identical to those 
in standard soft-kaon SU(2) HMrS$\chi$PT for $B\to K$ decays.  This is likely the reason that 
the standard NLO SU(2) expressions describe our data even at such large kaon energies. 
Reference~\cite{Colangelo:2012ew} found that the hard-pion theory can break down at 
three-loop level, but we only work at one-loop level here.

The NLO SU(2) HMrS$\chi$PT formulae for $B \to K$ decays take the form 
\begin{align}
    r_1^{1/2}f_{\parallel} & = \frac{g_\pi         \left[ C_\parallel^{(0)} \left(1 + \text{logs} \right) + 
        C_\parallel^{(1)} \chi_l + C_\parallel^{(2)} \chi_h         + C_\parallel^{(3)} \chi_E 
        +C_\parallel^{(4)} \chi_{a^2} + C_\parallel^{(5)} \chi_E^2\right]} 
{f_\pi r_1(E_K + \Delta_{B_{s0}^*})r_1} , 
    \label{eq:chiral.fpara} \\
    r_1^{-1/2}f_{\perp} & = \frac{g_\pi         \left[ C_\perp^{(0)} \left(1 + \text{logs} \right) + 
        C_\perp^{(1)} \chi_l + C_\perp^{(2)} \chi_h         + C_\perp^{(3)} \chi_E 
        +C_\perp^{(4)} \chi_{a^2} + C_\parallel^{(5)} \chi_E^2\right]} 
{f_\pi r_1(E_K + \Delta_{B_s^*})r_1} , 
    \label{eq:chiral.fperp}
\end{align}
where ``$\text{logs}$'' denotes nonanalytic functions of the light-quark mass 
and lattice spacing; the explicit expressions are given in 
Eqs.~(\ref{eq:app:su2_fpara_log}), (\ref{eq:app:su2_fperp_D}), and (\ref{eq:app:su2_fperp_log}).
The dimensionless expansion parameters $\chi_i$ in Eqs.~(\ref{eq:chiral.fpara}) 
and~(\ref{eq:chiral.fperp}) are
\begin{align}
    \chi_{l}   & = \frac{2\mu m_l}{8\pi^2f_\pi^2},\\
    \chi_{h}   & = \frac{2\mu m_h}{8\pi^2f_\pi^2},\\
    \chi_{a^2} & = \frac{a^2\overline{\Delta}}{8\pi^2f_\pi^2},\\
    \chi_E     & = \frac{\sqrt{2}E_K}{4\pi f_\pi},
\end{align}
where $a^2\overline{\Delta}$ is the averaged taste-symmetry breaking parameter, 
$a^2\overline{\Delta}\equiv\frac{1}{16}\sum_{\xi}a^2\Delta_{\xi}$ and
$\mu$ denotes the leading-order QCD LEC; see 
Eqs.~(\ref{eq:app:ps_mass_start})--(\ref{eq:app:ps_mass_end}) for 
the definition.
If HMrS$\chi$PT gives a good description of the data, we expect the 
$C^{(i)}$, $i>0$, to be of order unity. The SU(2) $\chi$PT formulae do not 
contain $m_h$ explicitly; however, the low-energy constants (LECs) depend
on $m_h$. Because the strange-quark masses on different ensembles are 
slightly different from each other, we include a term proportional to 
$\chi_h$ in the set of analytic terms to account for the leading 
strange-quark mass dependence of the LECs and enable an interpolation 
to the physical strange-quark mass.

Equations (\ref{eq:chiral.fpara}) and (\ref{eq:chiral.fperp}) each contain a pole in $E_K$. 
The poles appear at negative energy $-\Delta_{B_{s(0)}^*}$ with
\begin{equation}
    \Delta_{B_{s(0)}^*} \equiv \frac{M_{B^*_{s(0)}}^2-M_B^2-M_K^2}{2M_B}\approx M_{B_{s(0)}^*}-M_B.
\end{equation}
The pole arises from low-lying states with flavor content $\bar{b}s$ and quantum numbers that 
depend upon the form factor: for $f_\perp$ and $f_T$, the relevant $B_s^*$ meson has $J^P=1^-$, 
while for $f_\parallel$, the $B_{s0}^*$ state has $J^P=0^+$.  In the chiral-continuum fits, we 
fix $M_B$ to its experimentally measured  value 5.27958~GeV~\cite{Agashe:2014kda} (recall that 
we tuned the lattice $b$-quark mass using the experimental  $B_s$-meson mass.). 
We also use the experimentally measured value of the lowest-lying vector meson 
$M_{B_s^*} = 5.4154$~GeV~\cite{Agashe:2014kda}, which is stable apart from 
$B_s^*\to B_s\gamma$, for the pole position in the fits of $f_\perp$ and $f_T$ to Eq.~(\ref{eq:chiral.fperp}).
Although a scalar $B_{s0}^*$ state has not been observed in experiments, 
theoretical predictions estimate its mass to be just below the $B$-$K$ production 
threshold~\cite{Bardeen:2003kt,Green:2003zza}. Therefore, in the fit of $f_\parallel$, 
we use the prediction $M_{B^*_{s0}} = 5.711 (23)$~GeV from a recent three-flavor 
lattice-QCD calculation~\cite{Lang:2015hza} for the pole position in Eq.~(\ref{eq:chiral.fpara}). 

Following the approach of Refs.~\cite{Lepage:2001ym,Morningstar:2001je}, we constrain 
the parameters of the chiral-continuum fit with Bayesian priors and minimize the 
augmented $\chi^2_{\rm aug}$ defined in Eq~(\ref{eq:aug_chi2}).  
The chiral logarithms in Eqs.~(\ref{eq:chiral.fpara}) and~(\ref{eq:chiral.fperp}) 
depend upon the universal $B$-$B^*$-$\pi$ coupling $g_\pi$, which we constrain with 
a Gaussian prior of central value 0.45 and width 0.08.  This prior is 
consistent with a direct lattice calculation~\cite{Detmold:2011bp,Detmold:2012ge, Flynn:2015xna}, 
yet conservative enough to accommodate other lattice results~\cite{Flynn:2013kwa,Bernardoni:2014kla}. 
The chiral logarithms also depend on the mass splittings between mesons 
of different tastes and on the leading-order LEC $\mu$. These parameters depend only on the 
light-quark action, and we fix them to the values determined in the MILC light-pseudoscalar 
analysis~\cite{Bazavov:2009bb}; see Table~\ref{tab:QCD_para}.  In the $f_\parallel$ 
chiral-continuum extrapolation, we account for the uncertainty on the scalar $B^*$ mass by 
taking a generous prior width of three times the theoretical error reported in 
Ref.~\cite{Lang:2015hza}, or $\pm 69$~MeV.

We constrain the coefficients of the LO and NLO analytic terms $C^{(0)}$--$C^{(5)}$
using priors with central values zero and widths two.  To allow for higher-order 
contributions in the chiral expansion, we also include the complete  set of NNLO 
analytic terms.  These are proportional to  $\chi_l^2$, $\chi_l \chi_{a^2}$, 
$\chi_l \chi_{E}$, $\chi_l \chi_{E}^2$, $\chi_{a^2} \chi_{E}$,
$\chi_{a^2} \chi_{E}^2$, $\chi_E^3$, $\chi_E^4$, and $\chi_{a^2}^2$. 
We use prior central values of 0 with widths 1 for
the coefficients of the NNLO analytic terms.
The systematic error from truncating the chiral expansion 
will be discussed in Sec.\ref{sec:sys_errs}.

\begin{table}[tp]
    \centering
    \caption{Fixed parameters used in the chiral fit~\cite{Bailey:2014tva}.
        $\mu$ is the leading-order low-energy constant in QCD.
        $r_1^2 a^2 \Delta_\Xi$ and $r_1^2 a^2\delta_{V/A}$ are the 
      taste splittings and hairpin parameters for asqtad staggered 
      fermions.}
    \label{tab:QCD_para}
 \begin{tabular}{ccccccc}
  \hline  
  \hline  
  & ~\coarse~ & ~\fine~ & ~\superfine~ & ~\ultrafine~  & continuum\\
  \hline  
  \hline  
  ~~~~$r_1\mu$~~~~ & 6.831904 & 6.638563 & 6.486649 & 6.417427 & 6.015349 \\
  \hline  
  $r_1^2 a^2 \Delta_P (10^{-2})$ & 0 & 0 & 0 & 0 & 0 \\
  $r_1^2 a^2 \Delta_A (10^{-2})$ & 22.70460 & 7.469220 & 2.634800 & 1.040930 & 0 \\
  $r_1^2 a^2 \Delta_T (10^{-2})$ & 36.61620 & 12.37760 & 4.297780 & 1.697920 & 0 \\
  $r_1^2 a^2 \Delta_V (10^{-2})$ & 48.02591 & 15.93220 & 5.743780 & 2.269190 & 0 \\
  $r_1^2 a^2 \Delta_S (10^{-2})$ & 60.08212 & 22.06520 & 7.038790 & 2.780810 & 0 \\
  \hline  
  $r_1^2 a^2\delta_V$ & 0.00 & 0.00 & 0.00 & 0.00 & 0\\
  $r_1^2 a^2\delta_A$ & $-0.28$ & $-0.09$ & $-0.03$ & $-0.01$ & 0\\
  \hline  
  \hline
 \end{tabular}
\end{table}

Staggered $\chi$PT incorporates taste-breaking discretization effects from the 
light valence and sea quarks, but the lattice data also contain generic 
light-quark and gluon discretization effects as well as discretization effects 
from the heavy quark. 
We account for generic light-quark and gluon discretization errors by adding
the term $z \alpha_s (a\Lambda_\text{QCD})^2$ in the HMrS$\chi$PT 
formulae with coefficient prior central value zero and width one.
Similarly, to account for heavy-quark discretization effects in both the 
action and heavy-light currents, we add terms of order $a^2$ and 
$\alpha_s a$ with coefficients constrained by heavy-quark power counting~\cite{Oktay:2008ex}. 
At this order there are five functions ($f_B$, $f_Y$, $f_3$, $f_E$, $f_X$) 
that depend upon the bare heavy-quark mass; their explicit forms are given 
in Appendix A of Ref.~\cite{Bailey:2008wp}.
Dimensional analysis can be used to estimate the heavy-quark error
\begin{equation}
\texttt{error}_i \propto f_i(m_0a)(a\Lambda)^{{\rm dim}\mathcal{O}_i-4} ,
\end{equation}
where $f_i$ is related to the mismatch between coefficients of the continuum 
operators in the action and currents and their lattice counterparts, and 
$\Lambda$ is a typical QCD scale for heavy-light mesons
which we take to be $\Lambda=500$~MeV. 
As in Ref.~\cite{Bazavov:2011aa}, we add terms 
$z_i \times \texttt{error}_i$ to the HMrS$\chi$PT formulae for 
$f_{\parallel,\perp,T}$.
The priors on the $z_i$ have central values zero and widths 
equal to the square root of the number of times each function appears.
(See Appendix~A of Ref.~\cite{Bailey:2008wp}.)
Because the discretization errors are included via the constrained fit
in the chiral-continuum extrapolations, our results for the extrapolated form 
factors include the systematic uncertainties from light and heavy 
discretization effects.

\begin{figure}
\centering
\begin{minipage}[c]{0.49\textwidth}
\centering
    \includegraphics[width=0.9\textwidth]{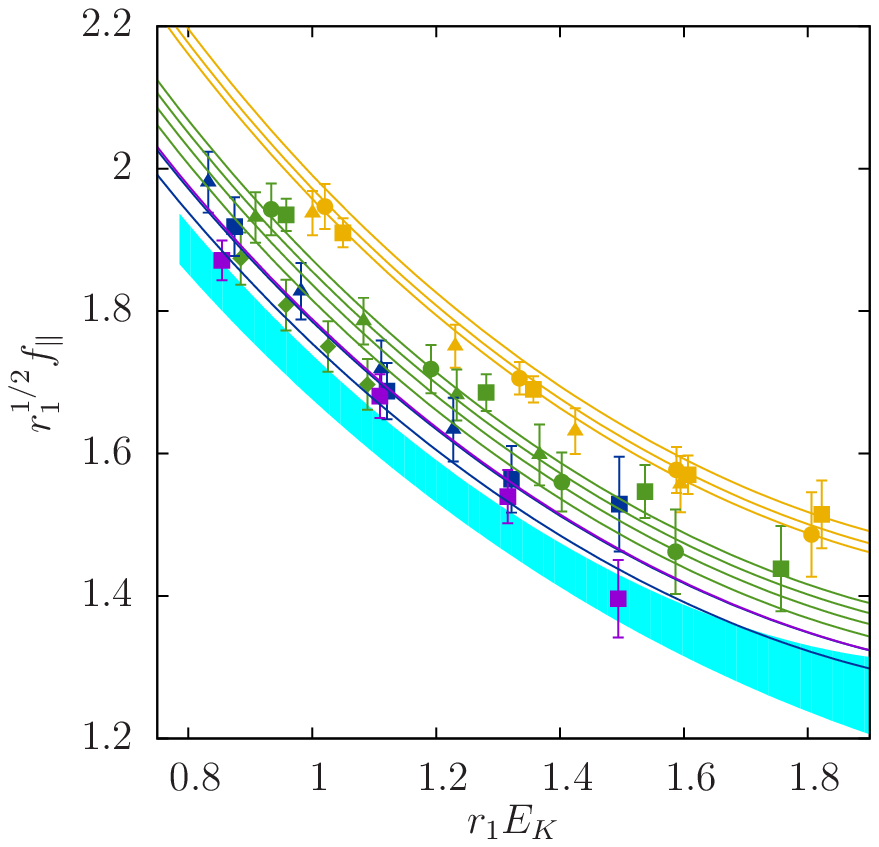}
\end{minipage}
\begin{minipage}[c]{0.49\textwidth}
\centering
    \includegraphics[width=0.85\textwidth]{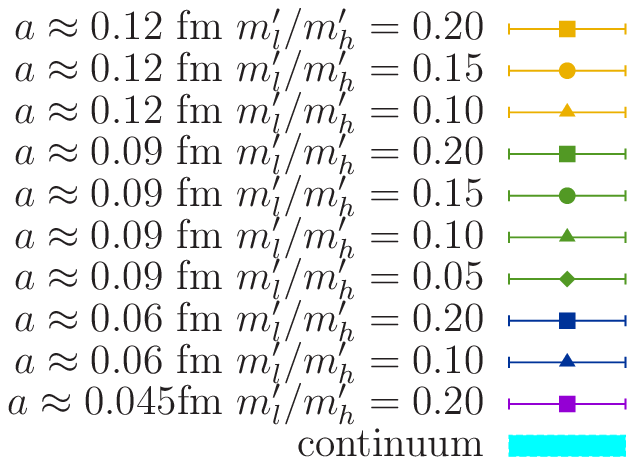}
\end{minipage}
\begin{minipage}[c]{0.49\textwidth}
\centering
    \includegraphics[width=0.9\textwidth]{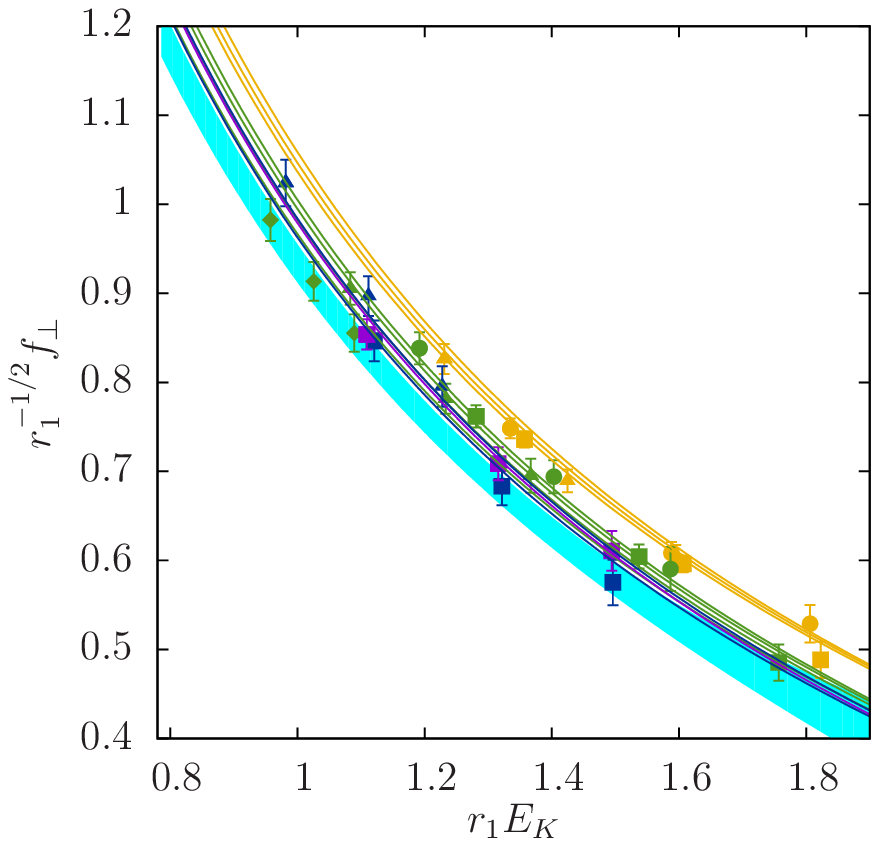}
\end{minipage}
\begin{minipage}[c]{0.49\textwidth}
\centering
    \includegraphics[width=0.9\textwidth]{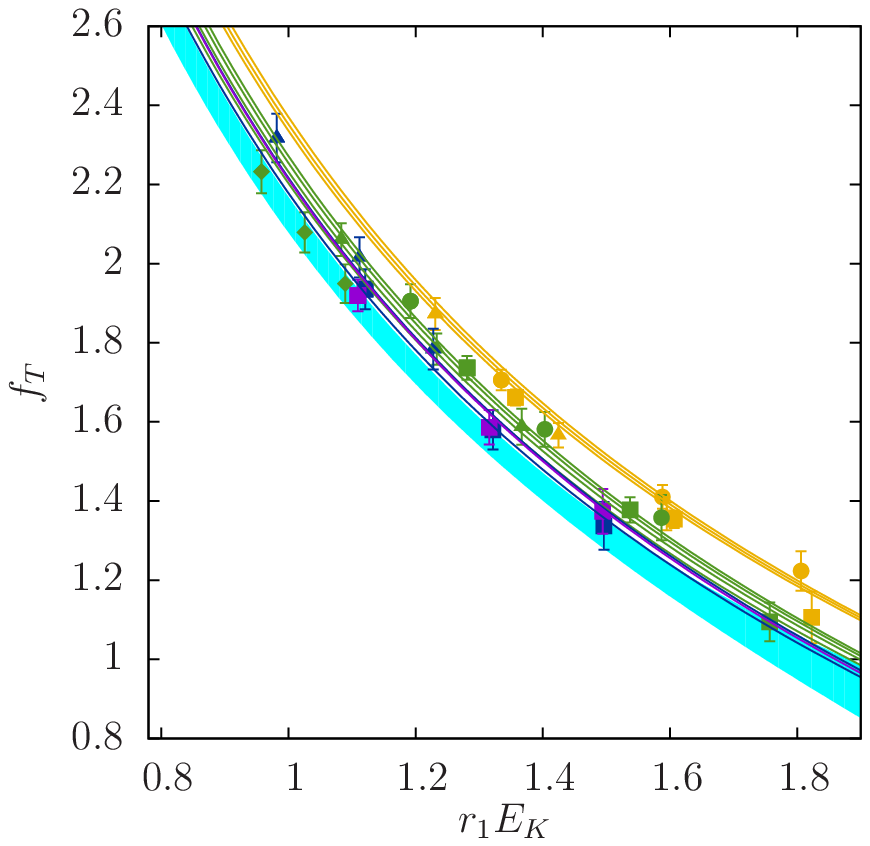}
\end{minipage}
    \caption{Chiral-continuum extrapolations of  $f_\parallel$ (upper left), $f_\perp$ (lower 
      left), and $f_T$ (lower right) using NLO SU(2) HMrS$\chi$PT plus NNLO analytical 
      terms. The squares, circles, triangles, and diamonds denote the 
        $m_l^\prime/m_h^\prime = 0.2$, 0.14, 0.1, and 0.05 data, respectively. 
        The colored fit lines correspond to the different lattice spacings as indicated 
        in the legend. The cyan band shows the continuum extrapolated curve with 
        statistical error, which includes the systematic uncertainties due to $g_\pi$, 
        and the heavy-quark, light-quark, and gluon discretization errors.
        Fit lines should pass through the data points of the corresponding color.}
    \label{fig:combined_su2_fpara_fperp_ft_chpt_fit}
\end{figure}

In summary, we use expressions derived in SU(2) HMrS$\chi$PT for the central 
chiral-continuum extrapolations of the form factors $f_\parallel$, $f_\perp$, 
and $f_T$; these are shown in Fig.~\ref{fig:combined_su2_fpara_fperp_ft_chpt_fit}.
The SU(2) theory describes our data well: the $p$~values of the fits are 0.91, 0.94, 
and 0.98 for $f_\parallel$, $f_\perp$, and $f_T$, respectively.
Our fit results for $g_\pi$ are 0.47(5), 0.46(4), and 0.47(3), respectively.
At this stage, only the statistical, $g_\pi$, chiral truncation, and discretization 
errors have been included. In the next section, we estimate the size of 
the remaining uncertainties before employing the $z$ expansion in 
Sec.~\ref{sec:zexp} to extend our results over the full kinematic range.

\section{Form-Factor Error Budget}
\label{sec:sys_errs}
In this section, we estimate the systematic errors in the 
form factors, discussing each source of uncertainty in a separate subsection.
We first discuss the error from the chiral-continuum extrapolation, which 
also includes heavy-quark, light-quark, and gluon discretization errors.  
We then discuss the remaining systematic uncertainties from the heavy-light 
current renormalization, lattice-scale determination, light- and 
strange-quark mass determinations, finite-volume effects, and
$b$-quark mass determination, discussing each in a separate subsection.
As discussed previously, the systematic errors from $g_\pi$ and 
heavy- and light-quark discretization effects are included in
the statistical errors of the chiral-continuum extrapolation result through the 
constrained fit. Finally, we visually summarize the error budgets for 
the three form factors as a function of $q^2$ in Fig.~\ref{fig:fp_f0_ft_errors}.

\subsection{Chiral-continuum extrapolation}

We use NLO SU(2) HMrS$\chi$PT supplemented by all possible NNLO analytic terms, 
as well as heavy-quark, light-quark, and gluon discretization terms,
in our preferred chiral extrapolations of $f_\perp$, $f_\parallel$, and $f_T$.

\begin{figure}[tp]
\centering
\begin{minipage}[c]{0.49\textwidth}
\centering
    \includegraphics[width=0.9\textwidth]{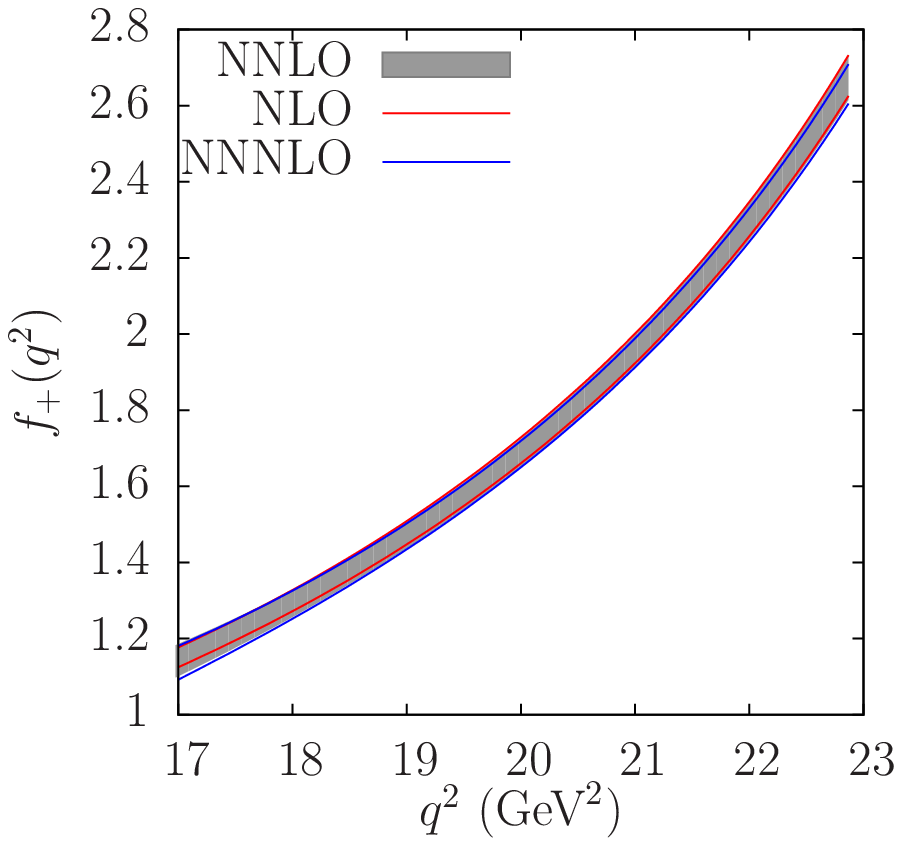}
\end{minipage}
\begin{minipage}[c]{0.49\textwidth}
\centering
    \includegraphics[width=0.9\textwidth]{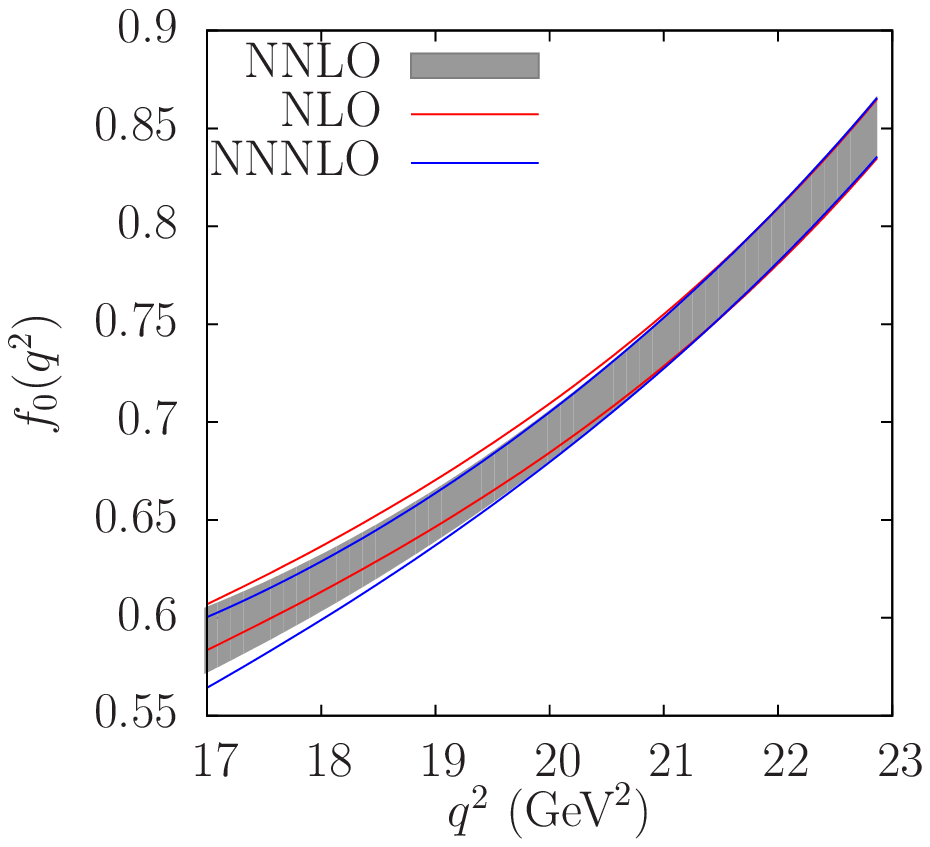}
\end{minipage}
\begin{minipage}[c]{0.49\textwidth}
\centering
    \includegraphics[width=0.9\textwidth]{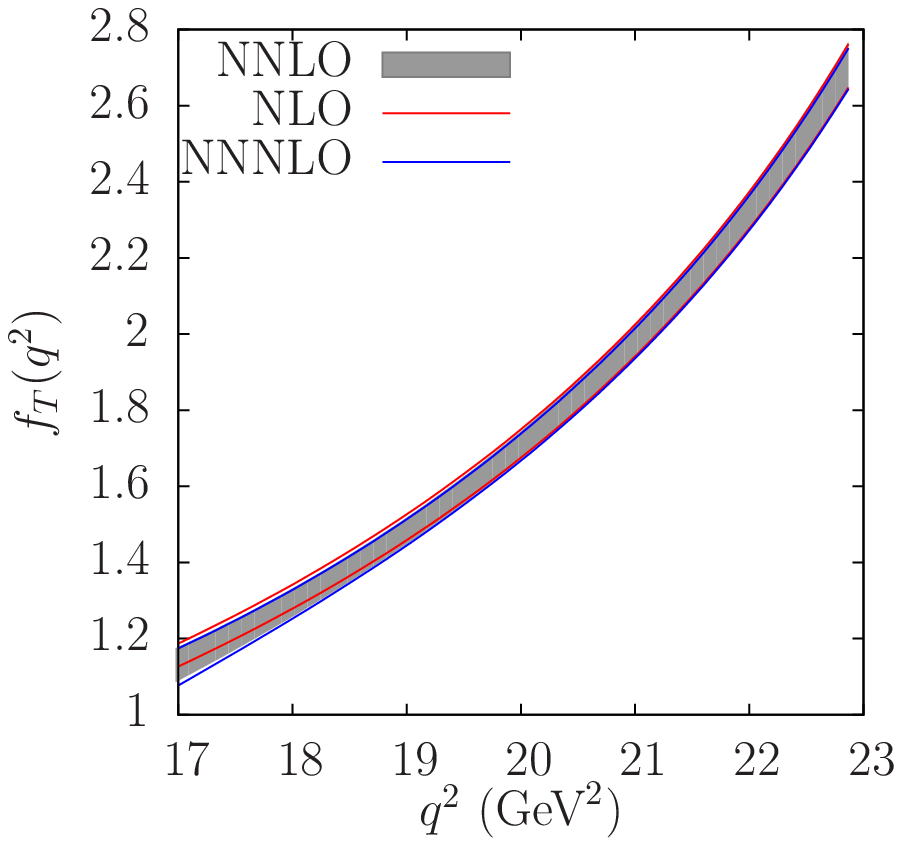}
\end{minipage}
    \caption{Chiral-continuum extrapolations with NLO, NNLO, or NNNLO analytic
      terms for $f_+$ (upper left), $f_0$ (upper right), and 
      $f_T$ (lower panel). In each plot, the grey band shows the statistical 
      error from the preferred NNLO SU(2) $\chi$PT. The red and blue
      lines show the error from the fits with NLO and NNNLO analytic 
      terms, respectively.}
    \label{fig:chipt_sys_err_truncation}
\end{figure}

First, to estimate truncation effects, we compare fit results using NLO 
HMrS$\chi$PT, our preferred fit function with NNLO analytic terms, and the 
same fit function with the addition of the complete set of NNNLO analytic 
terms in Fig.~\ref{fig:chipt_sys_err_truncation}. We see that the errors 
in the preferred fit with NNLO analytic terms are already saturated, since 
they are the same as the errors in the fit with NNNLO analytic terms. 
Hence,  truncation effects are included in the statistical fit errors 
from our preferred fit. 

\begin{figure}[tp]
\centering
\begin{minipage}[c]{0.49\textwidth}
\centering
    \includegraphics[width=0.9\textwidth]{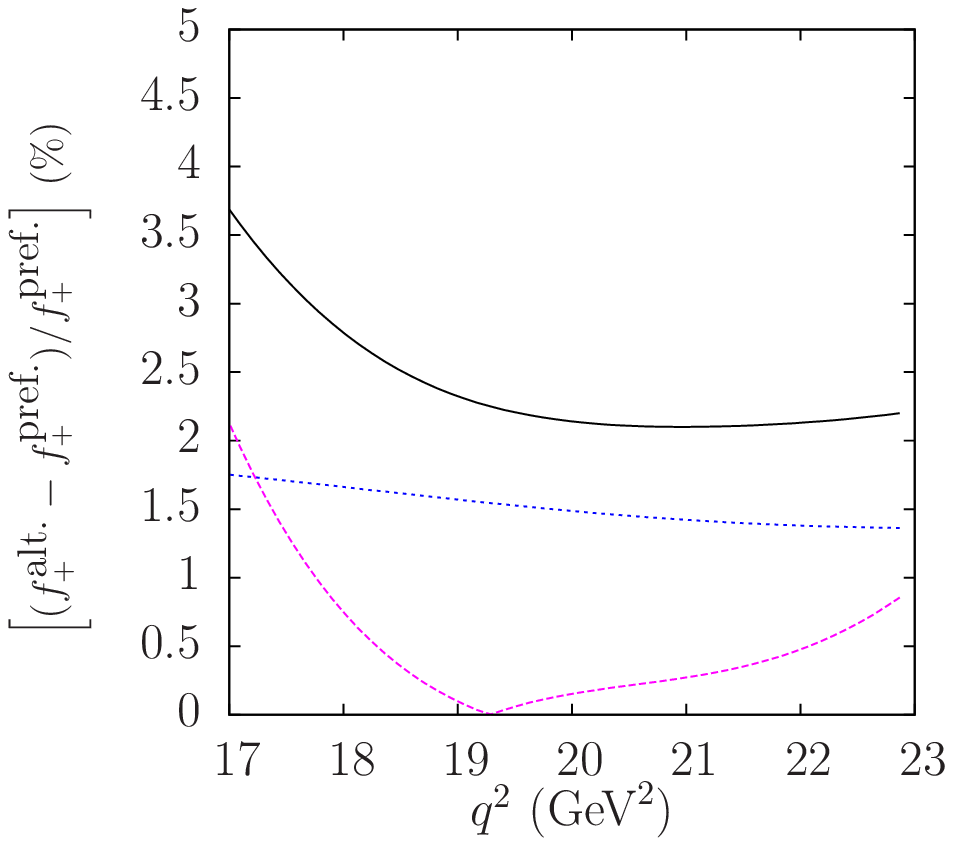}
\end{minipage}
\begin{minipage}[c]{0.49\textwidth}
\centering
    \includegraphics[width=0.5\textwidth]{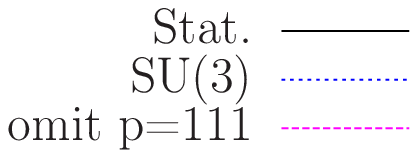}
\end{minipage}
\begin{minipage}[c]{0.49\textwidth}
\centering
    \includegraphics[width=0.9\textwidth]{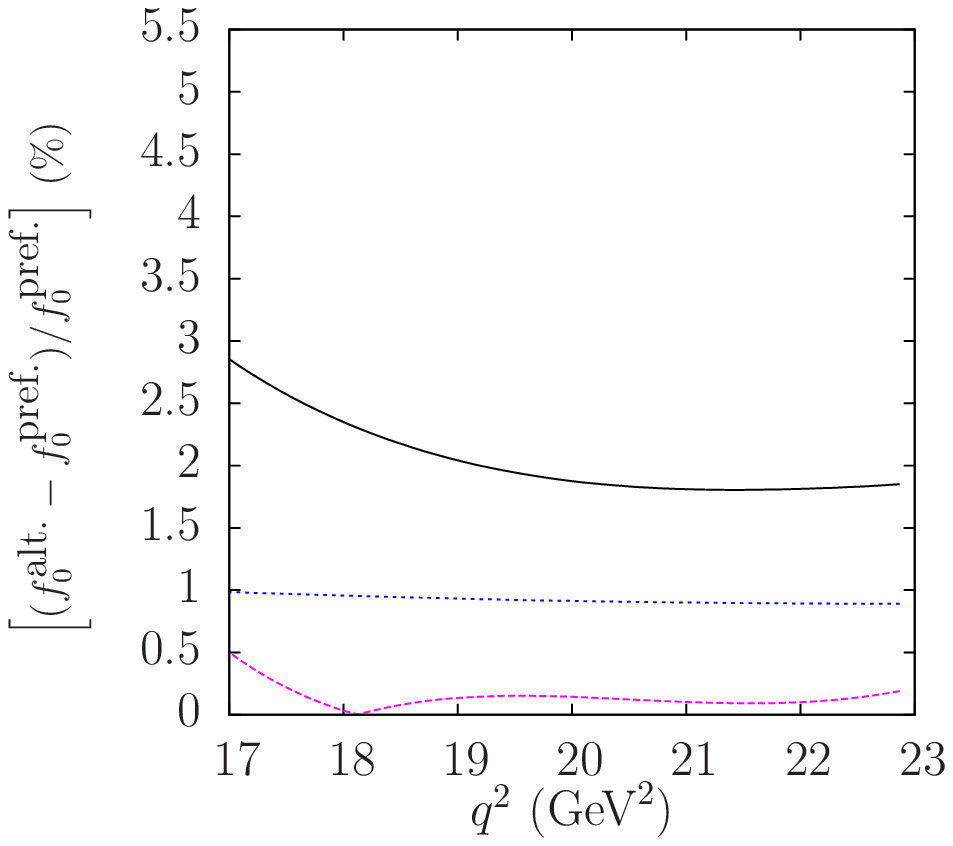}
\end{minipage}
\begin{minipage}[c]{0.49\textwidth}
\centering
    \includegraphics[width=0.9\textwidth]{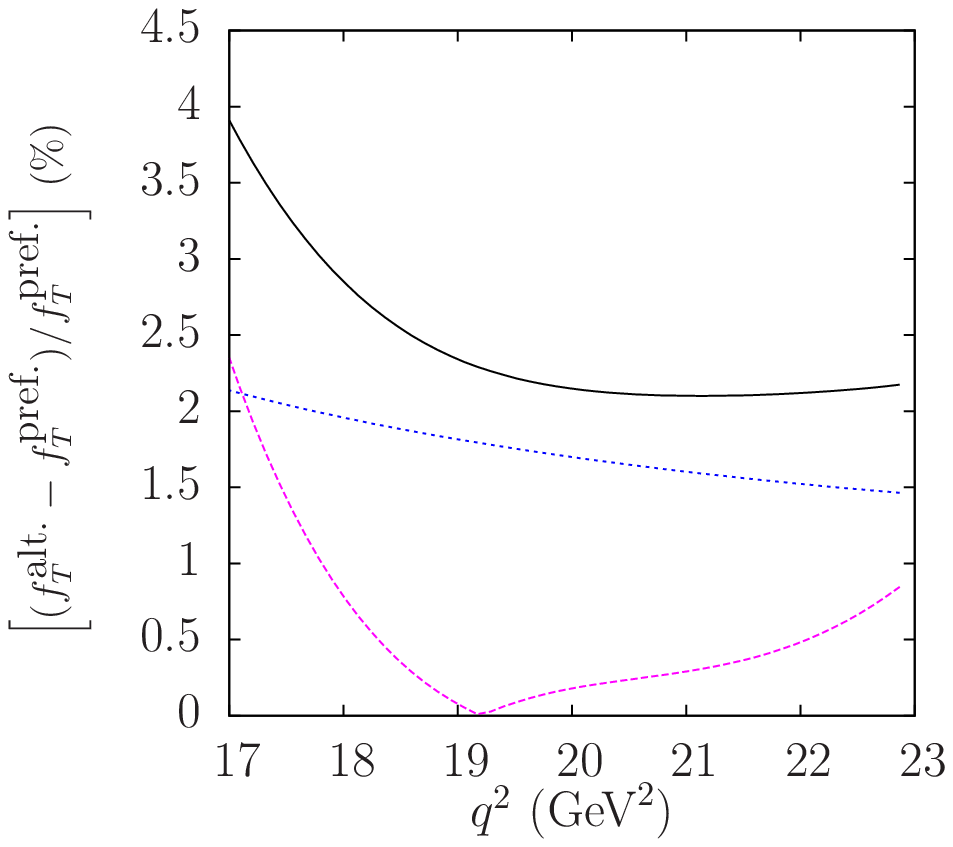}
\end{minipage}
    \caption{Deviations of alternate chiral-continuum extrapolations from the 
      central results for $f_+$ (upper left), $f_0$ (lower left), and 
      $f_T$ (lower right). In each plot, the black curve shows the statistical 
      error from the preferred NNLO SU(2) $\chi$PT. The blue and pink 
      lines show the \% difference from the central fit obtained by
      using SU(3) hard-kaon $\chi$PT and omitting $\bm{k}=2\pi(1,1,1)/L$ data,
      respectively.}
    \label{fig:chipt_sys_err}
\end{figure}

In addition, we also consider two alternative fit Ans{\"a}tze for
the chiral-continuum extrapolation.
First, we consider NLO SU(3) hard-pion 
HMrS$\chi$PT, which provides a good description of our data,
although the standard NLO SU(3) expressions do not.
We use the result from the SU(2) HMrS$\chi$PT fit as our preferred fit,
because the SU(2) theory converges faster than the SU(3) theory as studied in 
Ref.~\cite{Becirevic:2007dg}.  We compare the fit results from NLO hard-kaon SU(3) 
HMrS$\chi$PT plus NNLO analytical terms and our preferred fit, 
and find differences between the central values of about $1$-$2$\% 
for all form factors and $q^2$. 
Second, we consider the effect of the $E_K$ range of the lattice-QCD data to
the extrapolated continuum result by omitting the $\bm{k}=2\pi(1,1,1)/L$ data
from our fit. We find the differences are below $1$-$2$\%. 
Figure~\ref{fig:chipt_sys_err} summarizes the differences between the form 
factors obtained from the alternative chiral-continuum fits and the central 
results. Overall, the shifts of the continuum form-factor central values 
are within the quoted statistical errors of the preferred 
chiral fit that includes truncation effects.

\subsection{Heavy-light current renormalization}

To obtain the continuum form factors, we multiply the lattice form factors  
by the renormalization constant given in Eq.~(\ref{eq:rhodef}), 
using the values of $\rho_J$, $\ZVbb$, and $\ZVll$ listed in Tables~\ref{tab:Z}--\ref{tab:rho}. The statistical error on $\ZVll^{1/2}$ is about 0.2\%.
By using the jackknife blocks of \ZVbb\ calculated on the same ensembles, 
we incorporate the statistical error from $\ZVbb^{1/2}$ automatically in our fit results.

The $\rho_J$ are calculated at one-loop order in perturbation theory. They are close to unity 
by design, since they are defined as ratios of renormalization factors. Indeed their one-loop 
corrections are small, as shown in Table~\ref{tab:rho}. We estimate the error due to 
truncating the perturbative expansion as $2 \rho_{J, {\rm max}}^{[1]} \alpha_s^2$ 
in order to avoid sensitivity due to accidental cancellations. We obtain 
$\rho_{J, {\rm max}}^{[1]}$ as follows. For the scale-independent vector 
currents ($V^i$ and $V^4$), we simply look for the largest value of the
one-loop coefficients for both currents on all of the ensembles. We find that the 
spatial vector current has a larger one-loop coefficient with  $\rho_{V, {\rm max}}^{[1]} = 0.1$. 
We evaluate $\alpha_s$ at the $a \approx 0.06$~fm lattice spacing (the next to finest), 
which yields an error of 1\% for both components of the vector current.
For the scale-dependent tensor current the perturbative corrections include 
logarithmic contributions due to their anomalous dimension, which are responsible 
for the growth of $\rho_T$ towards smaller lattice spacings seen in Table~\ref{tab:rho} . 
In order to estimate the truncation error, we remove the effect of the anomalous 
dimension by setting $\mu=2/a$. We find that $\rho^{[1]}_{T,\text{max}}=0.2$, which 
corresponds to a truncation error of $2\%$ on $\rho_T$. In summary, we assign 
a perturbative truncation error of 1\% on $f_+,0$ and an error of 2\% on $f_T$.

\subsection{Scale uncertainty}
We use $r_1=0.3117(22)$~fm in the continuum from Ref.~\cite{Bazavov:2011aa} to  
convert lattice quantities
to physical units, where the quoted error includes both statistics and systematics.
We repeat our analysis varying $r_1$ by plus and minus one standard deviation from 
its central value and use the larger change of each form factor as an estimate 
of the systematic error due to the scale uncertainty. We find differences of less 
than 1\% for $f_\parallel$, $f_\perp$, and $f_T$ throughout the simulated $q^2$
region.

\subsection{Light- and strange-quark mass uncertainties}
After the chiral-continuum fit, we evaluate the form factors at the physical quark masses 
$r_1\hat{m}=0.000965(33)$ and $r_1m_s=0.0265(8)$ determined from the analysis of the
light pseudoscalar meson spectrum~\cite{Bazavov:2009bb, Bailey:2014tva}.
We vary the quark masses by plus and minus one standard deviation and find 
the differences in all three form factors due to changing $m_l$ and $m_s$ to 
be below 0.6\% in the simulated $q^2$ region.

\subsection{Finite-volume effects}
The lattices used in this work have finite spatial volumes with $M_\pi L \gtrsim 4$.
We estimate the size of finite-volume effects using HMrS$\chi$PT.
In chiral perturbation theory, finite volume contributions change 
loop-momentum integrals to sums which have been calculated in Refs.~\cite{Aubin:2007mc,Arndt:2004bg}.
We employ continuum integrals in the preferred chiral-continuum extrapolations.
To estimate the size of finite-volume effects, we evaluate the form factors with the 
LECs we obtain from the preferred chiral fits, and compare the results from the 
infinite-volume formulae and the finite-volume formulae on all ensembles used in this work.
We try both SU(2) HMrS$\chi$PT and SU(3) hard-kaon HMrS$\chi$PT.
We find that in all cases finite-volume effects are below 0.001\%.
Therefore, we neglect finite-volume effects in the total error budget.

\subsection{\texorpdfstring{\boldmath $b$-quark mass correction}{b-quark mass correction}}
We correct the form factors from the simulated $\kappa_b'$ to the physical $\kappa_b$
before we perform the chiral-continuum extrapolation.
Including these corrections accounts for the dominant effect from $b$-quark mistuning, 
but small errors in the form factors remain due to 
the uncertainties in the $\kappa_b$-correction factors.
The statistical errors in the slopes  $\frac{\partial{\ln{f}}}{\partial{\ln{m_2}}}$ are at
most about 10\% for $f_{\perp, T}$ at $2\pi(1,1,1)/L$, while the 
sizes of the $\kappa_b$ shifts applied to the data points are about 1\%--2\%.
We therefore take  the systematic error from the $\kappa_b$ correction to be 
2\% $\times$ 10\% = 0.2\%, which is conservative enough to accommodate the largest 
possible error in the shift.

\subsection{Summary of the systematic error budget}
\begin{figure}
\centering
\begin{minipage}[c]{0.49\textwidth}
\centering
    \includegraphics[width=0.9\textwidth]{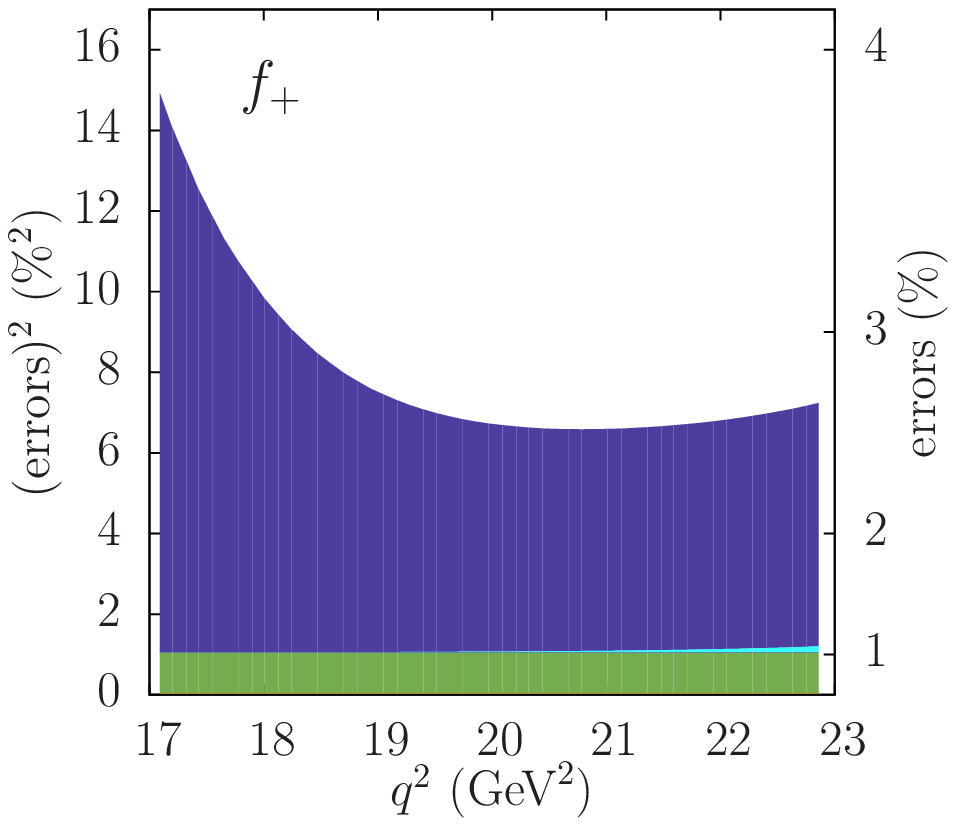}
\end{minipage}
\vspace*{1em}
\begin{minipage}[c]{0.49\textwidth}
\centering
    \includegraphics[width=0.8\textwidth]{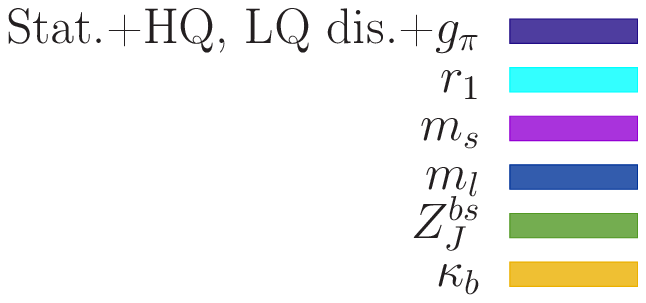}
\end{minipage}
\begin{minipage}[c]{0.49\textwidth}
\centering
    \includegraphics[width=0.9\textwidth]{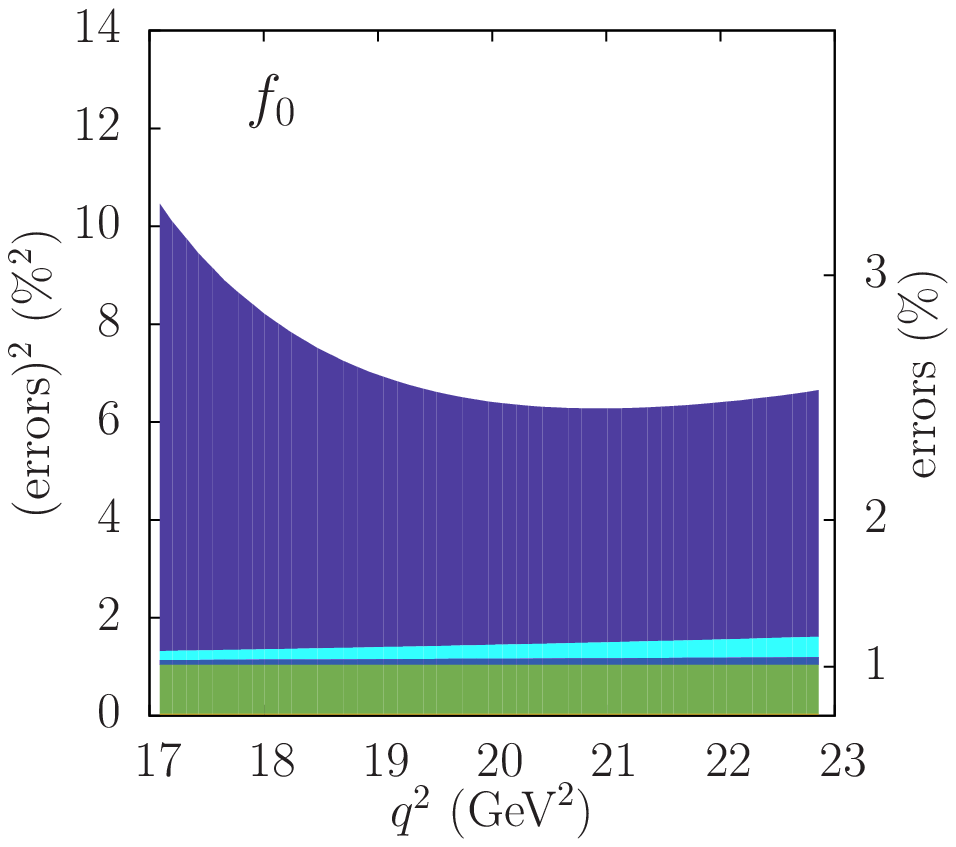}
\end{minipage}
\begin{minipage}[c]{0.49\textwidth}
\centering
    \includegraphics[width=0.9\textwidth]{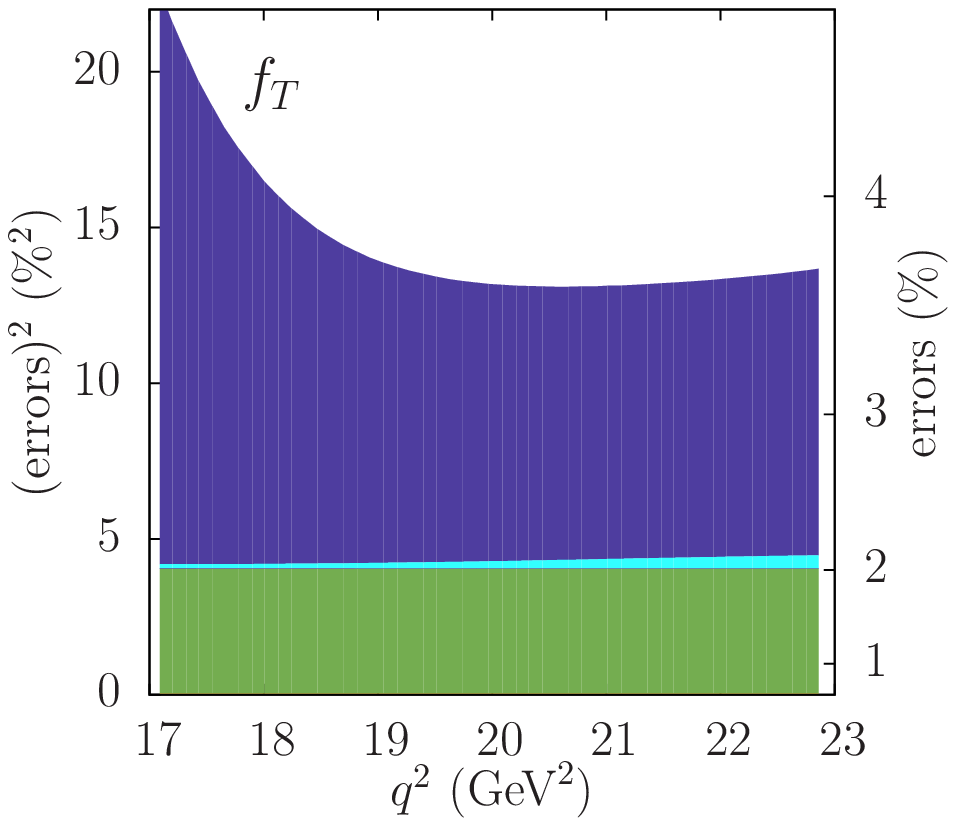}
\end{minipage}
    \caption{Statistical and systematic error contributions to $f_+$ (upper left), 
      $f_0$ (lower left),  and $f_T$ (lower right). The left vertical axis label shows the 
      squares of the errors added in quadrature, while the right vertical axis 
      label shows the errors themselves. The filled, stacked curves from bottom to top 
      show the total error when we add each individual source of error in quadrature
      one by one.}
    \label{fig:fp_f0_ft_errors}
\end{figure}

Figure~\ref{fig:fp_f0_ft_errors} visually summarizes the results for the statistical 
and systematic errors. For all three form factors, the combined chiral-continuum extrapolation 
error is the largest source of systematic uncertainty. The total errors in the form 
factors $f_+$, $f_0$, and $f_T$ are below 5\% for all $q^2>17~\text{GeV}^2$,
and are $\sim 3\%$ near $q^2_\text{max}$.
We quote numerical results for the form factors including all systematic errors 
over the entire $q^2$ range in the following section, after the $q^2$-extrapolation 
to the full kinematic range using the $z$~expansion.

\section{\texorpdfstring{\boldmath $z$~expansion of form factors}{z expansion of form factors}}
\label{sec:zexp}

The form factors obtained from the chiral-continuum fit are reliable for
high momentum transfer, $q^2 \gtrsim 17$~GeV$^2$.
We only simulated kaons with momenta up to $2\pi(1,1,1)/L$, because, at
higher momenta, the two- and three-point correlators become noisier and are
subject to larger discretization errors. Further, the HMrS$\chi$PT 
formalism used to take the continuum limit does not apply when $E_K$ is too 
large. In particular, for $E_K \gtrsim 1.2~\textrm{GeV}$ the 
expansion parameter $\chi_E \gtrsim 1$, so the terms analytic 
in $\chi_E^n$ increase with higher powers of $n$.
Because of these limitations, a way to extend the form factors to 
high kaon energy, or, equivalently, $q^2=0$, is needed.
In this paper, we follow Ref.~\cite{Boyd:1994tt}  and map $q^2$ to 
a new variable $z$ such that $|z|\leq1$.
Constraints from unitarity, analyticity, and heavy-quark physics ensure that 
the expansion of the form factors in terms of $z$ converges.  Thus we 
can use the $z$ expansion to obtain a model-independent parameterization 
of our form factors valid over the entire kinematic range.
This technique is now standard for analyzing $B\to\pi l\nu$ 
decays~\cite{Amhis:2012bh,Aoki:2013ldr,Agashe:2014kda}.

We first define the new variable $z$ via the conformal mapping~\cite{Boyd:1994tt}
\begin{align}
z(q^2,t_0)&
=\frac{\sqrt{t_+-q^2}-\sqrt{t_+-t_0}}{\sqrt{t_+-q^2}+\sqrt{t_+-t_0}} ,
\end{align}
where $t_\pm=(M_B\pm M_K)^2$ and $t_0$ is a free parameter that can be chosen to minimize 
$|z|$ for the semileptonic-decay region. In this work, we 
use $t_0=(M_B+M_K)(\sqrt{M_B}-\sqrt{M_K})^2$~\cite{Bourrely:2008za}, which maps the physical 
semileptonic decay region $0 \leq q^2 \leq 22.8$~GeV$^2$
to $|z|<0.15$.  The small range of $|z|$ helps control
the truncation error in the $z$~expansion.

Using the new variable $z$, we expand the form factors 
as~\cite{Bourrely:2008za}
\begin{align}
    f_+(q^2) &= \frac{1}{P_+(q^2)}\sum_{m=0}^{K-1} b^+_m \left[ 
        z^m -(-1)^{m-K}\frac{m}{K}z^K \right], 
    \label{eq:f+zexp} \\
    f_0(q^2) &= \frac{1}{P_0(q^2)}\sum_{m=0}^{K-1} b^0_m  z^m, 
    \label{eq:f0zexp} \\
    f_T(q^2) &= \frac{1}{P_T(q^2)}\sum_{m=0}^{K-1} b^T_m \left[ 
        z^m -(-1)^{m-K}\frac{m}{K}z^K \right], 
    \label{eq:fTzexp}
\end{align}
The function $P_{+,0,T}(q^2) = 1-q^2/M^2$ accounts for poles below and near the 
$B$-$K$ production threshold.  For the $z$ fits of $f_+$ and $f_T$, we fix the 
location of the vector $B_s^*$ pole to the measured value $M_{B_s^*} = 5.4154$~GeV~\cite{Agashe:2014kda}.
For the $f_0$ fit, we fix the location of the scalar $B_{s0}^*$ pole to the 
lattice-QCD prediction $M_{B_{s0}} = 5.711$~GeV from Ref.~\cite{Lang:2015hza}. 
We find that varying its location by three times the quoted theoretical 
error ($\pm 69$~MeV) does not change the extrapolated form factor.

The expression for $f_+$ in Eq.~(\ref{eq:f+zexp}) was derived by Bourrely, Caprini and Lellouch
in Ref.~\cite{Bourrely:2008za}, and is commonly called the BCL parameterization.
In the BCL expression for $f_+$ in Eq.~(\ref{eq:f+zexp}), the coefficient of the 
term proportional to $z^K$ is related to that of the lower-order terms.
This constraint is due to the conservation of momentum and the analyticity of the form
factors~\cite{Bourrely:2008za}. There is no analogous constraint for $f_0$.
We use the same expression for $f_T$ as for $f_+$ because they are proportional 
to each other at leading order in the heavy-quark expansion.  
These expressions were also used to analyze the lattice form factors for $B \to K l^+ l^-$ in
Refs.~\cite{Bouchard:2013eph,Bouchard:2013mia}.

Unitarity constrains the coefficients of the $z$~expansion such that
\begin{equation}
    \sum_{m,n=0}^{\infty}B_{mn}b_m b_n \lesssim 1 ,
    \label{eq:unitary.cons}
\end{equation}
where the values of $B_{mn}$  are calculated using the Taylor expansion of the function 
discussed in Ref.~\cite{Bourrely:2008za} and given in Table~\ref{tab:zexp.para}. 
We employ the same coefficients $B_{mn}$ for $f_T$ and $f_+$.
  The outer function $\phi$ defined in Ref.~\cite{Becher:2005bg}
  is used in the derivation of the $B_{mn}$.
Although the $\phi$ of $f_0$ in Ref.~\cite{Becher:2005bg} was derived
without a scalar pole, its form is not altered by the presence of the 
pole, because $|z|$ always equals $1$ on the unit circle.
In Ref.~\cite{Becher:2005bg}, Becher and Hill showed that, in the limit of large 
$b$-quark mass, the sizes of the $z$ coefficients for $f_+$ are even smaller than 
the expectation from~(\ref{eq:unitary.cons}).
Heavy-quark effective theory provides an estimate of the sum~\cite{Becher:2005bg}:
\begin{equation}
    \sum B_{mn}b_m b_n =\frac{1}{\pi}\int^{\infty}_{t_+} \frac{dt}{t-t_0}{\rm Im}
    \left(\sqrt{\frac{t_+-t_0}{t_+-t}} \right)|\phi_i(t)f_i(t)|^2 , 
    \label{eq:HQ.estimate}
\end{equation}
where $i=+$, $0$, or $T$, and the $\phi$ is an outer function.
To calculate the integral in Eq.~(\ref{eq:HQ.estimate}),
we need to know the form factors in the range $[t_+, \infty]$.
For $f_+$, we assume that $f_\perp$ gives the dominant contribution and has 
only the single $B_s^*$ pole.
Taking the limit $M_B \to \infty$ gives the following simple form for $f_+(q^2)$:
\begin{equation}
    f_+(q^2) \approx \frac{M_B}{\sqrt{2M_B}}f_\perp(E_K) \approx
        \frac{M_B}{\sqrt{2M_B}}\frac{C^{(0)}_\perp g_\pi}{f_\pi(E_K+\Delta_{B_s^*})} .
    \label{eq:HQ.integral}
\end{equation}
We then use our determination of $C_\perp^{(0)}$ from our preferred 
chiral-continuum fit to obtain the estimate 
\begin{align}
\sum B_{mn} b_m b_n & \approx 0.012 \ .
\end{align}
This result means that Eq.~(\ref{eq:unitary.cons}) is only a loose bound for $f_+$. 
In addition, it is consistent with a power-counting estimate~\cite{Becher:2005bg}, 
which anticipates $\sum B_{mn}b_mb_n$ to be of order $(\Lambda/m_b)^3$.
The analogous calculation for $f_T$ gives a similar result.
The analysis below will show that the heavy-quark (HQ) constraint on $f_+$ 
(and $f_T$), Eq.~(\ref{eq:HQ.integral}), together with the kinematic constraint, 
$f_0(0)=f_+(0)$, suffices to keep the $z$~fit under control.

We assume a log-normal distribution on $\sum B_{mn}b_m b_n$ to ensure that $\sum B_{mn}b_m b_n$ is always
positive. The contribution from this prior to the augmented $\chi^2$ is:
\begin{equation}
    \chi^2_{B_{mn}b_mb_n} = \frac{\left [ \ln(\sum B_{mn}b_mb_n)-\mu\right ]^2}{\sigma^2} ,
    \label{eq:log_normal_Bmnbmbn}
\end{equation}
where $\mu$ is the central value and $\sigma$ is the width of the prior.
For $f_+$ and $f_T$, we choose $\mu$ and $\sigma$ in Eq.~(\ref{eq:log_normal_Bmnbmbn}) 
as $\ln (0.02)$ and $\ln(\frac{0.07}{0.02})$.
This choice is conservative enough to accommodate the uncertainties in the estimates.

\begin{table}[tp]
 \centering
\caption{Lowest-order coefficients $B_{mn}$ for $B\to Kl^+l^-$ decay using $M_B=5.27958$~GeV, 
$M_K=0.497614$~GeV, and $t_0=(M_B+M_K)(\sqrt{M_B}-\sqrt{M_K})^2$.
The outer function used in the calculation is from Ref.~\cite{Arnesen:2005ez}
  with $\chi_{f_+} = 5.025 \times 10^{-4}$ and $\chi_{f_0} = 1.4575 \times 10^{-2}$.
  Although these $\chi_i$s are derived for the $B \to \pi l \nu$ process, the calculation
  in Ref.~\cite{Bharucha:2010im} shows the difference between $\chi_i$s of the
  $B \to K l^+l^-$ and $B \to \pi l \nu$ process is less than 10\%. Therefore,
  we quote the inputs from  Ref.~\cite{Arnesen:2005ez} to obtain these $B_{mn}$.
All $B_{mn}$ not listed here can be obtained from the relations
$B_{m(m+n)}=B_{0n}$ and $B_{mn}=B_{nm}$.}
 \label{tab:zexp.para}
 \begin{tabular}{ccccccc}
  \hline\hline
  & ~~~$B_{00}$~~~ & ~~~~$B_{01}$~~~ & ~~~$B_{02}$~~~ & ~~~~$B_{03}$~~~ & ~~~~$B_{04}$~~~ & ~~~$B_{05}$~~~ \\
  \hline  
  $f_{+,T}$ &  0.0161 & $-0.0003$  & $-0.0104$ & \mph0.0002 & \mph0.0022 & 0.0002 \\
  $f_{0}$   &  0.0921 & \mph0.0132 & $-0.0483$ & $-0.0168$  & $-0.0001$  & 0.0024  \\
  \hline\hline
 \end{tabular}
\end{table}

We first generate from the continuum, physical quark-mass limit of the chiral 
extrapolation a few synthetic data points in the energy range of the simulated
lattice data ($q^2 \gtrsim 16.8$~GeV$^2$). With the lattice spacing set 
to zero and  the quark masses fixed to their physical values in 
Eqs.~(\ref{eq:chiral.fpara})--(\ref{eq:chiral.fperp}), the physical form 
factors depend upon at most six independent functions of the kaon energy $E_K$.
These are proportional to $1/(E_K+\Delta_{B_s^*})$, $E_K^0$, $E_K$, $E_K^2$, 
$E_K^3$, and $E_K^4$. To the degree that the coefficients in front of these 
functions are correlated, the number of independent modes may be even fewer than six.
If we generate too many synthetic data points, the covariance matrix will be singular.
We therefore generate four synthetic data points each for $f_+$, $f_0$, and $f_T$ 
at $q^2 = (22.86, 21.13, 19.17, 17.09)~\text{GeV}^2$.
These cover the simulated lattice-momentum range and are approximately evenly 
spaced in $q^2$. We also fit with synthetic data from a smaller and larger 
range and find consistent results.

The full covariance matrix of the synthetic data points includes both the 
statistical and systematic error:
\begin{equation}
C_{mn}^{\rm full} = C_{mn}^{\rm stat} + C_{mn}^{\rm syst} \ ,
\end{equation}
where $m,n$ denote the four $q^2$ values.  The systematic error contribution is calculated
as
\begin{equation}
C_{mn}^{\rm syst} = \sum_i \sigma_m^i \sigma_n^i \,
\end{equation}
where the index $i$ runs over the sources of  systematic error discussed 
in Sec.~\ref{sec:sys_errs}. Because we assume that the systematic errors 
are 100\% correlated between $q^2$ values, all nontrivial correlations 
between points are due to statistical fluctuations of the chiral-continuum 
fit results.

\begin{figure}[tp]
\centering
    \includegraphics[scale=0.8]{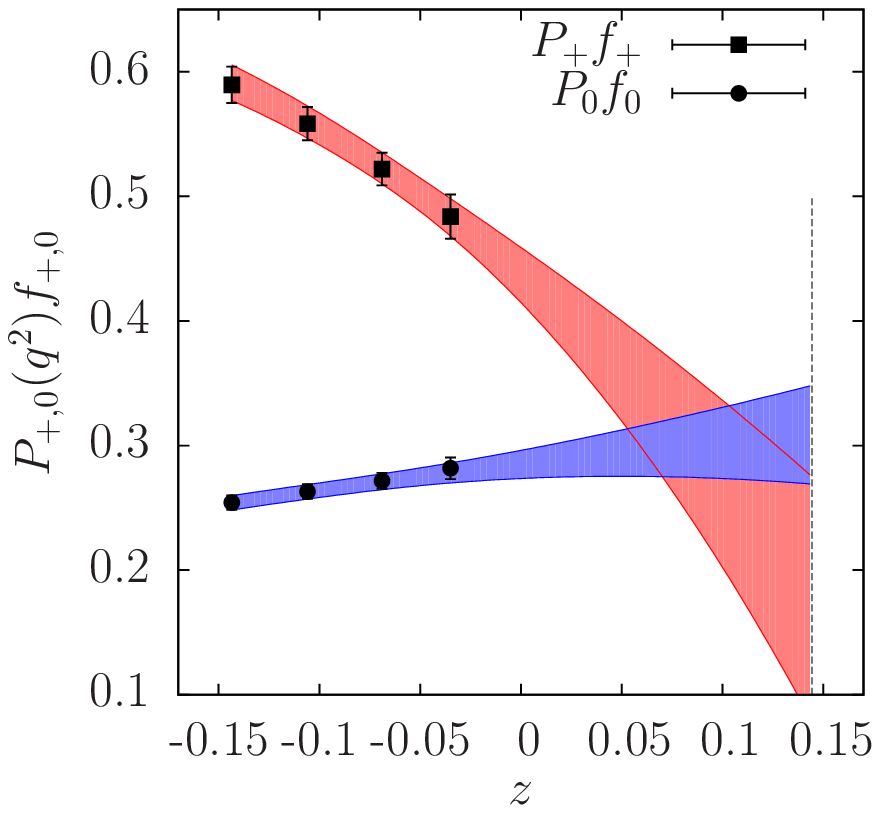}~~~~
    \includegraphics[scale=0.8]{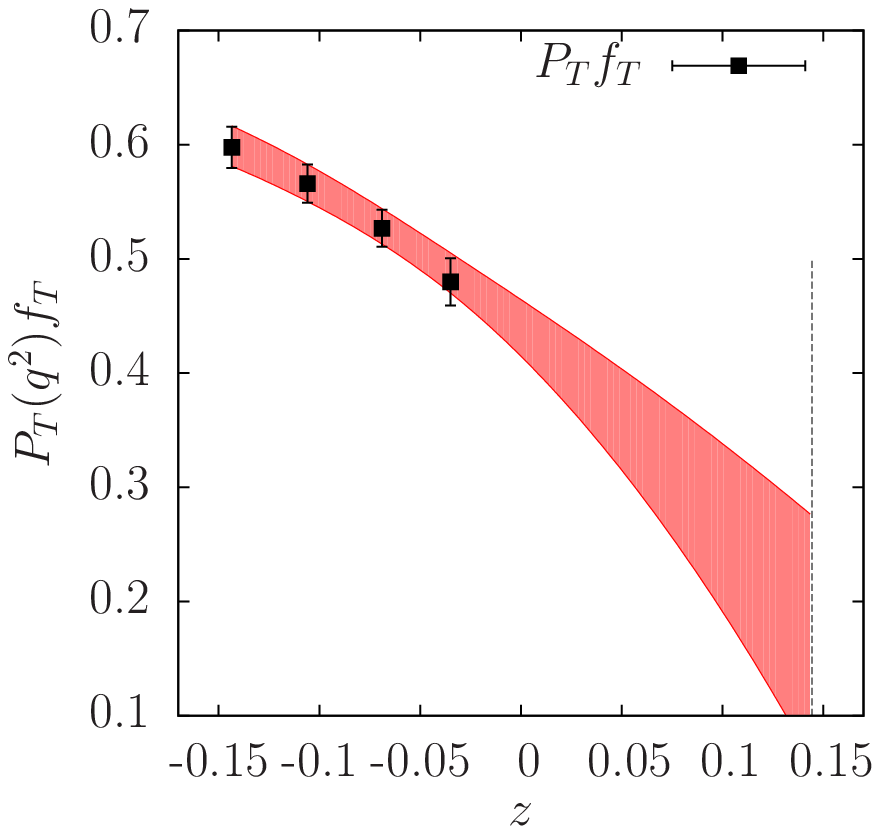}
    \includegraphics[scale=0.8]{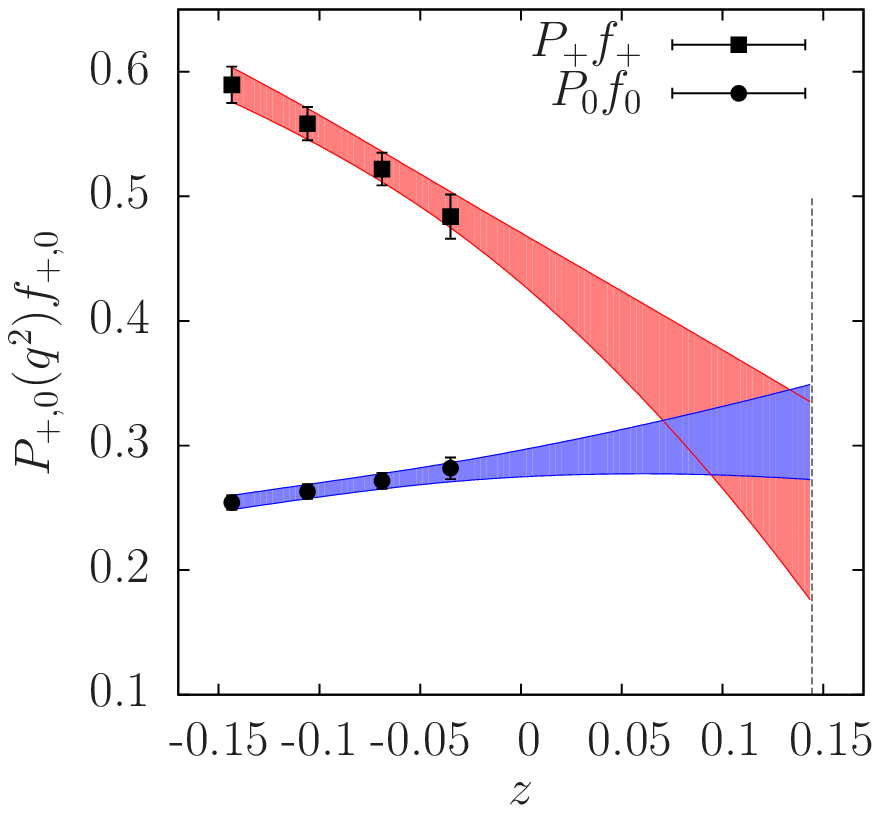}~~~~
    \includegraphics[scale=0.8]{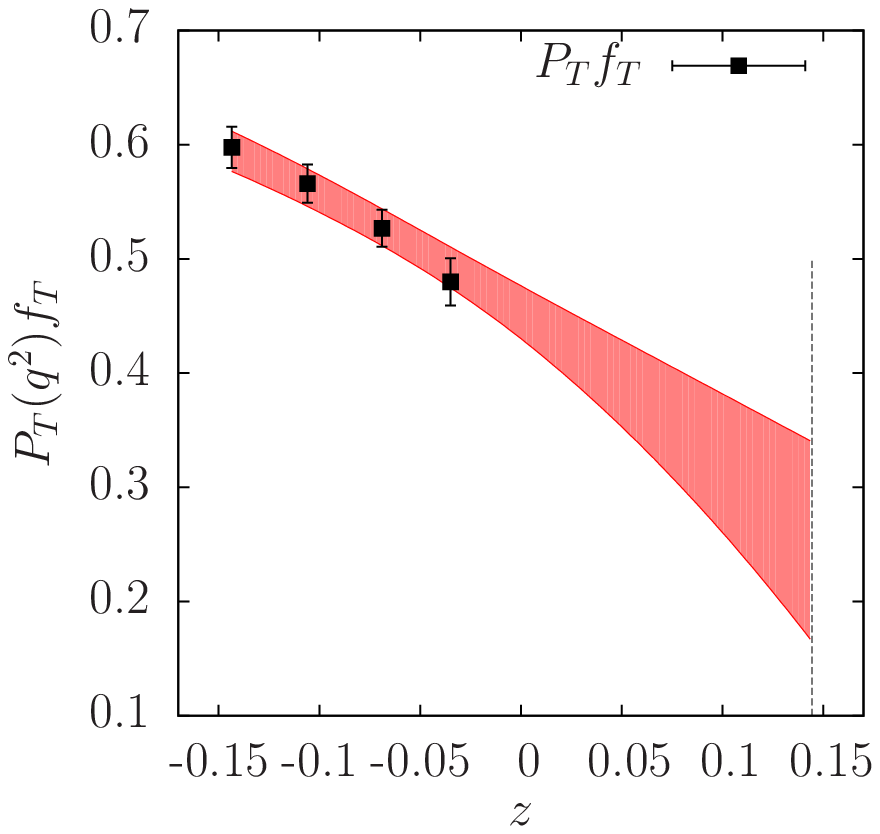}
    \caption{Separate $z$-expansion fits of $f_+$, $f_0$ (left) and $f_T$ (right) without 
      (upper) and with (lower) HQ constraints on the sum of coefficients for
      $f_+$ and $f_T$. The synthetic data points are 
      generated at large $q^2$ (small $z$) in the region of simulated lattice momenta.
      The kinematic condition $f_+(q^2=0) = f_0 (q^2=0)$ is satisfied better 
      when the HQ constraint
      is applied to $f_+$. (Recall that the factor $P_{+,0}= 1$ at $q^2=0$.)}
    \label{fig:zexp_sep_fpf0}
\end{figure}

We first fit $f_+$, $f_0$, and $f_T$ simultaneously in a combined fit using $K=3$ (three free 
parameters) in Eqs.~(\ref{eq:f+zexp})--(\ref{eq:fTzexp}) without any 
constraints on the coefficients.  
Table~\ref{tab:zfit_result} presents the results of these fits. 
We plot the fit results in Fig.~\ref{fig:zexp_sep_fpf0}. 
Although we do not impose the kinematic condition $f_+(q^2=0) = f_0(q^2=0)$, 
it is approximately satisfied with separate fits. Adding HQ constraints on the $f_+$ 
and $f_T$ fit makes the results even more consistent with the kinematic condition 
(see Fig.~\ref{fig:zexp_sep_fpf0}), and reduces the errors on $f_+$, $f_T$ at low $q^2$. 
We then fit $f_+$, $f_0$, and $f_T$ simultaneously with the kinematic constraint, and 
still including the HQ constraints on $f_+$ and $f_T$, which further decreases
the extrapolation error in the form factors at low~$q^2$. We
  implement the kinematic constraint by setting a prior of $f_+-f_0$ at
  $q^2$=0 with central value zero and width of 0.00001.

\begin{figure}[tp]
    \centering
    \includegraphics[scale=0.75]{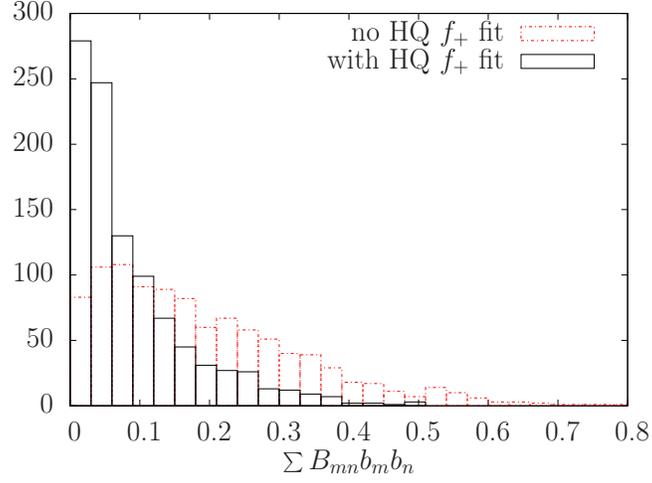} \\[1.2em]
    \includegraphics[scale=0.75]{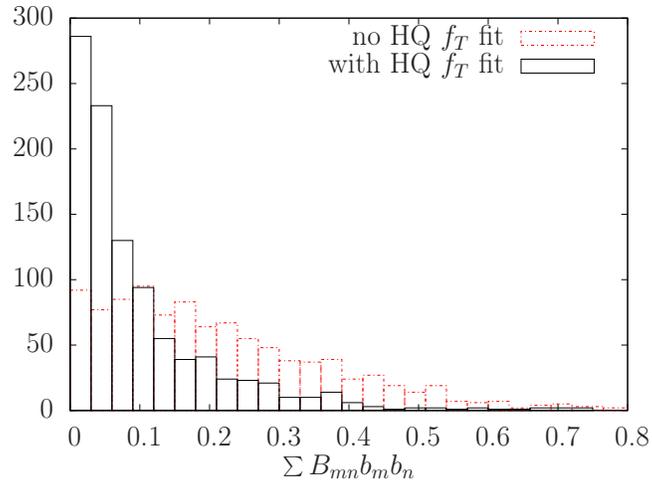}
    \caption{Histogram of the the sum of coefficients $B_{mn}b_mb_n$ for $f_+$ and $f_T$ from 
      fits with and without the HQ constraint. Use of the HQ constraint moves the
      distribution of $B_{mn}b_mb_n$ to smaller values.}
    \label{fig:Bmnbmbn_hist_fpfT}
\end{figure}

\begin{figure}[tp]
    \centering
    \includegraphics[scale=0.75]{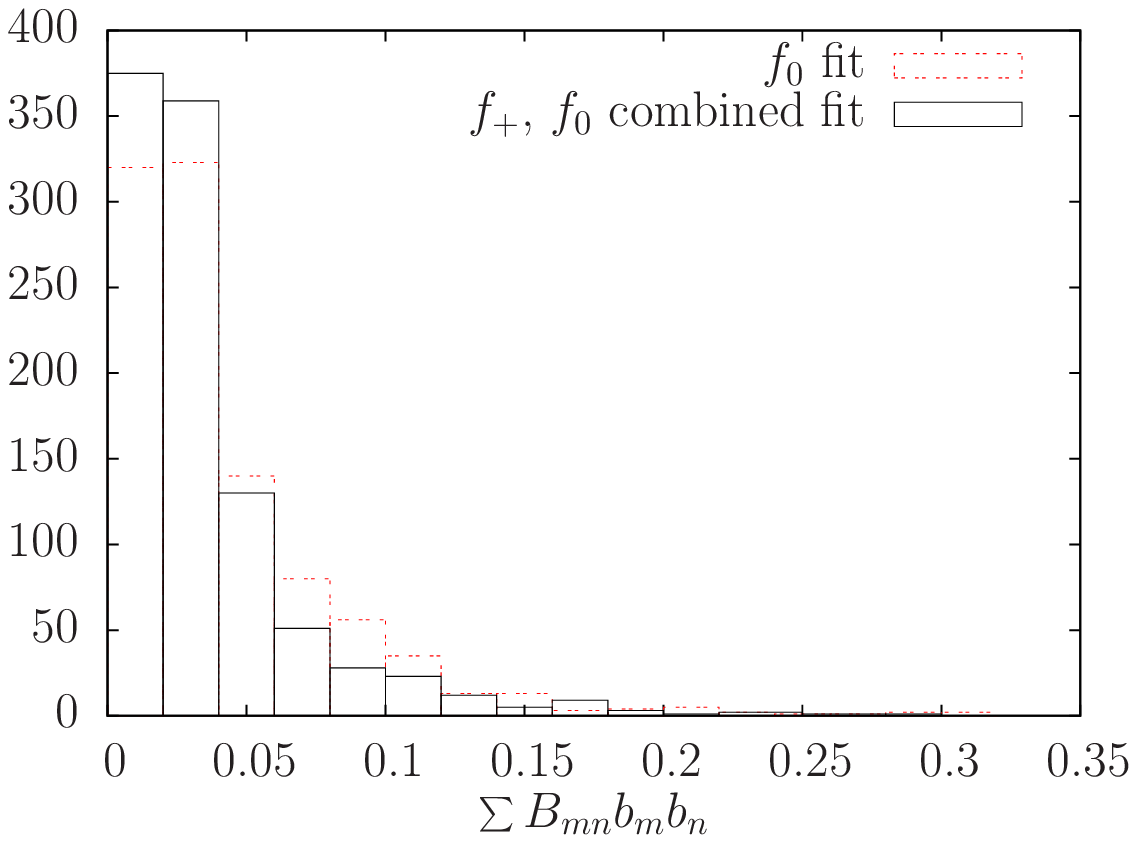}
    \caption{Histogram of the sum of coefficients $\sum B_{mn}b_mb_n$ for $f_0$ from 
      an independent fit and from a combined fit with $f_+$ that imposes the kinematic 
      constraint at $q^2=0$.     }
    \label{fig:Bmnbmbn_hist_f0}
\end{figure}

We show the $\sum B_{mn}b_mb_n$ bootstrap distribution of $f_+$ and $f_T$ from two fits 
with and without the HQ constraint in Fig.~\ref{fig:Bmnbmbn_hist_fpfT}. 
Adding the HQ constraint moves the distribution of $\sum B_{mn}b_mb_n$ to 
smaller values. 
We also compare the $\sum B_{mn}b_mb_n$ distribution of $f_0$ from two fits 
in Fig.~\ref{fig:Bmnbmbn_hist_f0}. One is a fit with $f_0$ only, the other is 
a combined $f_+$ and $f_0$ fit with the kinematic constraint. 
Adding the kinematic constraint decreases $\sum B_{mn}b_mb_n$ 
from the separate $f_0$ fit. Again, the result shows that the unitary constraint on 
the $\sum B_{mn}b_mb_n$ of $f_0$ is a loose bound.

We also check the truncation error by repeating the fit with
$K=4$. Because in $K=3$ fits, the coefficients $b^i_2$ are not well determined by 
data, and the results are zero within error, we add a prior of 0(2) on $b^i_4$ 
coefficients as in Ref.~\cite{Bouchard:2013eph} to control the fluctuations of 
the higher-order terms. 
All of the coefficients $b$ from fits with $K=3$ and $4$ are summarized in 
Table~\ref{tab:zexp.results}.
The results from different $K$ are consistent with each other.
The coefficients $b_{i,3}$ are zero within error and have little impact on 
the central value of the final result.  We therefore conclude that the $z$ truncation 
error is well controlled.

\begin{figure}[tp]
\centering
\includegraphics[scale=0.8]{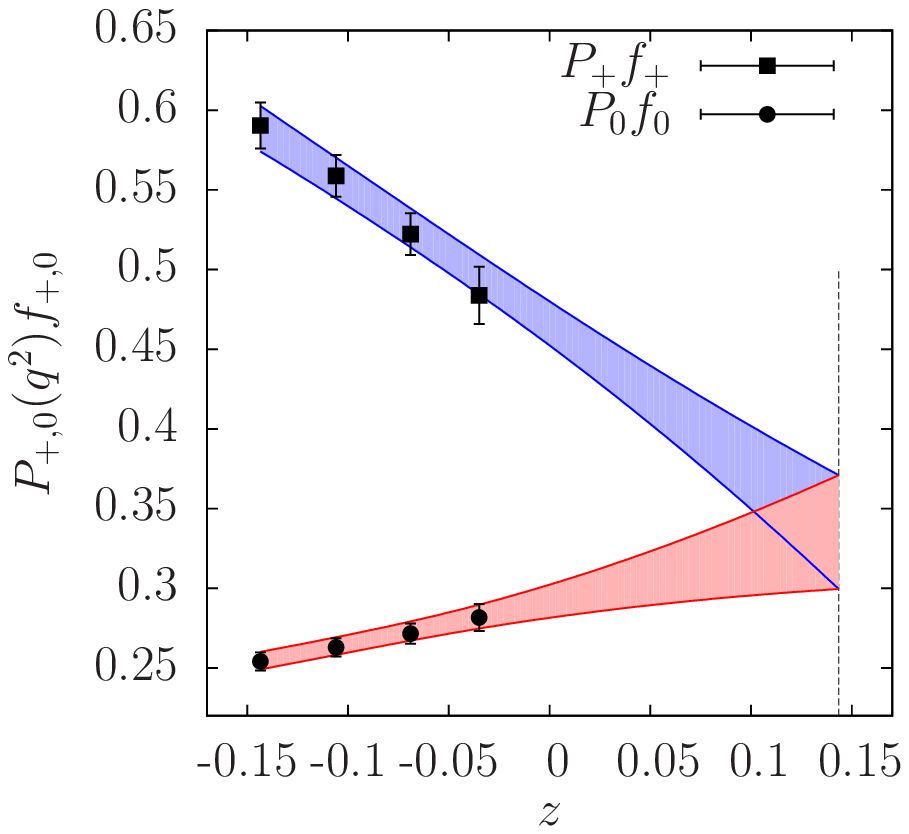}
\includegraphics[scale=0.8]{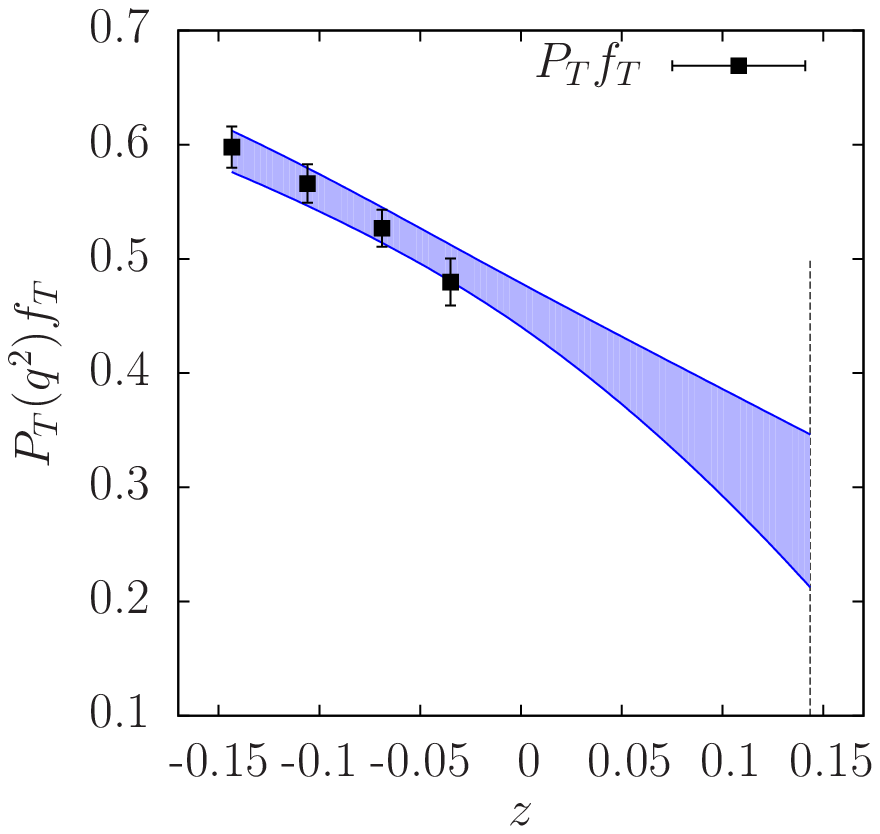}
\caption{$f_+$, $f_0$, and $f_T$ $z$-expansion fits. The synthetic data points are 
generated at large $q^2$ (small $z$) from LECs of the HMrS$\chi$PT fit result. The 
kinematic constraint $f_+(q^2=0)=f_0(q^2=0)$ is applied exactly in the combined 
$f_+$ and $f_0$ $z$-expansion fit. The vertical dashed lines correspond to $q^2$=0.  
We use three coefficients [$K=3$ in Eqs.~(\ref{eq:f+zexp})--(\ref{eq:fTzexp})] 
for $f_+$, $f_0$, and $f_T$.}
\label{fig:zexp_fvsz}
\end{figure}

\begin{figure}[tp]
\centering
\includegraphics[scale=0.8]{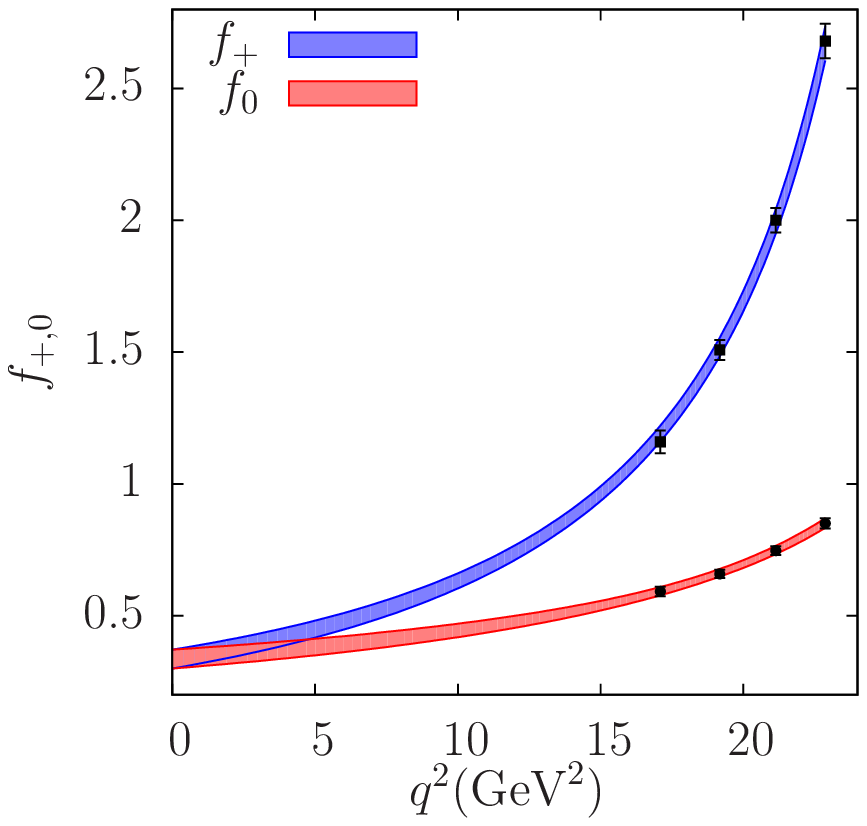}
\includegraphics[scale=0.8]{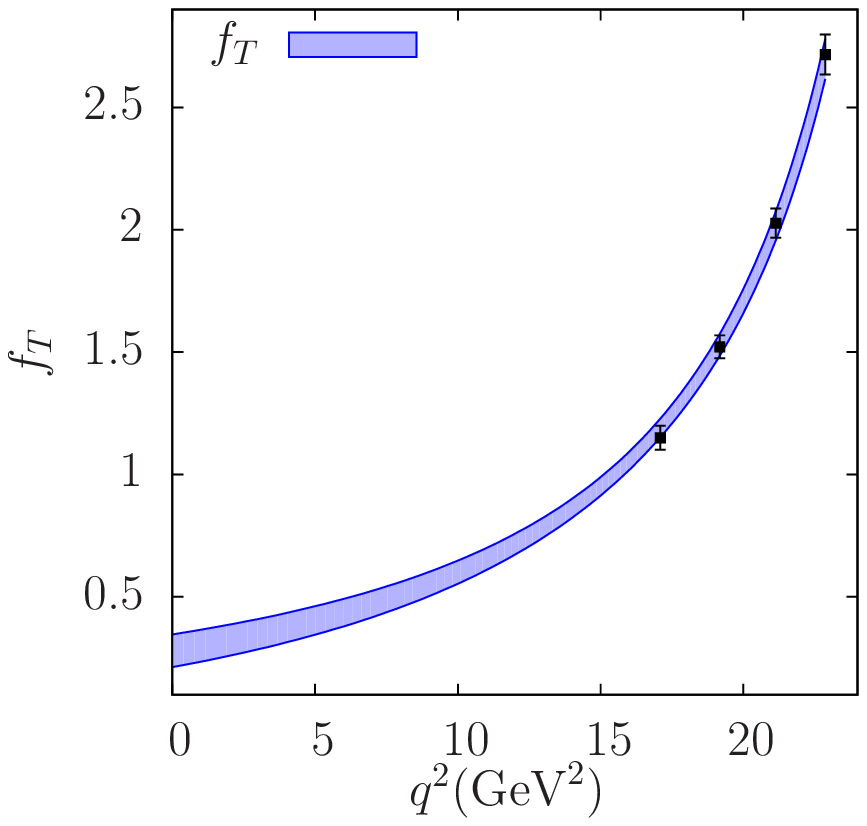}
\caption{$f_+$, $f_0$, and $f_T$ vs.~$q^2$ based on the $z$~expansion. 
The kinematic constraint $f_+(q^2=0)=f_0(q^2=0)$ is applied exactly in the fit. 
We use three coefficients [$K=3$ in Eqs.~(\ref{eq:f+zexp})--(\ref{eq:f+zexp})] 
for $f_+$, $f_0$, and $f_T$. }
\label{fig:zexp_fvsq2}
\end{figure}

We record our final, preferred results from $K=3$ $z$ fits including both 
the heavy-quark and kinematic constraints
in the third column of Table~\ref{tab:zexp.results}, and we give the
corresponding correlation matrix in Table~\ref{tab:zexp.results.covmat}.
Together with the pole masses (also in Table~\ref{tab:zexp.results}) and
Eqs.~(\ref{eq:f+zexp})--(\ref{eq:fTzexp}), this information allows the reader 
to reconstruct our form-factor results throughout the full kinematic 
range.  Our final form-factor results as a function of $z$ and $q^2$ are plotted in 
Figs.~\ref{fig:zexp_fvsz}--\ref{fig:zexp_fvsq2}.

\begin{table}[tp]
 \centering
 \caption{Results of $z$-expansion fits of the $B\to K$ form factors $f_+$ (top panel), 
   $f_0$ (middle panel), and $f_T$ (lower panel) using the
   formulae defined in Eqs.~(\ref{eq:f+zexp})--(\ref{eq:fTzexp}) with
   $t_0=(M_B+M_K)(\sqrt{M_B}-\sqrt{M_K})^2$~\cite{Bourrely:2008za},
   $M_{B_s^*}=5.4154$~GeV in $f_{+,T}$, $M_{B_{s0}^*}=5.711$~GeV in $f_{0}$, 
   $M_B$=5.27958~GeV, and $M_K$ = 0.497614 GeV~\cite{Agashe:2014kda}.}
 \label{tab:zexp.results}
 \begin{tabular}{c|c|ccc}
  \hline  
  \hline
  & ~~~unconstrained~~~ & \multicolumn{3}{c}{constrained} \\
  \hline
  & & HQ & \multicolumn{2}{c}{HQ + kinematic} \\
  \hline
  & ~~~~~$K=3$~~~~~ & ~~~~~~$K=3$~~~~~~ & ~~~~~~$K=3$~~~~~~ & ~~~~~~$K=4$~~~~~~ \\
  \hline
  \hline 
  $b^+_0$ &  0.437(22)   &  0.451(20)   &  0.466(14)    &  0.466(15)   \\
  $b^+_1$ & -1.41(33)    & -1.15(27)    & -0.89(13)     & -0.89(16)    \\
  $b^+_2$ & -2.5(1.4)    & -1.4(1.1)    & -0.21(55)     & -0.19(61)    \\
  $b^+_3$ & --           & --           & --            &  0.3(1.1)    \\
$\sum B_{mn}b_{m}b_{n}$ &  0.16 & 0.07  & 0.02          &  0.03        \\
  $f_+(0)$ & 0.18(10)      & 0.256(80)  & 0.335(36)     &  0.336(44)   \\
  \hline 
  $b^0_0$ &  0.285(11)   &  0.286(11)   &  0.292(10)    &  0.292(11)   \\
  $b^0_1$ &  0.19(14)    &  0.20(13)    &  0.28(12)     &  0.28(13)    \\
  $b^0_2$ & -0.17(49)    & -0.15(48)    &  0.15(44)     &  0.18(68)     \\
  $b^0_3$ & --           & --           &  --           &  0.2(1.7)     \\
$\sum B_{mn}b_{m}b_{n}$ & 0.02 & 0.02   & 0.02          &  0.02        \\
  $f_0(0)$ & 0.309(39)   & 0.311(38)    &  0.335(36)    &  0.336(44)     \\
  \hline 
  $b^T_0$ &  0.440(25)   & 0.453(23)    & 0.460(19)     &  0.459(20)   \\
  $b^T_1$ & -1.47(37)    & -1.17(30)    & -1.09(24)     & -1.11(24)    \\
  $b^T_2$ & -2.7(1.6)    & -1.4(1.2)    & -1.11(97)     & -1.15(95)     \\
  $b^T_3$ & --           & --           & --            & -0.2(1.1)     \\
$\sum B_{mn}b_{m}b_{n}$  & 0.18 & 0.07  & 0.05          &  0.05        \\
  $f_T(0)$ & 0.17(11)    & 0.254(87)    & 0.279(67)     &  0.276(68)   \\
  \hline
  $p$~value &  0.57      & 0.39         & 0.34          &  0.97        \\
  \hline
  \hline
 \end{tabular}
\label{tab:zfit_result}
\end{table}

\begin{table}[tp]
 \centering
\caption{The coefficients $b_i$ from the $z$-expansion fit (the first line) 
and their correlation matrix.
The upper index $+$, $0$, and $T$ denote the form factors $f_{+,0,T}$.
They are from the $z$-expansion fit formulae defined in Eqs.~(\ref{eq:f+zexp})--(\ref{eq:fTzexp}).
We use $t_0=(M_B+M_K)(\sqrt{M_B}-\sqrt{M_K})^2$~\cite{Bourrely:2008za}, $M_{B_s^*}=5.4154$~GeV in $f_{+,T}$,
$M_{B_{s0}^*}=5.711$~GeV in $f_{0}$, $M_B$=5.27958~GeV and $M_K$ = 0.497614 GeV~\cite{Agashe:2014kda}.}
 \label{tab:zexp.results.covmat}
 \begin{tabular}{cccccccccc}
  \hline  
  \hline
  &  ~~~$b^+_0$~~~ & ~~~$b^+_1$~~~ & ~~~$b^+_2$~~~ & ~~~$b^0_0$~~~ & ~~~$b^0_1$~~~ & ~~~$b^0_2$~~~ & ~~~$b^T_0$~~~ & ~~~$b^T_1$~~~ & ~~~$b^T_2$~~~ \\
  \hline
  Mean  & 0.466 & -0.885 & -0.213 & 0.292 & 0.281 & 0.150 & 0.460 & -1.089 & -1.114 \\          
  error & 0.014 &  0.128 &  0.548 & 0.010 & 0.125 & 0.441 & 0.019 &  0.236 & 0.971 \\
  \hline
  $b^+_0$ & 1     & 0.450 & 0.190 & 0.857 & 0.598 & 0.531 & 0.752 & 0.229 & 0.117 \\
  $b^+_1$ &       & 1     & 0.677 & 0.708 & 0.958 & 0.927 & 0.227 & 0.443 & 0.287 \\
  $b^+_2$ &       &       & 1     & 0.595 & 0.770 & 0.819 & -0.023 & 0.070 & 0.196 \\
  $b^0_0$ &       &       &       & 1     & 0.830 & 0.766 & 0.582 & 0.237 & 0.192 \\
  $b^0_1$ &       &       &       &       & 1     & 0.973 & 0.324 & 0.372 & 0.272 \\
  $b^0_2$ &       &       &       &       &       & 1     & 0.268 & 0.332 & 0.269 \\
  $b^T_0$ &       &       &       &       &       &       & 1     & 0.590 & 0.515 \\
  $b^T_1$ &       &       &       &       &       &       &       & 1     & 0.897 \\
  $b^T_2$ &       &       &       &       &       &       &       &       & 1     \\
  \hline
  \hline
 \end{tabular}
\end{table}

\section{Tests of QCD predictions for form-factor ratios}
\label{sec:QCDTests}

Because lattice-QCD calculations of the $B \to K$ semileptonic form factors have until recently 
been unavailable, theoretical calculations of $B\to Kl^+l^-$ observables sometimes use 
expectations from heavy-quark symmetries to relate them to others that can be constrained from 
experiment or computed with QCD models (see, e.g. Ref.~\cite{Bobeth:2011nj}).  Heavy-quark symmetry 
is also commonly used in phenomenological calculations of the related decays 
$B\to \pi l^+l^-$, $B\to K^{*}l^+l^-$, and 
$B\to K^*\gamma$~\cite{Grinstein:2004vb,Becher:2005fg,Ali:2007sj,Bobeth:2010wg,Beylich:2011aq,Bobeth:2011nj,Ali:2013zfa,Hou:2014dza}.  
Here we use our lattice-QCD form factors to directly test these heavy-quark 
symmetry relations in $B\to K$ decay at both high and low $q^2$.

\subsection{Low-recoil predictions from heavy-quark symmetry}
\label{sec:LowRecoil}

\begin{figure}
    \includegraphics[width=0.48\textwidth]{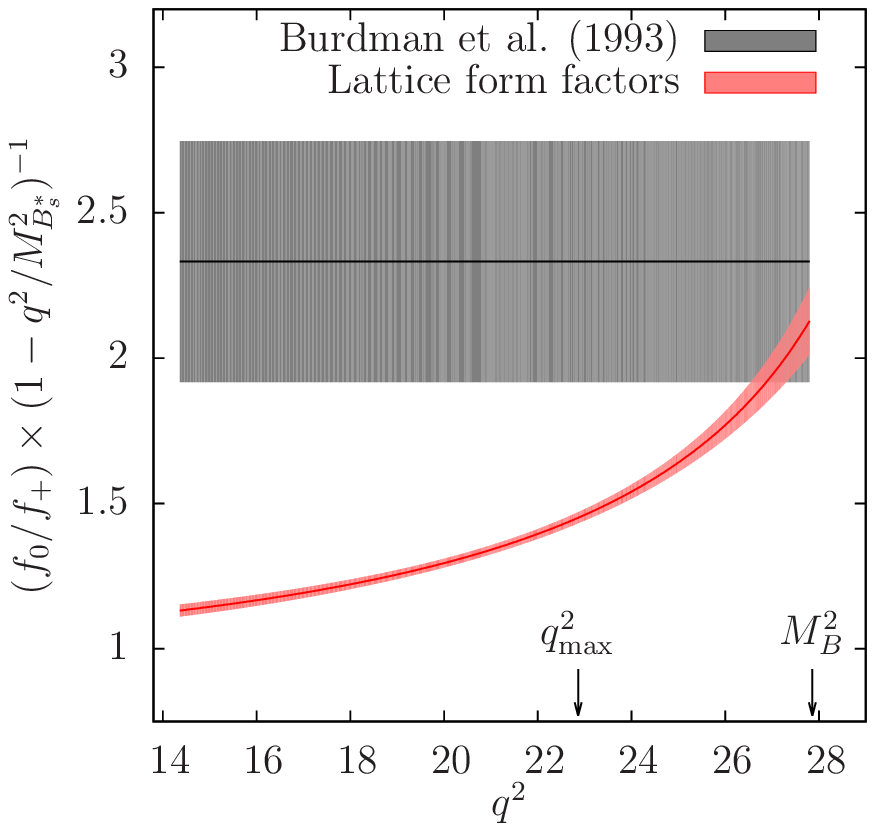} \hfill
    \includegraphics[width=0.48\textwidth]{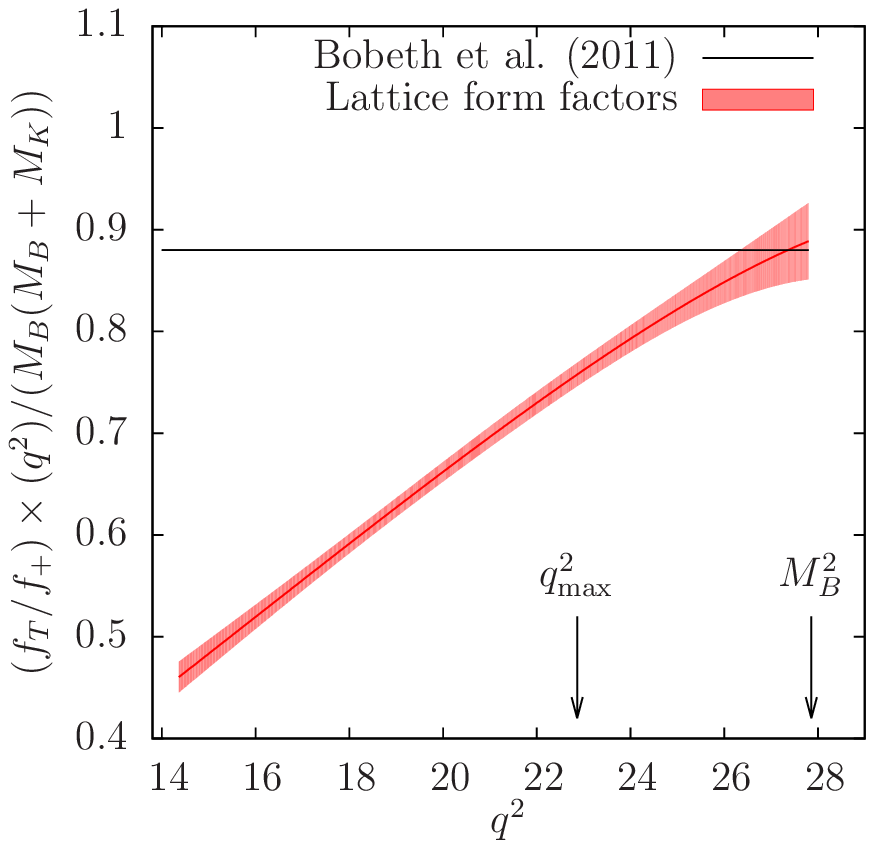}
    \caption{Comparison of lattice form-factor ratios with theoretical predictions from 
heavy-quark symmetry at low recoil. Left:  $(f_0 / f_+)/(1-q^2/M_{B_s^*}^2)^{-1}$ versus 
$q^2$ from lattice QCD (red curve with error band) and heavy-quark symmetry plus 
$\chi$PT~\cite{Burdman:1993es} (gray horizontal band). The width of the theoretical band 
includes the uncertainty on $g_{\pi} = 0.45(8)$ but no other theory errors.
Right:   $(f_T/f_+)\times (q^2)/(M_B(M_B + M_K))$   versus $q^2$ from lattice 
QCD (red curve with error band) and the improved Isgur-Wise relation~\cite{Bobeth:2011nj} 
(black horizontal line).}
    \label{fig:LowRecoil}
\end{figure}

In the soft-kaon ($E_K \ll M_B$) and chiral limits, the vector and scalar form factors can 
be related using heavy-quark  effective theory and chiral perturbation theory~\cite{Wise:1992hn,Burdman:1993es}:
\begin{equation}
        \lim_{q^2 \to M_B^2} \frac{f_0}{f_+} = \left(\frac{f_{B_s}}{f_{B^*_s}} \right) 
\frac{1-q^2/M_{B_s^*}^2}{g_{\pi}} + \order(\Lambda^2/m_b^2) \,,  \label{eq:f0f+lowrecoil} 
\end{equation}
where the decay-constant ratio accounts for heavy-quark corrections of $\order(1/m_b)$.  Heavy-quark 
spin symmetry relates the vector and tensor form factors in the soft-kaon limit as~\cite{Hewett:2004tv,Bobeth:2011nj}:
\begin{equation}
        \lim_{q^2 \to M_B^2}\frac{f_T}{f_+}(q^2, \mu)  = \kappa(\mu) \frac{M_B(M_B + M_K)}{q^2}
 + \order(\Lambda/m_b) \,,  \label{eq:fTf+lowrecoil} 
\end{equation}
where the scale-dependent coefficient $\kappa(\mu)$ incorporates corrections of $\order(\alpha_s^2)$ 
to the leading Isgur-Wise relation~\cite{Isgur:1990kf} and is given in Eq.~(2.5) of Ref.~\cite{Bobeth:2011nj}.   
We can estimate the size of higher-order corrections in the heavy-quark expansion from power counting.  
Taking $\Lambda = 500$~MeV and $m_b = 4.2$~GeV gives  $\Lambda/m_b \sim 12\%$ and $(\Lambda/m_b)^2 \sim 1\%$.  
Equations~(\ref{eq:f0f+lowrecoil}) and~(\ref{eq:fTf+lowrecoil}) also receive corrections from the kaon 
recoil energy that are of $\order(E_K/m_b)$.  For $q^2_{\rm max} \geq q^2 \geq 14 {\rm~GeV}^2$, this 
ratio varies from $12\% \leq E_K/m_b \leq 40\%$, so such corrections are expected to be significant 
even at low kaon recoil.
  
Figure~\ref{fig:LowRecoil}, left, compares the quantity $(f_0/f_+) \times (1-q^2/M_{B_s^*}^2)^{-1}$ 
obtained from our lattice form factors with the theoretical prediction Eq.~(\ref{eq:f0f+lowrecoil}).  
For the theoretical estimate, we take $f_{B^*_s}/f_{B_s} = 0.953(23)$ from the recent four-flavor  
lattice-QCD determination in Ref.~\cite{Colquhoun:2015oha} and $g_{\pi} = 0.45(8)$ as in our 
chiral-continuum fit. The width of the theoretical band is from the uncertainty on $g_{\pi}$, and does 
not include any other errors.  Figure~\ref{fig:LowRecoil}, right, compares the quantity 
$(f_T/f_+) \times (q^2)/(M_B(M_B + M_K))$ obtained from our  lattice form factors with the theoretical 
prediction Eq.~(\ref{eq:fTf+lowrecoil}) using $m_b = 4.18$~GeV and 
${\alpha_s}^{(4)}_{\overline{\rm MS}} (m_b) = 0.2268$, such that 
$\kappa(m_b) \approx 0.88$~\cite{Grinstein:2004vb,Bobeth:2011nj}.  We do not show any errors on 
the theoretical prediction.

The observed lattice form-factor ratios $f_0/f_+$ and $f_T/f_+$ at $q^2_{\rm max}$ are lower than 
the theoretical expectations by $38$\% and $15$\%, respectively; by $q^2 = 14.5$~GeV$^2$ the 
differences grow to $51$\% and $46$\%, respectively.   Although the observed disagreement with 
the theoretical expectation for the tensor form-factor ratio is large, it is within 
the size expected (from simple power counting) for higher-order corrections due to the kaon recoil energy.
The scalar form-factor ratio, however, differs from the theoretical expectation by a much larger amount.   
In Fig.~25 of Ref.~\cite{Lattice:2015tia} we compare the quantity $(f_0/f_+) \times (1-q^2/M_{B_s^*}^2)^{-1}$ 
for the related decay $B \to \pi l\nu$ with the heavy-quark prediction in the soft-pion limit.  
The observed agreement near $q^2_{\rm max}$ is better, which suggests that the discrepancy is indeed due to the light 
pseudoscalar-meson recoil energy, which is larger for $B\to K$ than for $B\to\pi$.  Thus our 
lattice form-factor results suggest that one should be cautious in using heavy-quark relations 
derived in the soft-pion/kaon limit for phenomenological predictions, especially for decays 
with $K$ or $K^*$ final-state mesons.

\subsection{Large-recoil predictions from QCD factorization}
\label{sec:LargeRecoil}

In the large-recoil limit ($E_K \gg M_K$), heavy-quark symmetry relates the vector, scalar, 
and tensor form factors to a single universal form factor~\cite{Charles:1998dr}:
\begin{align}
        \lim_{E_K \gg M_K} \frac{f_0}{f_+} & = \frac{2 E_K}{M_B} + \order(\Lambda/m_b), 
    \label{eq:f0f+highrecoil} \\
        \lim_{E_K \gg M_K} \frac{f_T}{f_+} & = \frac{M_B + M_K}{M_B}+ \order(\Lambda/m_b). 
    \label{eq:fTf+highrecoil} 
\end{align}
The $\order(\alpha_s)$ corrections to these leading large-recoil expressions were derived 
using QCD factorization (QCDF) in Ref.~\cite{Beneke:2000wa}, and the resulting expressions 
are given in Eqs.~(62)--(63) of that work.  Higher-order corrections in the heavy-quark 
expansion are expected to be about $\Lambda/m_b \sim 12\%$, while $\order(\alpha_s^2)$ 
corrections to the QCDF predictions from Ref.~\cite{Beneke:2000wa} are expected to be 
about 5\%. 

\begin{figure}
    \includegraphics[width=0.48\textwidth]{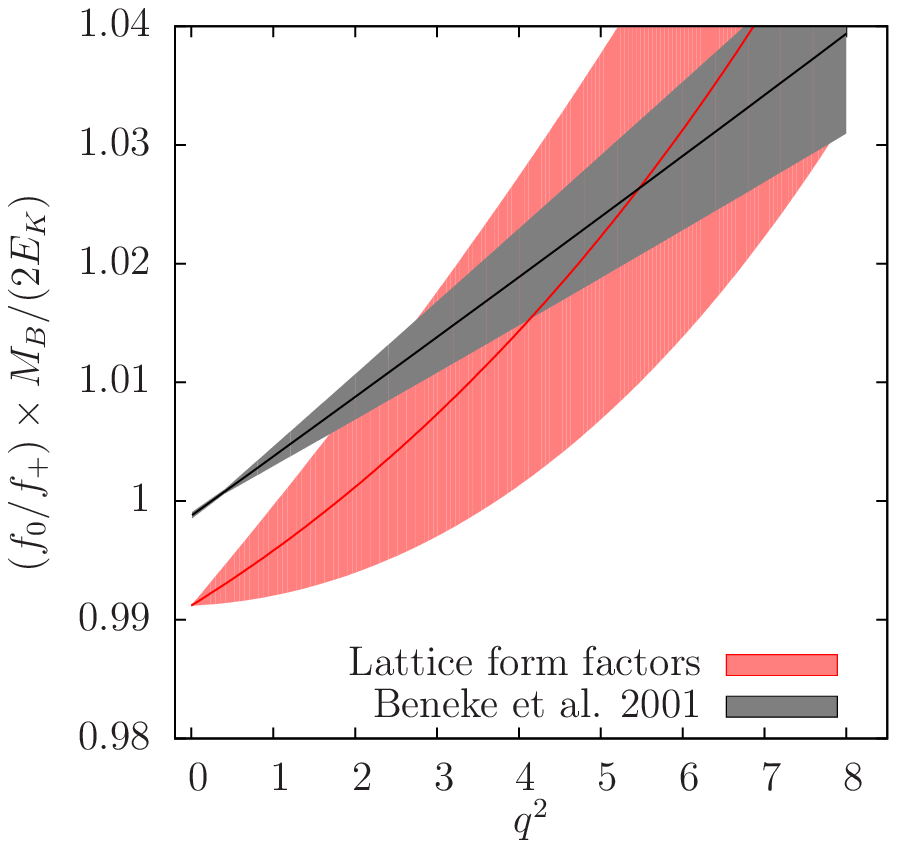} \hfill
    \includegraphics[width=0.48\textwidth]{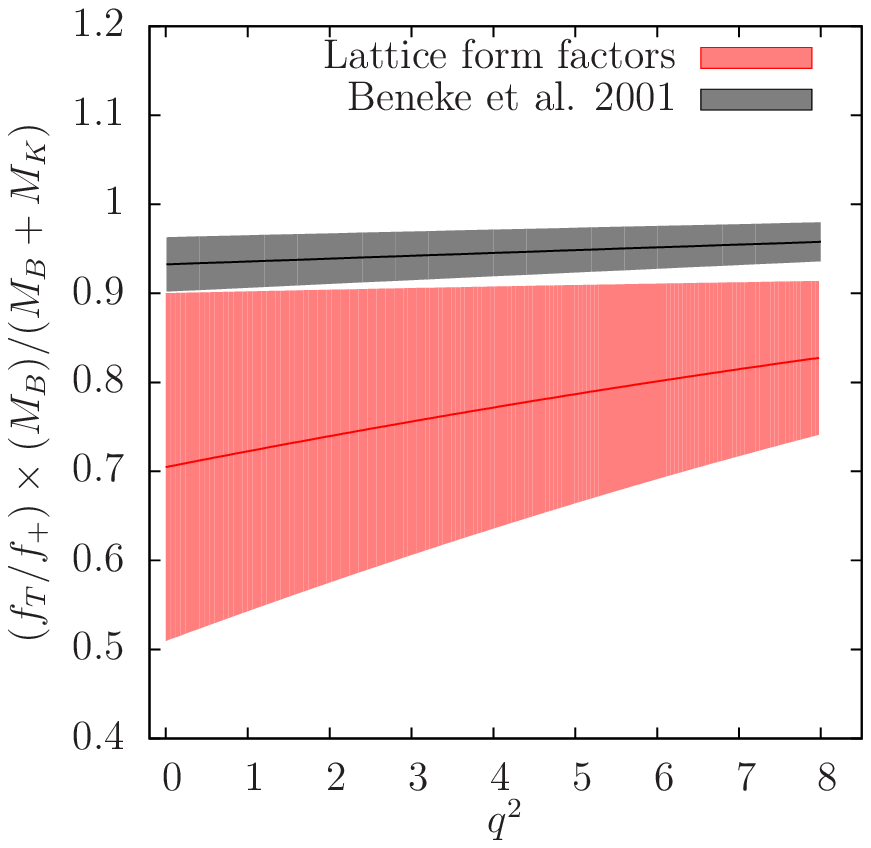}
    \caption{Comparison of lattice form-factor ratios with theoretical predictions from heavy-quark 
symmetry at large recoil. 
Left:  $(f_0 / f_+)\times M_B/(2E_K)$ versus $q^2$ from lattice QCD (red curve with error band) and 
theoretical prediction with $\order(\alpha_s)$ corrections~\cite{Beneke:2000wa} (gray curve with error band).  
Right:   $(f_T/f_+)\times (M_B)/(M_B + M_K)$ versus $q^2$ from lattice 
QCD (red curve with error band) and theoretical prediction 
with $\order(\alpha_s)$ corrections~\cite{Beneke:2000wa} (gray curve with error band).}
    \label{fig:HighRecoil}
\end{figure}

Figure~\ref{fig:HighRecoil} compares the lattice-form-factor ratios with the theoretical large-recoil 
predictions from Ref.~\cite{Beneke:2000wa}.  For the $\order(\alpha_s)$ corrections, we take the 
decay constants $f_B = 190.5(4.2)$~MeV from FLAG~\cite{Aoki:2013ldr} and $f_K = 156.2(7)$~MeV from the 
PDG~\cite{Agashe:2014kda}.  We take the first inverse moment of the $B$-meson distribution amplitude 
$\lambda^{-1}_{B}(2.2~{\rm GeV})=[0.51(12)~{\rm GeV}]^{-1}$ from LCSR~\cite{Ball:2006nr}, where 
the quoted theory error covers the spread of other determinations from QCD/light-cone sum rules and the 
operator-product expansion~\cite{Braun:2003wx,Khodjamirian:2005ea,Lee:2005gza}. We take the first and 
second moments of the kaon distribution amplitude $a_1^{K}(2~{\rm GeV}) = 0.061(4)$ 
and $a_2^{K}(2~{\rm GeV}) = 0.18(7)$ from a recent three-flavor lattice-QCD 
calculation~\cite{Arthur:2010xf}.  We use our own determination of $f_+(q^2=0) = 0.335(36)$.  
We take ${\alpha_s}^{(4)}_{\overline{\rm MS}} (m_b) = 0.2268$ as described above 
and ${\alpha_s}^{(4)}_{\overline{\rm MS}}(2.2~{\rm GeV}) = 0.279$~\cite{Chetyrkin:2000yt}.
The left panel of Fig.~\ref{fig:HighRecoil} shows the quantity 
$(f_0/f_+) \times (M_B) / (2E_K)$, while the right panel shows $(f_T/f_+) \times (M_B) / (M_B + M_K)$. 
The widths of the theoretical bands in Fig.~\ref{fig:HighRecoil} are from the 
uncertainty on $\lambda^{-1}_{B}$ and $f_+(q^2=0)$, and do not include any other errors.

For $(f_0 / f_+)\times M_B/(2E_K)$, the lattice-QCD result differs from the 
theoretical predictions by at most $1$\%, which is well within the expected 
size of heavy-quark corrections.  For $(f_T/f_+)\times (M_B)/(M_B + M_K)$, the 
lattice-QCD result is marginally consistent with the theoretical expectation of 
Ref.~\cite{Beneke:2000wa}. 
A more recent NNLO calculation within soft-collinear effective theory 
updates the large-recoil predictions to include $\order{(\alpha_s^2)}$ 
corrections~\cite{Bell:2010mg}.  The new $q^2=0$ result for 
$f_T / f_+ M_B / (M_B + M_K) = 0.817$ is in better agreement with 
the ratio obtained from lattice QCD. 
Overall, the uncertainty on the lattice-QCD tensor form-factor ratio at low 
$q^2$ is too large to draw any quantitative conclusions. (The vector and tensor 
form factors are not strongly correlated at low $q^2$.) Thus, while the 
scalar form-factor ratio suggests that the large-recoil predictions may 
be reliable, some caution is nevertheless warranted in their 
use for phenomenology given the limited number of tests they have undergone.

\section{Summary and Outlook}
\label{sec:conclusion}

As discussed in Sec.~\ref{sec:zexp}, Table~\ref{tab:zexp.results.covmat} presents our final results for the
form factors $f_+(q^2)$, $f_0(q^2)$, and $f_T(q^2)$ for the semileptonic process $B \to Kl^+l^-$.
These entries, which consist of the coefficients of the BCL $z$ expansion,
Eqs.~(\ref{eq:f+zexp})--(\ref{eq:fTzexp}), together with the correlations among them, can be used to
reconstruct our form factors with errors for all values of $0 \leq q^2 \leq q^2_\text{max}$.
This information can also be used to compute form-factor ratios and (differential) rates with squares of
linear combinations of the form factors.

\begin{figure}[b]
    \centering
    \includegraphics[scale=0.8]{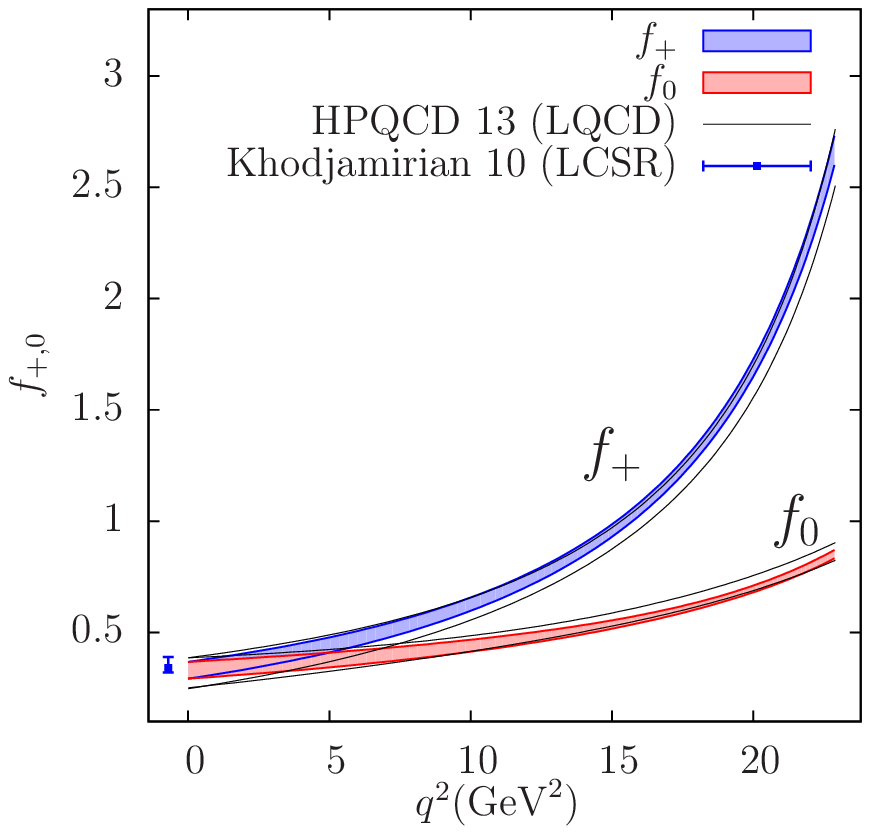} \quad
    \includegraphics[scale=0.8]{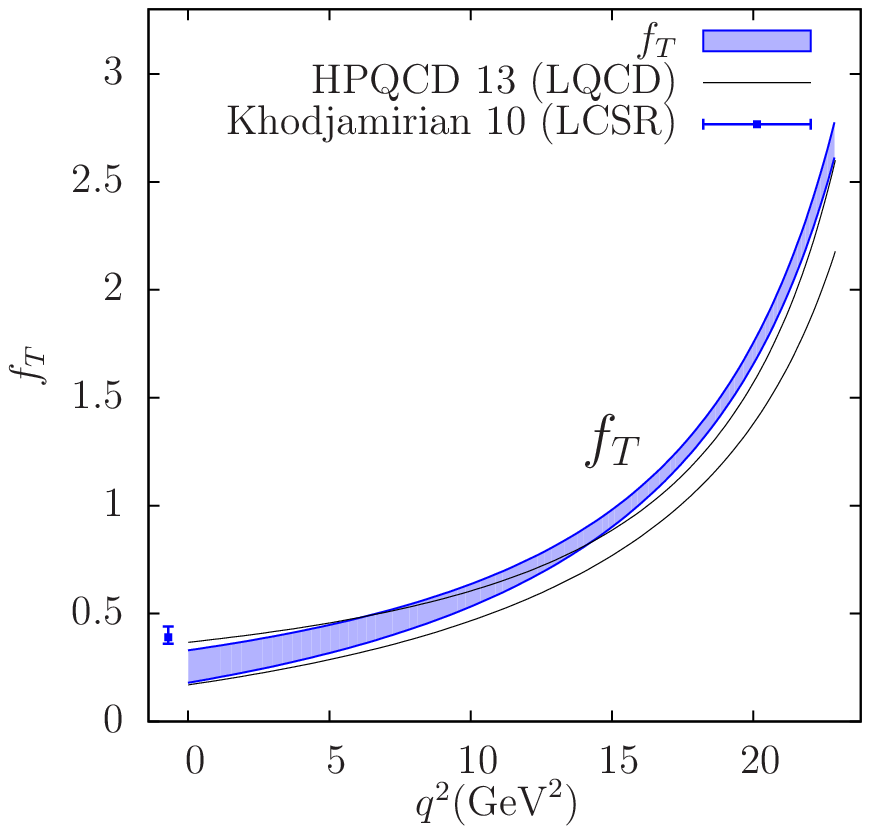}
    \caption{Our form factors compared with light-cone-sum-rule results~\cite{Khodjamirian:2010vf} and the 
        other unquenched lattice-QCD calculation~\cite{Bouchard:2013eph}.}
    \label{fig:ff_comparison}
\end{figure}

Figure~\ref{fig:ff_comparison} shows a comparison of our results with others in the literature.
At $q^2=0$, our result is consistent with a light-cone-sum-rule result from Khodjamirian
\emph{et~al.}~\cite{Khodjamirian:2010vf}.
For all~$q^2$, our results are consistent with the only other unquenched lattice-QCD 
calculation from the HPQCD Collaboration~\cite{Bouchard:2013eph}.
Our form factors are somewhat more precise than HPQCD's, especially at high~$q^2$, 
because we used more ensembles with finer lattice spacings and lighter quark masses.
The total errors, including both statistical and systematic errors, are less than 
4\% at high $q^2$, and at low $q^2$ about 10\% for $f_+$ and 30\% for $f_T$.

More generally, our results can be used to compute any $B \to K ll$ observable, including 
asymmetries and decay rates, for all possible dilepton final states ($l = \ell, \tau, \nu$), 
and even lepton-flavor-violating modes~\cite{Glashow:2014iga}.  We present a thorough 
analysis of observables for $B\to K$ semileptonic decays in a companion publication~\cite{Du:2015tda}, 
where we also present ratios of observables for $B\to K ll$ to $B\to \pi ll$ decay 
processes.  
The three form factors $f_+$, $f_0$, and $f_T$  
suffice to parameterize the factorizable hadronic contributions to 
$B\to K$ semileptonic decays in any extension of the Standard 
Model.  Other hadronic uncertainties, such as violations of 
quark-hadron duality due to intermediate charmonium resonances, 
must, of course, also be reliably estimated to obtain complete 
Standard-Model and new-physics predictions for $B\to K$ processes.
If deviations from the Standard Model are observed in any 
$B \to K ll$ decay channel, accurate results for the form factors will be essential 
to disentangling the underlying physics.

The main sources of uncertainty in our form factors are from the chiral-continuum extrapolation and extrapolation to
low~$q^2$.
We plan to reduce these uncertainties with newer gauge-field ensembles that are being generated by the MILC
Collaboration~\cite{Bazavov:2010ru,Bazavov:2012xda}.
These ensembles use the highly-improved staggered quark (HISQ) action for the light, strange, and charm
quarks. This action is designed to have smaller discretization effects which will help 
reduce the size of the continuum extrapolation errors~\cite{Follana:2006rc}.
In addition, the HISQ ensembles include ensembles with physical pion masses, which will eliminate the need
for the chiral extrapolation and the associated errors.
Indeed, these ensembles have already been used to improve the precision for kaon~\cite{Bazavov:2013maa}
and charmed-meson~\cite{Bazavov:2014wgs} physics.
In particular, we found with $D$- and $D_s$-meson decay constants~\cite{Bazavov:2014wgs} 
that an analysis with physical and unphysical quark masses provides better statistical precision with no 
penalty in systematic errors.

\acknowledgments

We thank Wolfgang Altmannshofer, Martin Beneke, and Enrico Lunghi 
for useful conversations about the form factors and phenomenology
 of $b\to s$ processes.
We thank Heechang Na for useful conversations about correlator fit techniques.
We thank Richard J. Hill for clarifying discussions about the unitarity constraints in the $z$-expansion.
% computing boilerplate
Computations for this work were carried out with resources provided by the USQCD Collaboration, the Argonne
Leadership Computing Facility, the National Energy Research Scientific Computing Center, and the Los Alamos
National Laboratory, which are funded by the Office of Science of the U.S.
Department of Energy; and with resources provided by the National Institute for Computational Science, the
Pittsburgh Supercomputer Center, the San Diego Supercomputer Center, and the Texas Advanced Computing Center,
which are funded through the National Science Foundation's Teragrid/XSEDE Program.
% university boilerplate
This work was supported in part by the U.S. Department of Energy under Grants
No.~DE-FG02-91ER40628 (C.B., J.K.),
No.~DE-FC02-12ER41879 (C.D., J.F., L.L.),
No.~DE-FG02-91ER40661 (S.G., R.Z.),
No. DE-SC0010120 (S.G.),
No. DE-FC02-06ER41443 (R.Z.),
No. DE-FG02-91ER40677 (A.X.K.,C.M.B.,D.D.,E.D.F.,R.D.J.), No. DE-FG02-13ER42001 (A.X.K.,D.D.)
No.~DE-FG02-ER41976 (D.T.);
No.~DE-SC0010114 (Y.L., Y.M.);
by the National Science Foundation under Grants
No.~PHY-0555243, No.~PHY-0757333, No.~PHY10-67881 (C.D., L.L.),
No.~PHY-1316748 (R.S.),
No. PHY-1212389 (R.Z.),
No.~PHY-1417805 (J.L., D. D.);
by the URA Visiting Scholars' program, (Y.L., Y.M., A.X.K., D.D., C.M.B.);
by the MICINN (Spain) under grant FPA2010-16696 and Ram\'on y Cajal program (E.G.);
by the Junta de Andaluc\'ia (Spain) under Grants No.~FQM-101 and 
No.~FQM-6552 (E.G.);
by European Commission (EC) under Grant No.~PCIG10-GA-2011-303781 (E.G.);
by the German Excellence Initiative and the European Union Seventh Framework 
Programme under grant agreement No.~291763 as well as the European Union's 
Marie Curie COFUND program (A.S.K.);
and  by the Basic Science Research Program of the National Research Foundation 
of Korea (NRF) funded by the Ministry of Education (No. 2014027937) and the 
Creative Research Initiatives Program (No. 2014001852) of the NRF grant funded 
by the Korean government (MEST) (J.A.B.).
% Brookhaven boilerplate
This manuscript has been co-authored by an employee of Brookhaven Science
Associates, LLC, under Contract No. DE-AC02-98CH10886 with the U.S. Department of Energy.
% Fermilab boilerplate
Fermilab is operated by Fermi Research Alliance, LLC, under Contract
No.~DE-AC02-07CH11359 with the United States Department of Energy.

\appendix

\section{\boldmath \texorpdfstring{$B\to Kll$}{B to KLL} form factors in SU(2) \texorpdfstring{S$\chi$PT}{SChPT}}
\label{app.su2}

We use expressions derived in 
HMrS$\chi$PT~\cite{Aubin:2007mc} as the low-energy effective theory of 
QCD in which the degrees of freedom are pions and kaons for our
chiral-continuum extrapolations.
SU(3) HMrS$\chi$PT~\cite{Becirevic:2002sc,Becirevic:2003ad,Aubin:2007mc} was 
applied to $B\to\pi l\nu$ semileptonic decays~\cite{Bailey:2008wp}.
More recently, SU(2) HM$\chi$PT~\cite{Becirevic:2006me, Becirevic:2007dg, DiVita:2011py} was also considered
as an alternative effective theory in studies of heavy meson physics.
We derive the SU(2) HMrS$\chi$PT formulae for form factors calculated with staggered quarks in this appendix.
These formulae can be used for $B\to \pi l \nu$, $B\to Kl^+l^-$ and $D$-meson semileptonic decays.
Our results are consistent with earlier studies of HM$\chi$PT for continuum QCD and Wilson
quarks~\cite{DiVita:2011py} after taking the continuum limit of the HMrS$\chi$PT expressions.
The differences in the detailed expressions can be absorbed into redefinitions of the scale or LECs.

\subsection{\boldmath\texorpdfstring{$f_\parallel$}{f\_parallel} and 
\texorpdfstring{$f_{\perp}$}{fperp} in SU(3) \texorpdfstring{HMrS$\chi$PT}{HMrSChPT}}

The SU(3) HMrS$\chi$PT expression of $f_\parallel$ is the same as in Eq.~(\ref{eq:chiral.fpara}).
We only list the expression of the log terms for unitary points 
($m_{l,h}=m_{l,h}^\prime=m_{l,s}$) here.

For the $B\to \pi$ process, the chiral logs in SU(3) HMrS$\chi$PT are given by~\cite{Aubin:2007mc}:
\begin{align}
\text{logs}^{B\to \pi}_{\parallel,\text{SU(3)}} = & \,
  \frac{1}{(4 \pi f)^2}\Biggl\{           \frac{1}{16}\sum_{\Xi}\biggl[
  \frac{1-3g^2_\pi}{2}\left[
    2I_1(m_{\pi,\Xi}) + I_1(m_{K,\Xi})
    \right]\hspace{5truecm}\nonumber\\&{}
  +2I_2(m_{\pi,\Xi},\vp) + I_2(m_{K,\Xi},\vp)
    \biggr] \nonumber\\
        & {}+\frac{1+3g^2_\pi}{4}\left[ I_1(m_{\pi,I})
  - \frac{1}{3}I_1(m_{\eta,I})\right]\nonumber\\
        & {}+
  \sum_{j\in\{\pi,\eta,\eta'\}}
   \biggl[a^2\delta'_V
    R^{[3,1]}_{j} \left(\{m_{\pi,V},m_{\eta,V},m_{\eta',V}\} ;
         \{m_{S,V}\}\right)\times\nonumber\\&{}\hspace*{2em}
  \left(\frac{3(g^2_\pi-1)}{2}
         I_1(m_{j,V})-2  I_2(m_{j,V},\vp)
         \right)\biggr] + [V\to A]\Biggr\}.
         \label{eq:B-to-pi-fv}
\end{align}
For the $B\to K$ process, the chiral logs in SU(3) HMrS$\chi$PT are given by:
\begin{eqnarray}
\text{logs}^{B\to K}_{\parallel, \text{SU(3)}} & = &
        \frac{1}{(4 \pi f)^2}\Biggl\{
\frac{1}{16}\sum_{\Xi}\biggl[
      \frac{2-3g^2_\pi}{2} I_1(m_{K,\Xi})
    - 3g^2_\pi I_1(m_{\pi,\Xi})
    + \frac{1}{2}I_1(m_{S,\Xi})\nonumber\\&&{}
        +2I_2(m_{K,\Xi},E) + I_2(m_{S,\Xi},E)
    \biggr]
    \nonumber\\& & {}
        - \frac{1}{2} I_1(m_{S,I})
        + \frac{3 g^2_\pi}{4} I_1(m_{\pi,I})
        + \frac{8-3g^2_\pi}{12} I_1(m_{\eta,I})
        + I_2(m_{\eta,I},E) - I_2(m_{S,I},E)
        \nonumber\\ & & {}
        + a^2\delta'_V \Biggl[
        \frac{I_1(m_{\eta',V}) - I_1(m_{\eta,V})
        +I_2(m_{\eta',V},E) - I_2(m_{\eta,V},E)}
        {m^2_{\eta',V} - m^2_{\eta,V}}
        \nonumber\\&&{}
        - \sum_{j\in\{S,\eta,\eta'\}}
    R^{[3,1]}_{j} \left(\{m_{S,V},m_{\eta,V},m_{\eta',V}\} ;
        \{m_{\pi,V}\}\right)\left(
        \frac{1}{2}I_1(m_{j,V}) + I_2(m_{j,V}, E)
        \right)\nonumber\\&&{}
        + \frac{3g^2_\pi}{2}\sum_{j\in\{\pi,\eta,\eta'\}}
    R^{[3,1]}_{j} \left(\{m_{\pi,V},m_{\eta,V},m_{\eta',V}\} ;
        \{m_{S,V}\}\right) I_1(m_{j,V}) \Biggr]
        \nonumber\\&&{}+ [V\to A]\Biggr\}\ . \label{eq:B-to-K-fv}
\end{eqnarray}
The masses and integrals that appear in Eqs.~(\ref{eq:B-to-pi-fv}) and
(\ref{eq:B-to-K-fv}) are as follows.
The flavor off-diagonal meson masses are:
\begin{align}
m_{\pi,\Xi}^2&= \mu(m_l+m_l)+a^2\Delta_\Xi  , \label{eq:app:ps_mass_start}\\
m_{K,\Xi}^2&= \mu(m_l+m_s)+a^2\Delta_\Xi ,\\
m_{S, \Xi}^2&= \mu(m_s+m_s)+a^2\Delta_\Xi ,
\end{align}
where $m_l$ and $m_s$ are sea-quark masses and the taste label
$\Xi$ has values $P$, $V$, $T$, $A$  and $I$.  
The masses of flavor-neutral mesons in the taste vector channel are~\cite{Aubin:2003mg}:
\begin{eqnarray}\label{eq:eigenvalues_V}
        m_{\pi^0,V}^2 & = & m_{U,V}^2 = m_{D,V}^2 = \mu(m_l+m_l)+a^2\Delta_V  ,  \\*
        m_{S,V}^2 & = & \mu(m_s+m_s)+a^2\Delta_V    , \\*
        m_{\eta,V}^2 & = & \frac{1}{2}\left( m_{U_V}^2 + m_{S_V}^2 +
        \frac{3}{4}a^2\delta'_V - Z
        \right)  ,\\*
        m_{\eta',V}^2 & = &  \frac{1}{2}\left( m_{U,V}^2 + m_{S,V}^2 +
        \frac{3}{4}a^2\delta'_V +  Z \right)  ,\\
         Z & \equiv &\sqrt{\left(m_{S,V}^2-m_{U,V}^2\right)^2
         - \frac{a^2\delta'_V}{2} 
        \left(m_{S,V}^2-m_{U,V}^2\right) +\frac{9(a^2\delta'_V)^2}{16}} .
\end{eqnarray}
The taste-axial case just requires substituting $A$ for $V$.  
For the taste-singlet case, we have:
\begin{eqnarray}\label{eq:eigenvalues_I}
        m_{\pi^0,I}^2 & = & m_{U,I}^2 = m_{D,I}^2 = \mu(m_l+m_l)+a^2\Delta_I  ,  \\*
        m_{S,I}^2 & = & \mu(m_h+m_h)+a^2\Delta_I   ,  \\*
        m_{\eta,I}^2 & = & \frac{m_{U,I}^2}{3}+
        \frac{2m_{S,I}^2}{3} ,\\  
        m_{\eta',I}^2 & = & m_0^2 . \label{eq:app:ps_mass_end}
\end{eqnarray}

The momentum integrals $I_1$ and $I_2$ that appear in the chiral log terms 
are defined as:
\begin{align}
I_1(m)  & =  m^2 \ln \left(\frac{m^2}{\Lambda ^2}\right) ,\\
I_2(m,\Delta)  & =  -2\Delta^2
        \ln \left(\frac{m^2}{\Lambda^2}\right)
        -4\Delta^2 F\left(\frac{m}{\Delta}\right)
        +2\Delta^2 ,\\
F(x)&=
\begin{cases}
 \sqrt{1-x^2}\tanh^{-1}{(\sqrt{1-x^2})} & 0 \le x \le 1 ,\\
 -\sqrt{x^2-1}\tan^{-1}{(\sqrt{x^2-1})} & x \ge 1 ,\\
 \end{cases}
\end{align}
where $\Lambda$ is the renormalization scale. 

Similarly, $f_\perp$ on the unitary points in NLO SU(3) HMrS$\chi$PT is 
given by~\cite{Aubin:2007mc}:
\begin{align}
f_\perp&=
\frac{C^{(0)}}{f}\left[ \frac{1}{E+\Delta^*_B+D} \right ] \nonumber \\ 
&+ \frac{C^{(0)} }{f(E+\Delta_{B^*})}( \text{logs}+C^{(1)} \chi_l+C^{(2)} 
\chi_s+C^{(3)} \chi_E+ C^{(4)}\chi_E^2+C^{(5)} \chi_{a^2}) , \label{eq:fperp-exp}
\end{align}
where $\Delta_{B_s^*} = M_{B^*_s} - M_{B}$. (The SU(3) expression has one extra chiral 
log term $D$ comparing with the 
SU(2) expression we used in our analysis.)
There are two chiral log related terms parameterized by
$D$ and logs in Eq.~(\ref{eq:fperp-exp}). 
For the $B\to \pi$ process, the SU(3) expressions are~\cite{Aubin:2007mc}:
\begin{eqnarray}
        D^{B\to\pi}_\text{SU(3)} & = &  - \frac{3g_\pi^2\vp}{(4 \pi f)^2}
  \Biggl\{ \frac{1}{16}\sum_{\Xi}
  \left[2J_1^{\rm sub}(m_{\pi,\Xi}, \vp )+
  J_1^{\rm sub}(m_{K,\Xi}, \vp) \right]
  \nonumber \\ &&{}
  -\frac{1}{2}J_1^{\rm sub}(m_{\pi,I}, \vp)
        +\frac{1}{6}J_1^{\rm sub}(m_{\eta,I}, \vp)
    \nonumber \\ &&{}+
        \sum_{j\in\{\pi,\eta,\eta'\}}
        \left[(- a^2\delta'_V)
        R^{[3,1]}_{j} \left(\{m_{\pi,V},m_{\eta,V},m_{\eta',V}\} ;
         \{m_{S,V}\}\right)
    J_1^{\rm sub}(m_{j,V}, \vp)\right] \nonumber\\
    &&{}+ \bigl[V\to A\bigr]
  \Biggr\} \ , \label{eq:B-to-pi-D}
\end{eqnarray}
and 
\begin{eqnarray}
\text{logs}^{B\to\pi}_{\perp, \text{SU(3)}} & = &
  \frac{1}{(4 \pi f)^2}\Biggl\{           \frac{1}{16}\sum_{\Xi}\left[
  -  \frac{1+3 g_\pi^2}{2}\left[ 2I_1(m_{\pi,\Xi})
  + I_1(m_{K,\Xi})\right]
  \right]\nonumber\\
  &&
   -\frac{1}{2}g^2_\pi J_1^{\rm sub}(m_{\pi,I}, \vp)
   +\frac{1}{6}g^2_\pi J_1^{\rm sub}(m_{\eta,I}, \vp)
  +\frac{1+3 g^2_\pi}{12}
  \biggl[3I_1(m_{\pi,I}) - I_1(m_{\eta,I}) \biggr]
  \nonumber\\&&
  {}+
  \sum_{j\in\{\pi,\eta,\eta'\}}
  \biggl[ a^2\delta'_V
    R^{[3,1]}_{j} \left(\{m_{\pi,V},m_{\eta,V},m_{\eta',V}\} ;
         \{m_{S,V}\}\right)\nonumber\\&&{}\times
    \left(
    g_\pi^2 J_1^{\rm sub}(m_{j,V}, \vp)
        +\frac{1+3g^2_\pi}{2}
    I_1(m_{j,V})\right) \biggr] + [V\to A]\Biggr\}\ . \label{eq:B-to-pi-fp}
\end{eqnarray}
For the $B\to K$ process, the SU(3) expressions are:
\begin{eqnarray}
        D^{B\to K}_\text{SU(3)} & = &  - \frac{3g_\pi^2(\vp)}{(4 \pi f)^2}
        \Biggl\{ \frac{1}{16}\sum_{\Xi}
        \left[2J_1^{\rm sub}(m_{K,\Xi}, \vp )+
        J_1^{\rm sub}(m_{S,\Xi}, \vp) \right]
        \nonumber \\ &&{}
        +\frac{2}{3}J_1^{\rm sub}(m_{\eta,I}, \vp)
        -J_1^{\rm sub}(m_{S,I}, \vp)
    \nonumber \\ &&{}+
        \sum_{j\in\{S,\eta,\eta'\}}
        \left[(- a^2\delta'_V)
        R^{[3,1]}_{j} \left(\{m_{S,V},m_{\eta,V},m_{\eta',V}\} ;
         \{m_{\pi,V}\}\right)
    J_1^{\rm sub}(m_{j,V}, \vp)\right] \nonumber\\
    &&{}+ \bigl[V\to A\bigr] \Biggr\} \ ,\label{eq:B-to-K-D}
\end{eqnarray}
and
\begin{eqnarray}
\text{logs}^{B\to K}_{\perp, \text{SU(3)}} & = &
        \frac{1}{(4 \pi f)^2}\Biggl\{           \frac{1}{16}\sum_{\Xi}\left[
        -  \frac{2+3 g_\pi^2}{2} I_1(m_{K,\Xi})
        -\frac{1}{2} I_1(m_{S,\Xi})
        -3g_\pi^2 I_1(m_{\pi,\Xi})
        \right]\hspace{1truecm}\nonumber\\&&
        -\frac{1}{3}g^2_\pi J_1^{\rm sub}(m_{\eta,I}, \vp)
        +\frac{3 g^2_\pi}{4}I_1(m_{\pi,I})
        - \frac{4+3 g^2_\pi}{12}I_1(m_{\eta,I})
        + \frac{1}{2}I_1(m_{S,I})
        \nonumber\\&&{}
        + a^2\delta'_V  \biggl[
        \frac{g_\pi^2}{m^2_{\eta',V}-m^2_{\eta,V}}
        \biggl(J_1^{\rm sub}(m_{\eta,V}, \vp)
        -J_1^{\rm sub}(m_{\eta',V}, \vp)
        \biggr)\nonumber\\&&{}
        +\frac{3g^2_\pi}{2}\sum_{j\in\{\pi,\eta,\eta'\}}
        R^{[3,1]}_{j} \left(\{m_{\pi,V},m_{\eta,V},m_{\eta',V}\} ;
        \{m_{S,V}\}\right) I_1(m_{j,V})
        \nonumber\\&&{}
        +\frac{1}{2}\sum_{j\in\{S,\eta,\eta'\}}
        R^{[3,1]}_{j} \left(\{m_{S,V},m_{\eta,V},m_{\eta',V}\} ;
        \{m_{\pi,V}\}\right) I_1(m_{j,V})
     \biggr] \nonumber \\
&&{} 
+ [V\to A]\Biggr\}\ . \label{eq:B-to-K-fp}
%%%
\end{eqnarray}
The definition of the meson mass terms and $I_1$ are the same as for
the $f_\parallel$ case. 
The $f_\perp$ expression has an extra function $J_1$, that is defined as:
\begin{align}
        J_1(m,\Delta)  & = 
        \left(-m^2 + \frac{2}{3}\Delta^2\right)
        \ln\left(\frac{m^2}{\Lambda^2}\right)
        +\frac{4}{3}(\Delta^2-m^2)
        F\left(\frac{m}{\Delta}\right)-
        \frac{10}{9}\Delta^2
        +\frac{4}{3}m^2  \label{eq:J1}\ , \\
        J_1^{\rm sub}(m,\Delta) &\equiv J_1(m,\Delta)
        - \frac{2\pi m^3}{3\Delta}\ .
\end{align}

\subsection{\boldmath\texorpdfstring{$f_\parallel$}{f\_parallel} and 
\texorpdfstring{$f_{\perp}$}{fperp} in SU(2) \texorpdfstring{HMrS$\chi$PT}{HMrSChPT}}

We derive the SU(2) formula for $f_\parallel$ and $f_\perp$ based on the SU(3) expression.
We also use the same expression for $f_T$ as for $f_\perp$ as discussed in Sec.~\ref{subsec:chiral_fit}.
To obtain the SU(2) limit of an SU(3) expression, we treat the strange-quark mass as infinitely heavy.
The SU(2) form does not contain $m_s$ explicitly, but all LECs depend implicitly on $m_s$.
Because our lattice data have slightly different $m_s$ on different ensembles, we keep the analytic term
which is proportional to $m_s$.
Next, we consider all terms in the SU(3) chiral log expression.
If a term is proportional to $m_s$ or $\ln m_s$ in the large $m_s$ limit, it is absorbed into the
redefinition of other LECs.
If a term is proportional to $1/m_s$ or $1/\ln m_s$ in the large $m_s$ limit, it does not appear in the
SU(2) expression.
We now derive the form of the chiral log terms in the SU(2) limit.

For the chiral log terms in $f_\parallel$, because we take $m_s$ to 
infinity, all $m_s$ related terms, such as $m_{K,\Xi}$, $m_{S,\Xi}$, $m_{\eta,I}$ and $m_{\eta^\prime,V/A/I}$, 
go to infinity. They are absorbed into LECs. Only $m_{\eta,V/A}$ is finite and 
goes like $\sqrt{m_U^2+\frac{\delta^\prime_{V/A}}{2}}$. 
We now consider all contributing chiral log terms.
\begin{itemize}
\item $I_1(m)$ goes like $m^2\ln m^2$, so only $I_1(M_\pi)$ survives.
\item $I_2(m,E)$ diverges as $2\pi mE$ when $m\to \infty$, so only $I_2(M_\pi,E)$ survives.
\item The ratio
\begin{equation}
  \frac{I_1(m_{\eta',V}) - I_1(m_{\eta,V})
    +I_2(m_{\eta',V},E) - I_2(m_{\eta,V},E)}
       {m^2_{\eta',V} - m^2_{\eta,V}}      , \nonumber
\end{equation}
diverges as $2\ln m_s$ at large $m_s$, so it is removed.
\item We find that
\begin{align}
    \lim_{m_s\to\infty}a^2\delta^\prime_{V/A} R^{[3,1]}_{j} \left(\{m_{S,V},m_{\eta,V},m_{\eta',V}\} ;
        \{m_{\pi,V}\}\right) =
\begin{cases}
 4, & j=S \\
-\frac{a^4\delta^{\prime 2}_{V/A}}{2m_S^4}=0, & j=\eta \\
-4, & j=\eta^\prime ,\\
 \end{cases} .
\end{align}
When this term multiplies $I_1$ or $I_2$, it is divergent as $m_s \to \infty$ for $j=S$ or $j=\eta^\prime$. 
For $j=\eta$, the $I_1$ and $I_2$ are finite, but the total contribution is zero as $m_s \to \infty$.
So these terms are removed.
\item We find that
\begin{align}
    \lim_{m_s\to\infty}a^2\delta^\prime_{V/A} R^{[3,1]}_{j} \left(\{m_{\pi,V},m_{\eta,V},m_{\eta',V}\} ;
        \{m_{S,V}\}\right) =
\begin{cases}
 2, & j=\pi \\
-2, & j=\eta \\
-\frac{a^4\delta^{\prime 2}_{V/A}}{4m_S^4}=0, & j=\eta^\prime \\
 \end{cases},
\end{align}
so only $j=\pi$ and $j=\eta$ terms contribute in the SU(2) theory.
\end{itemize}

In summary, for the $B\to \pi$ process, the chiral log in SU(2) HMrS$\chi$PT is
given by
\begin{align}
\text{logs}^{B\to\pi}_{\parallel, \text{SU(2)}} & =
  \frac{1}{(4 \pi f)^2}\left\{           \frac{1}{16}\sum_{\Xi}\left[
{(1-3g^2_\pi)}I_1(m_{\pi,\Xi})     
  +2I_2(m_{\pi,\Xi},\vp)     \right]
+\frac{1+3g^2_\pi}{4}I_1(m_{\pi,I}) \right. \nonumber\\
       & { } +
2\left[\frac{3(g^2_\pi-1)}{2}I_1(m_{\pi,V})-2  I_2(m_{\pi,V},\vp) \right] 
-2\left[\frac{3(g^2_\pi-1)}{2}I_1(m_{\eta,V})-2  I_2(m_{\eta,V},\vp) \right] \nonumber\\
       & { } +
 \left. [V\to A] \vphantom{\sum_{\Xi}\frac{1+3g^2_\pi}{4}} \right\}\ . 
\end{align}
For the $B\to K$ process, the chiral log in SU(2) HMrS$\chi$PT is given by
\begin{align}
\text{logs}^{B\to K}_{\parallel, \text{SU(2)}} & =
        \frac{1}{(4 \pi f)^2}\left\{
\frac{1}{16}\sum_{\Xi}\left[
    - 3g^2_\pi I_1(m_{\pi,\Xi}) \right]
        + \frac{3 g^2_\pi}{4} I_1(m_{\pi,I}) \right.
        \nonumber\\ & {}
        + \frac{3g^2_\pi}{2}\left[2I_1(m_{\pi,V})-2I_1(m_{\eta,V})\right] +
 \left. [V\to A] \vphantom{\sum_{\Xi}\frac{1+3g^2_\pi}{4}} \right\}\ . 
\label{eq:app:su2_fpara_log}
\end{align}

We then derive the expression for the  $f_\perp$ chiral log terms in SU(2) HMrS$\chi$PT. We use the 
same treatment of analytic terms as was done for $f_\parallel$. 
To calculate the SU(2) chiral log terms, we consider the large $m_s$ limit 
of $J_1$:
\begin{flalign}
& \lim_{m\to\infty}J_1(m,E)  \to  -m^2\ln{m^2}  \to -\infty  ,\\
& \lim_{m_s\to\infty}\frac{J_1^{\rm sub}(m_{\eta,V}, \vp)-J_1^{\rm sub}(m_{\eta',V}, \vp) }
{m^2_{\eta',V}-m^2_{\eta,V}}   \to  2\ln {m_s} \to \infty  .
\end{flalign}
So all $J_1$ related terms are absorbed into the redefinition of LECs and 
 disappear. 

Via a procedure similar to that for $f_\parallel$, we obtain the SU(2) chiral log terms in $f_\perp$ for the
$B\to \pi$ channel:
\begin{align}
        D^{B\to\pi}_\text{SU(2)} & =  - \frac{3g_\pi^2\vp}{(4 \pi f)^2}
  \left\{ \frac{1}{16}\sum_{\Xi}
  \left[2J_1^{\rm sub}(m_{\pi,\Xi}, \vp )\right]
  -\frac{1}{2}J_1^{\rm sub}(m_{\pi,I}, \vp) \right.
    \nonumber \\ & -
        \left.\left[
    2J_1^{\rm sub}(m_{\pi,V}, \vp)-2J_1^{\rm sub}(m_{\eta,V}, \vp)
\right] + [V\to A] \vphantom{\frac{1}{16}\sum_\Xi}
  \right\},
    \label{eq:app:su2_fperp_D_B2pi}
\end{align}
\begin{align}
\text{logs}^{B\to\pi}_{\perp, \text{SU(2)}} & =
  \frac{1}{(4 \pi f)^2}\Biggl\{           \frac{1}{16}\sum_{\Xi}\left[
  -  \frac{1+3 g_\pi^2}{2}\left[ 2I_1(m_{\pi,\Xi})\right]  \right]
   -\frac{1}{2}g^2_\pi J_1^{\rm sub}(m_{\pi,I}, \vp)
  \nonumber\\&
  +\frac{1+3 g^2_\pi}{12}
  \biggl[3I_1(m_{\pi,I}) \biggr]
  + \biggl[
    2\left(g_\pi^2 J_1^{\rm sub}(m_{\pi,V}, \vp) +\frac{1+3g^2_\pi}{2} I_1(m_{\pi,V})\right) \nonumber\\&
    -2\left(g_\pi^2 J_1^{\rm sub}(m_{\eta,V}, \vp) +\frac{1+3g^2_\pi}{2} I_1(m_{\eta,V})\right) 
    \biggr] + [V\to A]\Biggr\}.
    \label{eq:app:su2_fperp_log_B2pi}
\end{align}

Similarly, the SU(2) chiral log terms in $B\to K$ are:
\begin{equation}
    D^{B\to K}_\text{SU(2)} = 0 ,
    \label{eq:app:su2_fperp_D}
\end{equation}
\begin{align}
    \text{logs}^{B\to K}_{\perp, \text{SU(2)}} & =
        \frac{1}{(4 \pi f)^2}\Biggl\{ \frac{1}{16}\sum_{\Xi}\left[
        -3g_\pi^2 I_1(m_{\pi,\Xi})
        \right]
        +\frac{3 g^2_\pi}{4}I_1(m_{\pi,I})
        \nonumber\\&
        +\frac{3g^2_\pi}{2}
        \left[2I_1(m_{\pi,V})-2I_1(m_{\eta,V})\right]
        + [V\to A]\Biggr\}\ .
    \label{eq:app:su2_fperp_log}
\end{align}
Equations~(\ref{eq:app:su2_fperp_D_B2pi}), (\ref{eq:app:su2_fperp_log_B2pi}),
and~(\ref{eq:app:su2_fperp_log}) are written with a structure similar to
their SU(3) counterparts, which makes it easier to implement a unified 
computer code for the various choices of $\chi$PT studied in this paper.

\subsection{Form factors in hard-pion/kaon ChPT}
The hard-kaon (pion) continuum HM$\chi$PT for $B\to K$ and $B\to \pi$ semileptonic 
decays was derived in Refs.~\cite{Bijnens:2010ws,Bijnens:2010jg}. 
The pion or kaon with large $E$ is integrated out from the theory and its effects 
are absorbed into the LECs.
We derive the hard-kaon (pion) limit of the HMrS$\chi$PT in this section. We first study the asymptotic 
behavior of the integrals which contain $E_{\pi}$ or $E_K$. We find that
\begin{align}
I_2(m,E)& \to  A_0 E^2 \ln (E^2)+A_1 E^2 +A_2 \ln E -m^2 \ln (\frac{m^2}{\Lambda^2}) \ , \\
J_1^{\rm sub}(m,E)& \to  B_0 E^2 \ln (E^2)+B_1 E^2 +B_2 \ln E +B_3  \ ,
\end{align}
in the large $E$ limit, where the coefficients $A_i$ and $B_i$ are either constants or analytic 
functions of $m$. The divergent terms in the large $E$ limit  decouple from the 
expression. The analytic terms in $m$ are absorbed into the redefinition of the 
LECs. So the rules to derive the hard-kaon (pion) HMrS$\chi$PT are the following:
\begin{itemize}
\item{Replace the term $I_2(m,E)$ by $-I_1(m)$}
\item{Remove $J_1^{\rm sub}(m,E)$ term}
\end{itemize}
To compare our results with Refs.~\cite{Bijnens:2010ws,Bijnens:2010jg}, 
we set all taste splitting parameters, hairpin parameters and lattice spacings 
to zero. We then can reproduce the continuum 
hard-kaon (pion) HM$\chi$PT results.
\begin{align}
\text{logs}^{B\to \pi}_{\perp, \rm SU(3)} & = 
  -(\frac{3}{4}+\frac{9}{4}g^2_\pi)\frac{I_1(m_{\pi})}{(4 \pi f)^2}
  -(\frac{1}{2}+\frac{3}{2}g^2_\pi)\frac{I_1(m_{K})}{(4 \pi f)^2}
  -(\frac{1}{12}+\frac{1}{4}g^2_\pi)\frac{I_1(m_{\eta})}{(4 \pi f)^2}  \ , \\
\text{logs}^{B\to K}_{\perp, \rm SU(3)} & = 
  -(\frac{9}{4}g^2_\pi)\frac{I_1(m_{\pi})}{(4 \pi f)^2}
  -(1+\frac{3}{2}g^2_\pi)\frac{I_1(m_{K})}{(4 \pi f)^2}
  -(\frac{1}{3}+\frac{1}{4}g^2_\pi)\frac{I_1(m_{\eta})}{(4 \pi f)^2} \ , \\
D & = 0 \ .
\end{align}
Our derivation shows that $\text{logs}^{B\to \pi}_{\perp\rm SU(3)}=\text{logs}^{B\to \pi}_{\parallel\rm SU(3)}$
and $\text{logs}^{B\to K}_{\perp\rm SU(3)}=\text{logs}^{B\to K}_{\parallel\rm SU(3)}$ in the continuum,
which is also found in Refs.~\cite{Bijnens:2010ws,Bijnens:2010jg}.

\bibliographystyle{apsrev4-1}
\bibliography{paper}

%merlin.mbs apsrev4-1.bst 2010-07-25 4.21a (PWD, AO, DPC) hacked
%Control: key (0)
%Control: author (72) initials jnrlst
%Control: editor formatted (1) identically to author
%Control: production of article title (-1) disabled
%Control: page (0) single
%Control: year (1) truncated
%Control: production of eprint (0) enabled
\begin{thebibliography}{156}%
\makeatletter
\providecommand \@ifxundefined [1]{%
 \@ifx{#1\undefined}
}%
\providecommand \@ifnum [1]{%
 \ifnum #1\expandafter \@firstoftwo
 \else \expandafter \@secondoftwo
 \fi
}%
\providecommand \@ifx [1]{%
 \ifx #1\expandafter \@firstoftwo
 \else \expandafter \@secondoftwo
 \fi
}%
\providecommand \natexlab [1]{#1}%
\providecommand \enquote  [1]{``#1''}%
\providecommand \bibnamefont  [1]{#1}%
\providecommand \bibfnamefont [1]{#1}%
\providecommand \citenamefont [1]{#1}%
\providecommand \href@noop [0]{\@secondoftwo}%
\providecommand \href [0]{\begingroup \@sanitize@url \@href}%
\providecommand \@href[1]{\@@startlink{#1}\@@href}%
\providecommand \@@href[1]{\endgroup#1\@@endlink}%
\providecommand \@sanitize@url [0]{\catcode `\\12\catcode `\$12\catcode
  `\&12\catcode `\#12\catcode `\^12\catcode `\_12\catcode `\%12\relax}%
\providecommand \@@startlink[1]{}%
\providecommand \@@endlink[0]{}%
\providecommand \url  [0]{\begingroup\@sanitize@url \@url }%
\providecommand \@url [1]{\endgroup\@href {#1}{\urlprefix }}%
\providecommand \urlprefix  [0]{URL }%
\providecommand \Eprint [0]{\href }%
\providecommand \doibase [0]{http://dx.doi.org/}%
\providecommand \selectlanguage [0]{\@gobble}%
\providecommand \bibinfo  [0]{\@secondoftwo}%
\providecommand \bibfield  [0]{\@secondoftwo}%
\providecommand \translation [1]{[#1]}%
\providecommand \BibitemOpen [0]{}%
\providecommand \bibitemStop [0]{}%
\providecommand \bibitemNoStop [0]{.\EOS\space}%
\providecommand \EOS [0]{\spacefactor3000\relax}%
\providecommand \BibitemShut  [1]{\csname bibitem#1\endcsname}%
\let\auto@bib@innerbib\@empty
%</preamble>
\bibitem [{\citenamefont {Bailey}\ \emph {et~al.}(2009)\citenamefont {Bailey}
  \emph {et~al.}}]{Bailey:2008wp}%
  \BibitemOpen
  \bibfield  {author} {\bibinfo {author} {\bibfnamefont {J.~A.}\ \bibnamefont
  {Bailey}} \emph {et~al.} (\bibinfo {collaboration} {Fermilab Lattice and MILC
  Collaborations}),\ }\href {\doibase 10.1103/PhysRevD.79.054507} {\bibfield
  {journal} {\bibinfo  {journal} {Phys. Rev.}\ }\textbf {\bibinfo {volume}
  {D79}},\ \bibinfo {pages} {054507} (\bibinfo {year} {2009})},\ \Eprint
  {http://arxiv.org/abs/0811.3640} {arXiv:0811.3640 [hep-lat]} \BibitemShut
  {NoStop}%
%%CITATION = ARXIV:0811.3640;%%
\bibitem [{\citenamefont {Bailey}\ \emph
  {et~al.}(2015{\natexlab{a}})\citenamefont {Bailey} \emph
  {et~al.}}]{Lattice:2015tia}%
  \BibitemOpen
  \bibfield  {author} {\bibinfo {author} {\bibfnamefont {J.~A.}\ \bibnamefont
  {Bailey}} \emph {et~al.} (\bibinfo {collaboration} {Fermilab Lattice and MILC
  Collaboration}),\ }\href@noop {} {\  (\bibinfo {year}
  {2015}{\natexlab{a}})},\ \Eprint {http://arxiv.org/abs/1503.07839}
  {arXiv:1503.07839 [hep-lat]} \BibitemShut {NoStop}%
%%CITATION = ARXIV:1503.07839;%%
\bibitem [{\citenamefont {Bernard}\ \emph {et~al.}(2009)\citenamefont {Bernard}
  \emph {et~al.}}]{Bernard:2008dn}%
  \BibitemOpen
  \bibfield  {author} {\bibinfo {author} {\bibfnamefont {C.}~\bibnamefont
  {Bernard}} \emph {et~al.} (\bibinfo {collaboration} {Fermilab Lattice and
  MILC Collaborations}),\ }\href {\doibase 10.1103/PhysRevD.79.014506}
  {\bibfield  {journal} {\bibinfo  {journal} {Phys. Rev.}\ }\textbf {\bibinfo
  {volume} {D79}},\ \bibinfo {pages} {014506} (\bibinfo {year} {2009})},\
  \Eprint {http://arxiv.org/abs/0808.2519} {arXiv:0808.2519 [hep-lat]}
  \BibitemShut {NoStop}%
%%CITATION = ARXIV:0808.2519;%%
\bibitem [{\citenamefont {Bailey}\ \emph
  {et~al.}(2014{\natexlab{a}})\citenamefont {Bailey} \emph
  {et~al.}}]{Bailey:2014tva}%
  \BibitemOpen
  \bibfield  {author} {\bibinfo {author} {\bibfnamefont {J.~A.}\ \bibnamefont
  {Bailey}} \emph {et~al.} (\bibinfo {collaboration} {Fermilab Lattice and MILC
  Collaborations}),\ }\href {\doibase 10.1103/PhysRevD.89.114504} {\bibfield
  {journal} {\bibinfo  {journal} {Phys. Rev.}\ }\textbf {\bibinfo {volume}
  {D89}},\ \bibinfo {pages} {114504} (\bibinfo {year} {2014}{\natexlab{a}})},\
  \Eprint {http://arxiv.org/abs/1403.0635} {arXiv:1403.0635 [hep-lat]}
  \BibitemShut {NoStop}%
%%CITATION = ARXIV:1403.0635;%%
\bibitem [{\citenamefont {Bailey}\ \emph
  {et~al.}(2015{\natexlab{b}})\citenamefont {Bailey} \emph
  {et~al.}}]{Lattice:2015rga}%
  \BibitemOpen
  \bibfield  {author} {\bibinfo {author} {\bibfnamefont {J.~A.}\ \bibnamefont
  {Bailey}} \emph {et~al.} (\bibinfo {collaboration} {Fermilab Lattice and MILC
  Collaborations}),\ }\href@noop {} {\  (\bibinfo {year}
  {2015}{\natexlab{b}})},\ \Eprint {http://arxiv.org/abs/1503.07237}
  {arXiv:1503.07237 [hep-lat]} \BibitemShut {NoStop}%
%%CITATION = ARXIV:1503.07237;%%
\bibitem [{\citenamefont {Olive}\ \emph {et~al.}(2014)\citenamefont {Olive}
  \emph {et~al.}}]{Agashe:2014kda}%
  \BibitemOpen
  \bibfield  {author} {\bibinfo {author} {\bibfnamefont {K.~A.}\ \bibnamefont
  {Olive}} \emph {et~al.} (\bibinfo {collaboration} {Particle Data Group}),\
  }\href {\doibase 10.1088/1674-1137/38/9/090001} {\bibfield  {journal}
  {\bibinfo  {journal} {Chin. Phys.}\ }\textbf {\bibinfo {volume} {C38}},\
  \bibinfo {pages} {090001} (\bibinfo {year} {2014})}\BibitemShut {NoStop}%
%%CITATION = CHPHD,C38,090001;%%
\bibitem [{\citenamefont {Bailey}\ \emph
  {et~al.}(2015{\natexlab{c}})\citenamefont {Bailey} \emph
  {et~al.}}]{Bailey:2015nbd}%
  \BibitemOpen
  \bibfield  {author} {\bibinfo {author} {\bibfnamefont {J.~A.}\ \bibnamefont
  {Bailey}} \emph {et~al.} (\bibinfo {collaboration} {Fermilab Lattice,
  MILC}),\ }\href@noop {} {\  (\bibinfo {year} {2015}{\natexlab{c}})},\ \Eprint
  {http://arxiv.org/abs/1507.01618} {arXiv:1507.01618 [hep-ph]} \BibitemShut
  {NoStop}%
%%CITATION = ARXIV:1507.01618;%%
\bibitem [{\citenamefont {Bailey}\ \emph
  {et~al.}(2012{\natexlab{a}})\citenamefont {Bailey} \emph
  {et~al.}}]{Bailey:2012rr}%
  \BibitemOpen
  \bibfield  {author} {\bibinfo {author} {\bibfnamefont {J.~A.}\ \bibnamefont
  {Bailey}} \emph {et~al.} (\bibinfo {collaboration} {Fermilab Lattice and MILC
  Collaborations}),\ }\href {\doibase 10.1103/PhysRevD.85.114502} {\bibfield
  {journal} {\bibinfo  {journal} {Phys. Rev.}\ }\textbf {\bibinfo {volume}
  {D85}},\ \bibinfo {pages} {114502} (\bibinfo {year} {2012}{\natexlab{a}})},\
  \Eprint {http://arxiv.org/abs/1202.6346} {arXiv:1202.6346 [hep-lat]}
  \BibitemShut {NoStop}%
%%CITATION = ARXIV:1202.6346;%%
\bibitem [{\citenamefont {Bailey}\ \emph
  {et~al.}(2012{\natexlab{b}})\citenamefont {Bailey} \emph
  {et~al.}}]{Bailey:2012jg}%
  \BibitemOpen
  \bibfield  {author} {\bibinfo {author} {\bibfnamefont {J.~A.}\ \bibnamefont
  {Bailey}} \emph {et~al.} (\bibinfo {collaboration} {Fermilab Lattice and MILC
  Collaborations}),\ }\href {\doibase 10.1103/PhysRevLett.109.071802}
  {\bibfield  {journal} {\bibinfo  {journal} {Phys. Rev. Lett.}\ }\textbf
  {\bibinfo {volume} {109}},\ \bibinfo {pages} {071802} (\bibinfo {year}
  {2012}{\natexlab{b}})},\ \Eprint {http://arxiv.org/abs/1206.4992}
  {arXiv:1206.4992 [hep-ph]} \BibitemShut {NoStop}%
%%CITATION = ARXIV:1206.4992;%%
\bibitem [{\citenamefont {Hurth}\ and\ \citenamefont
  {Nakao}(2010)}]{Hurth:2010tk}%
  \BibitemOpen
  \bibfield  {author} {\bibinfo {author} {\bibfnamefont {T.}~\bibnamefont
  {Hurth}}\ and\ \bibinfo {author} {\bibfnamefont {M.}~\bibnamefont {Nakao}},\
  }\href {\doibase 10.1146/annurev.nucl.012809.104424} {\bibfield  {journal}
  {\bibinfo  {journal} {Annu. Rev. Nucl. Part. Sci.}\ }\textbf {\bibinfo
  {volume} {60}},\ \bibinfo {pages} {645} (\bibinfo {year} {2010})},\ \Eprint
  {http://arxiv.org/abs/1005.1224} {arXiv:1005.1224 [hep-ph]} \BibitemShut
  {NoStop}%
%%CITATION = ARXIV:1005.1224;%%
\bibitem [{\citenamefont {Antonelli}\ \emph {et~al.}(2010)\citenamefont
  {Antonelli} \emph {et~al.}}]{Antonelli:2009ws}%
  \BibitemOpen
  \bibfield  {author} {\bibinfo {author} {\bibfnamefont {M.}~\bibnamefont
  {Antonelli}} \emph {et~al.},\ }\href {\doibase 10.1016/j.physrep.2010.05.003}
  {\bibfield  {journal} {\bibinfo  {journal} {Phys. Rept.}\ }\textbf {\bibinfo
  {volume} {494}},\ \bibinfo {pages} {197} (\bibinfo {year} {2010})},\ \Eprint
  {http://arxiv.org/abs/0907.5386} {arXiv:0907.5386 [hep-ph]} \BibitemShut
  {NoStop}%
%%CITATION = ARXIV:0907.5386;%%
\bibitem [{\citenamefont {Aubert}\ \emph {et~al.}(2009)\citenamefont {Aubert}
  \emph {et~al.}}]{Aubert:2008ps}%
  \BibitemOpen
  \bibfield  {author} {\bibinfo {author} {\bibfnamefont {B.}~\bibnamefont
  {Aubert}} \emph {et~al.} (\bibinfo {collaboration} {BaBar Collaboration}),\
  }\href {\doibase 10.1103/PhysRevLett.102.091803} {\bibfield  {journal}
  {\bibinfo  {journal} {Phys. Rev. Lett.}\ }\textbf {\bibinfo {volume} {102}},\
  \bibinfo {pages} {091803} (\bibinfo {year} {2009})},\ \Eprint
  {http://arxiv.org/abs/0807.4119} {arXiv:0807.4119 [hep-ex]} \BibitemShut
  {NoStop}%
%%CITATION = ARXIV:0807.4119;%%
\bibitem [{\citenamefont {Wei}\ \emph {et~al.}(2009)\citenamefont {Wei} \emph
  {et~al.}}]{Wei:2009zv}%
  \BibitemOpen
  \bibfield  {author} {\bibinfo {author} {\bibfnamefont {J.-T.}\ \bibnamefont
  {Wei}} \emph {et~al.} (\bibinfo {collaboration} {Belle Collaboration}),\
  }\href {\doibase 10.1103/PhysRevLett.103.171801} {\bibfield  {journal}
  {\bibinfo  {journal} {Phys. Rev. Lett.}\ }\textbf {\bibinfo {volume} {103}},\
  \bibinfo {pages} {171801} (\bibinfo {year} {2009})},\ \Eprint
  {http://arxiv.org/abs/0904.0770} {arXiv:0904.0770 [hep-ex]} \BibitemShut
  {NoStop}%
%%CITATION = ARXIV:0904.0770;%%
\bibitem [{\citenamefont {Aaltonen}\ \emph
  {et~al.}(2011{\natexlab{a}})\citenamefont {Aaltonen} \emph
  {et~al.}}]{Aaltonen:2011cn}%
  \BibitemOpen
  \bibfield  {author} {\bibinfo {author} {\bibfnamefont {T.}~\bibnamefont
  {Aaltonen}} \emph {et~al.} (\bibinfo {collaboration} {CDF Collaboration}),\
  }\href {\doibase 10.1103/PhysRevLett.106.161801} {\bibfield  {journal}
  {\bibinfo  {journal} {Phys. Rev. Lett.}\ }\textbf {\bibinfo {volume} {106}},\
  \bibinfo {pages} {161801} (\bibinfo {year} {2011}{\natexlab{a}})},\ \Eprint
  {http://arxiv.org/abs/1101.1028} {arXiv:1101.1028 [hep-ex]} \BibitemShut
  {NoStop}%
%%CITATION = ARXIV:1101.1028;%%
\bibitem [{\citenamefont {Aaltonen}\ \emph
  {et~al.}(2011{\natexlab{b}})\citenamefont {Aaltonen} \emph
  {et~al.}}]{Aaltonen:2011qs}%
  \BibitemOpen
  \bibfield  {author} {\bibinfo {author} {\bibfnamefont {T.}~\bibnamefont
  {Aaltonen}} \emph {et~al.} (\bibinfo {collaboration} {CDF Collaboration}),\
  }\href {\doibase 10.1103/PhysRevLett.107.201802} {\bibfield  {journal}
  {\bibinfo  {journal} {Phys. Rev. Lett.}\ }\textbf {\bibinfo {volume} {107}},\
  \bibinfo {pages} {201802} (\bibinfo {year} {2011}{\natexlab{b}})},\ \Eprint
  {http://arxiv.org/abs/1107.3753} {arXiv:1107.3753 [hep-ex]} \BibitemShut
  {NoStop}%
%%CITATION = ARXIV:1107.3753;%%
\bibitem [{\citenamefont {Lees}\ \emph {et~al.}(2012)\citenamefont {Lees} \emph
  {et~al.}}]{Lees:2012tva}%
  \BibitemOpen
  \bibfield  {author} {\bibinfo {author} {\bibfnamefont {J.~P.}\ \bibnamefont
  {Lees}} \emph {et~al.} (\bibinfo {collaboration} {BaBar Collaboration}),\
  }\href {\doibase 10.1103/PhysRevD.86.032012} {\bibfield  {journal} {\bibinfo
  {journal} {Phys. Rev.}\ }\textbf {\bibinfo {volume} {D86}},\ \bibinfo {pages}
  {032012} (\bibinfo {year} {2012})},\ \Eprint {http://arxiv.org/abs/1204.3933}
  {arXiv:1204.3933 [hep-ex]} \BibitemShut {NoStop}%
%%CITATION = ARXIV:1204.3933;%%
\bibitem [{\citenamefont {Aaij}\ \emph {et~al.}(2012)\citenamefont {Aaij} \emph
  {et~al.}}]{Aaij:2012cq}%
  \BibitemOpen
  \bibfield  {author} {\bibinfo {author} {\bibfnamefont {R.}~\bibnamefont
  {Aaij}} \emph {et~al.} (\bibinfo {collaboration} {LHCb Collaboration}),\
  }\href {\doibase 10.1007/JHEP07(2012)133} {\bibfield  {journal} {\bibinfo
  {journal} {JHEP}\ }\textbf {\bibinfo {volume} {1207}},\ \bibinfo {pages}
  {133} (\bibinfo {year} {2012})},\ \Eprint {http://arxiv.org/abs/1205.3422}
  {arXiv:1205.3422 [hep-ex]} \BibitemShut {NoStop}%
%%CITATION = ARXIV:1205.3422;%%
\bibitem [{\citenamefont {Aaij}\ \emph {et~al.}(2013)\citenamefont {Aaij} \emph
  {et~al.}}]{Aaij:2012vr}%
  \BibitemOpen
  \bibfield  {author} {\bibinfo {author} {\bibfnamefont {R.}~\bibnamefont
  {Aaij}} \emph {et~al.} (\bibinfo {collaboration} {LHCb Collaboration}),\
  }\href {\doibase 10.1007/JHEP02(2013)105} {\bibfield  {journal} {\bibinfo
  {journal} {JHEP}\ }\textbf {\bibinfo {volume} {1302}},\ \bibinfo {pages}
  {105} (\bibinfo {year} {2013})},\ \Eprint {http://arxiv.org/abs/1209.4284}
  {arXiv:1209.4284 [hep-ex]} \BibitemShut {NoStop}%
%%CITATION = ARXIV:1209.4284;%%
\bibitem [{\citenamefont {Aaij}\ \emph {et~al.}(2014)\citenamefont {Aaij} \emph
  {et~al.}}]{Aaij:2014pli}%
  \BibitemOpen
  \bibfield  {author} {\bibinfo {author} {\bibfnamefont {R.}~\bibnamefont
  {Aaij}} \emph {et~al.} (\bibinfo {collaboration} {LHCb collaboration}),\
  }\href {\doibase 10.1007/JHEP06(2014)133} {\bibfield  {journal} {\bibinfo
  {journal} {JHEP}\ }\textbf {\bibinfo {volume} {1406}},\ \bibinfo {pages}
  {133} (\bibinfo {year} {2014})},\ \Eprint {http://arxiv.org/abs/1403.8044}
  {arXiv:1403.8044 [hep-ex]} \BibitemShut {NoStop}%
%%CITATION = ARXIV:1403.8044;%%
\bibitem [{\citenamefont {Hewett}\ \emph {et~al.}(2012)\citenamefont {Hewett},
  \citenamefont {Weerts} \emph {et~al.}}]{Hewett:2012ns}%
  \BibitemOpen
  \bibfield  {author} {\bibinfo {author} {\bibfnamefont {J.~L.}\ \bibnamefont
  {Hewett}}, \bibinfo {author} {\bibfnamefont {H.}~\bibnamefont {Weerts}},
  \emph {et~al.},\ }\href@noop {} {\emph {\bibinfo {title} {{Fundamental
  Physics at the Intensity Frontier}}}}\ (\bibinfo  {publisher} {{U.S.
  Department of Energy}},\ \bibinfo {year} {2012})\ \Eprint
  {http://arxiv.org/abs/1205.2671} {arXiv:1205.2671 [hep-ex]} \BibitemShut
  {NoStop}%
%%CITATION = ARXIV:1205.2671;%%
\bibitem [{\citenamefont {Bouchard}\ \emph
  {et~al.}(2013{\natexlab{a}})\citenamefont {Bouchard}, \citenamefont {Lepage},
  \citenamefont {Monahan}, \citenamefont {Na},\ and\ \citenamefont
  {Shigemitsu}}]{Bouchard:2013eph}%
  \BibitemOpen
  \bibfield  {author} {\bibinfo {author} {\bibfnamefont {C.}~\bibnamefont
  {Bouchard}}, \bibinfo {author} {\bibfnamefont {G.~P.}\ \bibnamefont
  {Lepage}}, \bibinfo {author} {\bibfnamefont {C.}~\bibnamefont {Monahan}},
  \bibinfo {author} {\bibfnamefont {H.}~\bibnamefont {Na}}, \ and\ \bibinfo
  {author} {\bibfnamefont {J.}~\bibnamefont {Shigemitsu}} (\bibinfo
  {collaboration} {HPQCD Collaboration}),\ }\href {\doibase
  10.1103/PhysRevD.88.079901, 10.1103/PhysRevD.88.054509} {\bibfield  {journal}
  {\bibinfo  {journal} {Phys. Rev.}\ }\textbf {\bibinfo {volume} {D88}},\
  \bibinfo {pages} {054509} (\bibinfo {year} {2013}{\natexlab{a}})},\ \Eprint
  {http://arxiv.org/abs/1306.2384} {arXiv:1306.2384 [hep-lat]} \BibitemShut
  {NoStop}%
%%CITATION = ARXIV:1306.2384;%%
\bibitem [{\citenamefont {Bouchard}\ \emph
  {et~al.}(2013{\natexlab{b}})\citenamefont {Bouchard}, \citenamefont {Lepage},
  \citenamefont {Monahan}, \citenamefont {Na},\ and\ \citenamefont
  {Shigemitsu}}]{Bouchard:2013mia}%
  \BibitemOpen
  \bibfield  {author} {\bibinfo {author} {\bibfnamefont {C.}~\bibnamefont
  {Bouchard}}, \bibinfo {author} {\bibfnamefont {G.~P.}\ \bibnamefont
  {Lepage}}, \bibinfo {author} {\bibfnamefont {C.}~\bibnamefont {Monahan}},
  \bibinfo {author} {\bibfnamefont {H.}~\bibnamefont {Na}}, \ and\ \bibinfo
  {author} {\bibfnamefont {J.}~\bibnamefont {Shigemitsu}} (\bibinfo
  {collaboration} {HPQCD Collaboration}),\ }\href {\doibase
  10.1103/PhysRevLett.112.149902, 10.1103/PhysRevLett.111.162002} {\bibfield
  {journal} {\bibinfo  {journal} {Phys. Rev. Lett.}\ }\textbf {\bibinfo
  {volume} {111}},\ \bibinfo {pages} {162002} (\bibinfo {year}
  {2013}{\natexlab{b}})},\ \Eprint {http://arxiv.org/abs/1306.0434}
  {arXiv:1306.0434 [hep-ph]} \BibitemShut {NoStop}%
%%CITATION = ARXIV:1306.0434;%%
\bibitem [{\citenamefont {Horgan}\ \emph {et~al.}(2014)\citenamefont {Horgan},
  \citenamefont {Liu}, \citenamefont {Meinel},\ and\ \citenamefont
  {Wingate}}]{Horgan:2013hoa}%
  \BibitemOpen
  \bibfield  {author} {\bibinfo {author} {\bibfnamefont {R.~R.}\ \bibnamefont
  {Horgan}}, \bibinfo {author} {\bibfnamefont {Z.}~\bibnamefont {Liu}},
  \bibinfo {author} {\bibfnamefont {S.}~\bibnamefont {Meinel}}, \ and\ \bibinfo
  {author} {\bibfnamefont {M.}~\bibnamefont {Wingate}},\ }\href {\doibase
  10.1103/PhysRevD.89.094501} {\bibfield  {journal} {\bibinfo  {journal} {Phys.
  Rev.}\ }\textbf {\bibinfo {volume} {D89}},\ \bibinfo {pages} {094501}
  (\bibinfo {year} {2014})},\ \Eprint {http://arxiv.org/abs/1310.3722}
  {arXiv:1310.3722 [hep-lat]} \BibitemShut {NoStop}%
%%CITATION = ARXIV:1310.3722;%%
\bibitem [{\citenamefont {Grinstein}\ \emph {et~al.}(1989)\citenamefont
  {Grinstein}, \citenamefont {Savage},\ and\ \citenamefont
  {Wise}}]{Grinstein:1988me}%
  \BibitemOpen
  \bibfield  {author} {\bibinfo {author} {\bibfnamefont {B.}~\bibnamefont
  {Grinstein}}, \bibinfo {author} {\bibfnamefont {M.~J.}\ \bibnamefont
  {Savage}}, \ and\ \bibinfo {author} {\bibfnamefont {M.~B.}\ \bibnamefont
  {Wise}},\ }\href {\doibase 10.1016/0550-3213(89)90078-3} {\bibfield
  {journal} {\bibinfo  {journal} {Nucl. Phys.}\ }\textbf {\bibinfo {volume}
  {B319}},\ \bibinfo {pages} {271} (\bibinfo {year} {1989})}\BibitemShut
  {NoStop}%
%%CITATION = NUPHA,B319,271;%%
\bibitem [{\citenamefont {Buras}\ \emph {et~al.}(1994)\citenamefont {Buras},
  \citenamefont {Misiak}, \citenamefont {{M\"unz}},\ and\ \citenamefont
  {Pokorski}}]{Buras:1993xp}%
  \BibitemOpen
  \bibfield  {author} {\bibinfo {author} {\bibfnamefont {A.~J.}\ \bibnamefont
  {Buras}}, \bibinfo {author} {\bibfnamefont {M.}~\bibnamefont {Misiak}},
  \bibinfo {author} {\bibfnamefont {M.}~\bibnamefont {{M\"unz}}}, \ and\
  \bibinfo {author} {\bibfnamefont {S.}~\bibnamefont {Pokorski}},\ }\href
  {\doibase 10.1016/0550-3213(94)90299-2} {\bibfield  {journal} {\bibinfo
  {journal} {Nucl. Phys.}\ }\textbf {\bibinfo {volume} {B424}},\ \bibinfo
  {pages} {374} (\bibinfo {year} {1994})},\ \Eprint
  {http://arxiv.org/abs/hep-ph/9311345} {arXiv:hep-ph/9311345 [hep-ph]}
  \BibitemShut {NoStop}%
%%CITATION = HEP-PH/9311345;%%
\bibitem [{\citenamefont {Bobeth}\ \emph {et~al.}(2000)\citenamefont {Bobeth},
  \citenamefont {Misiak},\ and\ \citenamefont {Urban}}]{Bobeth:1999mk}%
  \BibitemOpen
  \bibfield  {author} {\bibinfo {author} {\bibfnamefont {C.}~\bibnamefont
  {Bobeth}}, \bibinfo {author} {\bibfnamefont {M.}~\bibnamefont {Misiak}}, \
  and\ \bibinfo {author} {\bibfnamefont {J.}~\bibnamefont {Urban}},\ }\href
  {\doibase 10.1016/S0550-3213(00)00007-9} {\bibfield  {journal} {\bibinfo
  {journal} {Nucl. Phys.}\ }\textbf {\bibinfo {volume} {B574}},\ \bibinfo
  {pages} {291} (\bibinfo {year} {2000})},\ \Eprint
  {http://arxiv.org/abs/hep-ph/9910220} {arXiv:hep-ph/9910220 [hep-ph]}
  \BibitemShut {NoStop}%
%%CITATION = HEP-PH/9910220;%%
\bibitem [{\citenamefont {Altmannshofer}\ \emph {et~al.}(2009)\citenamefont
  {Altmannshofer} \emph {et~al.}}]{Altmannshofer:2008dz}%
  \BibitemOpen
  \bibfield  {author} {\bibinfo {author} {\bibfnamefont {W.}~\bibnamefont
  {Altmannshofer}} \emph {et~al.},\ }\href {\doibase
  10.1088/1126-6708/2009/01/019} {\bibfield  {journal} {\bibinfo  {journal}
  {JHEP}\ }\textbf {\bibinfo {volume} {0901}},\ \bibinfo {pages} {019}
  (\bibinfo {year} {2009})},\ \Eprint {http://arxiv.org/abs/0811.1214}
  {arXiv:0811.1214 [hep-ph]} \BibitemShut {NoStop}%
%%CITATION = ARXIV:0811.1214;%%
\bibitem [{\citenamefont {Sheikholeslami}\ and\ \citenamefont
  {Wohlert}(1985)}]{Sheikholeslami:1985ij}%
  \BibitemOpen
  \bibfield  {author} {\bibinfo {author} {\bibfnamefont {B.}~\bibnamefont
  {Sheikholeslami}}\ and\ \bibinfo {author} {\bibfnamefont {R.}~\bibnamefont
  {Wohlert}},\ }\href {\doibase 10.1016/0550-3213(85)90002-1} {\bibfield
  {journal} {\bibinfo  {journal} {Nucl. Phys.}\ }\textbf {\bibinfo {volume}
  {B259}},\ \bibinfo {pages} {572} (\bibinfo {year} {1985})}\BibitemShut
  {NoStop}%
%%CITATION = NUPHA,B259,572;%%
\bibitem [{\citenamefont {El-Khadra}\ \emph {et~al.}(1997)\citenamefont
  {El-Khadra}, \citenamefont {Kronfeld},\ and\ \citenamefont
  {Mackenzie}}]{ElKhadra:1996mp}%
  \BibitemOpen
  \bibfield  {author} {\bibinfo {author} {\bibfnamefont {A.~X.}\ \bibnamefont
  {El-Khadra}}, \bibinfo {author} {\bibfnamefont {A.~S.}\ \bibnamefont
  {Kronfeld}}, \ and\ \bibinfo {author} {\bibfnamefont {P.~B.}\ \bibnamefont
  {Mackenzie}},\ }\href {\doibase 10.1103/PhysRevD.55.3933} {\bibfield
  {journal} {\bibinfo  {journal} {Phys. Rev.}\ }\textbf {\bibinfo {volume}
  {D55}},\ \bibinfo {pages} {3933} (\bibinfo {year} {1997})},\ \Eprint
  {http://arxiv.org/abs/hep-lat/9604004} {hep-lat/9604004} \BibitemShut
  {NoStop}%
%%CITATION = HEP-LAT/9604004;%%
\bibitem [{\citenamefont {Lepage}\ \emph {et~al.}(1992)\citenamefont {Lepage},
  \citenamefont {Magnea}, \citenamefont {Nakhleh}, \citenamefont {Magnea},\
  and\ \citenamefont {Hornbostel}}]{Lepage:1992tx}%
  \BibitemOpen
  \bibfield  {author} {\bibinfo {author} {\bibfnamefont {G.~P.}\ \bibnamefont
  {Lepage}}, \bibinfo {author} {\bibfnamefont {L.}~\bibnamefont {Magnea}},
  \bibinfo {author} {\bibfnamefont {C.}~\bibnamefont {Nakhleh}}, \bibinfo
  {author} {\bibfnamefont {U.}~\bibnamefont {Magnea}}, \ and\ \bibinfo {author}
  {\bibfnamefont {K.}~\bibnamefont {Hornbostel}},\ }\href {\doibase
  10.1103/PhysRevD.46.4052} {\bibfield  {journal} {\bibinfo  {journal} {Phys.
  Rev.}\ }\textbf {\bibinfo {volume} {D46}},\ \bibinfo {pages} {4052} (\bibinfo
  {year} {1992})},\ \Eprint {http://arxiv.org/abs/hep-lat/9205007}
  {arXiv:hep-lat/9205007 [hep-lat]} \BibitemShut {NoStop}%
%%CITATION = HEP-LAT/9205007;%%
\bibitem [{\citenamefont {Boyd}\ \emph {et~al.}(1995)\citenamefont {Boyd},
  \citenamefont {Grinstein},\ and\ \citenamefont {Lebed}}]{Boyd:1994tt}%
  \BibitemOpen
  \bibfield  {author} {\bibinfo {author} {\bibfnamefont {C.~G.}\ \bibnamefont
  {Boyd}}, \bibinfo {author} {\bibfnamefont {B.}~\bibnamefont {Grinstein}}, \
  and\ \bibinfo {author} {\bibfnamefont {R.~F.}\ \bibnamefont {Lebed}},\ }\href
  {\doibase 10.1103/PhysRevLett.74.4603} {\bibfield  {journal} {\bibinfo
  {journal} {Phys. Rev. Lett.}\ }\textbf {\bibinfo {volume} {74}},\ \bibinfo
  {pages} {4603} (\bibinfo {year} {1995})},\ \Eprint
  {http://arxiv.org/abs/hep-ph/9412324} {hep-ph/9412324} \BibitemShut {NoStop}%
%%CITATION = HEP-PH/9412324;%%
\bibitem [{\citenamefont {Arnesen}\ \emph {et~al.}(2005)\citenamefont
  {Arnesen}, \citenamefont {Grinstein}, \citenamefont {Rothstein},\ and\
  \citenamefont {Stewart}}]{Arnesen:2005ez}%
  \BibitemOpen
  \bibfield  {author} {\bibinfo {author} {\bibfnamefont {M.~C.}\ \bibnamefont
  {Arnesen}}, \bibinfo {author} {\bibfnamefont {B.}~\bibnamefont {Grinstein}},
  \bibinfo {author} {\bibfnamefont {I.~Z.}\ \bibnamefont {Rothstein}}, \ and\
  \bibinfo {author} {\bibfnamefont {I.~W.}\ \bibnamefont {Stewart}},\ }\href
  {\doibase 10.1103/PhysRevLett.95.071802} {\bibfield  {journal} {\bibinfo
  {journal} {Phys. Rev. Lett.}\ }\textbf {\bibinfo {volume} {95}},\ \bibinfo
  {pages} {071802} (\bibinfo {year} {2005})},\ \Eprint
  {http://arxiv.org/abs/hep-ph/0504209} {hep-ph/0504209} \BibitemShut {NoStop}%
%%CITATION = HEP-PH/0504209;%%
\bibitem [{\citenamefont {Bourrely}\ \emph {et~al.}(2009)\citenamefont
  {Bourrely}, \citenamefont {Caprini},\ and\ \citenamefont
  {Lellouch}}]{Bourrely:2008za}%
  \BibitemOpen
  \bibfield  {author} {\bibinfo {author} {\bibfnamefont {C.}~\bibnamefont
  {Bourrely}}, \bibinfo {author} {\bibfnamefont {I.}~\bibnamefont {Caprini}}, \
  and\ \bibinfo {author} {\bibfnamefont {L.}~\bibnamefont {Lellouch}},\ }\href
  {\doibase 10.1103/PhysRevD.82.099902, 10.1103/PhysRevD.79.013008} {\bibfield
  {journal} {\bibinfo  {journal} {Phys. Rev.}\ }\textbf {\bibinfo {volume}
  {D79}},\ \bibinfo {pages} {013008} (\bibinfo {year} {2009})},\ \Eprint
  {http://arxiv.org/abs/0807.2722} {arXiv:0807.2722 [hep-ph]} \BibitemShut
  {NoStop}%
%%CITATION = ARXIV:0807.2722;%%
\bibitem [{\citenamefont {Bharucha}\ \emph {et~al.}(2010)\citenamefont
  {Bharucha}, \citenamefont {Feldmann},\ and\ \citenamefont
  {Wick}}]{Bharucha:2010im}%
  \BibitemOpen
  \bibfield  {author} {\bibinfo {author} {\bibfnamefont {A.}~\bibnamefont
  {Bharucha}}, \bibinfo {author} {\bibfnamefont {T.}~\bibnamefont {Feldmann}},
  \ and\ \bibinfo {author} {\bibfnamefont {M.}~\bibnamefont {Wick}},\ }\href
  {\doibase 10.1007/JHEP09(2010)090} {\bibfield  {journal} {\bibinfo  {journal}
  {JHEP}\ }\textbf {\bibinfo {volume} {1009}},\ \bibinfo {pages} {090}
  (\bibinfo {year} {2010})},\ \Eprint {http://arxiv.org/abs/1004.3249}
  {arXiv:1004.3249 [hep-ph]} \BibitemShut {NoStop}%
%%CITATION = ARXIV:1004.3249;%%
\bibitem [{\citenamefont {Zhou}\ \emph {et~al.}(2011)\citenamefont {Zhou} \emph
  {et~al.}}]{Zhou:2011be}%
  \BibitemOpen
  \bibfield  {author} {\bibinfo {author} {\bibfnamefont {R.}~\bibnamefont
  {Zhou}} \emph {et~al.} (\bibinfo {collaboration} {Fermilab Lattice and MILC
  Collaborations}),\ }\href@noop {} {\bibfield  {journal} {\bibinfo  {journal}
  {PoS}\ }\textbf {\bibinfo {volume} {{Lattice2011}}},\ \bibinfo {pages} {298}
  (\bibinfo {year} {2011})},\ \Eprint {http://arxiv.org/abs/1111.0981}
  {arXiv:1111.0981 [hep-lat]} \BibitemShut {NoStop}%
%%CITATION = ARXIV:1111.0981;%%
\bibitem [{\citenamefont {Zhou}\ \emph {et~al.}(2012)\citenamefont {Zhou} \emph
  {et~al.}}]{Zhou:2012sn}%
  \BibitemOpen
  \bibfield  {author} {\bibinfo {author} {\bibfnamefont {R.}~\bibnamefont
  {Zhou}} \emph {et~al.} (\bibinfo {collaboration} {Fermilab Lattice and MILC
  Collaborations}),\ }\href@noop {} {\bibfield  {journal} {\bibinfo  {journal}
  {PoS}\ }\textbf {\bibinfo {volume} {{Lattice2012}}},\ \bibinfo {pages} {120}
  (\bibinfo {year} {2012})},\ \Eprint {http://arxiv.org/abs/1211.1390}
  {arXiv:1211.1390 [hep-lat]} \BibitemShut {NoStop}%
%%CITATION = ARXIV:1211.1390;%%
\bibitem [{\citenamefont {Zhou}(2013)}]{Zhou:2013uu}%
  \BibitemOpen
  \bibfield  {author} {\bibinfo {author} {\bibfnamefont {R.}~\bibnamefont
  {Zhou}}\ }(\bibinfo {year} {2013})\ \Eprint {http://arxiv.org/abs/1301.0666}
  {arXiv:1301.0666 [hep-lat]} \BibitemShut {NoStop}%
%%CITATION = ARXIV:1301.0666;%%
\bibitem [{\citenamefont {{Be\v{c}irevi\'c}}\ \emph {et~al.}(2012)\citenamefont
  {{Be\v{c}irevi\'c}}, \citenamefont {Ko\v{s}nik}, \citenamefont {Mescia},\
  and\ \citenamefont {Schneider}}]{Becirevic:2012fy}%
  \BibitemOpen
  \bibfield  {author} {\bibinfo {author} {\bibfnamefont {D.}~\bibnamefont
  {{Be\v{c}irevi\'c}}}, \bibinfo {author} {\bibfnamefont {N.}~\bibnamefont
  {Ko\v{s}nik}}, \bibinfo {author} {\bibfnamefont {F.}~\bibnamefont {Mescia}},
  \ and\ \bibinfo {author} {\bibfnamefont {E.}~\bibnamefont {Schneider}},\
  }\href {\doibase 10.1103/PhysRevD.86.034034} {\bibfield  {journal} {\bibinfo
  {journal} {Phys. Rev.}\ }\textbf {\bibinfo {volume} {D86}},\ \bibinfo {pages}
  {034034} (\bibinfo {year} {2012})},\ \Eprint {http://arxiv.org/abs/1205.5811}
  {arXiv:1205.5811 [hep-ph]} \BibitemShut {NoStop}%
%%CITATION = ARXIV:1205.5811;%%
\bibitem [{\citenamefont {Ali}\ \emph {et~al.}(2014)\citenamefont {Ali},
  \citenamefont {Parkhomenko},\ and\ \citenamefont {Rusov}}]{Ali:2013zfa}%
  \BibitemOpen
  \bibfield  {author} {\bibinfo {author} {\bibfnamefont {A.}~\bibnamefont
  {Ali}}, \bibinfo {author} {\bibfnamefont {A.~Y.}\ \bibnamefont
  {Parkhomenko}}, \ and\ \bibinfo {author} {\bibfnamefont {A.~V.}\ \bibnamefont
  {Rusov}},\ }\href {\doibase 10.1103/PhysRevD.89.094021} {\bibfield  {journal}
  {\bibinfo  {journal} {Phys. Rev.}\ }\textbf {\bibinfo {volume} {D89}},\
  \bibinfo {pages} {094021} (\bibinfo {year} {2014})},\ \Eprint
  {http://arxiv.org/abs/1312.2523} {arXiv:1312.2523 [hep-ph]} \BibitemShut
  {NoStop}%
%%CITATION = ARXIV:1312.2523;%%
\bibitem [{\citenamefont {Bernard}\ \emph {et~al.}(2001)\citenamefont {Bernard}
  \emph {et~al.}}]{Bernard:2001av}%
  \BibitemOpen
  \bibfield  {author} {\bibinfo {author} {\bibfnamefont {C.~W.}\ \bibnamefont
  {Bernard}} \emph {et~al.},\ }\href {\doibase 10.1103/PhysRevD.64.054506}
  {\bibfield  {journal} {\bibinfo  {journal} {Phys. Rev.}\ }\textbf {\bibinfo
  {volume} {D64}},\ \bibinfo {pages} {054506} (\bibinfo {year} {2001})},\
  \Eprint {http://arxiv.org/abs/hep-lat/0104002} {hep-lat/0104002} \BibitemShut
  {NoStop}%
%%CITATION = HEP-LAT/0104002;%%
\bibitem [{\citenamefont {Aubin}\ \emph {et~al.}(2004)\citenamefont {Aubin}
  \emph {et~al.}}]{Aubin:2004wf}%
  \BibitemOpen
  \bibfield  {author} {\bibinfo {author} {\bibfnamefont {C.}~\bibnamefont
  {Aubin}} \emph {et~al.},\ }\href {\doibase 10.1103/PhysRevD.70.094505}
  {\bibfield  {journal} {\bibinfo  {journal} {Phys. Rev.}\ }\textbf {\bibinfo
  {volume} {D70}},\ \bibinfo {pages} {094505} (\bibinfo {year} {2004})},\
  \Eprint {http://arxiv.org/abs/hep-lat/0402030} {arXiv:hep-lat/0402030
  [hep-lat]} \BibitemShut {NoStop}%
%%CITATION = HEP-LAT/0402030;%%
\bibitem [{\citenamefont {{L\"uscher}}\ and\ \citenamefont
  {Weisz}(1985{\natexlab{a}})}]{Luscher:1984xn}%
  \BibitemOpen
  \bibfield  {author} {\bibinfo {author} {\bibfnamefont {M.}~\bibnamefont
  {{L\"uscher}}}\ and\ \bibinfo {author} {\bibfnamefont {P.}~\bibnamefont
  {Weisz}},\ }\href {\doibase 10.1007/BF01206178} {\bibfield  {journal}
  {\bibinfo  {journal} {Commun. Math. Phys.}\ }\textbf {\bibinfo {volume}
  {97}},\ \bibinfo {pages} {59} (\bibinfo {year}
  {1985}{\natexlab{a}})}\BibitemShut {NoStop}%
%%CITATION = CMPHA,97,59;%%
\bibitem [{\citenamefont {{L\"uscher}}\ and\ \citenamefont
  {Weisz}(1985{\natexlab{b}})}]{Luscher:1985zq}%
  \BibitemOpen
  \bibfield  {author} {\bibinfo {author} {\bibfnamefont {M.}~\bibnamefont
  {{L\"uscher}}}\ and\ \bibinfo {author} {\bibfnamefont {P.}~\bibnamefont
  {Weisz}},\ }\href {\doibase 10.1016/0370-2693(85)90966-9} {\bibfield
  {journal} {\bibinfo  {journal} {Phys. Lett.}\ }\textbf {\bibinfo {volume}
  {B158}},\ \bibinfo {pages} {250} (\bibinfo {year}
  {1985}{\natexlab{b}})}\BibitemShut {NoStop}%
%%CITATION = PHLTA,B158,250;%%
\bibitem [{\citenamefont {Hao}\ \emph {et~al.}(2007)\citenamefont {Hao},
  \citenamefont {von Hippel}, \citenamefont {Horgan}, \citenamefont {Mason},\
  and\ \citenamefont {Trottier}}]{Hao:2007iz}%
  \BibitemOpen
  \bibfield  {author} {\bibinfo {author} {\bibfnamefont {Z.}~\bibnamefont
  {Hao}}, \bibinfo {author} {\bibfnamefont {G.~M.}\ \bibnamefont {von Hippel}},
  \bibinfo {author} {\bibfnamefont {R.~R.}\ \bibnamefont {Horgan}}, \bibinfo
  {author} {\bibfnamefont {Q.~J.}\ \bibnamefont {Mason}}, \ and\ \bibinfo
  {author} {\bibfnamefont {H.~D.}\ \bibnamefont {Trottier}},\ }\href {\doibase
  10.1103/PhysRevD.76.034507} {\bibfield  {journal} {\bibinfo  {journal} {Phys.
  Rev.}\ }\textbf {\bibinfo {volume} {D76}},\ \bibinfo {pages} {034507}
  (\bibinfo {year} {2007})},\ \Eprint {http://arxiv.org/abs/0705.4660}
  {arXiv:0705.4660 [hep-lat]} \BibitemShut {NoStop}%
%%CITATION = ARXIV:0705.4660;%%
\bibitem [{\citenamefont {Blum}\ \emph {et~al.}(1997)\citenamefont {Blum} \emph
  {et~al.}}]{Blum:1996uf}%
  \BibitemOpen
  \bibfield  {author} {\bibinfo {author} {\bibfnamefont {T.}~\bibnamefont
  {Blum}} \emph {et~al.},\ }\href {\doibase 10.1103/PhysRevD.55.1133}
  {\bibfield  {journal} {\bibinfo  {journal} {Phys. Rev.}\ }\textbf {\bibinfo
  {volume} {D55}},\ \bibinfo {pages} {1133} (\bibinfo {year} {1997})},\ \Eprint
  {http://arxiv.org/abs/hep-lat/9609036} {hep-lat/9609036} \BibitemShut
  {NoStop}%
%%CITATION = HEP-LAT/9609036;%%
\bibitem [{\citenamefont {Lepage}(1998)}]{Lepage:1997id}%
  \BibitemOpen
  \bibfield  {author} {\bibinfo {author} {\bibfnamefont {G.~P.}\ \bibnamefont
  {Lepage}},\ }\href@noop {} {\bibfield  {journal} {\bibinfo  {journal} {Nucl.
  Phys. Proc. Suppl.}\ }\textbf {\bibinfo {volume} {60A}},\ \bibinfo {pages}
  {267} (\bibinfo {year} {1998})},\ \Eprint
  {http://arxiv.org/abs/hep-lat/9707026} {hep-lat/9707026} \BibitemShut
  {NoStop}%
%%CITATION = HEP-LAT/9707026;%%
\bibitem [{\citenamefont {{Laga\"e}}\ and\ \citenamefont
  {Sinclair}(1999)}]{Lagae:1998pe}%
  \BibitemOpen
  \bibfield  {author} {\bibinfo {author} {\bibfnamefont {J.~F.}\ \bibnamefont
  {{Laga\"e}}}\ and\ \bibinfo {author} {\bibfnamefont {D.~K.}\ \bibnamefont
  {Sinclair}},\ }\href {\doibase 10.1103/PhysRevD.59.014511} {\bibfield
  {journal} {\bibinfo  {journal} {Phys. Rev.}\ }\textbf {\bibinfo {volume}
  {D59}},\ \bibinfo {pages} {014511} (\bibinfo {year} {1999})},\ \Eprint
  {http://arxiv.org/abs/hep-lat/9806014} {hep-lat/9806014} \BibitemShut
  {NoStop}%
%%CITATION = HEP-LAT/9806014;%%
\bibitem [{\citenamefont {Lepage}(1999)}]{Lepage:1998vj}%
  \BibitemOpen
  \bibfield  {author} {\bibinfo {author} {\bibfnamefont {G.~P.}\ \bibnamefont
  {Lepage}},\ }\href {\doibase 10.1103/PhysRevD.59.074502} {\bibfield
  {journal} {\bibinfo  {journal} {Phys. Rev.}\ }\textbf {\bibinfo {volume}
  {D59}},\ \bibinfo {pages} {074502} (\bibinfo {year} {1999})},\ \Eprint
  {http://arxiv.org/abs/hep-lat/9809157} {arXiv:hep-lat/9809157} \BibitemShut
  {NoStop}%
%%CITATION = HEP-LAT/9809157;%%
\bibitem [{\citenamefont {Orginos}\ and\ \citenamefont
  {Toussaint}(1999)}]{Orginos:1998ue}%
  \BibitemOpen
  \bibfield  {author} {\bibinfo {author} {\bibfnamefont {K.}~\bibnamefont
  {Orginos}}\ and\ \bibinfo {author} {\bibfnamefont {D.}~\bibnamefont
  {Toussaint}} (\bibinfo {collaboration} {MILC Collaboration}),\ }\href
  {\doibase 10.1103/PhysRevD.59.014501} {\bibfield  {journal} {\bibinfo
  {journal} {Phys. Rev.}\ }\textbf {\bibinfo {volume} {D59}},\ \bibinfo {pages}
  {014501} (\bibinfo {year} {1999})},\ \Eprint
  {http://arxiv.org/abs/hep-lat/9805009} {hep-lat/9805009} \BibitemShut
  {NoStop}%
%%CITATION = HEP-LAT/9805009;%%
\bibitem [{\citenamefont {Orginos}\ \emph {et~al.}(1999)\citenamefont
  {Orginos}, \citenamefont {Toussaint},\ and\ \citenamefont
  {Sugar}}]{Orginos:1999cr}%
  \BibitemOpen
  \bibfield  {author} {\bibinfo {author} {\bibfnamefont {K.}~\bibnamefont
  {Orginos}}, \bibinfo {author} {\bibfnamefont {D.}~\bibnamefont {Toussaint}},
  \ and\ \bibinfo {author} {\bibfnamefont {R.~L.}\ \bibnamefont {Sugar}}
  (\bibinfo {collaboration} {MILC Collaboration}),\ }\href {\doibase
  10.1103/PhysRevD.60.054503} {\bibfield  {journal} {\bibinfo  {journal} {Phys.
  Rev.}\ }\textbf {\bibinfo {volume} {D60}},\ \bibinfo {pages} {054503}
  (\bibinfo {year} {1999})},\ \Eprint {http://arxiv.org/abs/hep-lat/9903032}
  {hep-lat/9903032} \BibitemShut {NoStop}%
%%CITATION = HEP-LAT/9903032;%%
\bibitem [{\citenamefont {Bernard}\ \emph
  {et~al.}(2000{\natexlab{a}})\citenamefont {Bernard} \emph
  {et~al.}}]{Bernard:1999xx}%
  \BibitemOpen
  \bibfield  {author} {\bibinfo {author} {\bibfnamefont {C.~W.}\ \bibnamefont
  {Bernard}} \emph {et~al.} (\bibinfo {collaboration} {MILC Collaboration}),\
  }\href {\doibase 10.1103/PhysRevD.61.111502} {\bibfield  {journal} {\bibinfo
  {journal} {Phys. Rev.}\ }\textbf {\bibinfo {volume} {D61}},\ \bibinfo {pages}
  {111502} (\bibinfo {year} {2000}{\natexlab{a}})},\ \Eprint
  {http://arxiv.org/abs/hep-lat/9912018} {hep-lat/9912018} \BibitemShut
  {NoStop}%
%%CITATION = HEP-LAT/9912018;%%
\bibitem [{\citenamefont {Bazavov}\ \emph
  {et~al.}(2010{\natexlab{a}})\citenamefont {Bazavov} \emph
  {et~al.}}]{Bazavov:2009bb}%
  \BibitemOpen
  \bibfield  {author} {\bibinfo {author} {\bibfnamefont {A.}~\bibnamefont
  {Bazavov}} \emph {et~al.},\ }\href {\doibase 10.1103/RevModPhys.82.1349}
  {\bibfield  {journal} {\bibinfo  {journal} {Rev. Mod. Phys.}\ }\textbf
  {\bibinfo {volume} {82}},\ \bibinfo {pages} {1349} (\bibinfo {year}
  {2010}{\natexlab{a}})},\ \Eprint {http://arxiv.org/abs/0903.3598}
  {arXiv:0903.3598 [hep-lat]} \BibitemShut {NoStop}%
%%CITATION = ARXIV:0903.3598;%%
\bibitem [{\citenamefont {Shamir}(2005)}]{Shamir:2004zc}%
  \BibitemOpen
  \bibfield  {author} {\bibinfo {author} {\bibfnamefont {Y.}~\bibnamefont
  {Shamir}},\ }\href {\doibase 10.1103/PhysRevD.71.034509} {\bibfield
  {journal} {\bibinfo  {journal} {Phys. Rev.}\ }\textbf {\bibinfo {volume}
  {D71}},\ \bibinfo {pages} {034509} (\bibinfo {year} {2005})},\ \Eprint
  {http://arxiv.org/abs/hep-lat/0412014} {hep-lat/0412014} \BibitemShut
  {NoStop}%
%%CITATION = HEP-LAT/0412014;%%
\bibitem [{\citenamefont {Shamir}(2007)}]{Shamir:2006nj}%
  \BibitemOpen
  \bibfield  {author} {\bibinfo {author} {\bibfnamefont {Y.}~\bibnamefont
  {Shamir}},\ }\href {\doibase 10.1103/PhysRevD.75.054503} {\bibfield
  {journal} {\bibinfo  {journal} {Phys. Rev.}\ }\textbf {\bibinfo {volume}
  {D75}},\ \bibinfo {pages} {054503} (\bibinfo {year} {2007})},\ \Eprint
  {http://arxiv.org/abs/hep-lat/0607007} {hep-lat/0607007} \BibitemShut
  {NoStop}%
%%CITATION = HEP-LAT/0607007;%%
\bibitem [{\citenamefont {Lee}\ and\ \citenamefont
  {Sharpe}(1999)}]{Lee:1999zxa}%
  \BibitemOpen
  \bibfield  {author} {\bibinfo {author} {\bibfnamefont {W.-J.}\ \bibnamefont
  {Lee}}\ and\ \bibinfo {author} {\bibfnamefont {S.~R.}\ \bibnamefont
  {Sharpe}},\ }\href {\doibase 10.1103/PhysRevD.60.114503} {\bibfield
  {journal} {\bibinfo  {journal} {Phys. Rev.}\ }\textbf {\bibinfo {volume}
  {D60}},\ \bibinfo {pages} {114503} (\bibinfo {year} {1999})},\ \Eprint
  {http://arxiv.org/abs/hep-lat/9905023} {hep-lat/9905023} \BibitemShut
  {NoStop}%
%%CITATION = HEP-LAT/9905023;%%
\bibitem [{\citenamefont {Bernard}(2006)}]{Bernard:2006zw}%
  \BibitemOpen
  \bibfield  {author} {\bibinfo {author} {\bibfnamefont {C.}~\bibnamefont
  {Bernard}},\ }\href {\doibase 10.1103/PhysRevD.73.114503} {\bibfield
  {journal} {\bibinfo  {journal} {Phys. Rev.}\ }\textbf {\bibinfo {volume}
  {D73}},\ \bibinfo {pages} {114503} (\bibinfo {year} {2006})},\ \Eprint
  {http://arxiv.org/abs/hep-lat/0603011} {hep-lat/0603011} \BibitemShut
  {NoStop}%
%%CITATION = HEP-LAT/0603011;%%
\bibitem [{\citenamefont {Bernard}\ \emph {et~al.}(2008)\citenamefont
  {Bernard}, \citenamefont {Golterman},\ and\ \citenamefont
  {Shamir}}]{Bernard:2007ma}%
  \BibitemOpen
  \bibfield  {author} {\bibinfo {author} {\bibfnamefont {C.}~\bibnamefont
  {Bernard}}, \bibinfo {author} {\bibfnamefont {M.}~\bibnamefont {Golterman}},
  \ and\ \bibinfo {author} {\bibfnamefont {Y.}~\bibnamefont {Shamir}},\ }\href
  {\doibase 10.1103/PhysRevD.77.074505} {\bibfield  {journal} {\bibinfo
  {journal} {Phys. Rev.}\ }\textbf {\bibinfo {volume} {D77}},\ \bibinfo {pages}
  {074505} (\bibinfo {year} {2008})},\ \Eprint {http://arxiv.org/abs/0712.2560}
  {arXiv:0712.2560 [hep-lat]} \BibitemShut {NoStop}%
%%CITATION = ARXIV:0712.2560;%%
\bibitem [{\citenamefont {{Prelov\v sek}}(2006)}]{Prelovsek:2005rf}%
  \BibitemOpen
  \bibfield  {author} {\bibinfo {author} {\bibfnamefont {S.}~\bibnamefont
  {{Prelov\v sek}}},\ }\href {\doibase 10.1103/PhysRevD.73.014506} {\bibfield
  {journal} {\bibinfo  {journal} {Phys. Rev.}\ }\textbf {\bibinfo {volume}
  {D73}},\ \bibinfo {pages} {014506} (\bibinfo {year} {2006})},\ \Eprint
  {http://arxiv.org/abs/hep-lat/0510080} {hep-lat/0510080} \BibitemShut
  {NoStop}%
%%CITATION = HEP-LAT/0510080;%%
\bibitem [{\citenamefont {Bernard}\ \emph {et~al.}(2007)\citenamefont
  {Bernard}, \citenamefont {DeTar}, \citenamefont {Fu},\ and\ \citenamefont
  {{Prelov\v sek}}}]{Bernard:2007qf}%
  \BibitemOpen
  \bibfield  {author} {\bibinfo {author} {\bibfnamefont {C.}~\bibnamefont
  {Bernard}}, \bibinfo {author} {\bibfnamefont {C.~E.}\ \bibnamefont {DeTar}},
  \bibinfo {author} {\bibfnamefont {Z.}~\bibnamefont {Fu}}, \ and\ \bibinfo
  {author} {\bibfnamefont {S.}~\bibnamefont {{Prelov\v sek}}},\ }\href
  {\doibase 10.1103/PhysRevD.76.094504} {\bibfield  {journal} {\bibinfo
  {journal} {Phys. Rev.}\ }\textbf {\bibinfo {volume} {D76}},\ \bibinfo {pages}
  {094504} (\bibinfo {year} {2007})},\ \Eprint {http://arxiv.org/abs/0707.2402}
  {arXiv:0707.2402 [hep-lat]} \BibitemShut {NoStop}%
%%CITATION = ARXIV:0707.2402;%%
\bibitem [{\citenamefont {Aubin}\ \emph {et~al.}(2008)\citenamefont {Aubin},
  \citenamefont {Laiho},\ and\ \citenamefont {Van~de Water}}]{Aubin:2008wk}%
  \BibitemOpen
  \bibfield  {author} {\bibinfo {author} {\bibfnamefont {C.}~\bibnamefont
  {Aubin}}, \bibinfo {author} {\bibfnamefont {J.}~\bibnamefont {Laiho}}, \ and\
  \bibinfo {author} {\bibfnamefont {R.~S.}\ \bibnamefont {Van~de Water}},\
  }\href {\doibase 10.1103/PhysRevD.77.114501} {\bibfield  {journal} {\bibinfo
  {journal} {Phys. Rev.}\ }\textbf {\bibinfo {volume} {D77}},\ \bibinfo {pages}
  {114501} (\bibinfo {year} {2008})},\ \Eprint {http://arxiv.org/abs/0803.0129}
  {arXiv:0803.0129 [hep-lat]} \BibitemShut {NoStop}%
%%CITATION = ARXIV:0803.0129;%%
\bibitem [{\citenamefont {Sharpe}\ and\ \citenamefont {Van~de
  Water}(2005)}]{Sharpe:2004is}%
  \BibitemOpen
  \bibfield  {author} {\bibinfo {author} {\bibfnamefont {S.~R.}\ \bibnamefont
  {Sharpe}}\ and\ \bibinfo {author} {\bibfnamefont {R.~S.}\ \bibnamefont
  {Van~de Water}},\ }\href {\doibase 10.1103/PhysRevD.71.114505} {\bibfield
  {journal} {\bibinfo  {journal} {Phys. Rev.}\ }\textbf {\bibinfo {volume}
  {D71}},\ \bibinfo {pages} {114505} (\bibinfo {year} {2005})},\ \Eprint
  {http://arxiv.org/abs/hep-lat/0409018} {hep-lat/0409018} \BibitemShut
  {NoStop}%
%%CITATION = HEP-LAT/0409018;%%
\bibitem [{\citenamefont {{D\"urr}}(2006)}]{Durr:2005ax}%
  \BibitemOpen
  \bibfield  {author} {\bibinfo {author} {\bibfnamefont {S.}~\bibnamefont
  {{D\"urr}}},\ }\href@noop {} {\bibfield  {journal} {\bibinfo  {journal}
  {PoS}\ }\textbf {\bibinfo {volume} {LAT2005}},\ \bibinfo {pages} {021}
  (\bibinfo {year} {2006})},\ \Eprint {http://arxiv.org/abs/hep-lat/0509026}
  {hep-lat/0509026} \BibitemShut {NoStop}%
%%CITATION = HEP-LAT/0509026;%%
\bibitem [{\citenamefont {Sharpe}(2006)}]{Sharpe:2006re}%
  \BibitemOpen
  \bibfield  {author} {\bibinfo {author} {\bibfnamefont {S.~R.}\ \bibnamefont
  {Sharpe}},\ }\href@noop {} {\bibfield  {journal} {\bibinfo  {journal} {PoS}\
  }\textbf {\bibinfo {volume} {LAT2006}},\ \bibinfo {pages} {022} (\bibinfo
  {year} {2006})},\ \Eprint {http://arxiv.org/abs/hep-lat/0610094}
  {hep-lat/0610094} \BibitemShut {NoStop}%
%%CITATION = HEP-LAT/0610094;%%
\bibitem [{\citenamefont {Kronfeld}(2007)}]{Kronfeld:2007ek}%
  \BibitemOpen
  \bibfield  {author} {\bibinfo {author} {\bibfnamefont {A.~S.}\ \bibnamefont
  {Kronfeld}},\ }\href@noop {} {\bibfield  {journal} {\bibinfo  {journal}
  {PoS}\ }\textbf {\bibinfo {volume} {LAT2007}},\ \bibinfo {pages} {016}
  (\bibinfo {year} {2007})},\ \Eprint {http://arxiv.org/abs/0711.0699}
  {arXiv:0711.0699 [hep-lat]} \BibitemShut {NoStop}%
%%CITATION = ARXIV:0711.0699;%%
\bibitem [{\citenamefont {Donald}\ \emph {et~al.}(2011)\citenamefont {Donald},
  \citenamefont {Davies}, \citenamefont {Follana},\ and\ \citenamefont
  {Kronfeld}}]{Donald:2011if}%
  \BibitemOpen
  \bibfield  {author} {\bibinfo {author} {\bibfnamefont {G.~C.}\ \bibnamefont
  {Donald}}, \bibinfo {author} {\bibfnamefont {C.~T.~H.}\ \bibnamefont
  {Davies}}, \bibinfo {author} {\bibfnamefont {E.}~\bibnamefont {Follana}}, \
  and\ \bibinfo {author} {\bibfnamefont {A.~S.}\ \bibnamefont {Kronfeld}}
  (\bibinfo {collaboration} {{HPQCD and Fermilab Lattice Collaborations}}),\
  }\href {\doibase 10.1103/PhysRevD.84.054504} {\bibfield  {journal} {\bibinfo
  {journal} {Phys. Rev.}\ }\textbf {\bibinfo {volume} {D84}},\ \bibinfo {pages}
  {054504} (\bibinfo {year} {2011})},\ \Eprint {http://arxiv.org/abs/1106.2412}
  {arXiv:1106.2412 [hep-lat]} \BibitemShut {NoStop}%
%%CITATION = ARXIV:1106.2412;%%
\bibitem [{\citenamefont {{MILC
  Collaboration}}(2015{\natexlab{a}})}]{asqtad_C0.2ms}%
  \BibitemOpen
  \bibfield  {author} {\bibinfo {author} {\bibnamefont {{MILC
  Collaboration}}},\ }\href@noop {} {\enquote {\bibinfo {title}
  {{asqtad.en06a}},}\ }\bibinfo {howpublished}
  {\href{http://dx.doi.org/10.15484/milc.asqtad.en06a/1178158}{10.15484/milc.asqtad.en06a/1178158}}
  (\bibinfo {year} {2015}{\natexlab{a}})\BibitemShut {NoStop}%
\bibitem [{\citenamefont {{MILC
  Collaboration}}(2015{\natexlab{b}})}]{asqtad_C0.2ms_b}%
  \BibitemOpen
  \bibfield  {author} {\bibinfo {author} {\bibnamefont {{MILC
  Collaboration}}},\ }\href@noop {} {\enquote {\bibinfo {title}
  {{asqtad.en06b}},}\ }\bibinfo {howpublished}
  {\href{http://dx.doi.org/10.15484/milc.asqtad.en06b/1178159}{10.15484/milc.asqtad.en06b/1178159}}
  (\bibinfo {year} {2015}{\natexlab{b}})\BibitemShut {NoStop}%
\bibitem [{\citenamefont {{MILC
  Collaboration}}(2015{\natexlab{c}})}]{asqtad_C0.15ms}%
  \BibitemOpen
  \bibfield  {author} {\bibinfo {author} {\bibnamefont {{MILC
  Collaboration}}},\ }\href@noop {} {\enquote {\bibinfo {title}
  {{asqtad.en05a}},}\ }\bibinfo {howpublished}
  {\href{http://dx.doi.org/10.15484/milc.asqtad.en05a/1178156}{10.15484/milc.asqtad.en05a/1178156}}
  (\bibinfo {year} {2015}{\natexlab{c}})\BibitemShut {NoStop}%
\bibitem [{\citenamefont {{MILC
  Collaboration}}(2015{\natexlab{d}})}]{asqtad_C0.1ms}%
  \BibitemOpen
  \bibfield  {author} {\bibinfo {author} {\bibnamefont {{MILC
  Collaboration}}},\ }\href@noop {} {\enquote {\bibinfo {title}
  {{asqtad.en04a}},}\ }\bibinfo {howpublished}
  {\href{http://dx.doi.org/10.15484/milc.asqtad.en04a/1178155}{10.15484/milc.asqtad.en04a/1178155}}
  (\bibinfo {year} {2015}{\natexlab{d}})\BibitemShut {NoStop}%
\bibitem [{\citenamefont {{MILC
  Collaboration}}(2015{\natexlab{e}})}]{asqtad_F0.2ms}%
  \BibitemOpen
  \bibfield  {author} {\bibinfo {author} {\bibnamefont {{MILC
  Collaboration}}},\ }\href@noop {} {\enquote {\bibinfo {title}
  {{asqtad.en15a}},}\ }\bibinfo {howpublished}
  {\href{http://dx.doi.org/10.15484/milc.asqtad.en15a/1178095}{10.15484/milc.asqtad.en15a/1178095}}
  (\bibinfo {year} {2015}{\natexlab{e}})\BibitemShut {NoStop}%
\bibitem [{\citenamefont {{MILC
  Collaboration}}(2015{\natexlab{f}})}]{asqtad_F0.2ms_b}%
  \BibitemOpen
  \bibfield  {author} {\bibinfo {author} {\bibnamefont {{MILC
  Collaboration}}},\ }\href@noop {} {\enquote {\bibinfo {title}
  {{asqtad.en15b}},}\ }\bibinfo {howpublished}
  {\href{http://dx.doi.org/10.15484/milc.asqtad.en15b/1178096}{10.15484/milc.asqtad.en15b/1178096}}
  (\bibinfo {year} {2015}{\natexlab{f}})\BibitemShut {NoStop}%
\bibitem [{\citenamefont {{MILC
  Collaboration}}(2015{\natexlab{g}})}]{asqtad_F0.2ms_c}%
  \BibitemOpen
  \bibfield  {author} {\bibinfo {author} {\bibnamefont {{MILC
  Collaboration}}},\ }\href@noop {} {\enquote {\bibinfo {title}
  {{asqtad.en15c}},}\ }\bibinfo {howpublished}
  {\href{http://dx.doi.org/10.15484/milc.asqtad.en15c/1178097}{10.15484/milc.asqtad.en15c/1178097}}
  (\bibinfo {year} {2015}{\natexlab{g}})\BibitemShut {NoStop}%
\bibitem [{\citenamefont {{MILC
  Collaboration}}(2015{\natexlab{h}})}]{asqtad_F0.15ms}%
  \BibitemOpen
  \bibfield  {author} {\bibinfo {author} {\bibnamefont {{MILC
  Collaboration}}},\ }\href@noop {} {\enquote {\bibinfo {title}
  {{asqtad.en14a}},}\ }\bibinfo {howpublished}
  {\href{http://dx.doi.org/10.15484/milc.asqtad.en14a/1178094}{10.15484/milc.asqtad.en14a/1178094}}
  (\bibinfo {year} {2015}{\natexlab{h}})\BibitemShut {NoStop}%
\bibitem [{\citenamefont {{MILC
  Collaboration}}(2015{\natexlab{i}})}]{asqtad_F0.1ms_a}%
  \BibitemOpen
  \bibfield  {author} {\bibinfo {author} {\bibnamefont {{MILC
  Collaboration}}},\ }\href@noop {} {\enquote {\bibinfo {title}
  {{asqtad.en13a}},}\ }\bibinfo {howpublished}
  {\href{http://dx.doi.org/10.15484/milc.asqtad.en13a/1178092}{10.15484/milc.asqtad.en13a/1178092}}
  (\bibinfo {year} {2015}{\natexlab{i}})\BibitemShut {NoStop}%
\bibitem [{\citenamefont {{MILC
  Collaboration}}(2015{\natexlab{j}})}]{asqtad_F0.1ms_b}%
  \BibitemOpen
  \bibfield  {author} {\bibinfo {author} {\bibnamefont {{MILC
  Collaboration}}},\ }\href@noop {} {\enquote {\bibinfo {title}
  {{asqtad.en13b}},}\ }\bibinfo {howpublished}
  {\href{http://dx.doi.org/10.15484/milc.asqtad.en13b/1178093}{10.15484/milc.asqtad.en13b/1178093}}
  (\bibinfo {year} {2015}{\natexlab{j}})\BibitemShut {NoStop}%
\bibitem [{\citenamefont {{MILC
  Collaboration}}(2015{\natexlab{k}})}]{asqtad_F0.05ms}%
  \BibitemOpen
  \bibfield  {author} {\bibinfo {author} {\bibnamefont {{MILC
  Collaboration}}},\ }\href@noop {} {\enquote {\bibinfo {title}
  {{asqtad.en12a}},}\ }\bibinfo {howpublished}
  {\href{http://dx.doi.org/10.15484/milc.asqtad.en12a/1178091}{10.15484/milc.asqtad.en12a/1178091}}
  (\bibinfo {year} {2015}{\natexlab{k}})\BibitemShut {NoStop}%
\bibitem [{\citenamefont {{MILC
  Collaboration}}(2015{\natexlab{l}})}]{asqtad_SF0.2ms_a}%
  \BibitemOpen
  \bibfield  {author} {\bibinfo {author} {\bibnamefont {{MILC
  Collaboration}}},\ }\href@noop {} {\enquote {\bibinfo {title}
  {{asqtad.en20a}},}\ }\bibinfo {howpublished}
  {\href{http://dx.doi.org/10.15484/milc.asqtad.en20a/1178036}{10.15484/milc.asqtad.en20a/1178036}}
  (\bibinfo {year} {2015}{\natexlab{l}})\BibitemShut {NoStop}%
\bibitem [{\citenamefont {{MILC
  Collaboration}}(2015{\natexlab{m}})}]{asqtad_SF0.2ms_b}%
  \BibitemOpen
  \bibfield  {author} {\bibinfo {author} {\bibnamefont {{MILC
  Collaboration}}},\ }\href@noop {} {\enquote {\bibinfo {title}
  {{asqtad.en20b}},}\ }\bibinfo {howpublished}
  {\href{http://dx.doi.org/10.15484/milc.asqtad.en20b/1178037}{10.15484/milc.asqtad.en20b/1178037}}
  (\bibinfo {year} {2015}{\natexlab{m}})\BibitemShut {NoStop}%
\bibitem [{\citenamefont {{MILC
  Collaboration}}(2015{\natexlab{n}})}]{asqtad_SF0.1ms_a}%
  \BibitemOpen
  \bibfield  {author} {\bibinfo {author} {\bibnamefont {{MILC
  Collaboration}}},\ }\href@noop {} {\enquote {\bibinfo {title}
  {{asqtad.en18a}},}\ }\bibinfo {howpublished}
  {\href{http://dx.doi.org/10.15484/milc.asqtad.en18a/1178033}{10.15484/milc.asqtad.en18a/1178033}}
  (\bibinfo {year} {2015}{\natexlab{n}})\BibitemShut {NoStop}%
\bibitem [{\citenamefont {{MILC
  Collaboration}}(2015{\natexlab{o}})}]{asqtad_SF0.1ms_b}%
  \BibitemOpen
  \bibfield  {author} {\bibinfo {author} {\bibnamefont {{MILC
  Collaboration}}},\ }\href@noop {} {\enquote {\bibinfo {title}
  {{asqtad.en18b}},}\ }\bibinfo {howpublished}
  {\href{http://dx.doi.org/10.15484/milc.asqtad.en18b/1178034}{10.15484/milc.asqtad.en18b/1178034}}
  (\bibinfo {year} {2015}{\natexlab{o}})\BibitemShut {NoStop}%
\bibitem [{\citenamefont {{MILC
  Collaboration}}(2015{\natexlab{p}})}]{asqtad_UF0.2ms}%
  \BibitemOpen
  \bibfield  {author} {\bibinfo {author} {\bibnamefont {{MILC
  Collaboration}}},\ }\href@noop {} {\enquote {\bibinfo {title}
  {{asqtad.en24a}},}\ }\bibinfo {howpublished}
  {\href{http://dx.doi.org/10.15484/milc.asqtad.en24a/1177873}{10.15484/milc.asqtad.en24a/1177873}}
  (\bibinfo {year} {2015}{\natexlab{p}})\BibitemShut {NoStop}%
\bibitem [{\citenamefont {Kronfeld}(2000)}]{Kronfeld:2000ck}%
  \BibitemOpen
  \bibfield  {author} {\bibinfo {author} {\bibfnamefont {A.~S.}\ \bibnamefont
  {Kronfeld}},\ }\href {\doibase 10.1103/PhysRevD.62.014505} {\bibfield
  {journal} {\bibinfo  {journal} {Phys. Rev.}\ }\textbf {\bibinfo {volume}
  {D62}},\ \bibinfo {pages} {014505} (\bibinfo {year} {2000})},\ \Eprint
  {http://arxiv.org/abs/hep-lat/0002008} {hep-lat/0002008} \BibitemShut
  {NoStop}%
%%CITATION = HEP-LAT/0002008;%%
\bibitem [{\citenamefont {Oktay}\ and\ \citenamefont
  {Kronfeld}(2008)}]{Oktay:2008ex}%
  \BibitemOpen
  \bibfield  {author} {\bibinfo {author} {\bibfnamefont {M.~B.}\ \bibnamefont
  {Oktay}}\ and\ \bibinfo {author} {\bibfnamefont {A.~S.}\ \bibnamefont
  {Kronfeld}},\ }\href {\doibase 10.1103/PhysRevD.78.014504} {\bibfield
  {journal} {\bibinfo  {journal} {Phys. Rev.}\ }\textbf {\bibinfo {volume}
  {D78}},\ \bibinfo {pages} {014504} (\bibinfo {year} {2008})},\ \Eprint
  {http://arxiv.org/abs/0803.0523} {arXiv:0803.0523 [hep-lat]} \BibitemShut
  {NoStop}%
%%CITATION = ARXIV:0803.0523;%%
\bibitem [{\citenamefont {Harada}\ \emph
  {et~al.}(2002{\natexlab{a}})\citenamefont {Harada}, \citenamefont
  {Hashimoto}, \citenamefont {Ishikawa}, \citenamefont {Kronfeld},
  \citenamefont {Onogi},\ and\ \citenamefont {Yamada}}]{Harada:2001fi}%
  \BibitemOpen
  \bibfield  {author} {\bibinfo {author} {\bibfnamefont {J.}~\bibnamefont
  {Harada}}, \bibinfo {author} {\bibfnamefont {S.}~\bibnamefont {Hashimoto}},
  \bibinfo {author} {\bibfnamefont {K.-I.}\ \bibnamefont {Ishikawa}}, \bibinfo
  {author} {\bibfnamefont {A.~S.}\ \bibnamefont {Kronfeld}}, \bibinfo {author}
  {\bibfnamefont {T.}~\bibnamefont {Onogi}}, \ and\ \bibinfo {author}
  {\bibfnamefont {N.}~\bibnamefont {Yamada}},\ }\href {\doibase
  10.1103/PhysRevD.65.094513, 10.1103/PhysRevD.71.019903} {\bibfield  {journal}
  {\bibinfo  {journal} {Phys. Rev.}\ }\textbf {\bibinfo {volume} {D65}},\
  \bibinfo {pages} {094513} (\bibinfo {year} {2002}{\natexlab{a}})},\ \Eprint
  {http://arxiv.org/abs/hep-lat/0112044} {hep-lat/0112044} \BibitemShut
  {NoStop}%
%%CITATION = HEP-LAT/0112044;%%
\bibitem [{\citenamefont {Harada}\ \emph
  {et~al.}(2002{\natexlab{b}})\citenamefont {Harada}, \citenamefont
  {Hashimoto}, \citenamefont {Kronfeld},\ and\ \citenamefont
  {Onogi}}]{Harada:2001fj}%
  \BibitemOpen
  \bibfield  {author} {\bibinfo {author} {\bibfnamefont {J.}~\bibnamefont
  {Harada}}, \bibinfo {author} {\bibfnamefont {S.}~\bibnamefont {Hashimoto}},
  \bibinfo {author} {\bibfnamefont {A.~S.}\ \bibnamefont {Kronfeld}}, \ and\
  \bibinfo {author} {\bibfnamefont {T.}~\bibnamefont {Onogi}},\ }\href
  {\doibase 10.1103/PhysRevD.65.094514} {\bibfield  {journal} {\bibinfo
  {journal} {Phys. Rev.}\ }\textbf {\bibinfo {volume} {D65}},\ \bibinfo {pages}
  {094514} (\bibinfo {year} {2002}{\natexlab{b}})},\ \Eprint
  {http://arxiv.org/abs/hep-lat/0112045} {arXiv:hep-lat/0112045 [hep-lat]}
  \BibitemShut {NoStop}%
%%CITATION = HEP-LAT/0112045;%%
\bibitem [{\citenamefont {Bailey}\ \emph
  {et~al.}(2014{\natexlab{b}})\citenamefont {Bailey}, \citenamefont {Jang},
  \citenamefont {Lee},\ and\ \citenamefont {Leem}}]{Bailey:2014jga}%
  \BibitemOpen
  \bibfield  {author} {\bibinfo {author} {\bibfnamefont {J.~A.}\ \bibnamefont
  {Bailey}}, \bibinfo {author} {\bibfnamefont {Y.-C.}\ \bibnamefont {Jang}},
  \bibinfo {author} {\bibfnamefont {W.}~\bibnamefont {Lee}}, \ and\ \bibinfo
  {author} {\bibfnamefont {J.}~\bibnamefont {Leem}} (\bibinfo {collaboration}
  {SWME Collaboration}),\ }\href@noop {} {\bibfield  {journal} {\bibinfo
  {journal} {PoS}\ }\textbf {\bibinfo {volume} {LATTICE2014}},\ \bibinfo
  {pages} {389} (\bibinfo {year} {2014}{\natexlab{b}})},\ \Eprint
  {http://arxiv.org/abs/1411.4227} {arXiv:1411.4227 [hep-lat]} \BibitemShut
  {NoStop}%
%%CITATION = ARXIV:1411.4227;%%
\bibitem [{\citenamefont {Sommer}(1994)}]{Sommer:1993ce}%
  \BibitemOpen
  \bibfield  {author} {\bibinfo {author} {\bibfnamefont {R.}~\bibnamefont
  {Sommer}},\ }\href {\doibase 10.1016/0550-3213(94)90473-1} {\bibfield
  {journal} {\bibinfo  {journal} {Nucl. Phys.}\ }\textbf {\bibinfo {volume}
  {B411}},\ \bibinfo {pages} {839} (\bibinfo {year} {1994})},\ \Eprint
  {http://arxiv.org/abs/hep-lat/9310022} {hep-lat/9310022} \BibitemShut
  {NoStop}%
%%CITATION = HEP-LAT/9310022;%%
\bibitem [{\citenamefont {Bernard}\ \emph
  {et~al.}(2000{\natexlab{b}})\citenamefont {Bernard} \emph
  {et~al.}}]{Bernard:2000gd}%
  \BibitemOpen
  \bibfield  {author} {\bibinfo {author} {\bibfnamefont {C.~W.}\ \bibnamefont
  {Bernard}} \emph {et~al.},\ }\href {\doibase 10.1103/PhysRevD.62.034503}
  {\bibfield  {journal} {\bibinfo  {journal} {Phys. Rev.}\ }\textbf {\bibinfo
  {volume} {D62}},\ \bibinfo {pages} {034503} (\bibinfo {year}
  {2000}{\natexlab{b}})},\ \Eprint {http://arxiv.org/abs/hep-lat/0002028}
  {hep-lat/0002028} \BibitemShut {NoStop}%
%%CITATION = HEP-LAT/0002028;%%
\bibitem [{\citenamefont {Bazavov}\ \emph {et~al.}(2012)\citenamefont {Bazavov}
  \emph {et~al.}}]{Bazavov:2011aa}%
  \BibitemOpen
  \bibfield  {author} {\bibinfo {author} {\bibfnamefont {A.}~\bibnamefont
  {Bazavov}} \emph {et~al.} (\bibinfo {collaboration} {Fermilab Lattice and
  MILC Collaborations}),\ }\href {\doibase 10.1103/PhysRevD.85.114506}
  {\bibfield  {journal} {\bibinfo  {journal} {Phys. Rev.}\ }\textbf {\bibinfo
  {volume} {D85}},\ \bibinfo {pages} {114506} (\bibinfo {year} {2012})},\
  \Eprint {http://arxiv.org/abs/1112.3051} {arXiv:1112.3051 [hep-lat]}
  \BibitemShut {NoStop}%
%%CITATION = ARXIV:1112.3051;%%
\bibitem [{\citenamefont {Arthur}\ \emph {et~al.}(2013)\citenamefont {Arthur}
  \emph {et~al.}}]{Arthur:2012opa}%
  \BibitemOpen
  \bibfield  {author} {\bibinfo {author} {\bibfnamefont {R.}~\bibnamefont
  {Arthur}} \emph {et~al.} (\bibinfo {collaboration} {RBC, UKQCD}),\ }\href
  {\doibase 10.1103/PhysRevD.87.094514} {\bibfield  {journal} {\bibinfo
  {journal} {Phys.Rev.}\ }\textbf {\bibinfo {volume} {D87}},\ \bibinfo {pages}
  {094514} (\bibinfo {year} {2013})},\ \Eprint {http://arxiv.org/abs/1208.4412}
  {arXiv:1208.4412 [hep-lat]} \BibitemShut {NoStop}%
%%CITATION = ARXIV:1208.4412;%%
\bibitem [{\citenamefont {Wingate}\ \emph {et~al.}(2003)\citenamefont
  {Wingate}, \citenamefont {Shigemitsu}, \citenamefont {Davies}, \citenamefont
  {Lepage},\ and\ \citenamefont {Trottier}}]{Wingate:2002fh}%
  \BibitemOpen
  \bibfield  {author} {\bibinfo {author} {\bibfnamefont {M.}~\bibnamefont
  {Wingate}}, \bibinfo {author} {\bibfnamefont {J.}~\bibnamefont {Shigemitsu}},
  \bibinfo {author} {\bibfnamefont {C.~T.~H.}\ \bibnamefont {Davies}}, \bibinfo
  {author} {\bibfnamefont {G.~P.}\ \bibnamefont {Lepage}}, \ and\ \bibinfo
  {author} {\bibfnamefont {H.~D.}\ \bibnamefont {Trottier}},\ }\href {\doibase
  10.1103/PhysRevD.67.054505} {\bibfield  {journal} {\bibinfo  {journal} {Phys.
  Rev.}\ }\textbf {\bibinfo {volume} {D67}},\ \bibinfo {pages} {054505}
  (\bibinfo {year} {2003})},\ \Eprint {http://arxiv.org/abs/hep-lat/0211014}
  {hep-lat/0211014} \BibitemShut {NoStop}%
%%CITATION = HEP-LAT/0211014;%%
\bibitem [{\citenamefont {El-Khadra}\ \emph {et~al.}(2001)\citenamefont
  {El-Khadra}, \citenamefont {Kronfeld}, \citenamefont {Mackenzie},
  \citenamefont {Ryan},\ and\ \citenamefont {Simone}}]{ElKhadra:2001rv}%
  \BibitemOpen
  \bibfield  {author} {\bibinfo {author} {\bibfnamefont {A.~X.}\ \bibnamefont
  {El-Khadra}}, \bibinfo {author} {\bibfnamefont {A.~S.}\ \bibnamefont
  {Kronfeld}}, \bibinfo {author} {\bibfnamefont {P.~B.}\ \bibnamefont
  {Mackenzie}}, \bibinfo {author} {\bibfnamefont {S.~M.}\ \bibnamefont {Ryan}},
  \ and\ \bibinfo {author} {\bibfnamefont {J.~N.}\ \bibnamefont {Simone}},\
  }\href {\doibase 10.1103/PhysRevD.64.014502} {\bibfield  {journal} {\bibinfo
  {journal} {Phys. Rev.}\ }\textbf {\bibinfo {volume} {D64}},\ \bibinfo {pages}
  {014502} (\bibinfo {year} {2001})},\ \Eprint
  {http://arxiv.org/abs/hep-ph/0101023} {hep-ph/0101023} \BibitemShut {NoStop}%
%%CITATION = HEP-PH/0101023;%%
\bibitem [{\citenamefont {Qiu}\ \emph {et~al.}(2011)\citenamefont {Qiu} \emph
  {et~al.}}]{Qiu:2011ur}%
  \BibitemOpen
  \bibfield  {author} {\bibinfo {author} {\bibfnamefont {S.-W.}\ \bibnamefont
  {Qiu}} \emph {et~al.} (\bibinfo {collaboration} {Fermilab Lattice and MILC
  Collaborations}),\ }\href@noop {} {\bibfield  {journal} {\bibinfo  {journal}
  {PoS}\ }\textbf {\bibinfo {volume} {LATTICE2011}},\ \bibinfo {pages} {289}
  (\bibinfo {year} {2011})},\ \Eprint {http://arxiv.org/abs/1111.0677}
  {arXiv:1111.0677 [hep-lat]} \BibitemShut {NoStop}%
%%CITATION = ARXIV:1111.0677;%%
\bibitem [{\citenamefont {El-Khadra}\ \emph {et~al.}(2007)\citenamefont
  {El-Khadra}, \citenamefont {{G\'amiz}}, \citenamefont {Kronfeld},\ and\
  \citenamefont {Nobes}}]{ElKhadra:2007qe}%
  \BibitemOpen
  \bibfield  {author} {\bibinfo {author} {\bibfnamefont {A.~X.}\ \bibnamefont
  {El-Khadra}}, \bibinfo {author} {\bibfnamefont {E.}~\bibnamefont
  {{G\'amiz}}}, \bibinfo {author} {\bibfnamefont {A.~S.}\ \bibnamefont
  {Kronfeld}}, \ and\ \bibinfo {author} {\bibfnamefont {M.~A.}\ \bibnamefont
  {Nobes}},\ }\href@noop {} {\bibfield  {journal} {\bibinfo  {journal} {PoS}\
  }\textbf {\bibinfo {volume} {LAT2007}},\ \bibinfo {pages} {242} (\bibinfo
  {year} {2007})},\ \Eprint {http://arxiv.org/abs/0710.1437} {arXiv:0710.1437
  [hep-lat]} \BibitemShut {NoStop}%
%%CITATION = ARXIV:0710.1437;%%
\bibitem [{\citenamefont {Lepage}\ and\ \citenamefont
  {Mackenzie}(1993)}]{Lepage:1992xa}%
  \BibitemOpen
  \bibfield  {author} {\bibinfo {author} {\bibfnamefont {G.~P.}\ \bibnamefont
  {Lepage}}\ and\ \bibinfo {author} {\bibfnamefont {P.~B.}\ \bibnamefont
  {Mackenzie}},\ }\href {\doibase 10.1103/PhysRevD.48.2250} {\bibfield
  {journal} {\bibinfo  {journal} {Phys. Rev.}\ }\textbf {\bibinfo {volume}
  {D48}},\ \bibinfo {pages} {2250} (\bibinfo {year} {1993})},\ \Eprint
  {http://arxiv.org/abs/hep-lat/9209022} {arXiv:hep-lat/9209022 [hep-lat]}
  \BibitemShut {NoStop}%
%%CITATION = HEP-LAT/9209022;%%
\bibitem [{\citenamefont {Mason}\ \emph {et~al.}(2005)\citenamefont {Mason}
  \emph {et~al.}}]{Mason:2005zx}%
  \BibitemOpen
  \bibfield  {author} {\bibinfo {author} {\bibfnamefont {Q.}~\bibnamefont
  {Mason}} \emph {et~al.} (\bibinfo {collaboration} {HPQCD Collaboration}),\
  }\href {\doibase 10.1103/PhysRevLett.95.052002} {\bibfield  {journal}
  {\bibinfo  {journal} {Phys. Rev. Lett.}\ }\textbf {\bibinfo {volume} {95}},\
  \bibinfo {pages} {052002} (\bibinfo {year} {2005})},\ \Eprint
  {http://arxiv.org/abs/hep-lat/0503005} {hep-lat/0503005} \BibitemShut
  {NoStop}%
%%CITATION = HEP-LAT/0503005;%%
\bibitem [{\citenamefont {El-Khadra}\ \emph {et~al.}()\citenamefont
  {El-Khadra}, \citenamefont {{G\'amiz}},\ and\ \citenamefont
  {Kronfeld}}]{rho_in_prep:2012}%
  \BibitemOpen
  \bibfield  {author} {\bibinfo {author} {\bibfnamefont {A.~X.}\ \bibnamefont
  {El-Khadra}}, \bibinfo {author} {\bibfnamefont {E.}~\bibnamefont
  {{G\'amiz}}}, \ and\ \bibinfo {author} {\bibfnamefont {A.~S.}\ \bibnamefont
  {Kronfeld}},\ }\href@noop {} {}\bibinfo {note} {{in preparation}}\BibitemShut
  {NoStop}%
\bibitem [{\citenamefont {Brodsky}\ \emph {et~al.}(1983)\citenamefont
  {Brodsky}, \citenamefont {Lepage},\ and\ \citenamefont
  {Mackenzie}}]{Brodsky:1982gc}%
  \BibitemOpen
  \bibfield  {author} {\bibinfo {author} {\bibfnamefont {S.~J.}\ \bibnamefont
  {Brodsky}}, \bibinfo {author} {\bibfnamefont {G.~P.}\ \bibnamefont {Lepage}},
  \ and\ \bibinfo {author} {\bibfnamefont {P.~B.}\ \bibnamefont {Mackenzie}},\
  }\href {\doibase 10.1103/PhysRevD.28.228} {\bibfield  {journal} {\bibinfo
  {journal} {Phys. Rev.}\ }\textbf {\bibinfo {volume} {D28}},\ \bibinfo {pages}
  {228} (\bibinfo {year} {1983})}\BibitemShut {NoStop}%
%%CITATION = PHRVA,D28,228;%%
\bibitem [{\citenamefont {Richardson}(1979)}]{Richardson:1978bt}%
  \BibitemOpen
  \bibfield  {author} {\bibinfo {author} {\bibfnamefont {J.~L.}\ \bibnamefont
  {Richardson}},\ }\href {\doibase 10.1016/0370-2693(79)90753-6} {\bibfield
  {journal} {\bibinfo  {journal} {Phys. Lett.}\ }\textbf {\bibinfo {volume}
  {B82}},\ \bibinfo {pages} {272} (\bibinfo {year} {1979})}\BibitemShut
  {NoStop}%
%%CITATION = PHLTA,B82,272;%%
\bibitem [{\citenamefont {Menscher}(2005)}]{Menscher:2005}%
  \BibitemOpen
  \bibfield  {author} {\bibinfo {author} {\bibfnamefont {D.~P.}\ \bibnamefont
  {Menscher}},\ }\emph {\bibinfo {title} {{Charmonium and Charmed Mesons with
  Improved Lattice QCD}}},\ \href@noop {} {Ph.D. thesis},\ \bibinfo  {school}
  {University of Illinois} (\bibinfo {year} {2005})\BibitemShut {NoStop}%
%%CITATION = PHLTA,B82,272;%%
\bibitem [{\citenamefont {Bernard}\ \emph {et~al.}(2011)\citenamefont {Bernard}
  \emph {et~al.}}]{Bernard:2010fr}%
  \BibitemOpen
  \bibfield  {author} {\bibinfo {author} {\bibfnamefont {C.}~\bibnamefont
  {Bernard}} \emph {et~al.} (\bibinfo {collaboration} {Fermilab Lattice and
  MILC Collaborations}),\ }\href {\doibase 10.1103/PhysRevD.83.034503}
  {\bibfield  {journal} {\bibinfo  {journal} {Phys. Rev.}\ }\textbf {\bibinfo
  {volume} {D83}},\ \bibinfo {pages} {034503} (\bibinfo {year} {2011})},\
  \Eprint {http://arxiv.org/abs/1003.1937} {arXiv:1003.1937 [hep-lat]}
  \BibitemShut {NoStop}%
%%CITATION = ARXIV:1003.1937;%%
\bibitem [{\citenamefont {Neil}\ \emph {et~al.}(2011)\citenamefont {Neil} \emph
  {et~al.}}]{Neil:2011ku}%
  \BibitemOpen
  \bibfield  {author} {\bibinfo {author} {\bibfnamefont {E.~T.}\ \bibnamefont
  {Neil}} \emph {et~al.} (\bibinfo {collaboration} {Fermilab Lattice and MILC
  Collaborations}),\ }\href@noop {} {\bibfield  {journal} {\bibinfo  {journal}
  {PoS}\ }\textbf {\bibinfo {volume} {LATTICE2011}},\ \bibinfo {pages} {320}
  (\bibinfo {year} {2011})},\ \Eprint {http://arxiv.org/abs/1112.3978}
  {arXiv:1112.3978 [hep-lat]} \BibitemShut {NoStop}%
%%CITATION = ARXIV:1112.3978;%%
\bibitem [{\citenamefont {Lepage}\ \emph {et~al.}(2002)\citenamefont {Lepage}
  \emph {et~al.}}]{Lepage:2001ym}%
  \BibitemOpen
  \bibfield  {author} {\bibinfo {author} {\bibfnamefont {G.~P.}\ \bibnamefont
  {Lepage}} \emph {et~al.},\ }\href {\doibase 10.1016/S0920-5632(01)01638-3}
  {\bibfield  {journal} {\bibinfo  {journal} {Nucl. Phys. Proc. Suppl.}\
  }\textbf {\bibinfo {volume} {106}},\ \bibinfo {pages} {12} (\bibinfo {year}
  {2002})},\ \Eprint {http://arxiv.org/abs/hep-lat/0110175} {hep-lat/0110175}
  \BibitemShut {NoStop}%
%%CITATION = HEP-LAT/0110175;%%
\bibitem [{\citenamefont {Morningstar}(2002)}]{Morningstar:2001je}%
  \BibitemOpen
  \bibfield  {author} {\bibinfo {author} {\bibfnamefont {C.}~\bibnamefont
  {Morningstar}},\ }\href@noop {} {\bibfield  {journal} {\bibinfo  {journal}
  {Nucl. Phys. Proc. Suppl.}\ }\textbf {\bibinfo {volume} {109A}},\ \bibinfo
  {pages} {185} (\bibinfo {year} {2002})},\ \Eprint
  {http://arxiv.org/abs/hep-lat/0112023} {hep-lat/0112023} \BibitemShut
  {NoStop}%
%%CITATION = HEP-LAT/0112023;%%
\bibitem [{\citenamefont {Aubin}\ and\ \citenamefont
  {Bernard}(2006)}]{Aubin:2005aq}%
  \BibitemOpen
  \bibfield  {author} {\bibinfo {author} {\bibfnamefont {C.}~\bibnamefont
  {Aubin}}\ and\ \bibinfo {author} {\bibfnamefont {C.}~\bibnamefont
  {Bernard}},\ }\href {\doibase 10.1103/PhysRevD.73.014515} {\bibfield
  {journal} {\bibinfo  {journal} {Phys. Rev.}\ }\textbf {\bibinfo {volume}
  {D73}},\ \bibinfo {pages} {014515} (\bibinfo {year} {2006})},\ \Eprint
  {http://arxiv.org/abs/hep-lat/0510088} {arXiv:hep-lat/0510088 [hep-lat]}
  \BibitemShut {NoStop}%
%%CITATION = HEP-LAT/0510088;%%
\bibitem [{\citenamefont {Aubin}\ and\ \citenamefont
  {Bernard}(2007)}]{Aubin:2007mc}%
  \BibitemOpen
  \bibfield  {author} {\bibinfo {author} {\bibfnamefont {C.}~\bibnamefont
  {Aubin}}\ and\ \bibinfo {author} {\bibfnamefont {C.}~\bibnamefont
  {Bernard}},\ }\href {\doibase 10.1103/PhysRevD.76.014002} {\bibfield
  {journal} {\bibinfo  {journal} {Phys. Rev.}\ }\textbf {\bibinfo {volume}
  {D76}},\ \bibinfo {pages} {014002} (\bibinfo {year} {2007})},\ \Eprint
  {http://arxiv.org/abs/0704.0795} {arXiv:0704.0795 [hep-lat]} \BibitemShut
  {NoStop}%
%%CITATION = ARXIV:0704.0795;%%
\bibitem [{\citenamefont {Bijnens}\ and\ \citenamefont
  {Jemos}(2010)}]{Bijnens:2010ws}%
  \BibitemOpen
  \bibfield  {author} {\bibinfo {author} {\bibfnamefont {J.}~\bibnamefont
  {Bijnens}}\ and\ \bibinfo {author} {\bibfnamefont {I.}~\bibnamefont
  {Jemos}},\ }\href {\doibase 10.1016/j.nuclphysb.2010.06.021,
  10.1016/j.nuclphysb.2010.10.024} {\bibfield  {journal} {\bibinfo  {journal}
  {Nucl. Phys.}\ }\textbf {\bibinfo {volume} {B840}},\ \bibinfo {pages} {54}
  (\bibinfo {year} {2010})},\ \Eprint {http://arxiv.org/abs/1006.1197}
  {arXiv:1006.1197 [hep-ph]} \BibitemShut {NoStop}%
%%CITATION = ARXIV:1006.1197;%%
\bibitem [{\citenamefont {Bijnens}\ and\ \citenamefont
  {Jemos}(2011)}]{Bijnens:2010jg}%
  \BibitemOpen
  \bibfield  {author} {\bibinfo {author} {\bibfnamefont {J.}~\bibnamefont
  {Bijnens}}\ and\ \bibinfo {author} {\bibfnamefont {I.}~\bibnamefont
  {Jemos}},\ }\href {\doibase 10.1016/j.nuclphysb.2010.12.012} {\bibfield
  {journal} {\bibinfo  {journal} {Nucl. Phys.}\ }\textbf {\bibinfo {volume}
  {B846}},\ \bibinfo {pages} {145} (\bibinfo {year} {2011})},\ \Eprint
  {http://arxiv.org/abs/1011.6531} {arXiv:1011.6531 [hep-ph]} \BibitemShut
  {NoStop}%
%%CITATION = ARXIV:1011.6531;%%
\bibitem [{\citenamefont {Colangelo}\ \emph {et~al.}(2012)\citenamefont
  {Colangelo}, \citenamefont {Procura}, \citenamefont {Rothen}, \citenamefont
  {Stucki},\ and\ \citenamefont {Tarrus~Castella}}]{Colangelo:2012ew}%
  \BibitemOpen
  \bibfield  {author} {\bibinfo {author} {\bibfnamefont {G.}~\bibnamefont
  {Colangelo}}, \bibinfo {author} {\bibfnamefont {M.}~\bibnamefont {Procura}},
  \bibinfo {author} {\bibfnamefont {L.}~\bibnamefont {Rothen}}, \bibinfo
  {author} {\bibfnamefont {R.}~\bibnamefont {Stucki}}, \ and\ \bibinfo {author}
  {\bibfnamefont {J.}~\bibnamefont {Tarrus~Castella}},\ }\href {\doibase
  10.1007/JHEP09(2012)081} {\bibfield  {journal} {\bibinfo  {journal} {JHEP}\
  }\textbf {\bibinfo {volume} {1209}},\ \bibinfo {pages} {081} (\bibinfo {year}
  {2012})},\ \Eprint {http://arxiv.org/abs/1208.0498} {arXiv:1208.0498
  [hep-ph]} \BibitemShut {NoStop}%
%%CITATION = ARXIV:1208.0498;%%
\bibitem [{\citenamefont {Bardeen}\ \emph {et~al.}(2003)\citenamefont
  {Bardeen}, \citenamefont {Eichten},\ and\ \citenamefont
  {Hill}}]{Bardeen:2003kt}%
  \BibitemOpen
  \bibfield  {author} {\bibinfo {author} {\bibfnamefont {W.~A.}\ \bibnamefont
  {Bardeen}}, \bibinfo {author} {\bibfnamefont {E.~J.}\ \bibnamefont
  {Eichten}}, \ and\ \bibinfo {author} {\bibfnamefont {C.~T.}\ \bibnamefont
  {Hill}},\ }\href {\doibase 10.1103/PhysRevD.68.054024} {\bibfield  {journal}
  {\bibinfo  {journal} {Phys. Rev.}\ }\textbf {\bibinfo {volume} {D68}},\
  \bibinfo {pages} {054024} (\bibinfo {year} {2003})},\ \Eprint
  {http://arxiv.org/abs/hep-ph/0305049} {hep-ph/0305049} \BibitemShut {NoStop}%
%%CITATION = HEP-PH/0305049;%%
\bibitem [{\citenamefont {Green}\ \emph {et~al.}(2004)\citenamefont {Green},
  \citenamefont {Koponen}, \citenamefont {McNeile}, \citenamefont {Michael},\
  and\ \citenamefont {Thompson}}]{Green:2003zza}%
  \BibitemOpen
  \bibfield  {author} {\bibinfo {author} {\bibfnamefont {A.~M.}\ \bibnamefont
  {Green}}, \bibinfo {author} {\bibfnamefont {J.}~\bibnamefont {Koponen}},
  \bibinfo {author} {\bibfnamefont {C.}~\bibnamefont {McNeile}}, \bibinfo
  {author} {\bibfnamefont {C.}~\bibnamefont {Michael}}, \ and\ \bibinfo
  {author} {\bibfnamefont {G.}~\bibnamefont {Thompson}} (\bibinfo
  {collaboration} {UKQCD Collaboration}),\ }\href {\doibase
  10.1103/PhysRevD.69.094505} {\bibfield  {journal} {\bibinfo  {journal} {Phys.
  Rev.}\ }\textbf {\bibinfo {volume} {D69}},\ \bibinfo {pages} {094505}
  (\bibinfo {year} {2004})},\ \Eprint {http://arxiv.org/abs/hep-lat/0312007}
  {hep-lat/0312007} \BibitemShut {NoStop}%
%%CITATION = HEP-LAT/0312007;%%
\bibitem [{\citenamefont {Lang}\ \emph {et~al.}(2015)\citenamefont {Lang},
  \citenamefont {Mohler}, \citenamefont {Prelovsek},\ and\ \citenamefont
  {Woloshyn}}]{Lang:2015hza}%
  \BibitemOpen
  \bibfield  {author} {\bibinfo {author} {\bibfnamefont {C.~B.}\ \bibnamefont
  {Lang}}, \bibinfo {author} {\bibfnamefont {D.}~\bibnamefont {Mohler}},
  \bibinfo {author} {\bibfnamefont {S.}~\bibnamefont {Prelovsek}}, \ and\
  \bibinfo {author} {\bibfnamefont {R.~M.}\ \bibnamefont {Woloshyn}},\
  }\href@noop {} {\  (\bibinfo {year} {2015})},\ \Eprint
  {http://arxiv.org/abs/1501.01646} {arXiv:1501.01646 [hep-lat]} \BibitemShut
  {NoStop}%
%%CITATION = ARXIV:1501.01646;%%
\bibitem [{\citenamefont {Detmold}\ \emph
  {et~al.}(2012{\natexlab{a}})\citenamefont {Detmold}, \citenamefont {Lin},\
  and\ \citenamefont {Meinel}}]{Detmold:2011bp}%
  \BibitemOpen
  \bibfield  {author} {\bibinfo {author} {\bibfnamefont {W.}~\bibnamefont
  {Detmold}}, \bibinfo {author} {\bibfnamefont {C.-J.~D.}\ \bibnamefont {Lin}},
  \ and\ \bibinfo {author} {\bibfnamefont {S.}~\bibnamefont {Meinel}},\
  }\href@noop {} {\bibfield  {journal} {\bibinfo  {journal} {Phys. Rev. Lett.}\
  }\textbf {\bibinfo {volume} {108}},\ \bibinfo {pages} {172003} (\bibinfo
  {year} {2012}{\natexlab{a}})},\ \Eprint {http://arxiv.org/abs/1109.2480}
  {arXiv:1109.2480 [hep-lat]} \BibitemShut {NoStop}%
%%CITATION = ARXIV:1109.2480;%%
\bibitem [{\citenamefont {Detmold}\ \emph
  {et~al.}(2012{\natexlab{b}})\citenamefont {Detmold}, \citenamefont {Lin},\
  and\ \citenamefont {Meinel}}]{Detmold:2012ge}%
  \BibitemOpen
  \bibfield  {author} {\bibinfo {author} {\bibfnamefont {W.}~\bibnamefont
  {Detmold}}, \bibinfo {author} {\bibfnamefont {C.-J.~D.}\ \bibnamefont {Lin}},
  \ and\ \bibinfo {author} {\bibfnamefont {S.}~\bibnamefont {Meinel}},\ }\href
  {\doibase 10.1103/PhysRevD.85.114508} {\bibfield  {journal} {\bibinfo
  {journal} {Phys. Rev.}\ }\textbf {\bibinfo {volume} {D85}},\ \bibinfo {pages}
  {114508} (\bibinfo {year} {2012}{\natexlab{b}})},\ \Eprint
  {http://arxiv.org/abs/1203.3378} {arXiv:1203.3378 [hep-lat]} \BibitemShut
  {NoStop}%
%%CITATION = ARXIV:1203.3378;%%
\bibitem [{\citenamefont {Flynn}\ \emph {et~al.}(2015)\citenamefont {Flynn}
  \emph {et~al.}}]{Flynn:2015xna}%
  \BibitemOpen
  \bibfield  {author} {\bibinfo {author} {\bibfnamefont {J.}~\bibnamefont
  {Flynn}} \emph {et~al.} (\bibinfo {collaboration} {RBC/UKQCD}),\ }\href@noop
  {} {\  (\bibinfo {year} {2015})},\ \Eprint {http://arxiv.org/abs/1506.06413}
  {arXiv:1506.06413 [hep-lat]} \BibitemShut {NoStop}%
%%CITATION = ARXIV:1506.06413;%%
\bibitem [{\citenamefont {Flynn}\ \emph {et~al.}(2014)\citenamefont {Flynn}
  \emph {et~al.}}]{Flynn:2013kwa}%
  \BibitemOpen
  \bibfield  {author} {\bibinfo {author} {\bibfnamefont {J.~M.}\ \bibnamefont
  {Flynn}} \emph {et~al.} (\bibinfo {collaboration} {RBC and UKQCD
  Collaborations}),\ }\href@noop {} {\bibfield  {journal} {\bibinfo  {journal}
  {PoS}\ }\textbf {\bibinfo {volume} {LATTICE2013}},\ \bibinfo {pages} {408}
  (\bibinfo {year} {2014})},\ \Eprint {http://arxiv.org/abs/1311.2251}
  {arXiv:1311.2251 [hep-lat]} \BibitemShut {NoStop}%
%%CITATION = ARXIV:1311.2251;%%
\bibitem [{\citenamefont {Bernardoni}\ \emph {et~al.}(2015)\citenamefont
  {Bernardoni}, \citenamefont {Bulava}, \citenamefont {Donnellan},\ and\
  \citenamefont {Sommer}}]{Bernardoni:2014kla}%
  \BibitemOpen
  \bibfield  {author} {\bibinfo {author} {\bibfnamefont {F.}~\bibnamefont
  {Bernardoni}}, \bibinfo {author} {\bibfnamefont {J.}~\bibnamefont {Bulava}},
  \bibinfo {author} {\bibfnamefont {M.}~\bibnamefont {Donnellan}}, \ and\
  \bibinfo {author} {\bibfnamefont {R.}~\bibnamefont {Sommer}} (\bibinfo
  {collaboration} {ALPHA Collaboration}),\ }\href {\doibase
  10.1016/j.physletb.2014.11.051} {\bibfield  {journal} {\bibinfo  {journal}
  {Phys. Lett.}\ }\textbf {\bibinfo {volume} {B740}},\ \bibinfo {pages} {278}
  (\bibinfo {year} {2015})},\ \Eprint {http://arxiv.org/abs/1404.6951}
  {arXiv:1404.6951 [hep-lat]} \BibitemShut {NoStop}%
%%CITATION = ARXIV:1404.6951;%%
\bibitem [{\citenamefont {{Be\v{c}irevi\'c}}\ \emph
  {et~al.}(2007{\natexlab{a}})\citenamefont {{Be\v{c}irevi\'c}}, \citenamefont
  {Fajfer},\ and\ \citenamefont {Kamenik}}]{Becirevic:2007dg}%
  \BibitemOpen
  \bibfield  {author} {\bibinfo {author} {\bibfnamefont {D.}~\bibnamefont
  {{Be\v{c}irevi\'c}}}, \bibinfo {author} {\bibfnamefont {S.}~\bibnamefont
  {Fajfer}}, \ and\ \bibinfo {author} {\bibfnamefont {J.~F.}\ \bibnamefont
  {Kamenik}},\ }\href@noop {} {\bibfield  {journal} {\bibinfo  {journal} {PoS}\
  }\textbf {\bibinfo {volume} {LAT2007}},\ \bibinfo {pages} {063} (\bibinfo
  {year} {2007}{\natexlab{a}})},\ \Eprint {http://arxiv.org/abs/0710.3496}
  {arXiv:0710.3496 [hep-lat]} \BibitemShut {NoStop}%
%%CITATION = ARXIV:0710.3496;%%
\bibitem [{\citenamefont {Arndt}\ and\ \citenamefont
  {Lin}(2004)}]{Arndt:2004bg}%
  \BibitemOpen
  \bibfield  {author} {\bibinfo {author} {\bibfnamefont {D.}~\bibnamefont
  {Arndt}}\ and\ \bibinfo {author} {\bibfnamefont {C.~J.~D.}\ \bibnamefont
  {Lin}},\ }\href {\doibase 10.1103/PhysRevD.70.014503} {\bibfield  {journal}
  {\bibinfo  {journal} {Phys. Rev.}\ }\textbf {\bibinfo {volume} {D70}},\
  \bibinfo {pages} {014503} (\bibinfo {year} {2004})},\ \Eprint
  {http://arxiv.org/abs/hep-lat/0403012} {hep-lat/0403012} \BibitemShut
  {NoStop}%
%%CITATION = HEP-LAT/0403012;%%
\bibitem [{\citenamefont {Amhis}\ \emph {et~al.}(2012)\citenamefont {Amhis}
  \emph {et~al.}}]{Amhis:2012bh}%
  \BibitemOpen
  \bibfield  {author} {\bibinfo {author} {\bibfnamefont {Y.}~\bibnamefont
  {Amhis}} \emph {et~al.} (\bibinfo {collaboration} {Heavy Flavor Averaging
  Group}),\ }\href@noop {} {\  (\bibinfo {year} {2012})},\ \Eprint
  {http://arxiv.org/abs/1207.1158} {arXiv:1207.1158 [hep-ex]} \BibitemShut
  {NoStop}%
%%CITATION = ARXIV:1207.1158;%%
\bibitem [{\citenamefont {Aoki}\ \emph {et~al.}(2014)\citenamefont {Aoki} \emph
  {et~al.}}]{Aoki:2013ldr}%
  \BibitemOpen
  \bibfield  {author} {\bibinfo {author} {\bibfnamefont {S.}~\bibnamefont
  {Aoki}} \emph {et~al.} (\bibinfo {collaboration} {Flavour Lattice Averaging
  Group}),\ }\href {\doibase 10.1140/epjc/s10052-014-2890-7} {\bibfield
  {journal} {\bibinfo  {journal} {Eur. Phys. J.}\ }\textbf {\bibinfo {volume}
  {C74}},\ \bibinfo {pages} {2890} (\bibinfo {year} {2014})},\ \Eprint
  {http://arxiv.org/abs/1310.8555} {arXiv:1310.8555 [hep-lat]} \BibitemShut
  {NoStop}%
%%CITATION = ARXIV:1310.8555;%%
\bibitem [{\citenamefont {Becher}\ and\ \citenamefont
  {Hill}(2006)}]{Becher:2005bg}%
  \BibitemOpen
  \bibfield  {author} {\bibinfo {author} {\bibfnamefont {T.}~\bibnamefont
  {Becher}}\ and\ \bibinfo {author} {\bibfnamefont {R.~J.}\ \bibnamefont
  {Hill}},\ }\href {\doibase 10.1016/j.physletb.2005.11.063} {\bibfield
  {journal} {\bibinfo  {journal} {Phys. Lett.}\ }\textbf {\bibinfo {volume}
  {B633}},\ \bibinfo {pages} {61} (\bibinfo {year} {2006})},\ \Eprint
  {http://arxiv.org/abs/hep-ph/0509090} {hep-ph/0509090} \BibitemShut {NoStop}%
%%CITATION = HEP-PH/0509090;%%
\bibitem [{\citenamefont {Bobeth}\ \emph {et~al.}(2012)\citenamefont {Bobeth},
  \citenamefont {Hiller}, \citenamefont {van Dyk},\ and\ \citenamefont
  {Wacker}}]{Bobeth:2011nj}%
  \BibitemOpen
  \bibfield  {author} {\bibinfo {author} {\bibfnamefont {C.}~\bibnamefont
  {Bobeth}}, \bibinfo {author} {\bibfnamefont {G.}~\bibnamefont {Hiller}},
  \bibinfo {author} {\bibfnamefont {D.}~\bibnamefont {van Dyk}}, \ and\
  \bibinfo {author} {\bibfnamefont {C.}~\bibnamefont {Wacker}},\ }\href
  {\doibase 10.1007/JHEP01(2012)107} {\bibfield  {journal} {\bibinfo  {journal}
  {JHEP}\ }\textbf {\bibinfo {volume} {1201}},\ \bibinfo {pages} {107}
  (\bibinfo {year} {2012})},\ \Eprint {http://arxiv.org/abs/1111.2558}
  {arXiv:1111.2558 [hep-ph]} \BibitemShut {NoStop}%
%%CITATION = ARXIV:1111.2558;%%
\bibitem [{\citenamefont {Grinstein}\ and\ \citenamefont
  {Pirjol}(2004)}]{Grinstein:2004vb}%
  \BibitemOpen
  \bibfield  {author} {\bibinfo {author} {\bibfnamefont {B.}~\bibnamefont
  {Grinstein}}\ and\ \bibinfo {author} {\bibfnamefont {D.}~\bibnamefont
  {Pirjol}},\ }\href {\doibase 10.1103/PhysRevD.70.114005} {\bibfield
  {journal} {\bibinfo  {journal} {Phys. Rev.}\ }\textbf {\bibinfo {volume}
  {D70}},\ \bibinfo {pages} {114005} (\bibinfo {year} {2004})},\ \Eprint
  {http://arxiv.org/abs/hep-ph/0404250} {arXiv:hep-ph/0404250 [hep-ph]}
  \BibitemShut {NoStop}%
%%CITATION = HEP-PH/0404250;%%
\bibitem [{\citenamefont {Becher}\ \emph {et~al.}(2005)\citenamefont {Becher},
  \citenamefont {Hill},\ and\ \citenamefont {Neubert}}]{Becher:2005fg}%
  \BibitemOpen
  \bibfield  {author} {\bibinfo {author} {\bibfnamefont {T.}~\bibnamefont
  {Becher}}, \bibinfo {author} {\bibfnamefont {R.~J.}\ \bibnamefont {Hill}}, \
  and\ \bibinfo {author} {\bibfnamefont {M.}~\bibnamefont {Neubert}},\ }\href
  {\doibase 10.1103/PhysRevD.72.094017} {\bibfield  {journal} {\bibinfo
  {journal} {Phys. Rev.}\ }\textbf {\bibinfo {volume} {D72}},\ \bibinfo {pages}
  {094017} (\bibinfo {year} {2005})},\ \Eprint
  {http://arxiv.org/abs/hep-ph/0503263} {arXiv:hep-ph/0503263 [hep-ph]}
  \BibitemShut {NoStop}%
%%CITATION = HEP-PH/0503263;%%
\bibitem [{\citenamefont {Ali}\ \emph {et~al.}(2008)\citenamefont {Ali},
  \citenamefont {Pecjak},\ and\ \citenamefont {Greub}}]{Ali:2007sj}%
  \BibitemOpen
  \bibfield  {author} {\bibinfo {author} {\bibfnamefont {A.}~\bibnamefont
  {Ali}}, \bibinfo {author} {\bibfnamefont {B.~D.}\ \bibnamefont {Pecjak}}, \
  and\ \bibinfo {author} {\bibfnamefont {C.}~\bibnamefont {Greub}},\ }\href
  {\doibase 10.1140/epjc/s10052-008-0623-5} {\bibfield  {journal} {\bibinfo
  {journal} {Eur. Phys. J.}\ }\textbf {\bibinfo {volume} {C55}},\ \bibinfo
  {pages} {577} (\bibinfo {year} {2008})},\ \Eprint
  {http://arxiv.org/abs/0709.4422} {arXiv:0709.4422 [hep-ph]} \BibitemShut
  {NoStop}%
%%CITATION = ARXIV:0709.4422;%%
\bibitem [{\citenamefont {Bobeth}\ \emph {et~al.}(2010)\citenamefont {Bobeth},
  \citenamefont {Hiller},\ and\ \citenamefont {van Dyk}}]{Bobeth:2010wg}%
  \BibitemOpen
  \bibfield  {author} {\bibinfo {author} {\bibfnamefont {C.}~\bibnamefont
  {Bobeth}}, \bibinfo {author} {\bibfnamefont {G.}~\bibnamefont {Hiller}}, \
  and\ \bibinfo {author} {\bibfnamefont {D.}~\bibnamefont {van Dyk}},\ }\href
  {\doibase 10.1007/JHEP07(2010)098} {\bibfield  {journal} {\bibinfo  {journal}
  {JHEP}\ }\textbf {\bibinfo {volume} {1007}},\ \bibinfo {pages} {098}
  (\bibinfo {year} {2010})},\ \Eprint {http://arxiv.org/abs/1006.5013}
  {arXiv:1006.5013 [hep-ph]} \BibitemShut {NoStop}%
%%CITATION = ARXIV:1006.5013;%%
\bibitem [{\citenamefont {Beylich}\ \emph {et~al.}(2011)\citenamefont
  {Beylich}, \citenamefont {Buchalla},\ and\ \citenamefont
  {Feldmann}}]{Beylich:2011aq}%
  \BibitemOpen
  \bibfield  {author} {\bibinfo {author} {\bibfnamefont {M.}~\bibnamefont
  {Beylich}}, \bibinfo {author} {\bibfnamefont {G.}~\bibnamefont {Buchalla}}, \
  and\ \bibinfo {author} {\bibfnamefont {T.}~\bibnamefont {Feldmann}},\ }\href
  {\doibase 10.1140/epjc/s10052-011-1635-0} {\bibfield  {journal} {\bibinfo
  {journal} {Eur. Phys. J.}\ }\textbf {\bibinfo {volume} {C71}},\ \bibinfo
  {pages} {1635} (\bibinfo {year} {2011})},\ \Eprint
  {http://arxiv.org/abs/1101.5118} {arXiv:1101.5118 [hep-ph]} \BibitemShut
  {NoStop}%
%%CITATION = ARXIV:1101.5118;%%
\bibitem [{\citenamefont {Hou}\ \emph {et~al.}(2014)\citenamefont {Hou},
  \citenamefont {Kohda},\ and\ \citenamefont {Xu}}]{Hou:2014dza}%
  \BibitemOpen
  \bibfield  {author} {\bibinfo {author} {\bibfnamefont {W.-S.}\ \bibnamefont
  {Hou}}, \bibinfo {author} {\bibfnamefont {M.}~\bibnamefont {Kohda}}, \ and\
  \bibinfo {author} {\bibfnamefont {F.}~\bibnamefont {Xu}},\ }\href {\doibase
  10.1103/PhysRevD.90.013002} {\bibfield  {journal} {\bibinfo  {journal} {Phys.
  Rev.}\ }\textbf {\bibinfo {volume} {D90}},\ \bibinfo {pages} {013002}
  (\bibinfo {year} {2014})},\ \Eprint {http://arxiv.org/abs/1403.7410}
  {arXiv:1403.7410 [hep-ph]} \BibitemShut {NoStop}%
%%CITATION = ARXIV:1403.7410;%%
\bibitem [{\citenamefont {Burdman}\ \emph {et~al.}(1994)\citenamefont
  {Burdman}, \citenamefont {Ligeti}, \citenamefont {Neubert},\ and\
  \citenamefont {Nir}}]{Burdman:1993es}%
  \BibitemOpen
  \bibfield  {author} {\bibinfo {author} {\bibfnamefont {G.}~\bibnamefont
  {Burdman}}, \bibinfo {author} {\bibfnamefont {Z.}~\bibnamefont {Ligeti}},
  \bibinfo {author} {\bibfnamefont {M.}~\bibnamefont {Neubert}}, \ and\
  \bibinfo {author} {\bibfnamefont {Y.}~\bibnamefont {Nir}},\ }\href {\doibase
  10.1103/PhysRevD.49.2331} {\bibfield  {journal} {\bibinfo  {journal} {Phys.
  Rev.}\ }\textbf {\bibinfo {volume} {D49}},\ \bibinfo {pages} {2331} (\bibinfo
  {year} {1994})},\ \Eprint {http://arxiv.org/abs/hep-ph/9309272}
  {arXiv:hep-ph/9309272 [hep-ph]} \BibitemShut {NoStop}%
%%CITATION = HEP-PH/9309272;%%
\bibitem [{\citenamefont {Wise}(1992)}]{Wise:1992hn}%
  \BibitemOpen
  \bibfield  {author} {\bibinfo {author} {\bibfnamefont {M.~B.}\ \bibnamefont
  {Wise}},\ }\href {\doibase 10.1103/PhysRevD.45.R2188} {\bibfield  {journal}
  {\bibinfo  {journal} {Phys. Rev.}\ }\textbf {\bibinfo {volume} {D45}},\
  \bibinfo {pages} {2188} (\bibinfo {year} {1992})}\BibitemShut {NoStop}%
%%CITATION = PHRVA,D45,2188;%%
\bibitem [{\citenamefont {Hewett}\ and\ \citenamefont
  {Hitlin}(2005)}]{Hewett:2004tv}%
  \BibitemOpen
  \bibinfo {editor} {\bibfnamefont {J.~L.}\ \bibnamefont {Hewett}}\ and\
  \bibinfo {editor} {\bibfnamefont {D.~G.}\ \bibnamefont {Hitlin}},\ eds.,\
  \href@noop {} {\emph {\bibinfo {title} {{The Discovery potential of a Super
  $B$ Factory}}}}\ (\bibinfo  {publisher} {SLAC},\ \bibinfo {address} {Menlo
  Park},\ \bibinfo {year} {2005})\ \Eprint
  {http://arxiv.org/abs/hep-ph/0503261} {arXiv:hep-ph/0503261 [hep-ph]}
  \BibitemShut {NoStop}%
%%CITATION = HEP-PH/0503261;%%
\bibitem [{\citenamefont {Isgur}\ and\ \citenamefont
  {Wise}(1990)}]{Isgur:1990kf}%
  \BibitemOpen
  \bibfield  {author} {\bibinfo {author} {\bibfnamefont {N.}~\bibnamefont
  {Isgur}}\ and\ \bibinfo {author} {\bibfnamefont {M.~B.}\ \bibnamefont
  {Wise}},\ }\href {\doibase 10.1103/PhysRevD.42.2388} {\bibfield  {journal}
  {\bibinfo  {journal} {Phys. Rev.}\ }\textbf {\bibinfo {volume} {D42}},\
  \bibinfo {pages} {2388} (\bibinfo {year} {1990})}\BibitemShut {NoStop}%
%%CITATION = PHRVA,D42,2388;%%
\bibitem [{\citenamefont {Colquhoun}\ \emph {et~al.}(2015)\citenamefont
  {Colquhoun} \emph {et~al.}}]{Colquhoun:2015oha}%
  \BibitemOpen
  \bibfield  {author} {\bibinfo {author} {\bibfnamefont {B.}~\bibnamefont
  {Colquhoun}} \emph {et~al.} (\bibinfo {collaboration} {HPQCD}),\ }\href@noop
  {} {\  (\bibinfo {year} {2015})},\ \Eprint {http://arxiv.org/abs/1503.05762}
  {arXiv:1503.05762 [hep-lat]} \BibitemShut {NoStop}%
%%CITATION = ARXIV:1503.05762;%%
\bibitem [{\citenamefont {Charles}\ \emph {et~al.}(1999)\citenamefont
  {Charles}, \citenamefont {Le~Yaouanc}, \citenamefont {Oliver}, \citenamefont
  {Pene},\ and\ \citenamefont {Raynal}}]{Charles:1998dr}%
  \BibitemOpen
  \bibfield  {author} {\bibinfo {author} {\bibfnamefont {J.}~\bibnamefont
  {Charles}}, \bibinfo {author} {\bibfnamefont {A.}~\bibnamefont {Le~Yaouanc}},
  \bibinfo {author} {\bibfnamefont {L.}~\bibnamefont {Oliver}}, \bibinfo
  {author} {\bibfnamefont {O.}~\bibnamefont {Pene}}, \ and\ \bibinfo {author}
  {\bibfnamefont {J.~C.}\ \bibnamefont {Raynal}},\ }\href {\doibase
  10.1103/PhysRevD.60.014001} {\bibfield  {journal} {\bibinfo  {journal} {Phys.
  Rev.}\ }\textbf {\bibinfo {volume} {D60}},\ \bibinfo {pages} {014001}
  (\bibinfo {year} {1999})},\ \Eprint {http://arxiv.org/abs/hep-ph/9812358}
  {arXiv:hep-ph/9812358 [hep-ph]} \BibitemShut {NoStop}%
%%CITATION = HEP-PH/9812358;%%
\bibitem [{\citenamefont {Beneke}\ and\ \citenamefont
  {Feldmann}(2001)}]{Beneke:2000wa}%
  \BibitemOpen
  \bibfield  {author} {\bibinfo {author} {\bibfnamefont {M.}~\bibnamefont
  {Beneke}}\ and\ \bibinfo {author} {\bibfnamefont {T.}~\bibnamefont
  {Feldmann}},\ }\href {\doibase 10.1016/S0550-3213(00)00585-X} {\bibfield
  {journal} {\bibinfo  {journal} {Nucl. Phys.}\ }\textbf {\bibinfo {volume}
  {B592}},\ \bibinfo {pages} {3} (\bibinfo {year} {2001})},\ \Eprint
  {http://arxiv.org/abs/hep-ph/0008255} {arXiv:hep-ph/0008255 [hep-ph]}
  \BibitemShut {NoStop}%
%%CITATION = HEP-PH/0008255;%%
\bibitem [{\citenamefont {Ball}\ and\ \citenamefont
  {Zwicky}(2006)}]{Ball:2006nr}%
  \BibitemOpen
  \bibfield  {author} {\bibinfo {author} {\bibfnamefont {P.}~\bibnamefont
  {Ball}}\ and\ \bibinfo {author} {\bibfnamefont {R.}~\bibnamefont {Zwicky}},\
  }\href {\doibase 10.1088/1126-6708/2006/04/046} {\bibfield  {journal}
  {\bibinfo  {journal} {JHEP}\ }\textbf {\bibinfo {volume} {0604}},\ \bibinfo
  {pages} {046} (\bibinfo {year} {2006})},\ \Eprint
  {http://arxiv.org/abs/hep-ph/0603232} {arXiv:hep-ph/0603232 [hep-ph]}
  \BibitemShut {NoStop}%
%%CITATION = HEP-PH/0603232;%%
\bibitem [{\citenamefont {Braun}\ \emph {et~al.}(2004)\citenamefont {Braun},
  \citenamefont {Ivanov},\ and\ \citenamefont {Korchemsky}}]{Braun:2003wx}%
  \BibitemOpen
  \bibfield  {author} {\bibinfo {author} {\bibfnamefont {V.~M.}\ \bibnamefont
  {Braun}}, \bibinfo {author} {\bibfnamefont {D.~Y.}\ \bibnamefont {Ivanov}}, \
  and\ \bibinfo {author} {\bibfnamefont {G.~P.}\ \bibnamefont {Korchemsky}},\
  }\href {\doibase 10.1103/PhysRevD.69.034014} {\bibfield  {journal} {\bibinfo
  {journal} {Phys. Rev.}\ }\textbf {\bibinfo {volume} {D69}},\ \bibinfo {pages}
  {034014} (\bibinfo {year} {2004})},\ \Eprint
  {http://arxiv.org/abs/hep-ph/0309330} {arXiv:hep-ph/0309330 [hep-ph]}
  \BibitemShut {NoStop}%
%%CITATION = HEP-PH/0309330;%%
\bibitem [{\citenamefont {Khodjamirian}\ \emph {et~al.}(2005)\citenamefont
  {Khodjamirian}, \citenamefont {Mannel},\ and\ \citenamefont
  {Offen}}]{Khodjamirian:2005ea}%
  \BibitemOpen
  \bibfield  {author} {\bibinfo {author} {\bibfnamefont {A.}~\bibnamefont
  {Khodjamirian}}, \bibinfo {author} {\bibfnamefont {T.}~\bibnamefont
  {Mannel}}, \ and\ \bibinfo {author} {\bibfnamefont {N.}~\bibnamefont
  {Offen}},\ }\href {\doibase 10.1016/j.physletb.2005.06.021} {\bibfield
  {journal} {\bibinfo  {journal} {Phys. Lett.}\ }\textbf {\bibinfo {volume}
  {B620}},\ \bibinfo {pages} {52} (\bibinfo {year} {2005})},\ \Eprint
  {http://arxiv.org/abs/hep-ph/0504091} {arXiv:hep-ph/0504091 [hep-ph]}
  \BibitemShut {NoStop}%
%%CITATION = HEP-PH/0504091;%%
\bibitem [{\citenamefont {Lee}\ and\ \citenamefont
  {Neubert}(2005)}]{Lee:2005gza}%
  \BibitemOpen
  \bibfield  {author} {\bibinfo {author} {\bibfnamefont {S.~J.}\ \bibnamefont
  {Lee}}\ and\ \bibinfo {author} {\bibfnamefont {M.}~\bibnamefont {Neubert}},\
  }\href {\doibase 10.1103/PhysRevD.72.094028} {\bibfield  {journal} {\bibinfo
  {journal} {Phys. Rev.}\ }\textbf {\bibinfo {volume} {D72}},\ \bibinfo {pages}
  {094028} (\bibinfo {year} {2005})},\ \Eprint
  {http://arxiv.org/abs/hep-ph/0509350} {arXiv:hep-ph/0509350 [hep-ph]}
  \BibitemShut {NoStop}%
%%CITATION = HEP-PH/0509350;%%
\bibitem [{\citenamefont {Arthur}\ \emph {et~al.}(2011)\citenamefont {Arthur}
  \emph {et~al.}}]{Arthur:2010xf}%
  \BibitemOpen
  \bibfield  {author} {\bibinfo {author} {\bibfnamefont {R.}~\bibnamefont
  {Arthur}} \emph {et~al.} (\bibinfo {collaboration} {RBC and UKQCD}),\ }\href
  {\doibase 10.1103/PhysRevD.83.074505} {\bibfield  {journal} {\bibinfo
  {journal} {Phys. Rev.}\ }\textbf {\bibinfo {volume} {D83}},\ \bibinfo {pages}
  {074505} (\bibinfo {year} {2011})},\ \Eprint {http://arxiv.org/abs/1011.5906}
  {arXiv:1011.5906 [hep-lat]} \BibitemShut {NoStop}%
%%CITATION = ARXIV:1011.5906;%%
\bibitem [{\citenamefont {Chetyrkin}\ \emph {et~al.}(2000)\citenamefont
  {Chetyrkin}, \citenamefont {Kuhn},\ and\ \citenamefont
  {Steinhauser}}]{Chetyrkin:2000yt}%
  \BibitemOpen
  \bibfield  {author} {\bibinfo {author} {\bibfnamefont {K.}~\bibnamefont
  {Chetyrkin}}, \bibinfo {author} {\bibfnamefont {J.~H.}\ \bibnamefont {Kuhn}},
  \ and\ \bibinfo {author} {\bibfnamefont {M.}~\bibnamefont {Steinhauser}},\
  }\href {\doibase 10.1016/S0010-4655(00)00155-7} {\bibfield  {journal}
  {\bibinfo  {journal} {Comput.Phys.Commun.}\ }\textbf {\bibinfo {volume}
  {133}},\ \bibinfo {pages} {43} (\bibinfo {year} {2000})},\ \Eprint
  {http://arxiv.org/abs/hep-ph/0004189} {arXiv:hep-ph/0004189 [hep-ph]}
  \BibitemShut {NoStop}%
%%CITATION = HEP-PH/0004189;%%
\bibitem [{\citenamefont {Bell}\ \emph {et~al.}(2011)\citenamefont {Bell},
  \citenamefont {Beneke}, \citenamefont {Huber},\ and\ \citenamefont
  {Li}}]{Bell:2010mg}%
  \BibitemOpen
  \bibfield  {author} {\bibinfo {author} {\bibfnamefont {G.}~\bibnamefont
  {Bell}}, \bibinfo {author} {\bibfnamefont {M.}~\bibnamefont {Beneke}},
  \bibinfo {author} {\bibfnamefont {T.}~\bibnamefont {Huber}}, \ and\ \bibinfo
  {author} {\bibfnamefont {X.-Q.}\ \bibnamefont {Li}},\ }\href {\doibase
  10.1016/j.nuclphysb.2010.09.022} {\bibfield  {journal} {\bibinfo  {journal}
  {Nucl. Phys.}\ }\textbf {\bibinfo {volume} {B843}},\ \bibinfo {pages} {143}
  (\bibinfo {year} {2011})},\ \Eprint {http://arxiv.org/abs/1007.3758}
  {arXiv:1007.3758 [hep-ph]} \BibitemShut {NoStop}%
%%CITATION = ARXIV:1007.3758;%%
\bibitem [{\citenamefont {Khodjamirian}\ \emph {et~al.}(2010)\citenamefont
  {Khodjamirian}, \citenamefont {Mannel}, \citenamefont {Pivovarov},\ and\
  \citenamefont {Wang}}]{Khodjamirian:2010vf}%
  \BibitemOpen
  \bibfield  {author} {\bibinfo {author} {\bibfnamefont {A.}~\bibnamefont
  {Khodjamirian}}, \bibinfo {author} {\bibfnamefont {T.}~\bibnamefont
  {Mannel}}, \bibinfo {author} {\bibfnamefont {A.~A.}\ \bibnamefont
  {Pivovarov}}, \ and\ \bibinfo {author} {\bibfnamefont {Y.-M.}\ \bibnamefont
  {Wang}},\ }\href {\doibase 10.1007/JHEP09(2010)089} {\bibfield  {journal}
  {\bibinfo  {journal} {JHEP}\ }\textbf {\bibinfo {volume} {1009}},\ \bibinfo
  {pages} {089} (\bibinfo {year} {2010})},\ \Eprint
  {http://arxiv.org/abs/1006.4945} {arXiv:1006.4945 [hep-ph]} \BibitemShut
  {NoStop}%
%%CITATION = ARXIV:1006.4945;%%
\bibitem [{\citenamefont {Glashow}\ \emph {et~al.}(2015)\citenamefont
  {Glashow}, \citenamefont {Guadagnoli},\ and\ \citenamefont
  {Lane}}]{Glashow:2014iga}%
  \BibitemOpen
  \bibfield  {author} {\bibinfo {author} {\bibfnamefont {S.~L.}\ \bibnamefont
  {Glashow}}, \bibinfo {author} {\bibfnamefont {D.}~\bibnamefont {Guadagnoli}},
  \ and\ \bibinfo {author} {\bibfnamefont {K.}~\bibnamefont {Lane}},\ }\href
  {\doibase 10.1103/PhysRevLett.114.091801} {\bibfield  {journal} {\bibinfo
  {journal} {Phys. Rev. Lett.}\ }\textbf {\bibinfo {volume} {114}},\ \bibinfo
  {pages} {091801} (\bibinfo {year} {2015})},\ \Eprint
  {http://arxiv.org/abs/1411.0565} {arXiv:1411.0565 [hep-ph]} \BibitemShut
  {NoStop}%
%%CITATION = ARXIV:1411.0565;%%
\bibitem [{\citenamefont {Du}\ \emph {et~al.}(2015)\citenamefont {Du},
  \citenamefont {El-Khadra}, \citenamefont {Gottlieb}, \citenamefont
  {Kronfeld}, \citenamefont {Laiho}, \citenamefont {Lunghi}, \citenamefont
  {Van~de Water},\ and\ \citenamefont {Zhou}}]{Du:2015tda}%
  \BibitemOpen
  \bibfield  {author} {\bibinfo {author} {\bibfnamefont {D.}~\bibnamefont
  {Du}}, \bibinfo {author} {\bibfnamefont {A.~X.}\ \bibnamefont {El-Khadra}},
  \bibinfo {author} {\bibfnamefont {S.}~\bibnamefont {Gottlieb}}, \bibinfo
  {author} {\bibfnamefont {A.~S.}\ \bibnamefont {Kronfeld}}, \bibinfo {author}
  {\bibfnamefont {J.}~\bibnamefont {Laiho}}, \bibinfo {author} {\bibfnamefont
  {E.}~\bibnamefont {Lunghi}}, \bibinfo {author} {\bibfnamefont {R.~S.}\
  \bibnamefont {Van~de Water}}, \ and\ \bibinfo {author} {\bibfnamefont
  {R.}~\bibnamefont {Zhou}},\ }\href@noop {} {\  (\bibinfo {year} {2015})},\
  \Eprint {http://arxiv.org/abs/1510.02349} {arXiv:1510.02349 [hep-ph]}
  \BibitemShut {NoStop}%
%%CITATION = ARXIV:1510.02349;%%
\bibitem [{\citenamefont {Bazavov}\ \emph
  {et~al.}(2010{\natexlab{b}})\citenamefont {Bazavov} \emph
  {et~al.}}]{Bazavov:2010ru}%
  \BibitemOpen
  \bibfield  {author} {\bibinfo {author} {\bibfnamefont {A.}~\bibnamefont
  {Bazavov}} \emph {et~al.} (\bibinfo {collaboration} {MILC Collaboration}),\
  }\href {\doibase 10.1103/PhysRevD.82.074501} {\bibfield  {journal} {\bibinfo
  {journal} {Phys. Rev.}\ }\textbf {\bibinfo {volume} {D82}},\ \bibinfo {pages}
  {074501} (\bibinfo {year} {2010}{\natexlab{b}})},\ \Eprint
  {http://arxiv.org/abs/1004.0342} {arXiv:1004.0342 [hep-lat]} \BibitemShut
  {NoStop}%
%%CITATION = ARXIV:1004.0342;%%
\bibitem [{\citenamefont {Bazavov}\ \emph {et~al.}(2013)\citenamefont {Bazavov}
  \emph {et~al.}}]{Bazavov:2012xda}%
  \BibitemOpen
  \bibfield  {author} {\bibinfo {author} {\bibfnamefont {A.}~\bibnamefont
  {Bazavov}} \emph {et~al.} (\bibinfo {collaboration} {MILC Collaboration}),\
  }\href {\doibase 10.1103/PhysRevD.87.054505} {\bibfield  {journal} {\bibinfo
  {journal} {Phys. Rev.}\ }\textbf {\bibinfo {volume} {D87}},\ \bibinfo {pages}
  {054505} (\bibinfo {year} {2013})},\ \Eprint {http://arxiv.org/abs/1212.4768}
  {arXiv:1212.4768 [hep-lat]} \BibitemShut {NoStop}%
%%CITATION = ARXIV:1212.4768;%%
\bibitem [{\citenamefont {Follana}\ \emph {et~al.}(2007)\citenamefont {Follana}
  \emph {et~al.}}]{Follana:2006rc}%
  \BibitemOpen
  \bibfield  {author} {\bibinfo {author} {\bibfnamefont {E.}~\bibnamefont
  {Follana}} \emph {et~al.} (\bibinfo {collaboration} {HPQCD Collaboration}),\
  }\href {\doibase 10.1103/PhysRevD.75.054502} {\bibfield  {journal} {\bibinfo
  {journal} {Phys. Rev.}\ }\textbf {\bibinfo {volume} {D75}},\ \bibinfo {pages}
  {054502} (\bibinfo {year} {2007})},\ \Eprint
  {http://arxiv.org/abs/hep-lat/0610092} {arXiv:hep-lat/0610092 [hep-lat]}
  \BibitemShut {NoStop}%
%%CITATION = HEP-LAT/0610092;%%
\bibitem [{\citenamefont {Bazavov}\ \emph
  {et~al.}(2014{\natexlab{a}})\citenamefont {Bazavov} \emph
  {et~al.}}]{Bazavov:2013maa}%
  \BibitemOpen
  \bibfield  {author} {\bibinfo {author} {\bibfnamefont {A.}~\bibnamefont
  {Bazavov}} \emph {et~al.} (\bibinfo {collaboration} {Fermilab Lattice and
  MILC Collaborations}),\ }\href {\doibase 10.1103/PhysRevLett.112.112001}
  {\bibfield  {journal} {\bibinfo  {journal} {Phys. Rev. Lett.}\ }\textbf
  {\bibinfo {volume} {112}},\ \bibinfo {pages} {112001} (\bibinfo {year}
  {2014}{\natexlab{a}})},\ \Eprint {http://arxiv.org/abs/1312.1228}
  {arXiv:1312.1228 [hep-ph]} \BibitemShut {NoStop}%
%%CITATION = ARXIV:1312.1228;%%
\bibitem [{\citenamefont {Bazavov}\ \emph
  {et~al.}(2014{\natexlab{b}})\citenamefont {Bazavov} \emph
  {et~al.}}]{Bazavov:2014wgs}%
  \BibitemOpen
  \bibfield  {author} {\bibinfo {author} {\bibfnamefont {A.}~\bibnamefont
  {Bazavov}} \emph {et~al.} (\bibinfo {collaboration} {Fermilab Lattice and
  MILC Collaborations}),\ }\href {\doibase 10.1103/PhysRevD.90.074509}
  {\bibfield  {journal} {\bibinfo  {journal} {Phys. Rev.}\ }\textbf {\bibinfo
  {volume} {D90}},\ \bibinfo {pages} {074509} (\bibinfo {year}
  {2014}{\natexlab{b}})},\ \Eprint {http://arxiv.org/abs/1407.3772}
  {arXiv:1407.3772 [hep-lat]} \BibitemShut {NoStop}%
%%CITATION = ARXIV:1407.3772;%%
\bibitem [{\citenamefont {{Be\v{c}irevi\'c}}\ \emph
  {et~al.}(2003{\natexlab{a}})\citenamefont {{Be\v{c}irevi\'c}}, \citenamefont
  {{Prelov\v sek}},\ and\ \citenamefont {Zupan}}]{Becirevic:2002sc}%
  \BibitemOpen
  \bibfield  {author} {\bibinfo {author} {\bibfnamefont {D.}~\bibnamefont
  {{Be\v{c}irevi\'c}}}, \bibinfo {author} {\bibfnamefont {S.}~\bibnamefont
  {{Prelov\v sek}}}, \ and\ \bibinfo {author} {\bibfnamefont {J.}~\bibnamefont
  {Zupan}},\ }\href {\doibase 10.1103/PhysRevD.67.054010} {\bibfield  {journal}
  {\bibinfo  {journal} {Phys. Rev.}\ }\textbf {\bibinfo {volume} {D67}},\
  \bibinfo {pages} {054010} (\bibinfo {year} {2003}{\natexlab{a}})},\ \Eprint
  {http://arxiv.org/abs/hep-lat/0210048} {hep-lat/0210048} \BibitemShut
  {NoStop}%
%%CITATION = HEP-LAT/0210048;%%
\bibitem [{\citenamefont {{Be\v{c}irevi\'c}}\ \emph
  {et~al.}(2003{\natexlab{b}})\citenamefont {{Be\v{c}irevi\'c}}, \citenamefont
  {{Prelov\v sek}},\ and\ \citenamefont {Zupan}}]{Becirevic:2003ad}%
  \BibitemOpen
  \bibfield  {author} {\bibinfo {author} {\bibfnamefont {D.}~\bibnamefont
  {{Be\v{c}irevi\'c}}}, \bibinfo {author} {\bibfnamefont {S.}~\bibnamefont
  {{Prelov\v sek}}}, \ and\ \bibinfo {author} {\bibfnamefont {J.}~\bibnamefont
  {Zupan}},\ }\href {\doibase 10.1103/PhysRevD.68.074003} {\bibfield  {journal}
  {\bibinfo  {journal} {Phys. Rev.}\ }\textbf {\bibinfo {volume} {D68}},\
  \bibinfo {pages} {074003} (\bibinfo {year} {2003}{\natexlab{b}})},\ \Eprint
  {http://arxiv.org/abs/hep-lat/0305001} {hep-lat/0305001} \BibitemShut
  {NoStop}%
%%CITATION = HEP-LAT/0305001;%%
\bibitem [{\citenamefont {{Be\v{c}irevi\'c}}\ \emph
  {et~al.}(2007{\natexlab{b}})\citenamefont {{Be\v{c}irevi\'c}}, \citenamefont
  {Fajfer},\ and\ \citenamefont {Kamenik}}]{Becirevic:2006me}%
  \BibitemOpen
  \bibfield  {author} {\bibinfo {author} {\bibfnamefont {D.}~\bibnamefont
  {{Be\v{c}irevi\'c}}}, \bibinfo {author} {\bibfnamefont {S.}~\bibnamefont
  {Fajfer}}, \ and\ \bibinfo {author} {\bibfnamefont {J.~F.}\ \bibnamefont
  {Kamenik}},\ }\href {\doibase 10.1088/1126-6708/2007/06/003} {\bibfield
  {journal} {\bibinfo  {journal} {JHEP}\ }\textbf {\bibinfo {volume} {06}},\
  \bibinfo {pages} {003} (\bibinfo {year} {2007}{\natexlab{b}})},\ \Eprint
  {http://arxiv.org/abs/hep-ph/0612224} {hep-ph/0612224} \BibitemShut {NoStop}%
%%CITATION = HEP-PH/0612224;%%
\bibitem [{\citenamefont {Di~Vita}\ \emph {et~al.}(2011)\citenamefont {Di~Vita}
  \emph {et~al.}}]{DiVita:2011py}%
  \BibitemOpen
  \bibfield  {author} {\bibinfo {author} {\bibfnamefont {S.}~\bibnamefont
  {Di~Vita}} \emph {et~al.},\ }\href@noop {} {\bibfield  {journal} {\bibinfo
  {journal} {PoS}\ }\textbf {\bibinfo {volume} {LAT2010}} (\bibinfo {year}
  {2011})},\ \Eprint {http://arxiv.org/abs/1104.0869} {arXiv:1104.0869
  [hep-lat]} \BibitemShut {NoStop}%
%%CITATION = 1104.0869;%%
\bibitem [{\citenamefont {Aubin}\ and\ \citenamefont
  {Bernard}(2003)}]{Aubin:2003mg}%
  \BibitemOpen
  \bibfield  {author} {\bibinfo {author} {\bibfnamefont {C.}~\bibnamefont
  {Aubin}}\ and\ \bibinfo {author} {\bibfnamefont {C.}~\bibnamefont
  {Bernard}},\ }\href {\doibase 10.1103/PhysRevD.68.034014} {\bibfield
  {journal} {\bibinfo  {journal} {Phys. Rev.}\ }\textbf {\bibinfo {volume}
  {D68}},\ \bibinfo {pages} {034014} (\bibinfo {year} {2003})}\BibitemShut
  {NoStop}%
%%CITATION = HEP-LAT/0304014;%%
\end{thebibliography}%
\end{document}